%% file: main.tex
    \include{preamble}

\documentclass[12pt]{report}
\usepackage{emlines,icthesis,epsf,uhead,rotate}
   \title{Modelling and Optimising $\protect{\rm \bf GaAs/
   Al_{x}\-Ga_{1-x}\-As}$ Multiple Quantum Well Solar Cells}
   \author {James P. Connolly}
   \submitdate{January 1997}
   
\begin{document}
   \setstretch{1.5}

   \maketitle
   \pagestyle{uheadings}
   \include{abstract}
   \include{acnow}
   \tableofcontents
   \listoffigures
   \listoftables

    \include{chap1}
    \include{chap2}
    \include{chap3}

    \include{chap4}
    \include{chap5}
    \include{chap6}

    \include{chap7}

    \include{chap8}

    \include{conclude}

    \appendix
    \include{app1}


\include{tbibli}
\end{document}

%% file: preamble.tex
\newcommand{\gaas}{$\rm GaAs$}
\newcommand{\algaas}{$\rm Al_{x}\-Ga_{1-x}\-As$}
\newcommand{\gaasalgaas}{$\rm GaAs/Al_{x}\-Ga_{1-x}\-As$}

\newcommand{\jsc}{$J_{sc}$}
\newcommand{\voc}{$V_{oc}$}
\newcommand{\pin}{p - i - n}
\newcommand{\pn}{p - n}
\newcommand{\figcap}[1]{\caption{\protect \footnotesize #1}}
\newcommand{\bigeps}[3]{\begin{figure*}[#1]\centerline{
   \epsfxsize=5.2in
   \epsffile{#2}}
   \figcap{#3}
   \end{figure*}}
\newcommand{\smalleps}[3]{\begin{figure*}[#1]\centerline{
   \epsfxsize=3.5in
   \epsffile{#2}}
   \figcap{#3}
   \end{figure*}}

%% file: abstract.tex
\begin{abstract}

The quantum well solar cell (QWSC) is a \pin\ solar cell with 
quantum wells in the intrinsic region. Previous work has shown that
QWSCs have a greater open circuit voltage (\voc) than would be
provided by a cell with the quantum well effective bandgap. This suggests
that the fundamental efficiency limits of QWSCs are greater than those of
single bandgap solar cells.

The following work investigates QWSCs in the \gaasalgaas\ materials
system. The design and optimisation of a QWSC in this system 
requires studies of the voltage and current dependencies on the 
aluminium fraction. QWSCs with different aluminium fractions have been
studied and show an increasing \voc\ with increasing barrier aluminium
composition.

The $QE$ however decreases with increasing aluminium composition. 
We develop a model of the $QE$ to test novel QWSC designs with a
view to minimising this problem.

This work concentrates on two design changes. The first deals
with compositionally graded structures in which the bandgap
varies with position. This bandgap variation introduces an
quasi electric field which can be used to increase minority
carrier collection in the low efficiency p and n layers.
This technique also increases the light flux reaching the
highly efficient depletion regions.

The second design change consists of coating the back of the cell 
with a mirror to exploit the portion of light which is not absorbed
on the first pass.

A model of the $QE$ of compositionally graded QWSC solar cells with
back surface mirrors is developed in order to analyse the effect of
these design changes.
These changes are implemented separately in a number of QWSC designs
and the resulting experimental data compared with the model. An optimised
design is then presented. 

\end{abstract}

%% file: acnow.tex
\begin{acknowledgements}

The work in this thesis would not have happened without the
support of a multitude of people. Keith Barnham has been a
tireless source of encouragement and ideas over the years.
Endless thanks to Jenny Nelson for always knowing what was
wrong with models, for her  encyclopaedic knowledge,
and being there to recommend films, or have them recommended.
Then there are all those who have passed through here, or
are still here, including Jenny Barnes, Ian Ballard, Alex
Zachariou, Paul Griffin, Ernest Tsui and Ned.

Next come the people who actually produced the solar cells in
this thesis. Christine Roberts was always ready to answer
constant questions about MBE machines and AlGaAs. John
Roberts' produced the grade which worked first time, and was 
full of suggestions for new material tests. Malcolm Pate 
produced innumerable devices over the years, including
the wafer squashed by the Royal Mail.

Then there are collaborators overseas. Antonio Marti Vega helped me
take some measurements which can't be fitted in. Two hundred pages
is quite enough! And Rosa Maria Fernandez Reyna in Madrid, who
explained them to me and showed me around Madrid, including some 
very nice cafes and cinemas.

Enrique Gr\"unbaum took TEM measurements which helped explain one
of the more mystifying samples in this project, along with SIMS
measurements by Federico Caccavale. The latter were enabled by Massimo
Mazzer who arranged the SIMS measurements in Padua, and always 
had ideas for everything, including jazz concerts.

And thanks to all the people who have made daily life about the
department more interesting, such as Jean Leveratt for 
chats on  level 8 at lunchtimes and solutions to all sorts of problems, 
Lydia, also of Level 8, and Betty Moynihan for help with the technicalities 
of writing up on level 3.

As for finance, I thank the Greenpeace Trust who made the project
possible, and who enabled me to go to the Solar World Congress.

Finally, there all the people I have known, or got to know over
the years. There are those
who are overseas - Chris and Olenka, Marc and Malar and the Smiths,
Claire and the Granjon Clan. All the Tuckmans and associates
in North London apart from Jo in Guatemala, Kerry and Alberta, 
Karen in the East, (big) Phil, (other) Phil,
Hiroe, Bruce who used to be in the West, Paul and Catherine in the
centre, Vince in Cambridge, the Marchants (Pete and Jo) and Kul.
Then theres Tanveer (I won, I think?), Tristan, Dorothy, Dave, 
Wing, thanks for the coffee Steve, and Rosa in Madrid. 
Thanks to everyone ... !

Finally, thanks to the family: Jim, Eva, Marie Pascal, Anne Catherine, 
Jeanne Chantal, Isabel and Oona for the support over the years, and for
coming over now and then  to visit. This thesis is dedicated 
to my parents.

\end{acknowledgements}

%% file: chap1.tex
\chapter{Introduction\label{secintro}}

\section{Current Energy Use and Alternatives}

The sun irradiates the planet
with approximately $4\times 10^{6}$ EJ per year \cite{markvart94}. 
When we compare this enormous energy flux with current human 
consumption of around 400 EJ per year, we find that we 
receive somewhat more than this amount every hour
in direct solar radiation.

With the exception of such sources as tidal, geothermal
and nuclear energy, all the energy we consume is ultimately
derived from the sun. It is largely harvested in a roundabout
manner. The most important is the burning of fossil fuels,
followed by nuclear and hydroelectric power.

The limited nature of the fossil fuel reserves however makes
the search for alternatives important. This is exacerbated
by recent concerns about global warming through the continued
emission of greenhouse gasses. The recently published results
of the United Nations Advisory Group on Climate Change
included the conclusions of a wide range of expert opinion
in this field. The conclusion was that some level of global
warming is expected, and indeed, is already measurable.

Currently exploited alternatives present problems. Nuclear
power, once hailed as the power source which would be too
cheap to meter, has experienced well publicised difficulties
related to the disposal of nuclear waste. 

Hydroelectric power has been
heavily exploited for decades, and in many countries is close to
capacity. Moreover, the environmental impact of such schemes has,
with hindsight, been found to be both significant and not always
beneficial.

Solar or photovoltaic (PV) power is an attractive alternative 
for a number of reasons.
PV installations can be designed on a small scale, point-of-use
basis. Such installations have a low land requirement and little
environmental impact. They do not require mechanical moving
parts, and are therefore potentially robust and reliable, and
subject to low maintainance costs. Photovoltaic module manufacturers 
already quote guaranteed lifetimes measured in decades. This compares
favourably with diesel generators, for example, or even other
renewable energy sources such thermal or wind schemes which require
moving parts and regular servicing.

Despite all these advantages, we see that photovoltaic power has yet
to make a significant impact worldwide. The
high capital cost associated with the production of the modules
is a significant problem. This is partly due to inertia
in the electricity generating industries and to institutional
barriers resulting from the interests of established industries.
Lowering the cost of alternative energy sources however clearly
remains a viable route to achieving sustainable energy consumption
patterns.

\section{Reducing the Cost of Photovoltaic Power}

There are three ways to reduce the cost of photovoltaic power.
One is to concentrate on designing low cost PV modules, at some
expense to the conversion efficiency. Larger areas will
therefore be necessary, but the reduction in final cost per
kilowatt may outweigh other considerations if the module
cost can be brought down far enough. Examples are amorphous
or poly-crystalline solar cells, and thin film solar cells
made from II-VI materials such as CdTe. This strategy is
currently the most successful for terrestrial PV applications.
Another example of this technique is the electrochemical
solar cell which has been reported by O'Regan and Gr\"atzel
\cite{oregan91}.

The second, related method is to reduce the cost of PV systems.
The cost of the solar cells in such a module may constitute
less than half the cost of the finished PV module. For example
the multicrystalline solar cells in Photowatt's PV  
modules \cite{franclieu95} comprise about 40\% of the cost of the 
finished module. 

Another approach is improve the efficiency of the solar cells
dramatically at the expense of cost. Again, the final module
cost may be attractive in dollars per watt terms if the increase
in output power more than compensates the increased module cost.
In some cases, it may be economically viable to increase the
light input to individual solar cells in a module by means
of light concentrators. This approach typically involves exotic 
structures, complex processing, and generally requires high 
quality material.

To date the low cost approach has delivered marketable PV modules,
whereas high efficiency cells are reserved for specialised
applications in which cost is a secondary issue, such as
space applications.

\section{Introduction to the Optimisation Program}

This work concerns the optimisation of a novel design of solar 
cell with the potential to reach high efficiencies through 
the use of novel physical processes. This design is the quantum 
well solar cell (QSWC). This is studied
here particularly with regard to the efficiency with which the
solar cell converts light into current, which is called the
quantum efficiency or $QE$.

Chapter \ref{secsolarcells} will give an introduction underlying 
semiconductor physics before defining important solar cell parameters.
A short review of some standard and high efficiency 
solar cell designs will be investigated, including the fundamental
principles behind the quantum well solar cell.

Chapter \ref{sectheory} outlines the theoretical basis for $QE$ calculations 
in general and describes the model we will use. This model requires 
knowledge of a large number of materials parameters, which are reviewed 
in chapter \ref{secparameters}.

Chapter \ref{secoptimisation} describes previous work on the optimisation
of the QWSC structure, and illustrates additional theoretical results 
from the model for a number of designs based on the QWSC. A number of
optimisations are presented. 

The practical aspects of growth together with a review of experimental
characterisation techniques are reviewed in chapter \ref{secexperiment}.
These include preliminary characterisation techniques which allow the
quality of individual growth and processing runs to be assessed.
Results for these preliminary techniques are given in chapter
\ref{secmivar}.

Chapter \ref{secqeresults} presents experimental measurements of
the $QE$ for a wide range of samples grown by two different techniques. 
These samples cover the optimisations suggested by the theory. The
experimental data are then compared with the modelling, and conclusions
drawn regarding the viability of the techniques proposed in
chapter \ref{secoptimisation}.

%% file: chap2.tex
\chapter{Introduction to Solar Cells\label{secsolarcells}}

In this chapter we introduce parameters such as short circuit
current, open circuit voltage and fill factor, and overall 
conversion efficiency, which are useful for characterising and 
optimising solar cells. The first sections summarise the basic 
semiconductor physics of semiconductor \pn\ junctions in equilibrium
and under illumination. These allows us to derive an expression
for the built in potential of a \pn\ junction.

The behaviour of a \pn\ junction under different applied bias conditions
is described as a basis for understanding the current voltage 
characteristic of a solar cell in the dark and under illumination. 
We describe the principles underlying the photovoltaic effect,
and can then define the solar cell parameters mentioned above.

Having defined these parameters, we can identify design requirements for
efficient solar cells. We review a number of high efficiency solar cell
designs in a range of materials. We introduce the quantum well solar cell
and explain why this design is expected to yield higher fundamental 
efficiencies limits than are expected from single band-gap solar cells.

\section{\label{secphysicsintro}Physics of Solar Cells}

In this section we introduce the fundamental physics of semiconductors
required to understand and optimise solar cell designs, and to define
all fundamental physical concepts and parameters which will be used 
in subsequent sections. The discussion is mainly based in the treatment 
given in Sze \cite{sze81}.

The crystal potential of a crystalline solid restricts the energies
which can be occupied by electrons into bands of electronic states.
The highest occupied band in energy in a semiconductor is the
valence band, and is completely filled by the outer valence 
electrons of the solid  at zero Kelvin.

The band immediately above it is
called the conduction band, and is empty at zero Kelvin. These two bands
are separated by a gap in energy where no electronic states exist.
This band-gap is principal bandgap of the semiconductor
$E_{g}$, and has important consequences for the electrical and optical
properties of these materials.

\subsection{Intrinsic Semiconductors in Equilibrium}

The probability that an electron in a solid should occupy a given
energy level is described by the Fermi-Dirac distribution
function $f(E)$ of the form
\begin{equation}
\label{fermidirac}
f(E)=\frac{1}{1+e^{(E-E_{F})/k_{B}T}}
\end{equation}
where $E_{F}$ is the Fermi level. Equation \ref{fermidirac} defines
it as the energy at which there is a 50\% chance of occupation by
an electron at any temperature. $k_{B}$ is Boltzmann's constant 
and $T$ is the temperature.

At room temperatures thermal excitation  in an undoped semiconductor
generates a finite equilibrium density of electron-hole 
pairs known as the intrinsic carrier concentration $n_{i}$. 
This quantity influences the properties of semiconductor
devices, and solar cells in particular.

The densities of holes $p_{0}$ and of electrons $n_{0}$ are
equal in an intrinsic semiconductor because each electron 
leaves exactly one hole behind in the
valence band. They are both equal to $n_{i}$. 
\begin{equation}
\label{electronsholesandni}
n_{0}=p_{0}=n_{i}
\end{equation}
The semiconductor in this case can conduct
current through the motion under an applied bias of the
electrons in the conduction band and of hole states in the valence
band. 

We now consider the magnitude of this intrinsic carrier concentration
by determining the number of electrons excited into the valence
band at a temperature $T$. This can be found by integrating the
product of the occupation probability $f(E)$ and the number of
available states $N(E)$. We consider only direct band gap
semiconductors for simplicity.

In this case, the number of available states near the
band-edges may be assumed to be parabolic. The relationship between
energy and electron momentum is analogous to the free-electron case,
with the electron mass $m_{e}$ replaced by an effective mass $m_{e}^{*}$,
to account for the crystal potential. It can be shown 
\cite{sze81} that the energy-momentum relationship 
yields the following density of states near the conduction 
band edge:
\begin{equation}
\label{electrondensityofstates}
N_{c}(E)=\frac{8\sqrt 2 \pi m_{e}^{*3/2}}{h^3}(E-E_{c})^{1/2}
\end{equation}
A similar relationship for the density of states near the valence
band edge can be found by substituting the hole effective mass $m_{h}^{*}$
mass for $m_{e}^{*}$ and integrating in energy from the valence band
edge to minus infinity:
from the valence band edge to infinity
\begin{equation}
\label{holedensityofstates}
N_{v}(E)=\frac{8\sqrt 2 \pi m_{h}^{*3/2}}{h^3}(E_{v}-E)^{1/2}
\end{equation}

For energies $E-E_{F}>3K_{B}T$, the Fermi-Dirac function can 
be approximated by
\begin{equation}
\label{approximatefermidirac}
f_{E}\simeq e^{-(E-E_{F})/k_{B}T}
\end{equation}
Integrating the product of equations \ref{electrondensityofstates}
and \ref{approximatefermidirac} over energies from the conduction
band edge to infinity then gives the number of conduction electrons
in the system:
\begin{equation}
\label{electrondensity}
n_{0}=N_{C}e^{(E_{F}-E_{c})/k_{B}T}
\end{equation}
where $N_{C}$ is known as the effective density of states in the
conduction band and is given by
\begin{equation}
\label{nc}
N_{C}=2\left( \frac{2\pi m_{e}^{*}k_{B}T}{h^{2}}\right)^{3/2}
\end{equation}
The number of holes is found by assuming a similar form to equation
\ref{approximatefermidirac} for $E_{F}-E>3K_{B}T$ and integrating
downwards in energy from the valence band edge. This gives
\begin{equation}
\label{holedensity}
p_{0}=N_{V}e^{(E_{v}-E_{F})/k_{B}T}
\end{equation}
where $N_{V}$ is the valence band effective density of states, given by
\begin{equation}
\label{nv}
N_{V}=2\left( \frac{2\pi m_{h}^{*}k_{B}T}{h^{2}}\right)^{3/2}
\end{equation}

Equations \ref{electrondensity} and \ref{holedensity} together with
equation \ref{electronsholesandni} yield the intrinsic carrier
concentration
\begin{equation}
\label{intrinsiccarrierconc}
n_{i}^{2}=p_{0}n_{0}=N_{C}N_{V}e^{-E_{g}/k_{B}T}
\end{equation}
The intrinsic carrier concentration in \gaas\ for example is
approximately $\rm{1.79 \times 10^{6}}$ $\rm {cm^{-3}}$ and a rather 
higher  $\rm{1.45 \times 10^{10}cm^{-3}}$ in Si at room temperature 
because of the smaller band gap of this material.

We can now calculate the Fermi level in an intrinsic semiconductor.
Equation \ref{electronsholesandni} shows that the ratio of
equations  \ref{electrondensity} and
\ref{holedensity} is equal to one 
Separating $E_{F}$ gives
\begin{equation}
\label{intrinsicfermilevel}
E_{F}=\frac{E_{v}+E_{c}}{2}+\frac{k_{B}T}{2}
\ln\left(\frac{N_{V}}{N_{C}}\right)
\end{equation}
The Fermi level is situated near the middle
of the bandgap, with a small offset caused by the difference between
the effective densities of states in the valence and conduction bands.

\subsection{Doped Semiconductors}

We now consider the case of a semiconductor which has been doped
with impurity atoms, starting with the case of a n type 
semiconductor. Donor or n type impurities
have valence electrons with energy levels located a few
meV below the conduction band edge. At non-zero temperatures, 
some of these donor atoms are
ionised by lattice vibrations and contribute extra electrons
to the conduction band edge. Charge neutrality is preserved, since
an ionised donor exists for every extra conduction band electron.
For most common semiconductor devices, the dopants are chosen so
that their ionisation energy is of the order or less that $3k_{B}T$, 
where $T$ is the operating temperature. In this case, the
donors are effectively all ionised, and the extra conduction
electron concentration is of the order of the doping density $N_{D}$.

The law of mass action states that the electron and hole concentrations
p and n are related as follows:
\begin{equation}
\label{massaction}
pn=n_{i}^{2}
\end{equation}
Useful doping densities in \gaasalgaas\ are typically in the range
$\rm 10^{16}$ - $10^{20} cm^{-3}$. We have seen previously that the $n_{i}$
is of the order of $\rm{10^{6}cm^{-3}}$ for \gaas\ at room temperature.
Since $n\simeq N_{D}$, it can be seen from equation \ref{massaction}.
that the hole concentration is typically many orders of magnitude
smaller than the electron concentration, and is given
approximately by
\begin{equation}
\label{minorityholeconcentration}
p_{n0}=\frac{n_{i}^{2}}{N_{D}}
\end{equation}
The electrons in n-type material are called majority carriers, and
the holes minority carriers. We use the subscript $n0$ to identify 
the minority hole concentration in n-type material in equilibrium.

A p-type semiconductor is one which has been doped with electron
acceptors with empty electron valence states a few meV above the
valence band. Electrons from the valence band are thermally excited
into these states at room temperatures, ionising the acceptor and leaving
a positive hole in the valence band. The majority carriers in this
case are holes, and their concentration is approximately equal to the
concentration $N_{A}$ of acceptor atoms. The minority electron
concentration is then given by
\begin{equation}
\label{minorityelectronconcentration}
n_{p0}=\frac{n_{i}^{2}}{N_{D}}
\end{equation}
where the subscript $p0$ identifies the electron minority concentration
in p-type material in equilibrium.

We can calculate the Fermi level in a n doped semiconductor from
equation \ref{electrondensity}, noting that $n\simeq N_{D}$.
This gives
\begin{equation}
\label{nfermilevel}
E_{F}=E_{c}-k_{B}T \ln\left(\frac{N_{C}}{N_{D}}\right)
\end{equation}
Equation \ref{nfermilevel} shows that as $N_{D}$ approaches the effective
conduction band density of states, the Fermi level increases towards
the conduction band edge. This reflects the increasing probability of
electronic states in the conduction band being occupied by electrons.

Similarly, the Fermi level in a p type semiconductor is
\begin{equation}
\label{pfermilevel}
E_{F}=E_{v}+k_{B}T \ln\left(\frac{N_{V}}{N_{A}}\right)
\end{equation}
The Fermi level for a p type semiconductor moves towards the valence
band edge as electrons are removed by acceptors, thereby decreasing the
probability of valence band states being occupied by electrons.

\section{The Photovoltaic Effect\label{secpveffect}}

In the preceding sections, we looked at the static behaviour of
electron and hole populations for semiconductors in equilibrium.
We now examine the behaviour of these populations when the system
is moved away from equilibrium under illumination, when electrical
power can be extracted via the photovoltaic effect.

The photovoltaic effect converts incident light into electrical 
power. The process consists of two stages. The first is to create
an excess minority carrier population by exciting electrons with a
light flux of intensity $F(\lambda)$, in photons per second per 
wavelength $\lambda$.
By excess minority carrier population we mean those minority carriers
which exceed the minority carrier population in equilibrium.
If the incident light has sufficient energy, an electron can absorb
a photon and transfer from a valence band state to a conduction band
state. The net effect is the creation of an electron-hole
pair. The excess minority carriers form a non equilibrium population
at a higher energy than the ground state of the system. 

The second stage of the photovoltaic process is to extract the
potential energy of this system. It is known as charge separation, and
consists of spatially separating the electron-hole pairs. This prevents
them  from recombining and losing their energy as heat or light, and
imposes motion in opposite directions which constitues the photocurrent.
This photocurrent can then be extracted through an external circuit
connected to the terminals of the solar cell.

\subsection{Light Absorption and Pair Generation\label{secintrogen}}

Minority carriers are generated when an electron absorbs a photon
and gains sufficient energy to be promoted to a conduction state,
leaving a positively charged hole state in the valence band. This
is illustrated in figure \ref{figgenerationthermalisation}. 

In p type material, electrons which absorb a photon with energy 
greater than the bandgap are promoted into minority carrier
states high in the conduction band.
They very rapidly lose this excess energy in collisions with the 
lattice, generating heat. This process, known as thermalisation,
ceases when they reach the band edge. The minority carriers then
remain in band edge states for a finite time known as the minority
electron lifetime $\tau_{n}$ before recombining with a majority 
hole in the valence band. The time taken to 
thermalise to the band-edge is much smaller than $\tau_{n}$,
and the carriers therefore build up a quasi-equilibrium thermal
population at the conduction band edge. 

Minority holes in n type material which absorb more energy than the 
bandgap are promoted to holes in the valence band, and thermalise up 
towards the valence band edge in a similar manner, 
again losing the excess energy as heat. They then remain at the band edge
for a minority hole recombination time $\tau_{p}$ before recombining
with a majority electron.

Thermalisation is an important energy loss mechanism in solar cells. It
ensures that the maximum amount of power delivered by a minority carrier
is proportional to the bandgap of the material, irrespective of the
energy of the photon providing the excitation.

\begin{figure}[t]
\begin{center}
\unitlength 1.00mm
\linethickness{0.4pt}
\begin{picture}(139.58,92.20)
\put(131.50,40.00){\makebox(0,0)[cc]{$E_{g}$}}
\put(131.50,62.00){\vector(0,1){0.2}}
\put(131.50,45.00){\line(0,1){17.00}}
\put(131.50,17.67){\vector(0,-1){0.2}}
\put(131.50,34.67){\line(0,-1){17.00}}
\bezier{48}(95.42,40.58)(100.42,37.25)(105.42,40.58)
\bezier{56}(105.42,40.58)(110.42,45.25)(115.42,40.25)
\bezier{32}(115.42,40.25)(119.09,38.58)(123.09,40.58)
\put(123.42,40.58){\vector(1,0){0.2}}
\put(122.75,40.58){\line(1,0){0.67}}
\put(35.25,91.00){\circle*{2.40}}
\put(35.25,11.50){\circle{3.40}}
\put(35.25,84.17){\vector(0,1){0.2}}
\put(35.25,18.50){\line(0,1){65.67}}
\bezier{36}(37.91,89.00)(38.91,83.00)(41.25,82.33)
\bezier{32}(41.25,82.33)(45.58,80.33)(44.91,76.67)
\bezier{28}(44.91,76.67)(42.25,72.00)(43.91,71.00)
\put(45.91,68.67){\vector(1,-1){0.2}}
\multiput(43.58,71.00)(0.12,-0.12){20}{\line(0,-1){0.12}}
\put(49.41,67.75){\circle*{2.40}}
\put(46.25,80.33){\makebox(0,0)[lc]{Thermalisation}}
\put(32.91,57.00){\makebox(0,0)[rc]{Photogeneration}}
\put(17.83,32.25){\makebox(0,0)[cc]{Incident photon}}
\put(77.50,17.00){\line(1,0){3.00}}
\put(70.41,17.00){\line(1,0){3.00}}
\put(63.75,17.00){\line(1,0){3.00}}
\put(57.08,17.00){\line(1,0){3.00}}
\put(50.41,17.00){\line(1,0){3.00}}
\put(43.75,17.00){\line(1,0){3.00}}
\put(37.08,17.00){\line(1,0){3.00}}
\put(30.41,17.00){\line(1,0){3.00}}
\put(23.75,17.00){\line(1,0){3.00}}
\put(17.08,17.00){\line(1,0){3.00}}
\put(10.41,17.00){\line(1,0){3.00}}
\put(7.50,16.25){\makebox(0,0)[rc]{$E_{F}$}}
\put(62.08,68.08){\circle*{2.40}}
\put(91.25,68.50){\circle*{2.40}}
\put(97.50,17.00){\line(1,0){3.00}}
\put(90.84,17.00){\line(1,0){3.00}}
\put(84.17,17.00){\line(1,0){3.00}}
\put(61.91,11.92){\circle{3.33}}
\put(91.08,11.92){\circle{3.33}}
\put(61.67,44.58){\vector(0,-1){0.2}}
\put(61.67,64.58){\line(0,-1){20.00}}
\put(61.67,39.58){\vector(0,1){0.2}}
\put(61.67,19.16){\line(0,1){20.42}}
\put(91.25,19.58){\vector(0,-1){0.2}}
\put(91.25,64.17){\line(0,-1){44.59}}
\put(93.75,56.67){\makebox(0,0)[lc]{Radiative}}
\bezier{48}(4.58,38.92)(9.58,35.58)(14.58,38.92)
\bezier{56}(14.58,38.92)(19.58,43.58)(24.58,38.58)
\bezier{32}(24.58,38.58)(28.25,36.92)(32.25,38.92)
\put(32.58,38.92){\vector(1,0){0.2}}
\put(31.91,38.92){\line(1,0){0.67}}
\put(58.34,42.08){\line(1,0){6.66}}
\put(10.42,14.17){\line(1,0){122.08}}
\put(10.83,65.83){\line(1,0){122.08}}
\put(110.42,32.92){\makebox(0,0)[cc]{Emitted photon}}
\put(139.58,65.83){\makebox(0,0)[lc]{$E_{c}$}}
\put(139.17,14.17){\makebox(0,0)[lc]{$E_{v}$}}
\put(104.58,17.00){\line(1,0){3.00}}
\put(124.58,17.00){\line(1,0){3.00}}
\put(117.92,17.00){\line(1,0){3.00}}
\put(111.25,17.00){\line(1,0){3.00}}
\put(64.17,56.67){\makebox(0,0)[lc]{Non radiative}}
\end{picture}
\figcap{Minority carrier photo generation, thermalisation and 
radiative and non radiative recombination mechanisms 
\label{figgenerationthermalisation}}
\end{center}
\end{figure}

The generation rate $G$ at a particular point is given by the absorption
coefficient $\alpha$ and the light flux $F$ 
\begin{equation}
G=\alpha F
\end{equation}

Direct and indirect semiconductors show distinctly different
optical and electrical behaviour. In direct material such as
\gaas\, the lowest electronic transition is one which does not
require a change of momentum. In this case the absorption
as a function of photon energy $E$ near the band edge is 
approximately given \cite{pankove71a} by
\begin{equation}
\label{directabsorption}
\alpha_{direct}(E)\sim (E-E_{g})^{1/2}f(E-E_{g})
\end{equation}
where $f(E-E_{g})$ is a step function with value 1 for $E>E_{g}$ and
zero for $E<E_{g}$.

Indirect material such as silicon or \algaas\ with an aluminium content
greater than about 45\% has an indirect bandgap. Light absorption
also requires phonon absorption or emission. As a result,
the absorption coefficient near the band-edge is significantly
smaller than in direct material and takes the form
\begin{equation}
\label{indirectabsorption}
\alpha_{indirect}\sim (E-E_{g})^{2}f(E-E_{g})
\end{equation}
The electron hole pairs generated by light absorption are free
to move through the crystal lattice until they either recombine,
losing the absorbed energy as heat or light, or are collected and
contribute to the photocurrent.

\subsection{Charge Separation\label{secchargesep}}

\begin{figure*}[t]
\setlength{\unitlength}{0.75mm}
\begin{picture}(160,100)(-12,-26)
\put(-10,-20){\vector(0,1){70}}
\put(-2,59){\makebox(0,0){{\em electron}}}
\put(-1.5,54){\makebox(0,0){{\em energy}}}
\put(-12,-18){\vector(1,0){180}}
\put(157,-26){\makebox(0,0){{\em position}}}
\put(-8,20){\line(1,0){28}}
\put(33,20){\line(1,0){30}}
\put(-8,48){\line(1,0){28}}
\put(33,48){\line(1,0){30}}
\multiput(-6,22)(10,0){3}{\line(1,0){5}}
\put(27,22){\makebox(0,0){$E_{F}$}}
\multiput(36,41)(10,0){3}{\line(1,0){5}}
\put(70,41){\makebox(0,0){$E_{F}$}}
\multiput(100,22)(10,0){6}{\line(1,0){5}}
\put(165,22){\makebox(0,0){$E_{F}$}}
\put(100,30){\vector(0,-1){8}}
\put(100,30){\vector(0,1){11}}
\put(98,41){\line(1,0){4}}
\put(98,22){\line(1,0){4}}
\put(105,30){\makebox(0,0){$V_{bi}$}}

\put(10,12){\makebox(0,0){p}}
\put(44,12){\makebox(0,0){n}}
\put(30,-24){\makebox(0,0){{\bf a)}}}
\put(124,-24){\makebox(0,0){{\bf b)}}}
\put(100,-5){\vector(1,0){10}}
\put(103,-8){\makebox(0,0){p}}
\put(165,-5){\vector(-1,0){10}}
\put(160,-8){\makebox(0,0){n}}
\put(140,-5){\vector(1,0){15}}
\put(140,-5){\vector(-1,0){30}}
\put(130,-8){\makebox(0,0){space charge}}
\put(133,-14){\makebox(0,0){region}}
\put(100,20){\line(1,0){10}}
\put(110,20){\line(6,-1){5}}
\put(114.93,19.17){\line(3,-1){5}}
\put(119.67,17.59){\line(2,-1){5}}
\put(124.6,15.0){\line(1,-1){10}}
\put(134.8,5){\line(2,-1){8}}
\put(142,1.42){\line(3,-1){7}}
\put(149,-0.9){\line(6,-1){6}}
\put(155,-1.9){\line(1,0){10}}
\put(100,48){\line(1,0){10}}
\put(110,48){\line(6,-1){5}}
\put(114.93,47.17){\line(3,-1){5}}
\put(119.67,45.59){\line(2,-1){5}}
\put(124.6,43){\line(1,-1){10}}
\put(134.8,33){\line(2,-1){8}}
\put(142,29.42){\line(3,-1){7}}
\put(149,27.1){\line(6,-1){6}}
\put(155,26.1){\line(1,0){10}}
\put(143,27){\makebox(0,0){{\small +}}}
\put(146,25){\makebox(0,0){{\small +}}}
\put(149,24){\makebox(0,0){{\small +}}}
\put(151.5,23.5){\makebox(0,0){{\small +}}}
\put(119,20){\makebox(0,0){{\Large -}}}
\put(122,19){\makebox(0,0){{\Large -}}}
\put(125,17){\makebox(0,0){{\Large -}}}
\put(128,14.5){\makebox(0,0){{\Large -}}}
\put(143,27){\makebox(0,0){{\Large o}}}
\put(146,25){\makebox(0,0){{\Large o}}}
\put(149,24){\makebox(0,0){{\Large o}}}
\put(119,20){\makebox(0,0){{\Large o}}}
\put(122,19){\makebox(0,0){{\Large o}}}
\put(125,17){\makebox(0,0){{\Large o}}}
\put(128,14.5){\makebox(0,0){{\Large o}}}
\put(151.5,23.5){\makebox(0,0){{\Large o}}}
\end{picture}
\figcap{\label{figpnjunction}p and n type semiconductors in
equilibrium a) before contact and b) after contact. The built in voltage 
$V_{bi}$ is equal to the difference between the Fermi levels in a), before
contact. Also shown are ionised donor (+) and acceptor (-) 
atoms which give rise to $V_{bi}$.}
\end{figure*}

Collection of photogenerated carriers requires a field to
separate electrons from holes, and to impose ordered motion
in the direction of the positive and negative terminals
of the solar cell. This constitutes the photocurrent.
                       
The charge separation is achieved by creating a junction region
in the solar cell with p type doped material on one side, and
n type dopants on the other. This results in a potential drop
across the structure. Electrons drift down the potential,
whereas holes drift up in the opposite direction. The precise origin of 
this separation potential is explained in more detail below.

The solar cell is divided into three regions. Charge separation
occurs mainly in the junction region, which we
will refer to as the space-charge region. In good material, the field
separates the electrons and holes with efficiencies approaching
100\%. 

There is no net electric field in the charge neutral sections of the
p and n regions adjoining the depletion layer, which we call the
p and n neutral regions.

Electrons and holes in these neutral regions move by diffusion.
Spatial variations in composition of doping may contribute a
small drift component. This will be investigated in chapter 
\ref{sectheory}. The diffusion process is much less efficient than the drift
collection which dominates in the space-charge region. Minority carriers
perform a random walk with a finite average lifetime and diffusion
length. A minority carrier which does not reach the interface with
the space-charge region eventually recombines with a majority
carrier and is lost. The collection of carriers generated
in the neutral regions is described in more detail in chapter
\ref{sectheory}.

Current generation in all regions of the cell therefore depends
on the presence of a built-in potential, which arises when a
junction is made between p and n type material.

When p and n type semiconductors are placed in electrical
contact, there is an imbalance between the large number of majority 
electrons on the n side and the thermal minority electron concentration
on the p side. A similar but opposite concentration gradient exists
for holes. These electron and hole gradients give rise to
a diffusion current of electrons from the n doped side 
of the junction to the p, and of holes in the opposite direction.

The electron current from the n side removes the electrostatic
shielding of the donor atoms, which become positively ionised.
Similarly, acceptor ions on the p side of the junction
are negatively ionised. The resulting distribution resembles a
parallel plate capacitor with positively charged donors on one
plate and negatively charged acceptors on the other. 
The region populated by ionised impurities is known as the
space-charge region, and is made up of the p and n depletion
layers. 

The electric field between the ionised dopants opposes the diffusion
current. In the absence of an external bias, the \pn\ junction
in the dark reaches an equilibrium in which the electrical field
cancels the diffusion potential. It can be shown
\cite{sze81} that this condition
is equivalent to a constant Fermi level throughout the system,
as illustrated in figure \ref{figpnjunction}b.

The thickness of these depletion layers can be found from charge
neutrality and assuming that the total potential dropped is equal
to the difference in Fermi levels between the two materials before
electrical contact.

This difference in Fermi levels $\Delta E_{F}$
is equal to $-qV_{bi}$ where $V_{bi}$ is the voltage dropped across 
the depleted region and is known as the built in voltage. It is
conveniently  expressed in terms of the intrinsic carrier concentration
$n_{i}$ and the dopant densities. Subtracting equation \ref{nfermilevel}
from equation \ref{pfermilevel} and substituting $n_{i}^{2}$
for the product $N_{C}N_{V}$ from equation \ref{intrinsiccarrierconc}
gives
\begin{equation}
\label{vbi}
V_{bi}=-\frac{\Delta E_{F}}{q}
=-\frac{k_{B}T}{q}\ln\left(\frac{N_{A}N_{D}}{n_{i}^{2}}\right)
\end{equation}
The built in voltage is negative, in the sense that it is opposite
to the standard terminal connection terminology, which is that the
n terminal is negative and the p positive.

For heavily doped high-bandgap materials, the Fermi levels in the p and
n neutral regions lie close to the conduction and valence band edges,
and $V_{bi}$ is only a few percent smaller than the bandgap of the 
semiconductor.

Since the space-charge region has essentially zero mobile charge
density in low injection conditions, we assume that all the
external bias is dropped across it, and that no net field
exists in the neutral sections.
The total potential drop across the space-charge region is therefore
given by the sum of the applied bias $V_{ext}$ and $V_{bi}$
\begin{equation}
\label{totaliregionpotential}
V=V_{ext}+V_{bi}
\end{equation}
This potential drop can also be derived by integrating Poisson's
equation across the space charge region of width 
$W=x_{w_{p}}+x_{w_{n}}$. 
\begin{equation}
\label{poisson}
V=-\int\!\!\int^{W}_{0}
(\frac{\rho}{\epsilon}) dx
\end{equation}
where $\epsilon$ is the permittivity of the material.
The charge density $\rho$ on both sides of the junction is given by 
$N_{A}$ and $N_{D}$.

Since there is no electrical field in the neutral sections of
the p and n layers, the sum of fixed positively and negatively
charges must cancel. The charge density in the p depletion layer
is negative and equal to $-qN_{A}$, where $q$ is the magnitude of 
the electronic charge. It is exactly opposite to $+qN_{D}$ in
the n depletion layer. 
If $x_{w_{p}}$ and $x_{w_{n}}$ are the p and
n layer depletion widths, this charge neutrality condition
can be written
\begin{equation}
\label{chargeneutrality}
x_{w_{p}}N_{A}=x_{w_{n}}N_{D}
\end{equation}
By solving equation \ref{poisson} subject to \ref{chargeneutrality}
and \ref{totaliregionpotential} we find that the depletion widths
(with $V_{bi}< 0$) are 
\begin{equation}
\label{pdepletion}
x_{w_{p}}=\frac{1}{N_{A}}
\sqrt{\frac{2\epsilon(-V_{bi}-V_{ext})}
{q(\frac{1}{N_{A}}+\frac{1}{N_{D}})}}
\end{equation}
\begin{equation}
\label{ndepletion}
x_{w_{n}}=\frac{1}{N_{D}}
\sqrt{\frac{2\epsilon(-V_{bi}-V_{ext})}
{q(\frac{1}{N_{A}}+\frac{1}{N_{D}})}}
\end{equation}
where $\epsilon$ is the permittivity of the material. We note that
the depletion widths increase as the applied bias becomes large
and negative, and tend to zero as the positive applied bias cancels
the built in potential.

\subsection{Current-Voltage Characteristic of a 
p-n Junction in the Dark\label{seciv}}

The current flowing through a solar cell is
the sum of the photocurrent and of a diode characteristic,
known as the dark current. The dark current component is
important in determining the open circuit voltage of the
cell. A good understanding of the dark current is therefore
important in optimising the overall efficiency of the
solar cell design.

Studies of dark current in \pn\ diodes were pioneered by
Shockley {em et al.\/} \cite{shockley50}. They established that an 
exponentially increasing current flows if the diode is driven into 
forward bias by making the p contact positive with respect to the n.
This is because the applied bias reduces the electric field in
the space charge region away from its equilibrium value. The
drift current no longer cancels the diffusion current, and
a net current flows.

A negative applied bias on ther other hand increases the field in 
the space charge region, inhibiting the diffusion of majority carriers
from the p and n sides of the junction.
The diffusion current therefore decreases and
no longer cancels the drift current due to generation-recombination
effects in the neutral regions. A small, voltage independent 
saturation current $J_{0}$ is seen 
to flow until the external applied bias exceeds a critical 
break-down voltage, at which the structure cannot sustain 
potential difference across the terminals and a sudden increase in
reverse current is observed.

Shockley {\em et al.\/} \cite{shockley50} initially 
modelled the dark current of an ideal diode 
in terms of semiconductor majority and minority carrier transport
equations. This considers only the diffusion currents of holes and
electrons from the neutral sections of the diode and neglects
the generation-recombination current in the space charge region. 
This approach yields the Shockley ideal diode equation
\begin{equation}
\label{idealdiode}
J_{ideal}(V)=J_{01}(e^{qV/k_{B}T}-1)
\end{equation}
where the saturation current $J_{01}$ is the ideal saturation current.
It can be expressed in terms of the electron and hole transport
parameters. It is a proportional to $n_{i}^{2}$ and is therefore 
given in Sze \cite{sze81} as
\begin{equation}
\label{approxsaturationcurrent}
J_{01}\simeq e^{E_{g}/k_{B}T}
\end{equation}
A more advanced analysis takes the generation-recombination current
in the space charge region into account. The discussion in Sze \cite{sze81}
based on analysis by Sah, Noyce and Shockley \cite{sah57}
shows that the electron and hole emission is dominated by recombination
centres situated at energies in the middle of the bandgap. The
generation-recombination current $J_{gr}$ can then be expressed in
terms of a recombination lifetime $\tau_{r}$, the total space-charge
region width $W$ and $n_{i}$:
\begin{equation}
\label{generationrecombination}
J_{gr}=\frac{qWn_{i}}{2\tau_{r}}\left(e^{qV/2k_{B}T}-1\right)
\end{equation}
The sum of equations \ref{idealdiode} and \ref{generationrecombination}
gives the Shockley-Read-Hall diode dark current $J_{D}$ 
\begin{equation}
\label{shrdarkcurrent}
J_{D}=J_{01}(e^{qV/k_{B}T}-1)+J_{02}(e^{qV/2k_{B}T}-1)
\end{equation}
Equation \ref{shrdarkcurrent} neglects the thermal contribution to
the dark current, which has a bandgap dependence of 
$\simeq e^{-E_{g}/k_{B}T}$ but is negligibly small in semiconductor 
structures suitable for solar cell applications.
The expression is frequently simplified as follows when modelling
data:
\begin{equation}
\label{shockley}
J_{D}\simeq J_{0}(e^{qV/nk_{B}T}-1)
\end{equation}
where $n$ is called the ideality factor.
It can be seen from equations \ref{idealdiode}
and \ref{generationrecombination} that $n=1$ for an ideal diode,
and $n=2$ for \pn\ junctions with strong generation-recombination
currents. Figure \ref{figdarkcurrent} shows a schematic of a diode
dark current in forward bias.

\begin{figure*}
\begin{center}
\unitlength 0.60mm
\linethickness{0.4pt}
\begin{picture}(164.58,115.42)
\put(164.17,20.42){\vector(1,0){0.2}}
\put(20.42,20.42){\line(1,0){143.75}}
\put(164.58,11.25){\makebox(0,0)[cc]{Voltage}}
\multiput(36.67,27.08)(0.31,0.12){195}{\line(1,0){0.31}}
\multiput(97.92,50.42)(0.22,0.12){28}{\line(1,0){0.22}}
\multiput(104.17,53.75)(0.18,0.12){32}{\line(1,0){0.18}}
\multiput(110.00,57.50)(0.14,0.12){21}{\line(1,0){0.14}}
\multiput(112.92,60.00)(0.12,0.12){28}{\line(1,0){0.12}}
\multiput(116.25,63.33)(0.12,0.14){317}{\line(0,1){0.14}}
\put(134.17,74.17){\makebox(0,0)[cc]{n=1}}
\put(78.75,35.83){\makebox(0,0)[cc]{n=2}}
\put(27.92,109.17){\vector(0,1){0.2}}
\put(27.92,12.50){\line(0,1){96.67}}
\put(27.92,115.42){\makebox(0,0)[cc]{Log($J_{D}$)}}
\end{picture}
\figcap{Semilog schematic of the forward bias Shockley dark current 
characteristic showing the generation-recombination regime (n=2)
and the diffusion regime (n=1) 
\label{figdarkcurrent}}
\end{center}
\end{figure*}

Diodes made from materials with small, direct band gaps tend to follow the 
ideal diode equation adequately, because the relatively large value 
of $n_{i}$ ensures that the diffusion currents from the neutral
regions dominate the dark current.

Diodes made from high bandgap materials on the other hand frequently
show two regimes. At high forward bias, the diffusion current dominates
and a factor $n=1$ is seen. For low forward bias, the reverse is true
and n tends to $2$.

\subsection{The p-n Junction in the Light\label{secsimplecells}}

We have mentioned before that the photocurrent from a solar cell is made 
up of two components which flow in the opposite way to a conventional
current. Minority carriers in the space-charge region are rapidly 
separated by the built in field, and swept to the interface with the 
neutral regions where they become majority carriers.

Minority carriers generated in the neutral regions must reach the
space-charge region before they can contribute to the photocurrent,
because no field exists in these regions. Transport is dominated
by carrier diffusion.

Both these cases are examined in detail in section \ref{sectheory}.
The photocurrent $J_{L}$ is expressed as the sum of diffusion components
from the p and n regions $J_{p}$ and $J_{n}$, and a drift
component from the space charge region $J_{i}$
\begin{equation}
J_{L}=J_{p}+J_{i}+J_{n}
\end{equation}
A significant decrease in the photocurrent is observed if the solar
cell is forced into forward bias of the order of the built in
potential, because charge separation in the space-charge region
is impaired. A small increase is seen in reverse bias as the space charge
region broadens because of the enhanced internal field. However,
in the operating range of high quality solar cells, the photocurrent
$J_{L}$ may be assumed to be constant with bias and equal to the short 
circuit value $\rm J_{sc}$.

The current-voltage characteristic of an illuminated solar cell $J(V)$
is then approximately given by sum of $J_{L}$ and the dark current $J_{D}$ 
of equation \ref{shockley}, noting that the photocurrent is
negative
\begin{equation}
\label{lightiv}
J=J_{0}(e^{qV/nk_{B}T}-1)-J_{L}
\end{equation}
A typical light current characteristic is given in figure \ref{figlightiv}. 

\begin{figure*}[t]
\begin{center}
\unitlength 0.60mm
\linethickness{0.4pt}
\begin{picture}(162.50,121.25)
\put(162.50,35.83){\vector(1,0){0.2}}
\put(30.00,35.83){\line(1,0){132.50}}
\put(56.67,115.00){\vector(0,1){0.2}}
\put(56.67,10.42){\line(0,1){104.58}}
\put(56.67,121.25){\makebox(0,0)[cc]{Current}}
\put(162.50,27.50){\makebox(0,0)[rc]{Voltage}}
\bezier{232}(39.58,99.58)(97.50,99.58)(97.50,99.58)
\bezier{64}(97.50,99.58)(105.42,98.75)(107.08,90.83)
\bezier{56}(107.08,90.83)(109.58,86.67)(110.42,77.08)
\bezier{64}(110.42,77.08)(111.67,70.00)(111.67,61.25)
\bezier{168}(111.67,61.25)(112.08,19.17)(112.08,19.17)
\bezier{232}(39.17,35.83)(97.08,35.83)(97.08,35.83)
\bezier{64}(97.08,35.83)(105.00,35.00)(106.67,27.08)
\bezier{56}(106.67,26.67)(109.17,22.50)(110.00,12.92)
\put(56.67,96.25){\line(1,0){47.08}}
\put(103.75,96.25){\line(0,-1){60.42}}
\put(78.75,77.50){\makebox(0,0)[cc]{{\bf \large Area=Pmax}}}
\put(83.33,20.42){\makebox(0,0)[rc]{$J_{D}$}}
\put(106.67,20.42){\vector(1,0){0.2}}
\put(88.33,20.42){\line(1,0){18.34}}
\put(126.25,84.17){\makebox(0,0)[lc]{$J$}}
\put(112.08,84.17){\vector(-1,0){0.2}}
\put(120.83,84.17){\line(-1,0){8.75}}
\put(72.08,112.92){\makebox(0,0)[lc]{$J_{sc}$}}
\put(59.58,102.08){\vector(-3,-4){0.2}}
\multiput(67.50,111.67)(-0.12,-0.15){66}{\line(0,-1){0.15}}
\put(115.00,40.00){\vector(-3,-4){0.2}}
\multiput(122.92,49.58)(-0.12,-0.15){66}{\line(0,-1){0.15}}
\put(126.25,52.92){\makebox(0,0)[lc]{$V_{oc}$}}
\put(54.58,94.17){\vector(1,1){0.2}}
\multiput(45.83,84.17)(0.12,0.14){73}{\line(0,1){0.14}}
\put(44.58,80.83){\makebox(0,0)[rc]{$J_{P}$}}
\put(101.67,38.75){\vector(3,-4){0.2}}
\multiput(93.33,49.58)(0.12,-0.15){70}{\line(0,-1){0.15}}
\put(92.08,52.92){\makebox(0,0)[rc]{$V_{P}$}}
\end{picture}
\figcap{\label{figlightiv} Light and dark currents of a solar cell,
showing the maximum power rectangle}
\end{center}
\end{figure*}

We now define some of the more useful quantities for solar cell
characterisation and optimisation. The open circuit
voltage $\rm V_{oc}$ is the forward bias at which the dark current exactly
cancels the photocurrent, and no net current flows. Setting $J=0$
in equation \ref{lightiv} and assuming that $J_{D}$ is dominated by
diffusion currents at such voltages such that $J_{gr}<\!<J_{ideal}$,
we obtain
\begin{equation}
\label{voc}
V_{oc}=\frac{k_{B}T}{q}\ln\left(\frac{J_{sc}}{J_{01}}\right)
\end{equation}
The maximum power which can be extracted from the
solar cell is given by the largest
rectangle which can be fitted under the curve, as shown
in the schematic figure \ref{figlightiv}. This rectangle 
defines the maximum power point, and the associated optimum
operational current and voltage $J_{P}$ and $V_{P}$. The
maximum power is then
\begin{equation}
\label{pmax}
P_{max}=J_{P}V_{P}
\end{equation}
It is useful to define a fill factor $\rm F\!F$ as follows
\begin{equation}
\label{ff}
F\!F=\frac{P_{max}}{V_{oc}J_{sc}}
\end{equation}
The overall efficiency of the cell $\eta$ is given by the ratio of
the the maximum power $P_{max}$ and incident power $P_{inc}$. It
is usually expressed as follows
\begin{equation}
\label{efficiency}
\eta=\frac{V_{oc}J_{sc}F\!F}{P_{inc}}
\end{equation}

\section{The p-i-n Solar Cell\label{secpincell}}

In the last sections we have defined the principal characteristics
of \pn\ junction solar cells. 
We saw in section \ref{secchargesep} that carriers
photogenerated in the space-charge region are collected with
efficiencies approaching 100\%. It is therefore clearly desirable
to extend the width of this layer. Although equations \ref{pdepletion}
and \ref{ndepletion} show that this can be achieved by reducing
the p and n doping densities, this technique has two unfortunate
consequences. The first is the reduction of the $\rm V_{oc}$
according to equation \ref{voc}. The second is an increased resistivity
in the neutral regions due to lower doping.

We now introduce a structure which circumvents these problems.
This is the \pin\ diode, which is commonly used for high efficiency 
photodetectors because the spectral response of the cell can be
optimised as a function of the response time. It serves as a useful
starting point for the design of an efficient quantum well solar
cell, as we shall see below.

\subsection{Space Charge Region of a p-i-n Diode\label{secpindepletion}}

The \pin\ solar cell consists of an ordinary \pn\ diode, with a layer
of nominally undoped material inserted between the p and n regions.
This layer is known as the intrinsic
region, of thickness $x_{i}$. The total space-charge region thickness
is then given by
\begin{equation}
\label{piniregion}
W_{i}=x_{w_{p}}+x_{i}+x_{w_{n}}
\end{equation}

\begin{figure*}[ht]
\begin{center}
\unitlength 0.70mm
\linethickness{0.4pt}
\begin{picture}(135.00,87.92)
\put(135.00,8.75){\makebox(0,0)[rc]{Depth}}
\put(14.58,87.92){\makebox(0,0)[cc]{Field}}
\put(38.33,7.50){\line(0,1){6.67}}
\put(38.33,20.42){\line(0,1){6.66}}
\put(38.33,33.33){\line(0,1){6.67}}
\put(38.33,46.25){\line(0,1){6.67}}
\put(38.33,59.17){\line(0,1){6.66}}
\put(38.33,72.08){\line(0,1){6.67}}
\put(97.50,7.50){\line(0,1){6.67}}
\put(97.50,20.42){\line(0,1){6.66}}
\put(97.50,33.33){\line(0,1){6.67}}
\put(97.50,46.25){\line(0,1){6.67}}
\put(97.50,59.17){\line(0,1){6.66}}
\put(97.50,72.08){\line(0,1){6.67}}
\put(14.17,79.58){\vector(0,1){0.2}}
\put(14.17,7.92){\line(0,1){71.66}}
\put(135.00,14.17){\vector(1,0){0.2}}
\put(7.92,14.17){\line(1,0){127.08}}
\put(25.00,6.67){\makebox(0,0)[cc]{{\bf p}}}
\put(66.67,6.67){\makebox(0,0)[cc]{{\bf i}}}
\put(114.59,7.09){\makebox(0,0)[cc]{{\bf n}}}
\multiput(21.25,14.17)(0.12,0.38){143}{\line(0,1){0.38}}
\put(38.33,68.75){\line(1,0){59.17}}
\multiput(97.50,68.75)(0.12,-0.24){226}{\line(0,-1){0.24}}
\multiput(30.83,14.17)(0.12,0.36){63}{\line(0,1){0.36}}
\put(38.33,36.67){\line(1,0){59.17}}
\multiput(97.50,36.67)(0.12,-0.21){105}{\line(0,-1){0.21}}
\put(66.67,72.08){\makebox(0,0)[cc]{A) High reverse bias}}
\put(66.67,40.42){\makebox(0,0)[cc]{B) Zero bias}}
\end{picture}
\figcap{\label{figlownbgpinfield}Electric field variation in an
ideal \pin\ diode with zero background doping at zero and reverse biasses}
\end{center}
\end{figure*}

Figure \ref{figlownbgpinfield} shows the electric field variation
in such a structure at reverse bias and at zero bias. The p and
n depletion regions decrease in size as the field drops. The
field in the $i$ region will only drop to zero when the $V_{ext}$
is equal and opposite to $V_{bi}$.

Although the intrinsic layer is nominally undoped, we always find
a residual or background doping density $N_{BG}$, which may be either
n or p type depending on the growth machine and growth conditions.
It is of the order $N_{BG}\sim 10^{14}\rm{cm^{-3}}$ in good material, but
may be as high as $10^{16}$ or even $10^{17}\rm cm^{-3}$.

The calculation of $W_{i}$ is similar to the calculation
outlined in section \ref{secchargesep}. The built-in potential
is given by equation \ref{vbi} and by Poisson's equation
\ref{poisson} as before. The main difference is that the double
integral is over the new space charge region thickness $W_{i}$,
and must include any residual space-charge due to 
the backgound doping $N_{BG}$. Figure \ref{figpinfield} shows the
variation of field as a function of position in a \pin\ structure
where the background doping is p type. Curve A shows the field
for a large negative applied bias. In the case of curve B, the magnitude
of the negative applied
bias is lower, and the depletion regions are smaller. Curve
C shows a case where the $i$ layer is not completely depleted. 

\begin{figure*}[ht]
\begin{center}
\unitlength 0.70mm
\linethickness{0.4pt}
\begin{picture}(135.00,87.92)
\put(135.00,8.75){\makebox(0,0)[rc]{Depth}}
\put(14.58,87.92){\makebox(0,0)[cc]{Field}}
\put(38.33,7.50){\line(0,1){6.67}}
\put(38.33,20.42){\line(0,1){6.66}}
\put(38.33,33.33){\line(0,1){6.67}}
\put(38.33,46.25){\line(0,1){6.67}}
\put(38.33,59.17){\line(0,1){6.66}}
\put(38.33,72.08){\line(0,1){6.67}}
\put(97.50,7.50){\line(0,1){6.67}}
\put(97.50,20.42){\line(0,1){6.66}}
\put(97.50,33.33){\line(0,1){6.67}}
\put(97.50,46.25){\line(0,1){6.67}}
\put(97.50,59.17){\line(0,1){6.66}}
\put(97.50,72.08){\line(0,1){6.67}}
\put(14.17,79.58){\vector(0,1){0.2}}
\put(14.17,7.92){\line(0,1){71.66}}
\multiput(28.75,14.17)(0.12,0.42){80}{\line(0,1){0.42}}
\multiput(38.33,47.50)(0.41,-0.12){146}{\line(1,0){0.41}}
\multiput(97.50,30.00)(0.12,-0.19){84}{\line(0,-1){0.19}}
\multiput(18.33,14.17)(0.12,0.35){171}{\line(0,1){0.35}}
\multiput(38.75,73.33)(0.46,-0.12){129}{\line(1,0){0.46}}
\multiput(97.50,57.92)(0.12,-0.18){244}{\line(0,-1){0.18}}
\multiput(34.58,14.17)(0.12,0.42){32}{\line(0,1){0.42}}
\multiput(38.33,27.50)(0.47,-0.12){112}{\line(1,0){0.47}}
\put(135.00,14.17){\vector(1,0){0.2}}
\put(7.92,14.17){\line(1,0){127.08}}
\put(25.00,6.67){\makebox(0,0)[cc]{{\bf\large p}}}
\put(66.67,6.67){\makebox(0,0)[cc]{{\bf\large i}}}
\put(116.67,6.67){\makebox(0,0)[cc]{{\bf\large n}}}
\put(66.25,23.33){\makebox(0,0)[cc]{C}}
\put(66.25,42.08){\makebox(0,0)[cc]{B}}
\put(66.25,69.17){\makebox(0,0)[cc]{A}}
\end{picture}
\figcap{\label{figpinfield}Electric field variation in a residually n
doped \pin\ diode for A high reverse bias B moderate reverse bias and
C an undepleted i layer}
\end{center}
\end{figure*}

For the \pin\ cell with a n type background doping shown in figure 
\ref{figpinfield}, the potential $V$ dropped across $W_{i}$ is again
the sum of external and built in potentials, and is defined
as before by equations \ref{vbi} and \ref{totaliregionpotential}. It
can also be calculated by integrating Poissons equation across
$W_{i}$, 
\begin{equation}
\label{vbipinptype}
V=-
\frac{qN_{A}x_{w_{p}}^{2}}{2\epsilon}-
\frac{qN_{A}x_{w_{p}}x_{i}}{\epsilon}+
\frac{qN_{BG}x_{i}^{2}}{2\epsilon}-
\frac{qN_{D}x_{w_{n}}^{2}}{2\epsilon}
\end{equation}
The charge neutrality condition of equation \ref{chargeneutrality} 
must be re-written to take the background doping into account:
\begin{equation}
\label{pinchargeneutralityptype}
-x_{w_{p}}N_{A}+x_{i}N_{BG}+x_{w_{n}}N_{D}=0
\end{equation}
Equations \ref{vbipinptype} and \ref{pinchargeneutralityptype}
define the depletion widths in terms a quadratic equation.
For the p side depletion layer of the n type background doped 
cell we have been discussing, this quadratic equation takes the form
\begin{equation}
\label{pinpdepletion}
\begin{array}{c}
x_{w_{p}}^{2}
\left[\frac{N_{A}^{2}}{2\epsilon}(\frac{1}{N_{A}} +\frac{1}{N_{D}})\right]
 +x_{w_{p}}
\left[\frac{N_{A}N_{BG}x_{i}}{\epsilon}
(\frac{1}{N_{BG}}-\frac{1}{N_{D}})\right] \\
+x_{i}^{2}
\left[\frac{N_{BG}^{2}}{2\epsilon}(\frac{1}{N_{D}}-\frac{1}{N_{BG}})\right]
+\frac{V}{q}
 =0 \\
\end{array}
\end{equation}
Similar expressions describe the n depletion width $x_{w_{n}}$. If
the intrinsic doping is n type, the sign of the term in $N_{BG}$
of equation \ref{vbipinptype} must be reversed.

\subsection{The Background Doping Problem in p-i-n Structures 
\label{secniproblem}}

In this section we consider the influence of the intrinsic region 
thickness $x_{i}$ and the background doping $N_{BG}$ on the performance 
of a \pin\ solar cell, according
to theory described by Paxman \cite{paxman92}. We then
introduce a method of calculating $N_{BG}$ from the dependence
of the photocurrent $J_{L}$ on voltage. 

A calculation of depletion widths as a function of intrinsic region 
width $x_{i}$ and $N_{BG}$ raises a number of potential problems with this 
type of structure. The n and p depletion layer thicknesses 
decrease as $x_{i}$ is made larger, thereby 
decreasing the overall space charge region width. Calculations 
however show that the decrease in n and p depletion layer 
thicknesses is negligible compared to the increase due to the 
intrinsic layer.

Another consequence of increasing the intrinsic region width
is a lower electric field in the space charge
region, which simply results from the fact that the constant built in
potential is dropped across a wider thickness. This sets a practical
limit on the amount of intrinsic material which can be inserted
between the doped layers. This is not usually a
serious limitation in practical solar cells.

Studies of the space charge region thickness dependence on background
doping reveal more challenging technological problems, as we shall 
see in the next section. In this case, the ionised background charge 
may cause the electric field to drop to zero in the intrinsic
region. Since $N_{BG}$ is always much smaller that $N_{A}$ and $N_{D}$,
Poisson's equation shows that the rate of reduction of the space charge 
region thickness with
increasingly positive voltage becomes much greater. Furthermore, 
cell resistance may increase because of the presence of an undepleted
and low doped thickness adjoining either the n or the p region.
Both these effects will reduce the solar cell efficiency.

We now consider the variation of photocurrent
$J_{L}$ for \pin\ cells with high and low values of $N_{BG}$.
In the case of low $N_{BG}$, a nearly constant photocurrent is seen into
forward bias. A residual voltage dependence does in principle
exist, and is due to the change in depletion widths as a function
of applied bias. The photocurrent in forward bias may drop because
of experimental considerations which will be considered in chapters
\ref{secexperiment} and \ref{secmivar}. In theory, however, it drops
at the point where the applied bias is equal and opposite to
$V_{bi}$. This removes the charge separating potential, and
the device no longer operates as a solar cell.

In the case of a cell with a high background
doping, the higher space charge density in the
intrinsic region imposes a significant electric field
gradient, as illustrated by the curves of
figure \ref{figpinfield}. In this case, the width of the
space charge region decreases slowly while the p and n depletion
widths are greater than zero, and $J_{L}$ is only weakly dependent
on the applied bias.

As the applied bias is increased, either the p or the n depletion
layer vanishes. At this point, there is still a field in the
space charge region, and charge separation still occurs. However,
the space charge region diminishes much more quickly with
increasing bias because of the lower doping density in the
space charge region. A significant decrease in $J_{L}$
with increasing voltage is suddenly seen.

The corresponding shoulder in the curve of $J_{L}$ versus
applied bias provides us with a useful tool for estimating
$N_{BG}$ in poor quality samples. At the shoulder, the \pin\ cell
is effectively behaving like a $\rm{p^{-}n}$cell with the
p type depletion layer given simply by the intrinsic region width,
and an effective p doping $N_{A}^{-}=N_{BG}^{p}$. We can therefore
substitute $x_{i}$ for $x_{w_{p}}$ and $N_{BG}^{p}$ for $N_{A}$
in equation \ref{pdepletion}, giving
\begin{equation}
\label{pnifrommivshoulder}
N_{BG}^{p}=\frac{N_{D}}{2}\left[
\sqrt{1+\frac{8\epsilon(-V_{bi}-V_{sh})}{qx_{i}^{2}N_{D}}}-1\right]
\end{equation}
where $V_{sh}$ is the measured voltage at the shoulder.
The equivalent relation for a n type $N_{BG}^{n}=N_{D}^{-}$ is given
by equation \ref{ndepletion}
\begin{equation}
\label{nnifrommivshoulder}
N_{BG}^{n}=\frac{N_{A}}{2}\left[
\sqrt{1+\frac{8\epsilon(-V_{bi}-V_{sh})}{qx_{i}^{2}N_{A}}}-1\right]
\end{equation}
Equations \ref{nnifrommivshoulder} and \ref{pnifrommivshoulder}
can be used to estimate the maximum tolerable background doping
for a given \pin\ solar cell design. For a typical structure with
a p type background doping,  
$N_{D}\simeq 10^{17}\rm{cm^{-3}}$ and an intrinsic layer
thickness $x_{i}\simeq 0.8\mu\rm{m}$ we find that the highest
background doping density permitting the cell to operate
adequately to voltages of about a volt is 
$N_{BG}^{p}\sim 5\times 10^{14}\rm{cm^{-3}}$

\section{Limiting Efficiency of Solar Cells\label{secefflimit}}

In the previous sections we have introduced the quantities describing
the efficiency of a solar cell. Equation \ref{efficiency} provides a
starting point for designing a high efficiency solar cell. 

High efficiencies require
large values of $\rm J_{sc}$, $\rm V_{oc}$ and $\rm F\!F$.
The series resistance of the cell is important in determining
$\rm F\!F$ and is ideally small. Although this can be achieved by
high doping levels, practical limits are set
by the degradation of transport parameters in heavily doped
samples. 

A high $\rm J_{sc}$ clearly depends on absorption of as much light as
possible. Equations \ref{directabsorption} and \ref{indirectabsorption}
show that a low bandgap increases absorption by reducing the cut-off
energy below which no photons are absorbed.

The requirements for a high $\rm V_{oc}$ can be found from equation
\ref{voc}. A fundamental requirement for a low $J_{D}$ and
hence a high $\rm V_{oc}$ is a low recombination.
Recombination can be reduced to some extent by improved cell 
fabrication techniques and passivation which reduce the
concentration and activity of non-radiative recombination
centres. The fundamental and unavoidable radiative limit
however guarantees a lower limit to the dark current.
We shall see in subsequent sections that \gaasalgaas\
solar cells are dominated by Shockley-Hall-Read non-radiative 
recombination mentioned in section \ref{seciv}.

We have seen that the the saturation current 
contains an exponential bandgap dependence of the form 
$J_{01}\sim e^{-E_{g}/k_{B}T}$ for the ideal diode.

The dependence of $\rm J_{sc}$ on $E_{g}$ is more complex. 
Experimental and theoretical studies (see chapters \ref{sectheory} 
and \ref{secqeresults}) however show that it decreases with increasing
bandgap. This can be partly explained by the change in absorption.
An increase in bandgap moves the absorption edge to higher
energies and decreases the absorption coefficient above that
threshold (see equations \ref{directabsorption} and 
\ref{indirectabsorption}).
The direct result is a reduction in minority carrier generation
and therefore a smaller light current. This decrease however
is weaker in character and magnitude than the exponential dependence
of the saturation current. The form of the $\rm V_{oc}$ therefore
reduces to
\begin{equation}
\label{vocegdependence}
V_{oc}\sim E_{g}\ln(J_{sc})
\end{equation}
A compromise therefore exists between the high bandgap requirement
for a high $\rm V_{oc}$ and the low bandgap for a high $\rm J_{sc}$, as
illustrated in figure \ref{figvocjsccompromise}.

\begin{figure*}[t]
\setlength{\unitlength}{0.75mm}
\begin{picture}(90,100)(-10,5)
\put(0,10){\vector(1,0){150}}
\put(142,6){\makebox(0,0){\it Voltage}}
\put(20,0){\vector(0,1){95}}
\put(18,98){\makebox(0,0){\it Light current}}
\put(49.7,70){\vector(-1,0){8}}
\put(59.7,70){\makebox(0,0){Low $E_{g}$}}
\put(99.7,55){\vector(-1,0){8}}
\put(109.7,55){\makebox(0,0){Ideal $E_{g}$}}
\put(119.7,20){\vector(-1,0){8}}
\put(129.7,20){\makebox(0,0){High $E_{g}$}}
\put(20,60){\line(1,0){64.6}}
\put(84.6,60){\line(0,-1){50}}
\put(50,48){\makebox{\bf \large Maximum}}
\put(57,40){\makebox{\bf \large power}}
\put(0,65){\line(1,0){70}}
\put(70,65){\line(6,-1){5}}
\put(74.93,64.17){\line(3,-1){5}}
\put(79.67,62.59){\line(2,-1){5}}
\put(84.6,60){\line(1,-1){5}}
\put(89.7,55){\line(1,-2){5}}
\put(95,45){\line(1,-6){7.5}}
\put(0,30){\line(1,0){90}}
\put(90,30){\line(6,-1){5}}
\put(94.93,29.17){\line(3,-1){5}}
\put(99.67,27.59){\line(2,-1){5}}
\put(104.6,25){\line(1,-1){5}}
\put(109.7,20){\line(1,-3){3}}
\put(112.5,11){\line(1,-6){1}}
\put(0,80){\line(1,0){20}}
\put(20,80){\line(6,-1){5}}
\put(24.93,79.17){\line(3,-1){5}}
\put(29.67,77.59){\line(2,-1){5}}
\put(34.6,75){\line(1,-1){5}}
\put(39.7,70){\line(1,-3){3}}
\put(42.5,61){\line(1,-4){3}}
\put(45.5,49){\line(1,-5){2}}
\put(47.5,39){\line(1,-6){5.5}}
\end{picture}
\figcap{\label{figvocjsccompromise} Light currents of solar cells
with low and high bandgaps under similar illumination, showing
the compromise between voltage and current. Also shown is an optimum
design with a manifestly greater maximum power rectangle.}
\end{figure*}

Limiting efficiencies for solar cells as a function of bandgap have
been calculated by Henry \cite{henry80}. The figures quoted are for
the AM1.5d standard solar spectrum. This spectrum corresponds to
direct sunlight at noon at a latitude of $48^{\circ}$. Other
standards include the AM0 spectrum, which is sunlight without
any atmospheric absorption.

An optimum of 30.6\% reported by Henry for materials 
with bandgaps in the region of 1.35eV,
which is near to the GaAs bandgap (1.424eV) for direct semiconductors,
and to Si (1.12eV) for indirect materials. 
Higher efficiencies can be achieved by 
concentrating the incident light, which increases both the short circuit
current and the open circuit voltage and increases the maximum efficiency
to about 37.2\%. The concentrator approach is limited 
by cell heating under the concentrated light. The increased operating
temperature increases the dark current, thereby reducing the open circuit
voltage. Cells for concentration must therefore be designed with this
effect in mind, or cooled. 

Cells for concentrator applications must also be designed with a
much lower series resistance, because of the higher current
densities. Although cell resistances can be reduced by increasing
the doping, this leads to a reduced minority carrier mobility
because of the increased number of scattering centres.

Cells with efficiencies in the high  20's
have been achieved in practice. Standard tabulations of world
record solar cell efficiencies under AM1.5 ilumination 
published by Green \cite{green96}
quote the best GaAs solar cell as having an efficiency of
$\sim25\%$, and $\sim$24\% for Si. Laboratory cells are therefore
close to achieving the fundamental efficiency limits, and increasingly 
complex designs are necessary to achieve higher efficiencies in single
bandgap systems.

Higher efficiency designs however, which are not subject to the same 
efficiency limits as single band-gap solar cells, have been suggested.
These are reviewed in the next section.

\subsection{Higher Efficiency Designs\label{sechighefficiencies}}

We have seen that the two fundamental factors limiting the maximum
efficiency of single bandgap solar cells are the trade-off between
$\rm V_{oc}$ and $\rm J_{sc}$, and the related problem of thermalisation
losses.

A solution to this trade-off is to combine several band gaps
in a single solar cell. Different parts of the incident spectrum are
absorbed in solar cell components with appropriate bandgaps. 
The component with the highest bandgap absorbs short wavelength
light and delivers current at a high voltage. 
The cell with the lowest bandgap absorbs long wavelength light
and operates at a lower voltage. The operational principles of the 
component cells remain those of single bandgap cells. The combination of
the different bandgaps however reduces thermalisation losses, 
resulting in higher efficiency limits. This concept is known as 
the rainbow or cascade cell. Henry \cite{henry80} predicts a 
maximum theoretical efficiency of around 75\%.

Unfortunately, this design suffers from serious technological
problems. If the cells are not connected in series, the output
voltage is that of the cell with the lowest bandgap. On the
other hand, connection in series requires that the current
outputof all the components be equal, because the output current
of the cell will be that of the worst component. 

Finally, series connection of cells is technologically
difficult. It can be achieved at high cost in complex
electrical connections, or during growth by the incorporation of tunnel
junctions.

A simplified approach is the
tandem cell. This exploits the same principle
as the rainbow cell but restricts the design to two materials,
thereby simplifying the current matching and electrical connection
problems. Henry \cite{henry80} predicts a fundamental efficiency
for this system of 50\%. A recent indium-gallium-phosphide/gallium
arsenide tandem cell reported by Bertness \cite{bertness94} has
achieved a record efficiency of 29.5\% in AM1.5g (global) 
illumination.

\section{Quantum Well Solar Cells\label{qwscintro}}

In previous sections we have described the operating principle
of the photovoltaic effect and given an overview of fundamental
loss mechanisms. We have reviewed a number of higher efficiency
designs which can minimise these losses and lead to higher
efficiencies.

We now present the main design studied in this thesis, which
is the quantum well solar cell (QWSC), first suggested by Barnham
\cite{barnham90}. The QWSC is a \pin\ cell
with quantum wells grown in the intrinsic region. This cell
is similar to tandem cells because it contains material with
two different bandgaps. 

Figure \ref{figian} shows a schematic of the QWSC. The structure
contains two bandgaps $E_{g}$ and $E_{a}$ which are the barrier
bandgap and quantum well bandgap respectively. The material behaves
like bulk material for electron and hole energies greater
than $E_{g}$ and like a quantum well or superlattice states
for energies below $E_{g}$.

For incident photons with energies greater than $E_{g}$,
the QWSC operates like a conventional single bandgap \pin\ cell. 

Photons with energies between $E_{g}$ and $E_{a}$ however are absorbed
in the quantum well states. Once excited into well states, they
can recombine radiatively or non radiatively or escape from the
well. The escape mechanisms are thermionic emission into a bulk
state, tunnelling,
or a combination of the two. Experimental and theoretical studies
of this system by Nelson \cite{nelson93} have shown that the escape
probability from \gaas\ quantum wells embedded in \algaas\ with a 
composition of 34\% is about $100\%$ at room temperatures. This
very high escape probability is seen even when the field across
the quantum well is low. Thermionic emission therefore is sufficient
to ensure high carrier escape probabilities from the wells at room
temperatures.

Further work by Barnham {\em et al.} \cite{barnham96} has shown that 
the output voltage is greater than would be expected from a single
junction cell with the quantum well bandgap $E_{a}$. This has
been observed in three different materials systems. In the
\algaas\ system, this work showed that increasing barrier
bandgaps $E_{g}$ give increased voltage enhancements over the
$V_{oc}$ expected from a single bandgap cell with the
quantum well effective bandgap $E_{a}$.

The QWSC design therefore departs from the analysis used by
Henry \cite{henry80} in his analysis of the fundamental efficiency
limit. Unlike the separate components of a single bandgap cell,
the QWSC operates in a physically different way to single bandgap
cells because of the minority carrier escape from the quantum
wells. Since it is in principle capable of generating the same
photocurrent as a single bandgap cell with a higher voltage,
the design has a higher fundamental efficiency limit than a
single bandgap cell.

The fundamental efficiency limits of QWSCs are not yet clearly 
understood. The review article by Nelson \cite{nelson96} gives an 
overview of the subject.

\begin{figure*}[ht]
\begin{center}
\unitlength 0.80mm
\linethickness{0.4pt}
\begin{picture}(117.55,91.67)
\put(78.58,59.00){\circle*{2.40}}
\put(46.25,33.83){\circle{3.40}}
\put(87.25,65.83){\circle*{2.40}}
\put(98.17,57.67){\line(1,0){3.00}}
\put(66.91,21.25){\circle{3.33}}
\put(78.41,24.92){\circle{3.33}}
\put(105.25,57.67){\line(1,0){3.00}}
\put(111.92,57.67){\line(1,0){3.00}}
\put(42.33,76.50){\line(0,-1){10.50}}
\put(42.33,66.00){\line(5,-1){7.67}}
\put(50.00,64.47){\line(0,1){9.53}}
\put(49.67,74.00){\line(4,-1){8.67}}
\put(65.67,69.33){\line(4,-1){8.67}}
\put(81.67,64.67){\line(4,-1){8.67}}
\put(58.33,71.83){\line(0,-1){10.50}}
\put(74.33,67.17){\line(0,-1){10.50}}
\put(58.33,61.33){\line(5,-1){7.67}}
\put(74.33,56.67){\line(5,-1){7.67}}
\put(66.00,59.80){\line(0,1){9.53}}
\put(82.00,55.13){\line(0,1){9.53}}
\put(90.33,62.42){\line(6,-1){5.67}}
\put(95.99,61.48){\line(1,0){19.00}}
\put(29.91,79.35){\line(-1,0){24.00}}
\put(42.33,67.33){\line(1,0){7.67}}
\put(42.67,71.00){\line(1,0){7.67}}
\put(58.33,63.33){\line(1,0){7.67}}
\put(58.33,67.00){\line(1,0){7.67}}
\put(74.33,59.00){\line(1,0){7.67}}
\put(74.33,62.67){\line(1,0){7.67}}
\put(46.41,71.08){\circle*{2.40}}
\put(18.16,49.58){\makebox(0,0)[cc]{Light}}
\bezier{48}(4.91,56.25)(9.91,52.91)(14.91,56.25)
\bezier{56}(14.91,56.25)(19.91,60.91)(24.91,55.91)
\bezier{32}(24.91,55.91)(28.58,54.25)(32.58,56.25)
\put(32.91,56.25){\vector(1,0){0.2}}
\put(32.24,56.25){\line(1,0){0.67}}
\put(87.67,65.33){\vector(1,0){4.67}}
\bezier{36}(78.67,59.33)(81.33,61.67)(76.33,63.67)
\bezier{28}(76.33,63.67)(75.00,66.00)(79.00,65.67)
\bezier{32}(79.00,65.67)(84.00,64.33)(86.67,65.33)
\put(49.67,29.00){\line(4,-1){8.67}}
\put(65.67,24.33){\line(4,-1){8.67}}
\put(81.67,19.67){\line(4,-1){8.67}}
\put(90.36,17.52){\line(6,-1){5.67}}
\put(96.02,16.58){\line(1,0){19.00}}
\put(30.00,34.22){\line(-1,0){24.00}}
\put(81.67,19.67){\line(0,1){6.67}}
\put(81.33,26.67){\line(-3,1){6.67}}
\put(74.67,22.00){\line(0,1){6.67}}
\put(65.67,24.67){\line(0,1){6.67}}
\put(65.33,31.67){\line(-3,1){6.67}}
\put(58.67,27.00){\line(0,1){6.67}}
\put(49.67,29.00){\line(0,1){6.67}}
\put(49.33,36.00){\line(-3,1){6.67}}
\put(42.67,31.33){\line(0,1){6.67}}
\put(42.67,33.67){\line(1,0){7.00}}
\put(58.67,29.67){\line(1,0){7.00}}
\put(74.67,25.00){\line(1,0){7.00}}
\bezier{40}(78.67,23.00)(79.67,17.67)(75.67,20.33)
\bezier{16}(75.67,20.33)(72.33,20.67)(72.00,19.67)
\bezier{28}(72.00,19.67)(71.67,17.33)(68.67,20.33)
\put(65.33,21.33){\vector(-1,0){3.67}}
\put(46.33,35.33){\vector(0,1){33.00}}
\put(62.33,63.33){\vector(0,-1){33.67}}
\put(27.08,36.67){\line(1,0){3.00}}
\put(20.41,36.67){\line(1,0){3.00}}
\put(13.74,36.67){\line(1,0){3.00}}
\put(7.08,36.67){\line(1,0){3.00}}
\put(32.67,71.00){\vector(1,0){7.00}}
\put(29.33,71.00){\makebox(0,0)[rc]{Energy level}}
\put(18.33,6.00){\makebox(0,0)[cc]{p}}
\put(62.00,6.00){\makebox(0,0)[cc]{$\bf i$}}
\put(106.33,6.00){\makebox(0,0)[cc]{$\bf n$}}
\put(67.33,6.00){\vector(1,0){27.67}}
\put(109.33,6.00){\vector(1,0){6.67}}
\put(103.33,6.00){\vector(-1,0){5.33}}
\put(56.33,6.00){\vector(-1,0){24.67}}
\put(21.67,6.00){\vector(1,0){6.67}}
\put(14.67,6.00){\vector(-1,0){8.67}}
\put(85.00,40.00){\makebox(0,0)[cc]{$E_{a}$}}
\put(84.67,44.33){\vector(0,1){14.67}}
\put(84.67,35.67){\vector(0,-1){10.67}}
\put(55.00,91.67){\makebox(0,0)[cc]{Quantum wells}}
\put(32.67,36.67){\line(1,0){0.33}}
\put(35.33,36.67){\line(1,0){0.33}}
\put(38.00,36.67){\line(1,0){0.33}}
\put(40.67,36.67){\line(1,0){0.33}}
\put(43.33,36.67){\line(1,0){0.33}}
\put(46.00,36.67){\line(1,0){0.33}}
\put(48.67,36.67){\line(1,0){0.33}}
\put(51.33,36.67){\line(1,0){0.33}}
\put(54.00,36.67){\line(1,0){0.33}}
\put(56.67,36.67){\line(1,0){0.33}}
\put(59.33,36.67){\line(1,0){0.33}}
\put(62.00,36.67){\line(1,0){0.33}}
\put(64.67,36.67){\line(1,0){0.33}}
\put(67.33,36.67){\line(1,0){0.33}}
\put(70.00,36.67){\line(1,0){0.33}}
\put(72.67,36.67){\line(1,0){0.33}}
\put(75.33,36.67){\line(1,0){0.33}}
\put(78.00,36.67){\line(1,0){0.33}}
\put(80.67,36.67){\line(1,0){0.33}}
\put(83.33,36.67){\line(1,0){0.33}}
\put(86.00,36.67){\line(1,0){0.33}}
\put(88.67,36.67){\line(1,0){0.33}}
\put(91.33,36.67){\line(1,0){0.33}}
\put(94.00,36.67){\line(1,0){0.33}}
\put(96.67,36.67){\line(1,0){0.33}}
\put(99.33,36.67){\line(1,0){0.33}}
\put(102.00,36.67){\line(1,0){0.33}}
\put(104.67,36.67){\line(1,0){0.33}}
\put(107.33,36.67){\line(1,0){0.33}}
\put(110.00,36.67){\line(1,0){0.33}}
\put(112.67,36.67){\line(1,0){0.33}}
\put(108.33,47.67){\makebox(0,0)[cc]{$V_{ext}$}}
\put(108.00,52.33){\vector(0,1){4.33}}
\put(108.00,44.33){\vector(0,-1){6.33}}
\put(44.67,51.00){\makebox(0,0)[rc]{A}}
\put(61.00,47.67){\makebox(0,0)[rc]{B}}
\put(56.67,87.00){\vector(1,-3){4.33}}
\put(53.33,86.67){\vector(-2,-3){6.44}}
\put(8.67,25.00){\makebox(0,0)[lc]{A) Generation}}
\put(8.67,17.33){\makebox(0,0)[lc]{B) Recombination}}
\put(92.67,82.67){\makebox(0,0)[cc]{Thermal escape}}
\put(88.67,70.33){\vector(-1,-4){0.2}}
\multiput(90.33,79.00)(-0.12,-0.62){14}{\line(0,-1){0.62}}
\multiput(42.35,76.39)(-0.31,0.12){17}{\line(-1,0){0.31}}
\put(37.13,78.37){\line(0,1){0.00}}
\multiput(37.13,78.37)(-0.83,0.11){9}{\line(-1,0){0.83}}
\multiput(42.63,31.22)(-0.31,0.12){17}{\line(-1,0){0.31}}
\put(37.41,33.20){\line(0,1){0.00}}
\multiput(37.41,33.20)(-0.83,0.11){9}{\line(-1,0){0.83}}
\put(117.52,37.28){\makebox(0,0)[cc]{$E_{g}$}}
\put(117.55,61.51){\vector(0,1){0.2}}
\put(117.55,41.28){\line(0,1){20.23}}
\put(117.55,16.68){\vector(0,-1){0.2}}
\put(117.55,33.08){\line(0,-1){16.40}}
\end{picture}
\figcap{Schematic QWSC structure, absorption, recombination and escape
mechanisms\label{figian}}
\end{center}
\end{figure*}

\section{The QWSC in the AlGaAs System}

Work by Paxman \cite{paxman92} has demonstrated efficient carrier
escape from QWSC structures in \algaas\ with $x\simeq 30\%$. This
resulted in a sample grown by Tom Foxon at Phillips Laboratories,
Redhill, which was 14.2\% efficient in AM1.5. This sample showed
a good $V_{oc}$ of 1.07V, and a fill factor $F\!F=77\%$. More detail 
concerning optimisations carried out in this work will be given in 
chapter \ref{secoptimisation}.

The work by Paxman however showed scope for improvement in the 
$J_{sc}$ of QWSCs in \algaas. The advantages of this material 
include extensive characterisation and a material which is lattice
matched for compositions ranging from \gaas\ to $AlAs$. The lattice
spacing in \gaas\ and AlAs is only different by about 0.14\% according
to Adachi \cite{adachi85}.

The material however suffers from poor transport characteristics.
As we shall see in chapter \ref{secparameters}, minority
carrier transport is poor, and the $J_{sc}$ of single bandgap
\algaas\ cells decreases rapidly with increasing aluminium composition
x.

We conclude that QWSCs in the \algaas\ system mainly suffer
from poor minority carrier characteristics, which result in
a low $J_{sc}$. Work presented in further chapters will address
this problem with theoretical and experimental analysis of
design changes made to the basic QWSC design.

\section{Conclusions}

This chapter has reviewed the operating principles of solar
cells and defined the main parameters allowing characterisation
of their performance.

We have seen that single bandgap solar cells are limited to
efficiencies in unconcentrated sunlight of about 30\%. by 
fundamental loss mechanisms. Although designs such as the
cascade cell and variations on this design are capable of
reaching higher efficiencies, they are subject to complex
technological problems because they combine two or more
component solar cells in one design. Moreover, the approach
represents a technological solution rather than one which
introduces fundamental changes to solar cell operation.

The QWSC solar cell however is governed by physically different
efficiency limits, because it has a $V_{oc}$ which is higher than
that of a single bandgap cell with the quantum well effective
bandgap. This principle has been demonstrated in three
different materials systems.

The \algaas\ system shows a number of advantages
for quantum well applications. Unfortunately, it suffers from poor 
minority carrier transport properties, which degrade with increasing 
aluminium fraction. This results in a poor short circuit current, 
and poor efficiencies for solar cells fabricated from this material. 

We conclude that the voltage enhancement aspect of designing an
efficient QWSC in the \algaas\ cell has been demonstrated, but
that the $J_{sc}$ must be increased. The following chapters will
present a theoretical and experimental analysis of this problem.

%% file: chap3.tex
\chapter{\label{sectheory}Quantum efficiency and short 
circuit current models}

\section{Introduction}

Optimisation of the short circuit current (\jsc) requires
an understanding of the spectral response of the cell at zero
applied bias. In this chapter we give a brief survey of previous
theoretical work on such models, before going on to describe a 
model of the \jsc\ for \gaasalgaas\ QWSCs with position dependent
transport and optical properties. Values of the materials parameters
are discussed in chapter \ref{secparameters}.

\subsection{Review of Previous Work\label{seclittreview}}

Modelling of \algaas\ solar cells has largely been confined
to theoretical studies of cases which can be solved analytically.
Hovel \cite{hovel75} gives a good review of solar cell physics,
including discussions of a range of analytical and numerical
models.

The theoretical model by Konagai \cite{konagai76} considers a 
p \algaas\ - p \gaas\- n \gaas\ compositionally
graded p layers grown epitaxially on GaAs. He does not
consider the position dependence of the transport parameters,
and uses an analytical expression for the absorption
coefficient. The effect of the grade is expressed via
the quasi-electric minority carrier field it produces in
the graded region. The field is taken as
the gradient of the bandgap due to the compositional
change, and is discussed in more detail in chapter
\ref{secparameters}. This work suggests that the grade is
especially useful in reducing the minority carrier recombination
at the surface of the cell. Efficiencies of the order of 20\%
are predicted under air mass zero (AM0).

Hutchby \cite{hutchby76} has studied a similar design,
but for an n \algaas\ - n \gaas\- p \gaas\ design.
This work is again purely theoretical, and makes similar
assumptions by assuming that the minority
carrier transport is constant. The absorption coefficient
is a semi-empirical fit to data. The quasi-electric field in
this model is again taken as proportional to the band-gap 
gradient, and the minority carrier transport parameters
are assumed to be essentially independent of the aluminium 
composition. The theoretical efficiency in this case is 17.7\% for
an AM0 spectrum.

Sutherland \cite{sutherland76} has presented a theoretical study
which uses semi-empirical parameterisations to model the position
dependence of the materials parameters. He finds good agreement
with the earlier studies, predicting a peak efficiency of about
20\% for a n \algaas\ - n \gaas\- p \gaas\ design. 

Hamaker \cite{hamaker85} has presented a theoretical model
of a n -- p and p -- n \algaas\ cell with graded emitter
layers. This is the most accurate model published to date,
and uses semi-empirical parametrisations combined with
data for the materials parameters. An optimisation in
terms of doping and compositional profiles is included.
The predicted efficiencies are similar to those predicted
by the previous studies. Very little difference however
is reported between the different designs studied.

Of the work mentioned above, the Hamaker \cite{hamaker85} 
model is closest to the one we shall present in this chapter. 
We note however that the models, with the exception of
Hamaker \cite{hamaker85} have used analytical approximations
for the materials parameters and generally only consider the
position dependence of the optical parameters.

The models mentioned here are not compared with experimental
data. In the following sections we will present a model similar
to that of Hamaker \cite{hamaker85} but which is equipped to
deal with quantum wells and compositional grading throughout
the structure. The model furthermore will be designed to
reproduce experimental data, and the relevant fitting
parameters identified. The main methodology has been described
by Connolly \cite{connolly95}.

\section{Preliminary definitions}

We start by introducing the external quantum efficiency or QE.
It describes the spectral efficiency of
the cell independently of the particular incident spectrum and
is defined as the probability of an incident photon of wavelength
$\lambda$ generating an electron-hole pair which contributes to
the \jsc. It can be expressed as the ratio of the number of
steady state current carriers $J_{sc}/q$ collected under an illumination of
wavelength $\lambda$ to the number of photons incident on the top
surface of the cell $F(\lambda)$ where $q$ is the magnitude of
the electronic charge.
\begin{equation}
\label{qe}
QE(\lambda)=\frac{J_{sc}(\lambda)}{qF(\lambda)}
\end{equation}
We also define the internal $QE_{int}$, which is defined in the
same way, but where the flux $F$ is replaced by the flux actually
transmitted into the solar cell. The two differ only by a factor
dependent on the surface reflectivity.

The model calculates the $QE$, which can then be used to predict
the \jsc\ for a given cell design and illumination by multiplying 
the QE by the appropriate spectrum.

The wide range of QWSC and \pin\ solar cell designs considered
in the characterisation and optimisation program requires a
general approach to each part of the calculation. These designs
include cells with compositionally graded neutral and space
charge regions. These will be referred to as graded cells.
Cells without these compositional changes are referred to
as ungraded cells. We also consider cells which operate as
Fabry-Perot cavities because they feature a mirror which is
deposited on the back surface of the cell. These will be referred
to as mirror backed cells.

The first problem we consider in this chapter is the variation of
the light intensity in the solar cell as a function of position,
wavelength, and front and back surface reflectivities. The
generation rate can then be calculated from the absorption coefficient
and the light intensity at any particular position and wavelength.

We then describe the photocurrent calculation in the two
cases of depleted space-charge regions and neutral regions.
In a general \pin\ cell, the space-charge region is made up
of the p depletion region, the nominally undoped
i region, and the n region depletion layer. Under the depletion
approximation described below, the photocurrent calculation
reduces to an integral of the generation rate over the depletion
region and wavelength.

The second case deals with the photocurrent from the electrostatically
neutral sections of the n and p layers.
Solutions to the resulting transport equations are first discussed for
ungraded cells where materials parameters are constant, and an
analytical solution can be found.

In the case of graded cells however an analytical solution
cannot in general be found and the transport equations must be
solved numerically. A standard difference method for this case is
presented in the final part of the chapter.

\begin{figure*}
\vspace{-3ex}
\setlength{\unitlength}{0.88mm}
\begin{picture}(100,100)(-20,-8)

\put(8,87){\makebox(0,0){\small window}}
\put(30,76){\makebox(0,0){\small graded \normalsize $p$}}
\put(65,57){\makebox(0,0){$i$}}
\put(97,40){\makebox(0,0){$n$}}

\put(24,-6){\vector(-1,0){16}}
\put(25,-8){\makebox(0,0)[bl]{$x_{p}$}}
\put(30,-6){\vector(1,0){15}}
\put(39,-0.5){\makebox(0,0)[bl]{$x_{w_{p}}$}}
\put(58,-6){\vector(-1,0){13}}
\put(60,-8){\makebox(0,0)[bl]{$x_{i}$}}
\put(65,-6){\vector(1,0){12}}
\put(76.7,-0.5){\makebox(0,0)[bl]{$x_{w_{n}}$}}
\put(90,-6){\vector(-1,0){13}}
\put(92,-8){\makebox(0,0)[bl]{$x_{n}$}}
\put(100,-6){\vector(1,0){7}}

\put(0,-3){\vector(1,0){115}}
\put(115,-8){\makebox(0,0){Depth}}
\put(0,-3){\vector(0,1){95}}
\put(-9.5,82){\makebox(0,0){electron}}
\put(-9,77){\makebox(0,0){energy}}

\put(8,-4){\line(0,1){2}}
\put(39,-4){\line(0,1){2}}
\put(45,-4){\line(0,1){2}}
\put(77,-4){\line(0,1){2}}
\put(82,-4){\line(0,1){2}}
\put(107,-4){\line(0,1){2}}


\put(9.2,82.9){\line(-1,0){8}}
\put(9.2,74.9){\line(0,1){8}}
\put(24.3,72.4){\line(-6,1){15}}
\put(39,67.4){\line(-3,1){15}}
\put(39,67.4){\line(5,-2){4}}
\put(53,59.3){\line(-3,2){10}}
\put(53,59.3){\line(0,-1){10}}
\put(57,46.5){\line(-3,2){4}}
\put(57,46.5){\line(0,1){10}}
\put(61,54){\line(-3,2){4}}
\put(61,54){\line(0,-1){10}}
\put(65,41.4){\line(-3,2){4}}
\put(65,41.3){\line(0,1){10}}
\put(69,48.7){\line(-3,2){4}}
\put(69,48.7){\line(0,-1){10}}
\put(73,36.1){\line(-3,2){4}}
\put(73,36.1){\line(0,1){10}}
\put(83,39.4){\line(-3,2){10}}
\put(82.98,39.4){\line(3,-1){4}}
\put(87,38){\line(1,0){20}}


\put(39,31.4){\line(-1,0){37}}
\put(39,31.4){\line(3,-1){4}}
\put(53,23.3){\line(-3,2){10}}
\put(53,29.3){\line(0,-1){6}}
\put(57,26.5){\line(-3,2){4}}
\put(57,20.5){\line(0,1){6}}
\put(61,18){\line(-3,2){4}}
\put(61,24){\line(0,-1){6}}
\put(65,21.4){\line(-3,2){4}}
\put(65,15.3){\line(0,1){6}}
\put(69,12.7){\line(-3,2){4}}
\put(69,18.7){\line(0,-1){6}}
\put(73,16.1){\line(-3,2){4}}
\put(73,10.1){\line(0,1){6}}
\put(83,3.4){\line(-3,2){10}}
\put(82.98,3.4){\line(3,-1){4}}
\put(87,2){\line(1,0){20}}

\multiput(3,34)(10,0){11}{\line(1,0){5}}
\put(113,34){\makebox(0,0){$E_{F}$}}

\end{picture}
\figcap{Example generalised QWSC band diagramme, including position
dependent bandgap in the p type layer and high aluminium fraction
window layer
\label{figcelldesign}}
\end{figure*}

\section{Calculation of light intensity in a QSWC\label{secfluxcalc}}

\subsection{Definitions}

In this section we derive the light intensity in a solar cell
of total thickness $d$ and in which the optical parameters may
be position dependent. We define the optical parameters required
to calculate the light intensity for cells where interference
effects may be significant. We consider only the normal incidence
case. A comprehensive discussion is given in Macleod 
\cite{macleod69}.

The first phenomenon experienced by a beam of light of
intensity ${\cal I}_{i}$ which is incident on
a solar cell is reflection. We quantify this by the front
surface reflectivity $R$ which is the fraction of light
intensity of wavelength $\lambda$ reflected by that
surface.

For graded material the absorption coefficient varies
with distance. In this case the intensity at a particular point
depends on the integral of the absorption which the light has
suffered up to that point, minus the fraction of light reflected 
at the front surface.
\begin{equation}
\label{absorptiondefinition}
{\cal I}(x,\lambda)={\cal I}_{i}(\lambda)(1-R(\lambda))
e^{\int_{0}^{x}\alpha(x',\lambda)dx'}
\end{equation}


This expression allows us to calculate the light intensity for cases 
where light only makes one pass. By this we mean that
the amount of light internally reflected from the back surface
of the cell is negligible.

This is not the case for cells which are coated on the back surface
with a mirror and which leave a significant amount of light unabsorbed
on the first pass.
In this case, successive reflected beams suffer phase changes with
distance and upon reflection at the front and back surfaces.
The superposition of beams of light with different phases leads
to interference phenomena. To take these effects into account,
we must derive the intensity from the complex amplitude of the
light. The electric field component of the light $\cal E$ 
is a convenient quantity to use and we refer to it 
as the amplitude for brevity. It is of the same form as $\cal I$ 
and is characterised by a complex decay constant $k$. The
real part of $k$ characterises the exponential decrease in electric field
strength due to light absorption, and the imaginary part reflects
the change in phase of the electric field with distance. $\cal E$
is then
\begin{equation}
\label{electricfieldamplitude}
{\cal E}(x,\lambda)={\cal E}_{0}(\lambda) e^{-\int_{0}^{x} k(x',\lambda)dx'}
\end{equation}
where ${\cal E}_{0}$ is the amplitude at position zero.

The light intensity $\cal I$ is proportional the product of the electric
field amplitude and its complex conjugate \cite{macleod69}:
\begin{equation}
\label{fieldtointensity}
{\cal I}(x,\lambda)=\epsilon v {\cal E}{\cal E}^{*}
\end{equation}
where ${\cal E}^{*}$ denotes complex conjugate of ${\cal E}$, 
and $\epsilon$ and $v$ are respectively the permittivity
and the speed of light at position $x$. 

${\cal E}_{0}$ is related to the incident flux ${\cal I}_{i}$
by substituting the value of ${\cal I}(0,\lambda)$ from equation 
\ref{absorptiondefinition} into equation \ref{fieldtointensity}.
Since the phase changes in the cell are all relative to the phase
of the incident light, we choose to set this incident phase to zero.
${\cal E}_{0}$ is then real and of the form
\begin{equation}
\label{fieldatsurface}
{\cal E}_{0}=\sqrt{\frac{(1-R){\cal I}_{i}}{\epsilon v}}
\end{equation}
The total amplitude $\cal E$ is found by summing the
amplitude contributions from successive reflections
${\cal E}_{j}$, where $i$ is the order of reflection.
\begin{equation}
\label{sumelectricfield}
{\cal E}(x,\lambda)={\cal E}_{0} \sum_{j=1}^{\infty}{\cal E}_{j}(x,\lambda)
\end{equation}
The dimensionless components ${\cal E}_{j}$ are a measure of the amplitude
of the $j^{\rm th}$ amplitude component relative to the
amplitude ${\cal E}_{0}$ transmitted throught the
front surface of the cell, and are calculated below.

Equations \ref{sumelectricfield} and \ref{fieldtointensity} together
define the intensity as a function of distance and wavelength in terms
of the amplitude components ${\cal E}_{0}$.

\subsection{Phase and reflectance}

Before calculating the components ${\cal E}_{j}$ we must relate
the reflectivity and absorption coefficient to the corresponding
quantities for the amplitude. Both of these can be
derived from inspection of equation \ref{fieldtointensity}.

When a normally incident beam of light of intensity $\cal I$ is 
internally reflected, the intensity is reduced by a factor $R$. 
Consistency with equation \ref{fieldtointensity}
demands that the corresponding amplitude should be
modified by a factor $\sqrt R$. A factor $e^{i\psi}$ describes
the phase change upon reflection at the back surface,
where $i=\sqrt{-1}$ and $\psi$ is the phase change.
The complex reflectance is therefore the square
root of the $R$ multiplied by the $\psi$.
\begin{equation}
\label{reflectance}
r=e^{i \psi}\sqrt{R}
\end{equation}
The relationship between amplitude and intensity in equation
\ref{fieldtointensity} together with equations \ref{absorptiondefinition}
and \ref{electricfieldamplitude} shows us that the real part of $k$
is simply half the absorption coefficient $\alpha$. However,
it is a complex quantity and may have an associated phase.

The phase of an electromagnetic wave changes by a factor $2\pi$ over
a distance of one wavelength. Furthermore,
the wavelength of light in a medium with refractive index $n$
is $\lambda/n$. The phase change is therefore simply the ratio
$2\pi i n/\lambda$. The phase change over a distance
$x$ will be the integral of this phase over distance.

The complex decay constant is then the sum of $\alpha/2$ and
the phase
\begin{equation}
\label{extinctioncoeff}
k(x,\lambda)=\frac{\alpha}{2}-\frac{2 \pi\ i n}{\lambda}
\end{equation}

\subsection{Generation Rate\label{sectheorygenerationdefs}}

We can now explicitly write down the successive terms in
the sum of amplitudes as a function of position $x$.
The first is the simply the transmitted beam and starts
with the relative amplitude of 1.
\begin{equation}
\label{amp1}
{\cal E}_{1}(x)=e^{-\int_{0}^{x}kdx'}
\end{equation}
The initial amplitude of the second beam is equal to the limit
of ${\cal E}_{1}$ at the back mirror, but with a reflectance
factor. We introduce the back surface reflectance $r_{b}$ and it's
associated phase change $\psi_{b}$ which is related
to the back surface internal reflectivity $R_{b}$ as 
outlined above. The amplitude of the second beam is then dependent
on an integral over $-dx$ from the back surface $d$
to the position $x$. This is the same as integrating 
forwards from $x$ to $d$ over $+dx$. Component ${\cal E}_{2}$ is
then
\begin{equation}
{\cal E}_{2}(x)=e^{-\int_{0}^{d}kdx'}r_{b}e^{-\int_{x}^{d}{k(x')dx'}}
\end{equation}
If $r_{f}$ is the front surface internal reflectance, with associated
phase $\psi_{f}$ and reflectivity $R_{f}$, subsequent
reflections take the form
\begin{equation}
\label{amp3}
\begin{array}{l l l}
{\cal E}_{3}(x) & = & e^{-2\int_{0}^{d}kdx'}r_{f}r_{b}
e^{-\int_{0}^{x}{k(x')dx'}} \\
{\cal E}_{4}(x) & = & e^{-3\int_{0}^{d}kdx'}r_{f}r_{b}^{2}
e^{-\int_{x}^{d}{k(x')dx'}} \\
{\cal E}_{5}(x) & = & \cdots \\
\end{array}
\end{equation}
Rewriting the even reflections in terms of an integral from $0$ to $d$
minus an integral from $0$ to $x$ and factorising by the extra term in
$r_{b}$, we obtain two
geometric series which differ only by a factor relating
to the amplitude attenuation over one pass and the back surface
reflectance. It has the limit
\begin{equation}
\label{amplitudesum}
{\cal E}={\cal E}_{0}\left[e^{-\int_{0}^{x}kdx'}
+\frac{r_{b}e^{-2\int_{0}^{d}kdx'}} {e^{-\int_{0}^{x}kdx'}}\right]
\times
\left[\frac{1}{1-r_{f}r_{b}e^{-2\int_{0}^{d}kdx'}}\right]
\end{equation}
The intensity in the solar cell can be written independently of
the factor $\epsilon v$ as follows
\begin{equation}
\label{intensityexpression}
{\cal I}(x,\lambda)={\cal I}_{i}(1-R)\left(
\frac{{\cal E E}^{*} }{{\cal E}_{0}{\cal E}_{0}^{*}}
\right)
\end{equation}
We emphasize here that this treatment consitutes a first approximation
to the problem, since it does not consider the internal reflections
taking place at the quantum well interfaces. The reflectivity at these
surfaces, however, is not great, since the change in refractive index
between \gaas\ and \algaas\ is not large for the barrier compositions
considered. A brief calculation using optical parameters reviewed in 
chapter \ref{secparameters} indicates that the reflectivity in the
worst case of a 45\% aluminium barrier is only 0.2\%. On the other
hand, a large number of wells may make this effect significant. Further
discussion of this point is given with respect to experimental data in
chapter \ref{secqeresults}.

The generation rate $G$ is then readily given by the product of 
$\alpha$ and $I$ as follows
\begin{equation}
G(x,\lambda)=\alpha(x,\lambda){\cal I}(x,\lambda)
\end{equation}

\section{Photocurrent\label{secphotocurrent}}

Having derived an expression for the generation rate we now
investigate the mechanisms which transform these photo-generated
carriers into a useful electric current.

In general, the depleted or space-charge region of the 
QWSC includes the p depletion
region of width $x_{w_{p}}$ (see figure \ref{figcelldesign}), the 
intrinsically doped i region (width $x_{i}$), and the n
region depletion layer (width $x_{w_{n}}$). We have seen in section
\ref{secpincell} that the i region may not be fully depleted either
if the forward applied bias is larger than the magnitude of $V_{bi}$, 
or if the background doping level is high. In this case, the electric 
field drops to zero within the intrinsic region.

If, however, the background doping level is low, a field is
maintained across the i region in the moderate forward
bias regime and carrier transport is drift dominated. 
Under the depletion approximation described below,
carrier recombination in the space-charge region is negligible
and charge collection efficiencies are 100\%.

In the neutral window, p and n regions, transport is dominated by
diffusion and recombination is not negligible. The photocurrent
from these layers is found by solving minority carrier transport
equations. Previous work by Braun \cite{braun90} has shown that the high
band-gap window layer does not contribute a measurable photocurrent
and may be treated simply as a non-contributing surface absorber.

In the following sections the photocurrent calculations for space 
charge and neutral regions are discussed for
the general case of a \pin\ structure with position dependent 
parameters. We then examine the simplified case of an ungraded 
structure and the manner in which certain parameters can be 
determined from this simplified case. We then  develop a numerical 
method to solve the more general graded case.

\section{Intrinsic region photocurrent}

Calculation of the photocurrent $J_{i}$ from depletion layer is
based on the depletion approximation.  This assumes that minority carriers 
generated or injected into the i region are swept across 
it to the boundary with the depletion edge where they become majority 
carriers. That is, electrons are swept towards the n interface,
whilst holes are swept to the p interface with no loss.
This is a good approximation for cells in which a field is maintained
across the i region. For cells with either compensated or low background
doping, this condition is maintained into moderate forward bias.
$J_{i}$ can then be found from the generation rate.

The photocurrent contribution from the quantum wells is calculated 
assuming an escape efficiency of 100\%. Once the carriers have escaped from 
the wells, they behave like other minority carriers. Previous work 
by Nelson \cite{nelson93} has shown this to be true for \gaas\ wells in 
\algaas\ with  an aluminium fraction of 30\%.

The photocurrent density $J_{i}$  from the space charge region
is then simply given by the integral of the generation rate
\begin{equation} 
\label{icurrent}
J_{i}(\lambda)=\int_{x_{p}-x_{w_{p}}}^{x_{p}+x_{i}+x_{w_{n}}} 
G(x,\lambda) dx 
\end{equation}
Although in general an analytical form for $J_{i}$ can only be found 
for ungraded solar cells, the calculation is identical for the 
graded case. The quantum well absorption coefficient is described
by Paxman \cite{paxman93}.

\section{p and n Region Contributions \label{secpphotocurrent}}

The photocurrent from neutral p and n regions is found from
current and continuity relations outlined below. Since the method
in both cases is identical, the following discussion concentrates
on the p for clarity.

\subsection{Drift and Diffusion Currents}

Minority carrier currents can be caused by drift or diffusion.
The former results from the acceleration of carriers in an 
electric field, whereas the latter is the consequence of a
carrier concentration gradient.
In the low injection limit and for moderate electric fields,
the net current  flowing at depth $x$ in the p layer 
$j_{p}(x)$ is the sum of these two currents \cite{sze81}.
\begin{equation}
\label{currentgeneral}
j_{p}(x)=q\mu_{n}(x) E(x) n_{p}(x) + qD_{n}(x)\frac{dn_{p}}{dx} 
\end{equation}
where $\mu_{n}$ is the minority electron mobility, $D_{n}$ the electron 
diffusivity and $E$ the quasi-electric field. These parameters
are position dependent in the general case. We note here that some
workers use the terminology $j_{p}$ as the minority hole current in
the n layer. We choose to define it in this way in order to
keep the same subscript between the excess minority carrier concentration
$n_{p}$ and the corresponding current $j_{p}$.

The total  minority electron carrier concentration $n_{p}$ is
defined as the sum of the equilibrium carrier concentration
$n_{p_{0}}$ and the excess minority carrier concentration
$n$ as  follows:
\begin{equation}
\label{totalmincarr}
n_{p}=n_{p_{0}}+n
\end{equation}
No net current flows in equilibrium, and equation \ref{currentgeneral}
gives
\begin{equation}
\label{nocurrentindark}
q\mu_{n}(x) E(x) n_{p_{0}}(x) + qD_{n}(x)\frac{dn_{p_{0}}}{dx}=0 
\end{equation}                                                                                                                                                                                                                                                 
Subtracting \ref{nocurrentindark} from \ref{currentgeneral}, we obtain  
\begin{equation}
\label{pcurrent}
j_{p}(x)=q^{2}\frac{D_{n}(x)E(x)}{k_{B}T} n(x) + qD_{n}(x)\frac{dn}{dx} 
\end{equation}
where we have used the Einstein relation $\mu_{n}=qD_{n}/k_{B}T$.
This is valid for equilibrium \cite{rosenberg78} where the diffusion 
current due to the electron and hole carrier concentration gradients 
can be equated with the drift current flowing in the opposite direction.
We assume that in the low field diffusion dominated regime the mobility 
and diffusion coefficient are equal to their
equilibrium values. This is confirmed by low field mobility
measurements by Walukiewicz \cite{walukiewicz92}

Equation \ref{pcurrent} allows us to calculate the photocurrent in
terms of the excess minority carrier concentration $n$ only, because
the intrinsic carrier concentration gives no net current.

The total current contribution of the p layer is the current injected from
the p layer into the i region. In the depletion approximation, $n$ is 
zero at the edge of the space charge. In this case this is at the junction
between the electrostatically neutral section of the p region and the 
p depletion layer. Equation \ref{pcurrent} then becomes
\begin{equation}
\label{pphotocurrent}
J_{p}= \left(qD_{n}\frac{dn}{dx}\right)_{x=x_{w_{p}}}
\end{equation}
Calculating $J_{p}$ therefore reduces to the problem of determining 
the gradient $dn/dx$ at the depletion edge from the current
and continuity equations.

\subsection{Current and Continuity\label{seccurrentcontinuity}}

A minority electron in the p layer can be generated, drift, diffuse or
recombine with a hole. The rate of change of current as a function
of position is given by the difference between generation and 
recombination rates:
\begin{equation}
\label{continuity}
\frac{dJ_{p}}{dx}=-qG(x,\lambda)+q \frac{n}{\tau} 
\end{equation}
In \algaas\, the recombination time $\tau$ is dominated by 
non-radiadive Shockley-Read-Hall recombination 
mentioned earlier in section \ref{seciv}.
If we differentiate equation \ref{pcurrent}
and substitute for $\tau$ using $L_{n}=\sqrt{\tau D_{n}}$,
we obtain the following 
differential equation after dividing throughout by $D_{n}$
\begin{equation}
\label{differentialeqgeneral}
		\frac{d^{2}n}{dx^{2}}
		+(\frac{qE}{k_{B}T}+\frac{D'_{n}}{D_{n}})\frac{dn}{dx}
		+(\frac{qED'_{n}}{D_{n}k_{B}T}+
		\frac{qE'}{k_{B}T}-L_{n}^{-2})n+\frac{G}{D_{n}}=0 
\end{equation}
where $D'_{n}=dD_{n}/dx$.
In order to clarify subsequent discussions, we re-write equation
\ref{differentialeqgeneral} as follows
\begin{equation}
\label{differentialeq}
		 \frac{d^{2}n}{dx^{2}}
		 +b(x)\frac{dn}{dx}
		 +c(x) n+g=0 
\end{equation}
where functions $b$ and $c$ and the generation function $g$ 
are defined by comparison with equation \ref{differentialeqgeneral}. 
Given suitable boundary conditions, equation \ref{differentialeq} 
defines the minority carrier density as a function of position. 

\subsection{Transport Equations for n Layers}

For completeness we briefly indicate the differences between the
transport equations for p layer and n layers.
For an n type semiconductor, the minority carriers are holes
with a positive charge of $q$. The drift current takes the
same form, but the diffusion current changes sign. Equation
\ref{currentgeneral} becomes
\begin{equation}
j_{n}(x)=q\mu_{n}(x) E(x) p_{n}(x) - qD_{p}(x)\frac{dp_{n}}{dx} 
\end{equation}
The differential equation \ref{differentialeqgeneral} however 
does not change except the the hole diffusion length $L_{p}$
substituted for $L_{n}$, and $D_{p}$ and $p_{n}$ are substituted
for $D_{n}$ and $n_{p}$ respectively.

Since the solution to the problem is identical in both cases, we shall
not consider the n layer contribution explicitly in the following
sections in order to avoid unecessary repetition.

\subsection{General Analytical Solution
\label{analyticalsolutionmethods}}

Equation \ref{differentialeq} is an inhomogeneous differential
equation which has a standard solution described for example by Boas
\cite{boas82}. A study
of this method however increases our understanding of the influence
of the parameters governing the behaviour of the solution. 
In the dark, equation \ref{differentialeq} reduces to
\begin{equation}
\label{darkdiffeq}
		 \frac{d^{2}n}{dx^{2}}
		 +b(x)\frac{dn}{dx}
		 +c(x) n=0
\end{equation}
A general solution to this homogeneous second order
linear differential equation exists of the form
\begin{equation}
\label{darksoln}
n_{dark}=a_{1}U_{1}(x)+a_{2}U_{2}(x)
\end{equation}
where $a_{1}$ and $a_{2}$ are constants fixed by two boundary conditions
and $U_{1}$, $U_{2}$ are functions of position.

Following the methods outlined in Boas \cite{boas82}, we postulate 
that the solution to the inhomogeneous case \ref{differentialeq} 
is of the form
\begin{equation}
U_{3}=A(x)U_{1}+B(x)U_{2}
\end{equation}
where $A$ and $B$ are two arbitrary functions which can be found 
with standard solution methods, and take the form
\begin{equation}
\label{aandbfunctions} 
\begin{array}{l}
A(x) = \int^{x}{\frac{g U_{2}}{U_{1}U'_{2}-U'_{1}U_{2}}dx'} \\
B(x) = -\int^{x}{\frac{g U_{1}}{U_{1}U'_{2}-U'_{1}U_{2}}dx'} \\
\end{array}
\end{equation}
These two functions together with the dark solution form the particular
integral solution
\begin{equation}
\label{generalparticularsolution}
U_{3}=
U_{1}\int^{x}\frac{gU_{2}}
{U'_{1}U_{2}-U_{1}U'_{2}}dx'
-U_{2}\int^{x}\frac{gU_{1}}
{U'_{1}U_{2}-U_{1}U'_{2}}dx'
\end{equation}

The general solution is then
\begin{equation}
\label{generalsolution}
n(x)=a_{1}U_{1}+a_{2}U_{2}+U_{3}
\end{equation}
The constants $a_{1}$ and $a_{2}$ are found as described in the next
section. In general, analytical solutions cannot be found and we resort
to numerical methods of solving equation \ref{differentialeq}
described below.

\subsection{Boundary Conditions
\label{secboundaryconditions}}

The first boundary condition is that the surface current flow
is equal to a recombination current characterised by a
recombination velocity $S_{n}$
\begin{equation}
\label{bc1}
J_{p}(0)=qS_{n}n(0)
\end{equation}
The surface recombination may be either due to interface recombination
centres at an abrupt junction between layers with different aluminium
compositions or may be a surface recombination associated with defects
present at the interface between the semiconductor and the air
(Hovel \cite{hovel75}).

Real devices may depart from this behaviour because of increased
recombination at the front contacts \cite{fonash81} or damage to the 
front surface. Both these effects may lead to very high recombination
velocities, and a possible reduction of the p layer thickness.
This is discussed further in section \ref{secsn}.

Substituting equation \ref{pcurrent} for the current and the general
solution \ref{generalsolution} for $n$ into the previous expression
we obtain
\begin{equation}
\label{explicitbc1}
\begin{array}{l}
a_{1}\left( D_{n}U'_{1}(0)+(\frac{qD_{n}E}{k_{B}T}-S_{n})U_{1}(0)\right) \\
+ a_{2}\left(D_{n}U'_{2}(0)+(\frac{qD_{n}E}{k_{B}T}-S_{n})U_{2}(0) \right)\\
=U_{3}(0)\left(S_{n}-\frac{qD_{n}E}{k_{B}T}-D_{n}U'_{3}(0)\right) \\
\end{array}
\end{equation}
The second boundary condition states that the minority carrier 
concentration at the boundary with the  depletion region is zero, 
according to
the depletion approximation
\begin{equation}
\label{bc2}
n_{x_{w_{p}}}=0
\end{equation}
Susbituting for $n$ from equation \ref{generalsolution} we obtain
\begin{equation}
\label{explicitbc2}
a_{1}U_{1}(x_{w_{p}})+a_{2}U_{2}(x_{w_{p}})=-U_{3}(x_{w_{p}})
\end{equation}
The constants $a_{1}$ and $a_{2}$ are then obtained from the two
coupled equations \ref{explicitbc1} and \ref{explicitbc2}.

Two different cases requiring two methods of solving equation
\ref{differentialeq} are presented in the following sections.
The first covers the case of an ungraded solar cell, for
which there is no electric field in the neutral regions and materials
parameters can be assumed to be constant. A hypothetical cell with
constant materials parameters but with an effective electric field
applying only to minority carriers is examined as a preliminary to
the general compositionally varying case.

The second case is that of compositionally
graded QWSCs, for which all the parameters in equation
\ref{differentialeq} are position dependent. The same
boundary conditions are applicable in both cases.

\section{Constant Transport Parameters\label{sechomogeneoussolution}}

In this section we consider the p layer $QE$ of ungraded cells 
with and without effective fields. In this case, the parameters 
$L_{n}$ and $D_{n}$ are constant. The original work is described
in the recent review by Nelson \cite{nelson96}. 

\subsection{Zero Field\label{sectheoryzeroe}}

For a layer with a zero field, a trial solution $e^{kx}$
yields the functions
\begin{equation}
\label{homou1}
U_{1}=e^{x/L_{n}}
\end{equation}
\begin{equation}
\label{homou2}
U_{1}=e^{-x/L_{n}}
\end{equation}
Constants $a_{1}$ and $a_{2}$ are obtained from equations
\ref{explicitbc1} and \ref{explicitbc2}, thereby defining
the minority carrier profile. The photocurrent from 
the p layer is then found from equation \ref{pphotocurrent}.

Appendix \ref{secapp1} gives more detail, and shows that 
the photocurrent for these ungraded structures is only 
weakly dependent on the diffusion coefficient and the recombination
velocity if
\begin{equation}
\label{snlnbalance}
\frac{1}{L_{n}} \gg \frac{S_{n}}{D_{n}}
\end{equation}
The appendix further shows that the $QE$ in the limit of 
$L_{n}<x_{w_{p}}$ near the band edge is well approximanted
by
\begin{equation}
\label{lowlnqelimit}
QE \simeq L_{n}\alpha e^{-\alpha x_{w_{p}}}
\end{equation}
This result agrees with a brief treatment given by Ludowise
\cite{ludowise84}, and reviewed in more detail by Hovel \cite{hovel75}.
The \algaas\ structures we will study generally obey this limit
for acceptable p layer thicknesses. The photocurrent near the
band edge is dominated by p and i contributions. Since, as we have
seen, the i region photocurrent is well understood, $L_{n}$ can
be estimated by modelling the $QE$ in this wavelength range.

Appendix \ref{secapp1} further shows that the $QE$ at short wavelengths
remains sensitive to the recombination velocity, and tends to a
limit determined primarily by the surface recombination velocity and
the absorption coefficient. For diffusion lengths which are of the order
or greater than the p layer thickness, a simple limit is found,
depending solely on absorption coefficient and recombination velocity.
Since the absorption coefficient varies slowly at such wavelengths,
the overall $QE$ becomes weakly dependent on wavelength. This behaviour
is primarily observed in solar cells made from materials with high
absorption coefficients at short wavelengths, such as \gaas.

\subsection{Non Zero Effective Fields\label{sectheorynonzeroe}}

A brief examination of the hypothetical case of an ungraded
cell with electric fields present in the neutral regions is
useful here. Detail is again provided in appendix \ref{secapp1}.
We now define the effective field ${\cal E}_{f}$ due to the 
compositional grade, in units of per metre, as follows:
\begin{equation}
\label{egfielddefinition}
{\cal E}_{f}=\frac{qE}{k_{B}T}
\end{equation}

A trial solution for cells with ${\cal E}_{f}\neq 0$ yields the solutions
\begin{equation}
U_{1}=e^{k_{1}x}
\end{equation}
\begin{equation}
U_{2}=e^{k_{2}x}
\end{equation}
with
\begin{equation}
\label{k1withfield}
k_{1}=  +\sqrt{{\cal E}_{f}^{2}+L^{-2}}-{\cal E}_{f} 
\end{equation}
\begin{equation}
\label{k2withfield}
k_{2}=  -\sqrt{{\cal E}_{f}^{2}+L^{-2}}-{\cal E}_{f}
\end{equation}
Since ${\cal E}_{f} <0$, parameter $k_{1}$ tends to 
$2\mid\! {\cal E}_{f}\! \mid$ 
for large fields, whereas $k_{2}$ tends to zero. The exponentially growing
solution is therefore favoured, and the resulting minority carrier
distribution shifted in a positive direction towards the depletion
layer, increasing minority carrier collection.

We now examine the effect of the field on the surface boundary
condition. Appendix \ref{secapp1} shows that the condition
of \ref{snlnbalance} is replaced by
\begin{equation}
\label{fieldsnbalance}
\mid \!{\cal E}_{f}\!\mid \gg \frac{S_{n}}{D_{n}}
\end{equation}
In this limit, the $QE$ is insensitive to the surface boundary
condition and the surface recombination velocity in particular.
This limit is readily achieved even for high surface recombination
velocities. We shall see in section \ref{secparameters} that it
may fail however if the recombination velocity is unusually high.
This may occur if recombination at the contacts is large, or 
if the surface quality is poor.

We conclude however that the presence of a field renders the
$QE$ relatively insensitive to the surface recombination.

We shall see in subsequent sections that the real picture is somewhat
less clear in the case of a full numerical solution to 
equation \ref{differentialeqgeneral}. We shall see however in 
the modelling section that cells with electric fields are much less
sensitive to the magnitude of the surface recombination velocity,
as carriers in a well designed cell tend to diffuse in the direction
of the greater effective diffusion length.

\section{Graded QWSCs and Numerical Solutions
\label{secnumericalsolutions}}

Analytical solutions to equation \ref{generalsolution} generally cannot
be found if the parameters $b$ and $c$ are position dependent.
We now develop a numerical scheme to solve equation 
\ref{differentialeq} in this case.

We express the first and second differentials of $n$ 
using standard centred differencing described by Potter \cite{potter79}
as follows:
\begin{equation}
\label{numericaldiffs}
\begin{array}{l l}
n'_{k}= &\frac{1}{2\Delta}(n_{k+1}-n_{k-1}) \\
n''_{k}= &\frac{1}{\Delta^{2}}(n_{k+1}-2n_{k}+n_{k-1}) \\
\end{array}
\end{equation}
where $k$ is the position index and $\Delta$ the grid spacing of
a one dimensional grid of length $N$.
Substituting these expressions into equation \ref{differentialeq}
we obtain
\begin{equation}
\label{differenceeq}
n_{k-1}(\frac{1}{\Delta^{2}}-\frac{b_{k}}{2\Delta})
+n_{k}(c_{k}-\frac{2}{\Delta^{2}})
+n_{k+1}(\frac{1}{\Delta^{2}}+\frac{b_{k}}{2\Delta})
=-g_{k}
\end{equation}
The surface boundary condition equation \ref{bc1} can be written as
\begin{equation}
\label{diffbc1}
n_{1}( \frac{qE}{k_{B}T}-\frac{S_{n}}{D_{n}}-\frac{1}{\Delta})
+(\frac{n_{2}}{\Delta})=0
\end{equation}
The second boundary condition equation \ref{bc2} becomes
\begin{equation}
\label{diffbc2}
n_{N}=0
\end{equation}
Equations \ref{differenceeq}, \ref{diffbc1} and \ref{diffbc2} can be
expressed as the following matrix equation 
\begin{equation}
\label{numericalmatrix}
MN=G
\end{equation}
This equation is then solved by standard Gaussian elimination with
backsubstitution. A good review of this technique is given in
Press \cite{press86}

Although more complex methods were investigated, this simple
method was chosen because trials with a number of different 
numerical schemes shows that this method is preferable to methods 
which are nominally quicker and avoid explicitly calculating the 
inverse matrix. 
This is mainly because the matrix $M$ need only be inverted once,
but is typically used about six hundred times to cover the wavelength
range we are considering. For the matrices we consider, which may have
a dimension of a few hundreds, we observe that the calculations involved
in solving equation \ref{numericalmatrix} are the most time
consuming.

\subsection{Surface Boundary Condition in Compositionally Graded Cells}

The surface boundary condition is expressed by the topmost
row of matrix $M$, and is identical to the boundary condition
in the analytical case. 
We conclude that compositionally graded designs remain insensitive
to the surface recombination velocity.

More detailed conclusions regarding the general behaviour of the $QE$ 
of such cells are best illustrated by numerical calculation, and
are investigated further in chapter \ref{secoptimisation}.

\subsection{Numerical Accuracy} 

To determine the numerical accuracy of the scheme we express 
the minority carrier concentration as a Fourier series
\begin{equation}
\label{fouriern}
n=\sum_{j=0}^{\infty}a_{j}e^{ik_{j}x}
\end{equation}
where $k_{j}$ is the wavevector of the $j^{\rm th}$ mode.
The exact differential of an arbitrary mode $e^{ikx}$ is
\begin{equation}
n'=aikn
\end{equation}
When we use our difference expression of \ref{numericaldiffs}
for the same differential we obtain
\begin{equation}
\begin{array}{c c c}
n'& = & \frac{a}{2\Delta}(e^{ik(x+\Delta)}-e^{ik(x-\Delta)}) \\
  & = & \frac{an}{2\Delta}(2i \sin(k\Delta))
\end{array}
\end{equation}
If we substitute the Taylor series expansion for the sine term
this gives
\begin{equation}
\begin{array}{c c c}
n' & = & \frac{ain}{2\Delta}(2k\Delta
	 -\frac{(k\Delta)^{3}}{3}+O(k\Delta)^{5}) \\
   & = & aikn(1-\frac{(k\Delta)^{2}}{6}+O(k\Delta)^{4}) \\
\end{array}
\end{equation}
where $O(k\Delta)^{4}$ refers to fourth order terms. 
A similar treatment for the second differential gives the
analytical result
\begin{equation}
n''=-ak^{2}n
\end{equation}
whereas the numerical expression is
\begin{equation}
n''=-ak^{2}u(1-\frac{(k\Delta)^{2}}{24}+O(k\Delta)^{5})
\end{equation}
The error is given by the difference between the analytical expressions
and the numerical result in both cases and depends on $\Delta^{2}$. 
This scheme is therefore second order accurate.

We further note that the error also depends
on the square of the wavevector $k$ and will increase rapidly as $k$
becomes comparable with $\Delta$. A large value of $k$ is equivalent
to a minority carrier density which is varying rapidly in space.
We must therefore be careful of trusting the results
if the composition changes very rapidly over a small distance,
and this design is not suitable for neutral regions with sharp
discontinuities in any of the material parameters.
This is unlikely to cause problems since sharp interfaces are
generally avoided in the neutral regions because of the
increased interface recombination they cause. 

\subsection{Comparisons of Analytical and Numerical Solutions
\label{secnumericalchecks}}

In this section we brifely consider the stability of the
numerical solution. We then continue to compare numerical
and analytical calculation methods in order to test the numerical
method.

\subsubsection{Stability of Boundary Value Problems}
  
Unlike initial value problems, boundary value calculations are
not prone to numerical instability. This is because the solution
is constrained at the two end points.

However, rounding errors in the manipulation of large matrices can
grow and swamp the real solution. However, this is rarely a
problem in sparse matrices such as ours where the number
of operations is relatively low. Furthermore, Potter 
\cite{potter79} states that the use
of double precision can circumvent this problem for
medium size matrices where $N$ is of the order of
some hundreds. 

Furthermore, Gaussian elimination with back-substitution is
more stable than many more complex methods \cite{press86}.

\subsubsection{Zero Field}

In this section the results of the numerical model are compared 
to those of the analytical model in two cases where this is possible.

Figure \ref{figcheckhomoqwsc} shows a comparison of the internal p layer
$QE$ spectra predicted by analytical and numerical methods for an
typical highly doped p layer which is $0.15\mu m$ thick and has 
a composition of 30\% aluminium. The calculation uses a grid 
spacing of 400 points for the p layer.

We note that the numerical method slightly underestimates
the $QE$ at short wavelengths. This is a result of increased numerical
inaccuracy due to increased position dependence of both generation
rate and minority carrier concentration.

\smalleps{htbp}
{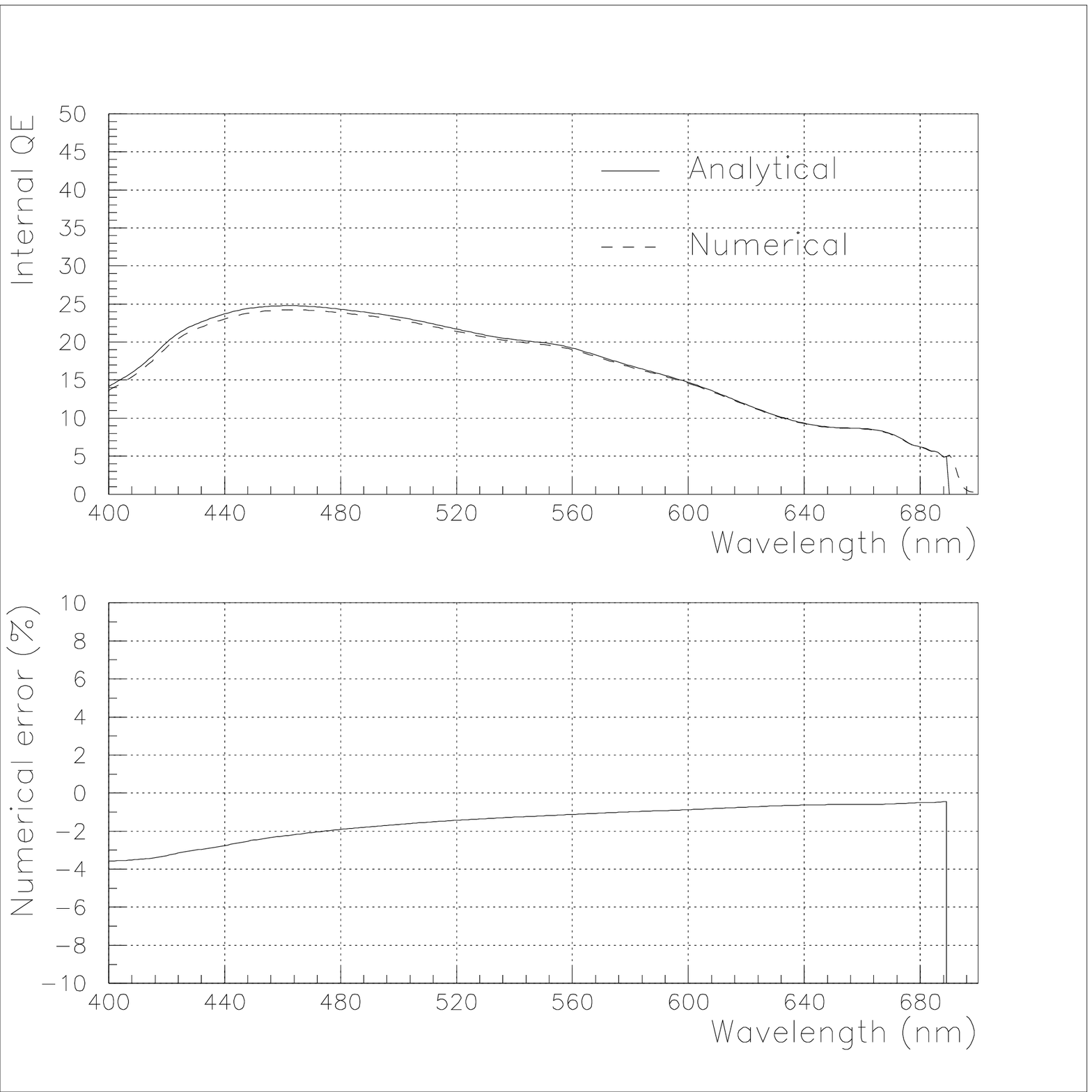}
{Numerical and analytical internal $QE$ plots for an ungraded
30\% aluminium QWSC \label{figcheckhomoqwsc}}

\smalleps{htbp}
{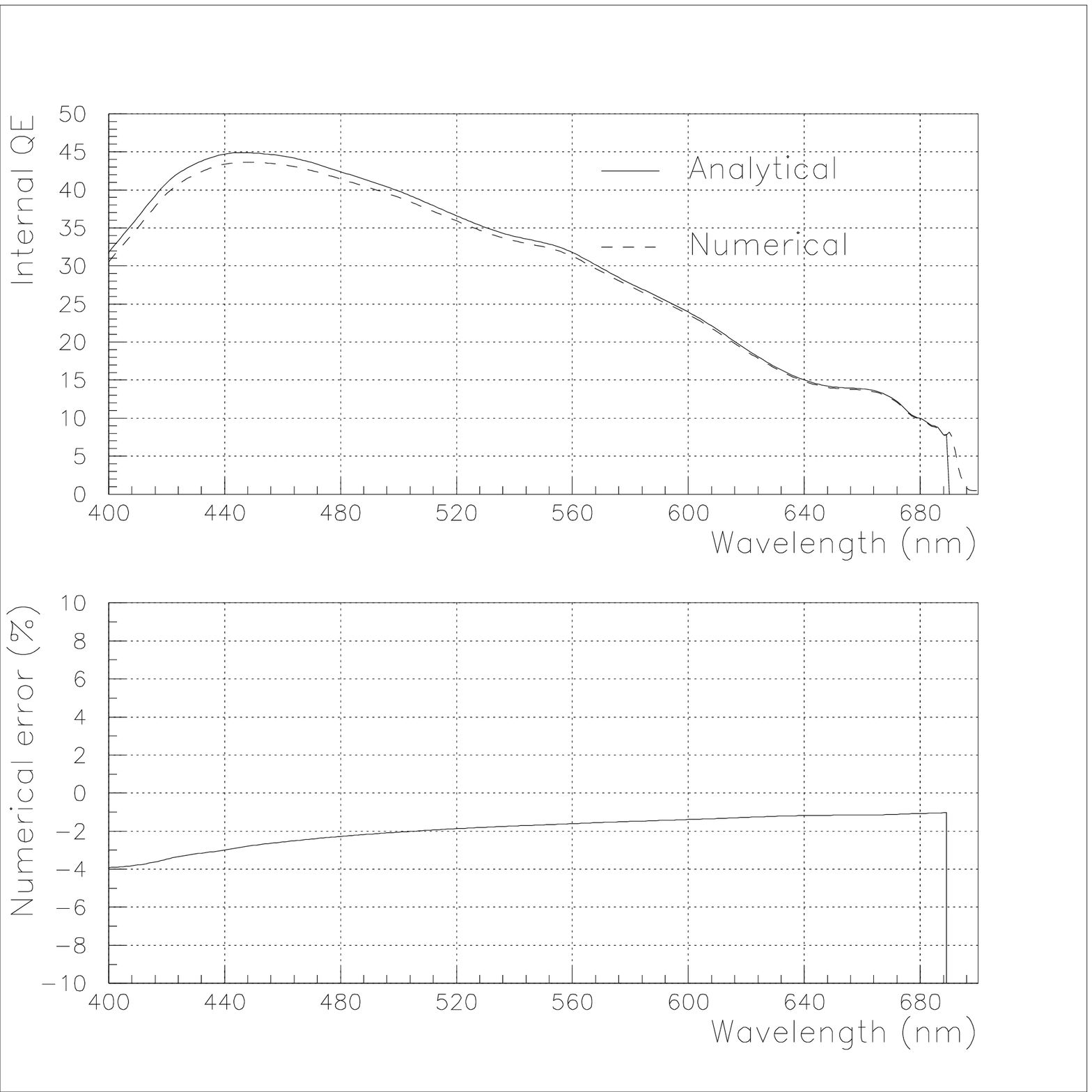}
{Numerical and analytical $QE$ plots for an ungraded
30\% aluminium QWSC with electric field \label{figcheckfieldqwsc}}

\subsubsection{Cells With Fields}

No analytical method exists for a graded p layer,
making precise evaluation of the performance of the numerical
model difficult. As a first approximation, we compare results
for the hypothetical case of ungraded cells with an effective 
field in the neutral regions.

Figure \ref{figcheckfieldqwsc} compares the analytical and numerical
results for a QWSC identical to that of the previous subsection,
but with an effective field of $-10^{6}Vm^{-1}$ in the p layer.
The results are again very similar, with the numerical model again
underestimating the $QE$ slightly at short wavelengths.

The error in the numerical case is marginally greater than in the
ungraded case, but still amounts only about one percent in
the predicted $QE$ at short wavelengths. It is again weakly 
sensitive to the grid spacing.

\subsubsection{Conclusions}

For the general case of an graded structure, we
have no analytical model model for comparison. Analysis
of the closest approximation however shows an error
which is about a a fifth of the experimental uncertainty 
at short wavelengths. The model furthermore is relatively 
insensitive to the number of grid-points used.

We conclude that the principal source of modelling error
lies in the physical assumptions underlying the model and
in the values of physical parameters, rather than in the
numerical method. This is investigated further by comparison
with experimental data in chapter \ref{secqeresults}.

\section{Conclusion}

In this chapter we have described a method of calculating the
light intensity for a solar cell with varying optical parameters
and with or without multiple internal reflections for a normally
incident spectrum.

Having calculated the generation rate from the light intensity,
we have looked at methods of calculating the resulting photocurrent.
These methods fall into two broad categories.
The first deals with ungraded cells where we find that
an analytical solution to the minority carrier concentration exists.
The QE in this case is only weakly dependent on the
diffusion coefficient for low recombination velocity.
We have found that is it possible to estimate the diffusion
length by modelling the QE in cells subject to this limit.

Brief examination of the abstract case of an ungraded solar cell
with effective fields in the neutral layers shows that such cells
are significantly less sensitive to surface recombination.

This useful property is also true for compositionally varying cells.
We have described a numerical scheme for calculating the $QE$ of 
such structures.
The second case, which deals with graded cells, requires a
numerical solution method and a good understanding of the dependence 
of transport parameters on aluminium content.

%% file: chap4.tex

\chapter{\label{secparameters}Model Parameters and Program Design}

In this chapter we describe the way a QWSC is represented by the model.
We then identify the main adjustable parameters used to fit experimental
data. We give a review of the current state of knowledge of these and
non-adjustable parameters used by the model.

\section{Program design}

The FORTRAN program SOLAL closely follows the progression
outlined in section \ref{sectheory}. The program design is
modular and menu driven, and interfaces with the PAW
graphics program. It calculates both analytical and
numerical $QE$ spectra.

The main design parameters are summarised in table 
\ref{tblgrowthparameters}. The anti-reflection (AR) coat is
represented by a wavelength dependent reflectivity, which
may be arbitrarily specified by the user. Three standard
reflectivities are provided. These are a free \gaas\ surface,
and two standard reflectivities based on data from Eindhoven
and Sheffield. The latter two will be discussed in chapter
\ref{secmivar}.

\begin{table}[t]
\begin{center}
\begin{tabular}{| c | c |} \hline
 Surface & GaAs reflectivity \\
 & {\em or} SiN reflectivity \\
 & {\em or} MgF:ZnS reflectivity \\
 & {\em or} Measured reflectivity data \\ \hline

Non-contributing & Layer thicknesses \\ \cline{2-2}
surface layers & Linear compositional profiles of the layers \\ \cline{2-2}
& Number of layers \\ \hline

p and n layers & Layer thicknesses \\ \cline{2-2}
 & Arbitrary composition profile \\ \cline{2-2}
 & Constant doping \\ \hline

Bulk i region & Number of strips \\ \cline{2-2}
 & Thickness of strips \\ \cline{2-2}
 & Linear compositional profiles of strips \\ \hline

MQW i region & Total i region width \\ \cline{2-2}
 & Well width \\ \cline{2-2}
 & Barrier composition \\ \cline{2-2}
 & Number of wells \\ \hline 
\end{tabular}
\figcap{List of the main QWSC growth and design 
parameters.\label{tblgrowthparameters}} 
\end{center}
\end{table}

\subsubsection{Sample structure}

The solar cells are represented 
as a series of \algaas\ layers which may have linearly varying 
aluminium composition. These are split into surface layers 
which do not contribute to the $QE$, and contributing p, 
n, and i layers. Most of our designs have a single surface 
layer, which is the \algaas\ window.

The p layer can be subdivided into several strips, each with
a linearly varying composition. The only recombination velocity
included in the calculation is that between the AR coat or GaAs
surface and the p layer. The doping however is assumed
constant. Depletion widths are included in the calculation,
and the effective aluminium fraction at the edge of the depletion
width is readjusted in graded structures to match that at the
edge of the calculated depletion width. The treatment for
the n layer is identical, with the difference that the
surface or interface recombination velocity is replaced
by a back surface recombination velocity.

The i layer may be of two types. The first is a conventional i
layer, which may be composed of a number of linearly graded strips.
A constant background doping level is assumed throughout.
The second type of i layer is one which contains quantum wells.
This layer may not be compositionally graded except for the
sharp interfaces between quantum wells and barriers. Details
of the quantum well modelling parameters have been supplied 
by Paxman \cite{paxman93}. The parameters in this case are the
total i region width, the barrier composition, the number of 
wells and the well thickness.

\begin{table}[t]
\begin{center}
\begin{tabular}{| c | c |} \hline
 Ungraded p  & $\left. \begin{array}{l}
 S_{n} \\ D_{n} \\ \end{array} \right\}                 
 \begin{array}{l} \mbox{Recombination parameter} \\
 {\cal S}=S_{n}/D_{n} \\ \end{array}$
 \\ \cline{2-2} 
 &  Diffusion length $L_{n}$ \\ \cline{2-2} 
\hline
 Graded p  & $\left. \begin{array}{l}
 S_{n} \\ D_{n}(x=0) \\ \end{array} \right\}                 
 \begin{array}{l} \mbox{Recombination parameter} \\
 {\cal S}=S_{n}/D_{n}(x=0) \\ \end{array}$
 \\ \cline{2-2} 
 &  Diffusion length $L_{n}(x)$ \\ \cline{2-2} 
 &  Diffusivity $D_{n}(x)$ \\
 \hline
\end{tabular}
\figcap{Transport parameters for the p layer required by the model
\label{tbltransportparameters}}
\end{center}
\end{table}

\subsubsection{Minority carrier transport in neutral layers}

The minority carrier transport parameters required to calculate the
$QE$ of the neutral p and n sections vary according to the type of
structure being considered. They are summarised for graded and ungraded
structures in table \ref{tbltransportparameters}.

For ungraded structures, the ratio of surface recombination velocity
and diffusivity are specified for each layer, together with the minority 
carrier diffusion length.

Graded structures require a parametrisation of diffusion length and
diffusivity as a function of position $x$.

These parameters are reviewed in the following sections. The
parametrisations used are reviewed in chapter 
\ref{secoptimisation}.

\section{Design and Growth Parameters}

In this section we review the parameters 
summarised in tables \ref{tblgrowthparameters}
and \ref{tbltransportparameters}. These parameters fall into two 
categories, the first of which are functions of cell design and 
growth. Examples are the dimensions of the cell layers, their 
composition and doping levels. Knowledge of these parameters 
is usually good because the QWSC fabrication process 
uses highly accurate epitaxial methods.

The second category contains parameters which are a function of
growth method, growth run and processing. Examples include
surface recombination, transport parameters and optical
properties. These parameters are largely taken from existing
data in the literature.

Tables \ref{tblgrowthparameters} and \ref{tbltransportparameters}
show the most important modelling parameters for
a graded QWSC with an anti reflection (AR) coat. 
Dimensions such p,i and n layer
thicknesses, and doping levels are generally well known. 

The quantum well width is frequently found to depart from
the design specification by a few percent. However, this does
not affect the overall electrical characteristics of the QWSC
and simply restricts or extends the QW absorption edge in the
infrared. Such deviations from the specifications are discussed
further with respect to modelling in chapter \ref{secqeresults}.

The background doping level $N_{BG}$ in the intrinsic region, however, 
is frequently subject to large uncertainty. This is largely because
residual impurity levels in the solid or gas sources are poorly
known. In addition, the net background doping level is a function
of reciprocal compensation by donor and acceptor impurities. The
final background doping level may also be affected by unpredictable
passivation effects during processing. 

As detailed in chapter \ref{secsolarcells}, a large value of 
$N_{BG}$ degrades the
performance of the QWSC because the built-in electric field drops
to zero in the intrinsic region.

The QE measurements however are taken when the photocurrent is
maximised. In this case, the only effect of $N_{BG}$ on the
photocurrent is to influence the width of the depletion
layers. A more detailed discussion is given with experimental
data in chapter \ref{secmivar}.

\begin{table}
\begin{center}
\begin{tabular}{| c | c |} \hline
AR coat & $\sim$70A SiN, measured reflectivity \\ \hline
Window layer & 300A $\rm Al_{0.67}Ga_{0.33}As$ \\ \hline
p layer & 0.15$\rm \mu m$ $\rm Al_{0.4}Ga_{0.6}As$ \\
 & {\em graded to} $\rm Al_{0.2}Ga_{0.8}As$ \\
 & Carbon doping $\rm 1.34\times 10^{18}cm^{-3}$ \\ \hline
i region & 30 GaAs wells of width 87A \\ 
 & 60A barriers, composition $\rm Al_{0.2}Ga_{0.8}As$ \\
 & Well absorption $\sim$1\% per level per well \\ \hline
n region & 0.6$\rm \mu m$ $\rm Al_{0.2}Ga_{0.8}As$ \\
 & Si doping $\rm 6\times 10^{17}cm^{-3}$ \\ \hline
Metallic back mirror & Reflectivity $\sim$95\% \\
 & across entire wavelength range \\ \hline
\end{tabular}
\figcap{Typical design parameter set for an anti-reflection coated
20\% Al 30 well QWSC with a graded p layer and a back mirror
\label{tblgeneralqwscparameters}}
\end{center}
\end{table}

\section{Optical Parameters}

\subsection{Incident Flux}

The $QE$ calculation is independent of the incident flux.
Estimated values of the short circuit current given in
subsequent chapters are given for the AM1.5 solar flux
which was described in section \ref{secefflimit}.

\subsection{Absorption Coefficient\label{secabsorption}}

Connolly \cite{connolly91} has shown that the QE is very
sensitive to the absorption coefficient $\alpha$, and has
proposed an absorption model based on accurate \gaas\
tabulations published by Aspnes \cite{aspnes89}. The absorption
model uses a non linear wavelength dependent energy shift which
depends on the aluminium composition $x$ to fit the \gaas\ 
absorption to data for \algaas\ published in \cite{aspnes86}.
This approach only gives the absorption of \algaas\ for direct
aluminium compositions.

Subsequent work in \cite{paxman93} 
fits an exponential tail below the direct absorption 
threshold for indirect aluminium fractions to account for the weaker 
indirect absorption. Since the indirect and direct absorption
routines were initially slightly ill matched, this work introduces 
a further continuous transition between the two parametrisations.

Connolly \cite{connolly91} has also indicated
that the absorption coefficient near the band edge is subject to
significant uncertainty for aluminium fractions
in the region of 30\%. This is seen both in the experimental data 
tabulations and in a range of models which are reviewed in the
reference. 

We conclude that the absorption coefficient remains a source of 
systematic error near the band edge. Since, however, the absorption
model uses the most recent tabulated absorption data available,
it is difficult to reduce this source of uncertainty.

\subsection{Reflectivity}

Given the very small size of typical QWSC mesa-photodiode test devices,
direct reflectivity measurements are difficult. In order to make such
measurements possible, pieces of wafer were exposed to all processing
steps at the same time as the material which was used to fabricate
devices, with the exception of mesa-etching and metallisation.
This is discussed in more detail in section \ref{secmivar}

These pieces of AR coated wafer are not available for devices processed
before 1995. For such samples, an average over previous processing
runs is used in the modelling.

The \gaas\ reflectivity was obtained from published data \cite{aspnes89}.
Measurements on \gaas\ wafers has produced close agreement.

Further reflectivity for the double-layer MgF:SiN  coating deposited
and characterised in Eindhoven Technical University was available
for devices processed in Eindhoven.

The reflectivity of the metallic back-mirror is given by Hass 
\cite{hass55}.
It is found to be very nearly constant across the wavelength range
of interest for gold and aluminium. The front surface reflectivity
was modelled as a fraction of the measured surface reflectivity. The
corresponding fraction is the prime parameter used in modelling
the amplitude of Fabry-Perot peaks in mirror-backed devices, and varies
from about 70\% to 100\% of the measured reflectivity. This factor is
described in more detail in chapter \ref{secqeresults}.

The front surface phase change $\psi_{f}$ for internal reflection in a 
dielectric material at an interface with air is given by Macleod 
\cite{macleod69} as zero. The phase change is $\pi$ if the interface is 
with a material with a higher refractive index. Work by Whitehead 
\cite{whitehead90} on multiple quantum well optical modulators 
indicates that the phase change  $\psi_{b}$ at the metallic 
back mirror is variable and device dependent. We shall see in 
chapter \ref{secqeresults} that different devices
processed from the same wafer may have different modelled values of
$\psi_{b}$. More recent discussions with the author \cite{whitehead94} 
have suggested that this quantity can be used as a fitting parameter
to adjust the frequency of the Fabry-Perot oscillations seen in
the $QE$ of mirror backed devices.

\subsection{Refractive index}

Reliable ellipsometric measurements of the refractive index of \algaas\ as 
a function of energy have been tabulated by Aspnes \cite{aspnes89} for 
eleven aluminium compositions for aluminium mole fractions ranging from 
$0$ to $1$. For intermediate energies and compositions, a linear 
interpolation is used.

\section{Transport parameters}

We have seen that good agreement exists on the optical materials
parameters of \algaas, and that consistent values are available in
the literature

Most of the minority transport parameters however are sensitive to device
history. Growth conditions between different runs on the same
machine may differ significantly due to relatively small
variations in the composition of the gas or solid sources used in the
epitaxial growth. Furthermore, post-growth processing, involving device
exposure to etches, plasmas and heat treatments, may also affect transport.
Examples include accidental passivation, diffusion of impurities
and a modification of the recombination velocity.
Finally, a number of these parameters are dependent on the device
structure.
The following section explores current knowledge of these parameters
in the literature.

\subsection{Band gap Parametrisations and the Quasi-Electric Field}

We saw in chapter \ref{secsolarcells} that
an externally applied bias in a \pin\  structure is entirely
dropped across the intrinsic region for moderate applied
bias where the majority carrier profile is not significantly
affected by the external voltage. That is, no externally applied
bias is dropped across the neutral sections of the n and p
layers.

The effective electric field acting on minority carriers in n and
p type graded structures is examined by Sutherland \cite{sutherland76},
Hutchby \cite{hutchby76} and is reviewed by Hamaker \cite{hamaker85}.
It is generally referred to as the quasi field.
This electric field is sometimes taken to be proportional to the 
gradient of the electron affinity as in Sutherland \cite{sutherland76}. 
Other workers however express it as the gradient of the bandgap.
The review by Hamaker \cite{hamaker85} explains the origin of
the field and expresses it in terms of doping, bandgap and Fermi
level gradients. In samples designed in the present work, 
the nominal doping is constant. We will not therefore consider
fields due to doping gradients.

Sassi \cite{sassi83} has estimated the field due to the
Fermi level gradient as less than 2\% of the field due to
the bandgap gradient. Sassi therefore neglects the gradient
of the Fermi level. Taking the example of the minority electron 
quasi field $E_{e}$ in p type material, this gives
\begin{equation}
E_{e}=-\frac{dE_{g}}{dx}
\end{equation}
where $E_{g}$ is in electronvolts and $x$ the position. This assumption
of low Fermi level gradient is especially true in heavily doped material
such as we use in our solar cell designs. It is the most commonly found
parametrisation for the quasi-field acting on minority carriers, and is
used for example in work by Konagai \cite{konagai76},
Hutchby \cite{hutchby76} and Sassi \cite{sassi83}. We shall therefore
use this method in this study.

The corresponding hole quasi field in electronvolts for
n type material is
\begin{equation}
E_{h}=+\frac{dE_{g}}{dx}
\end{equation}
We shall not consider graded n layers in the \algaas\ system,
and will revert to the nomenclature of chapter \ref{sectheory},
which refers to $E_{e}$ as $E$.

An added complication is the transition from direct to indirect
material in \algaas. The review article by Pollack \cite{pollack92} 
put this transition at a composition in the region of $45\%$ aluminium.

Two expressions are given for $E_{g}$
in the literature. We write these in terms of a composition dependent
step function $\Theta$ for convenience, which has value 1 below
the direct-indirect transition, and 0 above.
The bandgap is derived from expressions by Casey and Panish 
\cite{casey78} for the direct regime, and Hutchby \cite{hutchby76}
for the indirect material.
\begin{equation}
\label{bandgap}
E_{g}=\Theta(1.424+1.247X)+(1-\Theta)(1.92+0.17X+0.07X^{2})
\end{equation}

\subsection{Minority carrier lifetime} 

Although the model outlined in section \ref{secphotocurrent}
does not explicitly require knowledge of the minority carrier
lifetime $\tau$, this parameter is mentioned here for 
completeness. It is related to the minority carrier
diffusivity and the diffusion length by
\begin{equation}
\tau_{n}=\frac{L_{n}^{2}}{D_{n}}
\end{equation}
for electrons. A similar expression holds for holes.
Minority carrier lifetimes in \gaasalgaas\ have been measured
for a range of aluminium compositions.
The most common methods are time resolved photo-luminescence (TRPL)
(Ahrenkiel \cite{ahrenkiel91}) and zero-field time-of-flight (ZFTOF)
(Zarem \cite{zarem89}) techniques. This data is summarised and
discussed in the review article by Ahrenkiel \cite{ahrenkiel92}.
Unfortunately, little data is available for compositions above
about 40\% aluminium. The large scatter observed in experimental
data indicates that the lifetime is dominated by defect-related
Shockley-Hall-Read recombination processes. It is therefore sensitive
to material quality and according to Ahrenkiel may vary by orders
of magnitude compared with the published values in the review
article.

\subsection{Mobility and Diffusion Coefficient}

We have seen that the model does now explicitly depend on the
magnitude of the diffusivity. The functional form of this
parameter however is important in determining the $QE$ of graded
structures.

Values of the diffusion coefficients $D_{p}$ and $D_{n}$ for minority 
holes and electrons are not available in the literature, and, like the 
diffusion length values, must be obtained indirectly.

The review article by Ahrenkiel \cite{ahrenkiel92} 
states that the diffusivity is most conveniently
found by extrapolation from the electron and hole
majority Hall mobilities $\mu_{e}$ and $\mu_{h}$ 
in the low field, low injection limit where 
the Einstein relationship is valid. The two quantities are
then approximately related as follows
\begin{equation}
\begin{array}{l l}
D_{p}\simeq\frac{k_{B}T\mu_{h}}{q} & $(holes)$ \\
D_{n}\simeq\frac{k_{B}T\mu_{e}}{q} & $(electrons)$ \\
\end{array}
\end{equation}
Low field hole and electron mobilities are supplied in the review 
by Walukiewicz \cite{walukiewicz92}. A wide spread is observed,
indicating that the mobility is material dependent.
Both electron and hole mobilities decrease sharply from GaAs to 
$\rm Al_{0.5}Ga_{0.5}As$. For aluminium compositions greater than 
about 50\%, both mobilities appear essentially constant.

This behaviour is essentially reproduced by a parametrisation
of the diffusivity as a function of composition and doping
presented by Hamaker \cite{hamaker85}.
The Hamaker paper does not give details, but states that the
parametrisation is based on experimental data. We use this
as a guide to diffusivity values. More detail concerning this
point is provided in chapter \ref{secoptimisation}.

\subsection{Diffusion length}

The diffusion lengths $L_{n}$ and $L_{p}$ in \gaasalgaas\ are 
sensitive to growth method, device structure and aluminium 
composition. As we shall see in chapter \ref{secoptimisation},
they are significant QE fitting parameters. We review here
some of the expected values in  \algaas. Specific values of
this parameter will be referred to with respect to modelling in
chapter \ref{secqeresults}.

Ahrenkiel \cite{ahrenkiel92} recommends extracting $L$ for both
minority holes and electrons from the 
minority carrier lifetime and the diffusion coefficient discussed 
above through the relation
\begin{equation}
\label{diffusionlengthequation}
L_{n}=\sqrt{D_{n}\tau_{n}}
\end{equation}
for minority electrons. A similar expression holds for minority holes.
These values can only provide a rough starting point for
modelling, since the diffusion length is strongly device dependent.

Ahrenkiel reports that the diffusion length dependence on aluminium 
fraction is similar to that of the mobility. It generally decreases with 
increasing aluminium fraction over the direct region, but is found to 
increase near the direct/indirect transition region at about 40\% aluminium.
It remains approximately constant for indirect material, at a value
which is however only a fraction of the \gaas\ diffusion length.

Hamaker \cite{hamaker85} gives a parametrisation similar to that
which was discussed for the diffusivity in the previous section.
We shall compare values predicted by Hamaker with values estimated
from the modelling of $QE$ spectra in chapter \ref{secqeresults}.

Gr\"unbaum \cite{grunbaum95} has measured minority hole diffusion
lengths in some of our \algaas\ material using Electron Beam Induced
Current techniques. This technique however was only applied to one
sample relevant to this study. It is unfortunately currently incapable
of the resolution required to measure the more significant minority
electron diffusion length $L_{n}$. Moreover, the predicted $L_{p}$
values are rather large.

\subsection{Recombination velocity\label{secsn}}

Chapter \ref{sectheory} showed that surface or interface recombination
velocities in our material cannot be separated from the diffusivity,
which was reviewed earlier in this chapter. Modelled values of
the surface recombination parameter $\cal S$ however indicate whether or
not our values are in qualitative agreement with those reported
in the literature.

An unfortunate result of any heterojunction in a semiconductor
structure is the presence of interface recombination centres. These
may result for a variety of reasons. For heterojunctions these 
include lattice mismatch and abrupt changes in growth conditions.

Two more cases occur. The first is the junction between a semiconductor
and air. Recombination velocities in this case are particularly high.
Measurements presented by Fonash \cite{fonash81} quote values up to
three orders of magnitude greater than for heterojunctions.
The third case is that of a semiconductor and an anti-reflection
coating, such as deposited SiN. 

Measurements are carried out by time resolved photoluminescence 
(TRPL) studies of double heterojunction devices. The active region 
is grown with a lower bandgap, and its thickness varied. The variation of
the PL signal decay time as a function of active layer thickness can
then give an indirect measurement of the interface recombination
velocity. This method has been used by 'tHooft and Van Opdorp
\cite{thooft86}. Results are included in the review article
by Timmons \cite{timmons92}.

Devices which have suffered surface damage as a result of 
handling or processing may however deviate from this picture.
Furthermore, Fonash \cite{fonash81} has indicated that
recombination at the contacts increases the
effective surface recombination. This effect
can be reduced by the use of appropriate contacts. Selective
ohmic contacts, for example, present a potential barrier 
to minority carriers, and inhibit minority carrier recombination at
the contacts. They are, however, ohmic to majority carriers 
\cite{fonash81}.

Theoretical studies however such as those reviewed in chapter
\ref{sectheory} usually use values which are one to two
orders of magnitude greater than these. The largest reported value for
interface recombination in \algaas\ is quoted by Timmons \cite{timmons92}
as $3\times 10^{4}\rm cm/s$. Surface recombination velocities of
$4\times 10^{5}\rm cm/s$ have been observed in $\rm Al_{0.08}Ga_{0.92}As$
material exposed to the air.

Details concerning the recombination velocity and the way this
parameter is combined with the diffusivity are given in 
chapter \ref{secoptimisation} which  deals with modelling
techniques.

\subsection{Discussion}

In this chapter we have briefly described the method used to
design the modelling program, before discussing the parameters
required to model the $QE$ of real solar cells.
We have seen that these parameters can be  divided into two broad 
categories. The first are optical, compositional and geometrical 
parameters and are resonably well known.

The second category involves the transport parameters. These
are generally dependent on sample growth and device processing, 
and are poorly defined. Values quoted in the literature are 
variable and show a large spread.
However, experimental and parametrised values quoted in the 
literature provide bounds within which we can expect our values to lie. 

The treatment of the modelling techniques associated with
using mutually consistent values requires simultaneous
examination of the modelling results. These considerations
are desribed in chapters \ref{secoptimisation} and
\ref{secqeresults}.

%% file: chap5.tex
\chapter{$\bf QE$ Modelling and Optimisation\label{secoptimisation}}

The reliability of the analytical method has been established
in previous work by modelling published data by Bernd Braun 
\cite{braun90}. The first 
section of this chapter summarises the main results of this
work.

We go on to investigate a number of theoretical predictions
made by the model for a number of promising structures. 
All results apply to the QWSC design shown
in table \ref{tblbaseqwscdesign} and to \pin\ control samples which
are identical except for the lack of wells in the intrinsic region.

The optimisation is broadly separated into two cases according 
to incident photon energy. The first is the case of photons with 
energies above the bulk bandgap $E_{b2}$, which corresponds to 
aluminium fraction $X_{b2}$ (see table \ref{tblbaseqwscdesign}).

The $QE$ in this range is mainly determined by the minority carrier
transport efficiency of the p layer. Graded structures are further
characterised by the aluminium fraction $X_{b1}$ at the top of the
grade, and its associated bandgap $E_{b1}$.

The separate optimisation of the quantum well $QE$ applies
to incident photons with energies between the quantum well 
bandgap $E_{a}$ and the lower bulk bandgap $E_{b2}$ and its 
associated aluminium composition $X_{b2}$. This optimisation is
mainly dependent on the absorptivity of the quantum wells
and the light levels in the i region.


\sbox{51}{\begin{tabular}{c} {\bf Layer} \\ {\bf type} \\ \end{tabular}}
\sbox{52}{\begin{tabular}{c} $ \rm \bf Al_{x}Ga_{1-x}As$
\\ {\bf composition x (\%)} \\ \end{tabular}}
\sbox{53}{$30\times\left\{\begin{array}{l}\mbox{60A barrier} \\
\mbox{87A well} \\ \end{array} \right.$}
\sbox{54}{\begin{tabular}{l}
 Wells: $x=0$ \\ Barriers: $x=30 $ \\ \end{tabular}}
\sbox{55}{grade $\left\{\begin{array}{l}\mbox{$x=X_{b1}$} \\ 
 \mbox{$x=X_{b2}$} \\ \end{array} \right.$}
\sbox{56}{\begin{tabular}{r}
 GaAs  \\
 substrate 
 \end{tabular}}

\begin{table*}
\begin{center}
\begin{tabular}{| c | c | c | c |} \hline
\usebox{51}    & {\bf Thickness} &{\bf Doping} & \usebox{52} \\ \hline \hline
cap            & 400\AA          & $p^{+}$      & $x=0$       \\ \hline
window         & 300\AA          & $p^{+}$      & $x= 67$     \\ \hline
p layer        & 0.15$\mu m$     & $p$          & \usebox{55} \\ \hline
i buffer       & 400\AA          & $i$          & $x= X_{b1}$     \\ \hline
i layer        & \usebox{53} & $i$              & \usebox{54} \\ \hline
i buffer       & 400\AA          & $i$          & $x= X_{b1}$     \\ \hline
n layer        & 0.6$\mu m$      & $n$          &  $x= X_{b1}$    \\ \hline
GaAs substrate &                 & $n^{+}$      & $x=0$       \\ \hline
\end{tabular}
\figcap{Basic QWSC growth menu \label{tblbaseqwscdesign}}
\end{center}
\end{table*}

\section{Ungraded Cells\label{sechomooptimisation}}

\subsection{Previous Work}

\subsubsection{Photocurrent and Voltage Enhancement}

Early work presented by Mark Paxman {\it et al.} in reference 
\cite{paxman93} showed that the short circuit current  of a QWSC is 
greatly enhanced over that of a control structure which is identical, 
but with the well material replaced by an equal thickness of barrier 
material.

\bigeps{htbp}{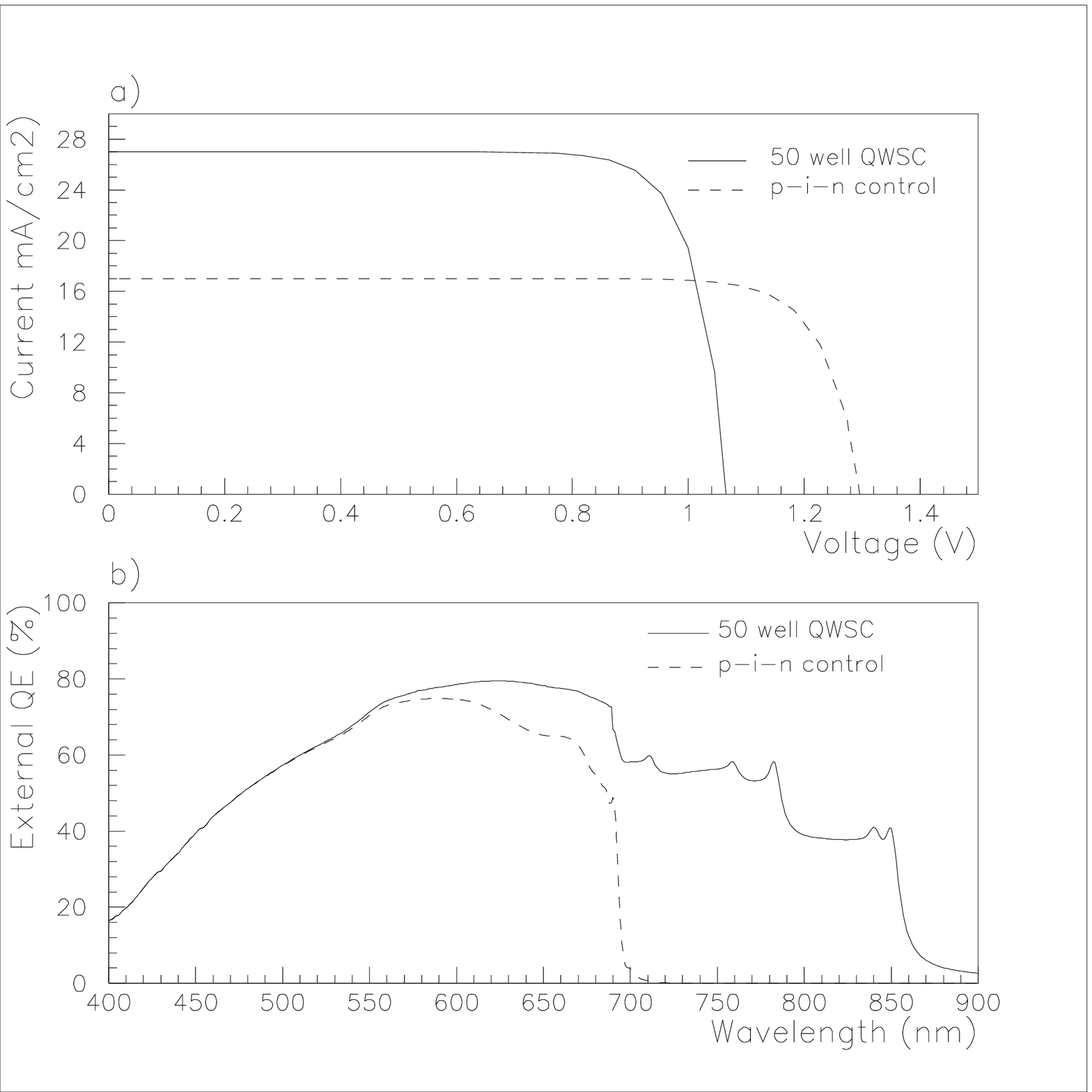}
{Modelled a) light IV curves and b) $QE$ spectra for a 50 well
QWSC and \pin\ control pair based on a successful device.
\label{figjscenhancement}}

Further work showed that the \voc\ for the QWSC drops relative
to that of the control sample, but remains greater than the
\voc\ which would be expected from a cell with the quantum 
well effective bandgap $E_{a}$.

One of the samples with fifty wells showed a \voc\ of 1.07V,
which is greater than that of the worlds' best GaAs cell 
(reference \cite{green96}). The modelled light IV curves and $QE$
spectra for this sample are illustrated in figure 
\ref{figjscenhancement}. The large increase in \jsc\
more than compensates for the decrease in \voc\, leading
to an overall efficiency enhancement in the QWSC relative 
to the \pin\ control.

A set of identical ungraded samples 
were designed with the help of the optimisation
program we described in chapter \ref{sectheory} and grown
by MBE. They feature different values 
of $X_{b2}$ but identical geometries. We concluded that the 
\voc\ increases with $X_{b2}$. Unfortunately, the material 
quality and $QE$ decrease with increasing
$X_{b2}$. We concluded that a value of $X_{b2}$ in the region
of 30\% is a reasonable compromise between the requirements
of good material quality and high \voc.

\subsubsection{Anti Reflection (AR) Coating}

The reflectivity of the standard single layer SiN AR coat has been 
optimised in reference \cite{paxman93}. 
Further reductions in reflectivity across the wavelength 
range of interest are possible with more sophisticated multilayer
AR coats. 

This work, however, is concerned mainly with the internal $QE$ of
QWSC structures, and has used the recommended SiN AR coat of
thickness $\sim 70nm$ throughout.

\subsubsection{Short Wavelength Optimisation}

Previous work has recommended a number of optimisations which
may be carried out. This work concluded that thinning the p layer
increases the minority carrier collection efficiency, particularly
at short wavelengths. It also transmits
more light to the i layer. Both effects enhance the $QE$.
This method, however, is limited by the increase in series
resistance as the p layer is thinned. This is particularly damaging
for concentrator applications.

The addition of a window layer has been found to reduce recombination
at the front of the p layer. This is due both to the replacement of
a surface recombination velocity by a lower interface recombination
between the p layer and the window, and the reflection of minority
carriers away from the surface by the bandgap discontinuity at the
interface.

Although this method successfully reduces the recombination velocity,
some of this advantage is lost because even thin window layers absorb
a significant fraction of incident light at short wavelengths. The
previous work has shown that the window layer is very inefficient
at converting this absorbed radiation to photocurrent, and 
essentially contributes no photocurrent.

Use of a very high aluminium fraction for the window layer can
reduce the window absorptivity, but introduces new difficulties.
This is due to an increased recombination velocity for high
window aluminium fraction, caused both by an increased lattice
mis-match at the interface and lower window material quality.

The techniques above have been successfully demonstrated in
reference \cite{paxman93}. However, we conclude that there is still 
much room for improvement at short wavelengths, which requires more 
complex structures.

\subsubsection{Long Wavelength Optimisation}

As shown indicated by Paxman \cite{paxman93}, the long wavelength
$QE$ can be increased simply by introducing more wells in
the i region. This can be done by reducing the barrier thickness.
This approach however is expected to reduce the \voc\ of
the cell. 

Increasing the width of the i layer is an
alternative route. We saw however in section \ref{secniproblem}
that this leads to a reduced $QE$ in forward bias and a
degradation of the fill factor $F\!F$. Furthermore, a more
fundamental limit is imposed by the increase in dark current 
for wider i layers.

Previous work resulted in a QWSC with fifty wells, which
nevertheless failed to absorb approximately half the incident
light at energies below $E_{b2}$. Failing material with lower 
background doping, more complex techniques are required to 
increase the $QE$ in the long wavelength range.

\subsubsection{Conclusions}

We have seen that the short circuit current enhancement techniques
presented by Paxman \cite{paxman93} are limited by a range of physical and 
materials considerations. 

Short wavelength efficiency enhancements in ungraded \gaasalgaas\
solar cells are limited by series resistance, window absorptivity and
poor material quality at high aluminium fractions.
                                  
At longer wavelengths, the background doping sets a limit on
the number of wells which may be inserted in the structure, and
hence the total absorptivity of the wells.

In both wavelength ranges therefore, there is potential for 
optimisation by improving p layer transport and by increasing
the absorptivity of the i layer.

\section{Further QE Enhancements Above the Barrier Bandgap}

Chapter \ref{sectheory} shows that a variation of material
composition with position can significantly alter the optical
and transport properties of the QWSC. This section investigates
the effect of such changes on light intensity and generation rate
as a function of position and incident light wavelength. 
The consequences for minority transport
are covered before reviewing the minority carrier density and
$QE$ optimisation of graded structures.

\subsection{Light Absorption and Minority Carrier Generation}

\bigeps{htbp}{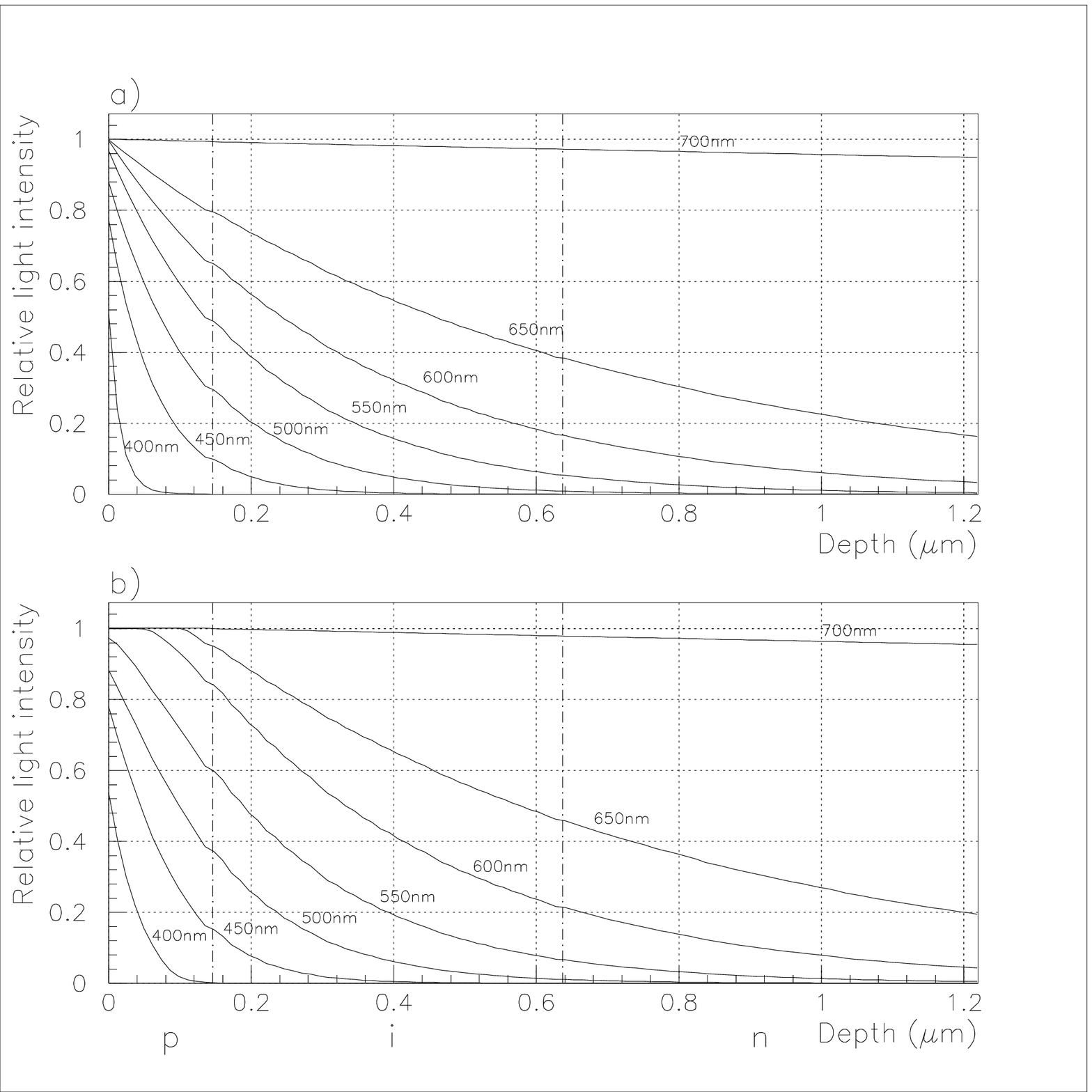}
{Relative light intensities in a a) ungraded \pin\ cell and b) a \pin\
cell with a p layer graded from $X_{b1}=67\%$ to $X_{b2}=30\%$
\label{figfluxcomparison}}
\bigeps{htbp}{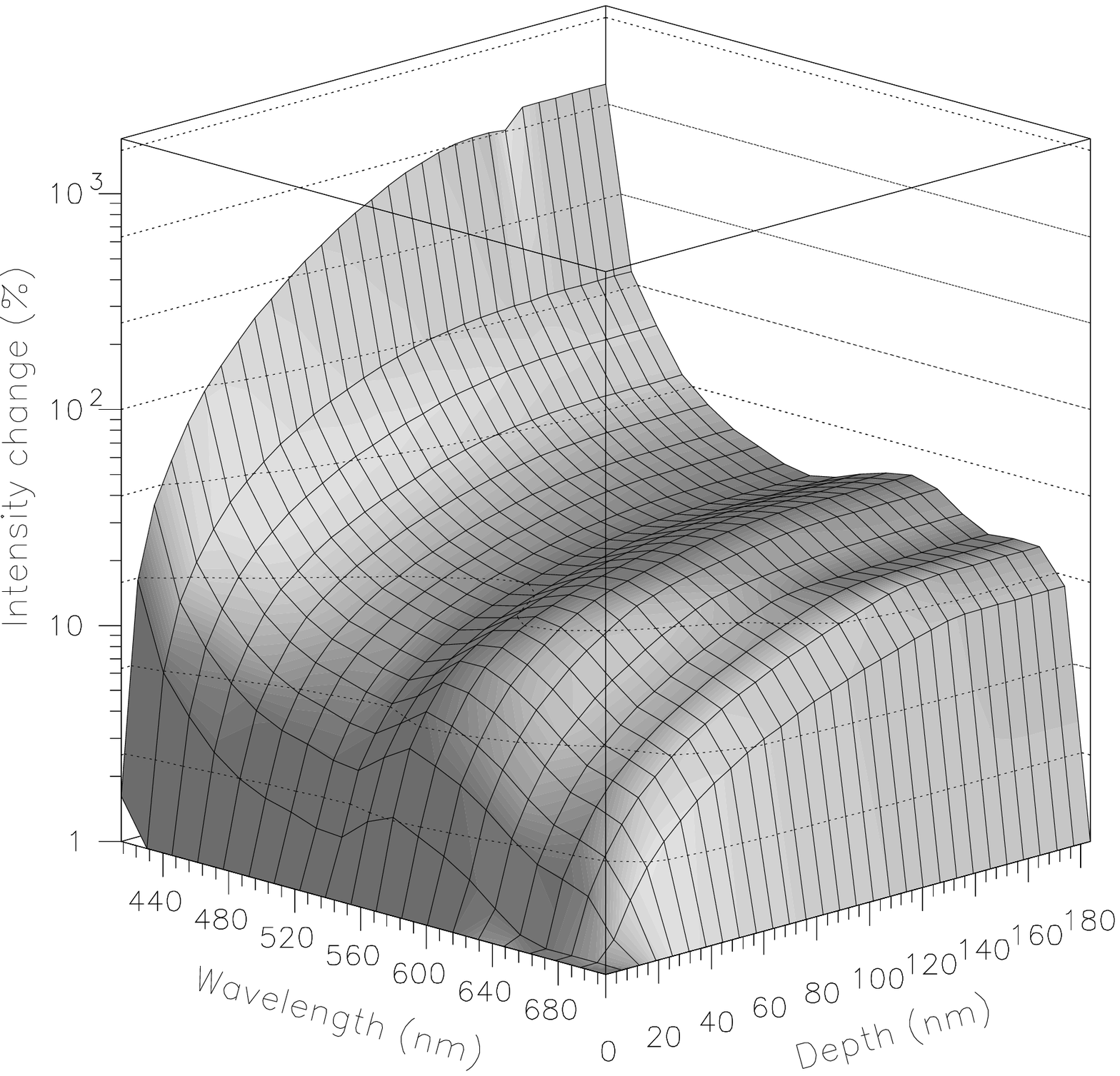}
{Fractional increase in light intensity in a \pin\ cell with a 
p layer graded from $X_{b1}=67\%$ to $X_{b2}=30\%$ compared to 
an ungraded cell
\label{figfluxratio}}

\subsubsection{Light Intensity}

In chapters \ref{sectheory} and \ref{secparameters} we saw that
\algaas\ is direct for aluminium fractions up to approximately
45\%. Direct materials are characterised by high absorption
coefficients. Unlike indirect materials such as silicon, direct
materials can absorb most of the incident light spectrum
in a relatively short depth. However, given the poor materials
characteristics of \algaas, it is desirable to increase light
levels deeper in the cell so that minority carriers are generated
preferentially near or in the depleted layers where they may be
collected by the built in potential. This may be achieved by
compositional grading.

The relative light intensity for typical ungraded and graded
\pin\ cells is shown for a range of wavelengths in figure 
\ref{figfluxcomparison}. The intensity immediately starts to decrease
in the ungraded cell for all wavelengths above the bandgap. At 
very short wavelengths, no light reaches the i region.

In the graded sample for long wavelengths such as 600nm, absorption in the 
p layer only becomes significant at a depth of about 70nm, at which
point the bandgap decreases below the corresponding absorption
threshold. A constant plateau in light intensity is seen from
zero to 70nm because of the lack of absorption.

Figure \ref{figfluxratio} gives the difference in light levels between 
ungraded and graded cells, expressed as a percentage of the light level
in the ungraded cell.
Higher light levels are seen for all wavelengths in the graded design. 
The gain in intensity increases with depth until the junction with
the i layer at a depth of 150nm. For greater depths, no further
increase is seen since the absorption coefficients of the i and n
layer are identical for the two cell designs used in this example.

The gain in light level is strongest at short wavelengths because
the difference in the absorptivities of the graded and ungraded
p layers is greatest for high energies. The calculations 
of intensity at 400nm represented in figure \ref{figfluxcomparison}
show that $\sim 95\%$ of the incident light is absorbed
in the first 40nm of an ungraded cell, whereas $\sim 85\%$ is 
absorbed in the same depth for the graded structure. Whilst the light
intensity in the i and n layers appears much higher in the graded
structure relative to the ungraded cell, the absolute light 
level remains very low. The absolute light intensity in the i and n 
layers in the short wavelength range therefore remains
very low.

Improvement in the short wavelength range will be due to
an improved p layer response due to a better distribution of
light, as we shall see in subsequent sections.

For wavelengths between approximately 450nm and the bandedge,
increases of between 10\% and 40\% are seen throughout the cell.
This is particularly significant for the i layer, which converts
absorbed light into photocurrent with nearly 100\% efficiency. 
For energies very close to $E_{b2}$, the intensity enhancement 
tends to zero since neither cell absorbs strongly.

\bigeps{htbp}{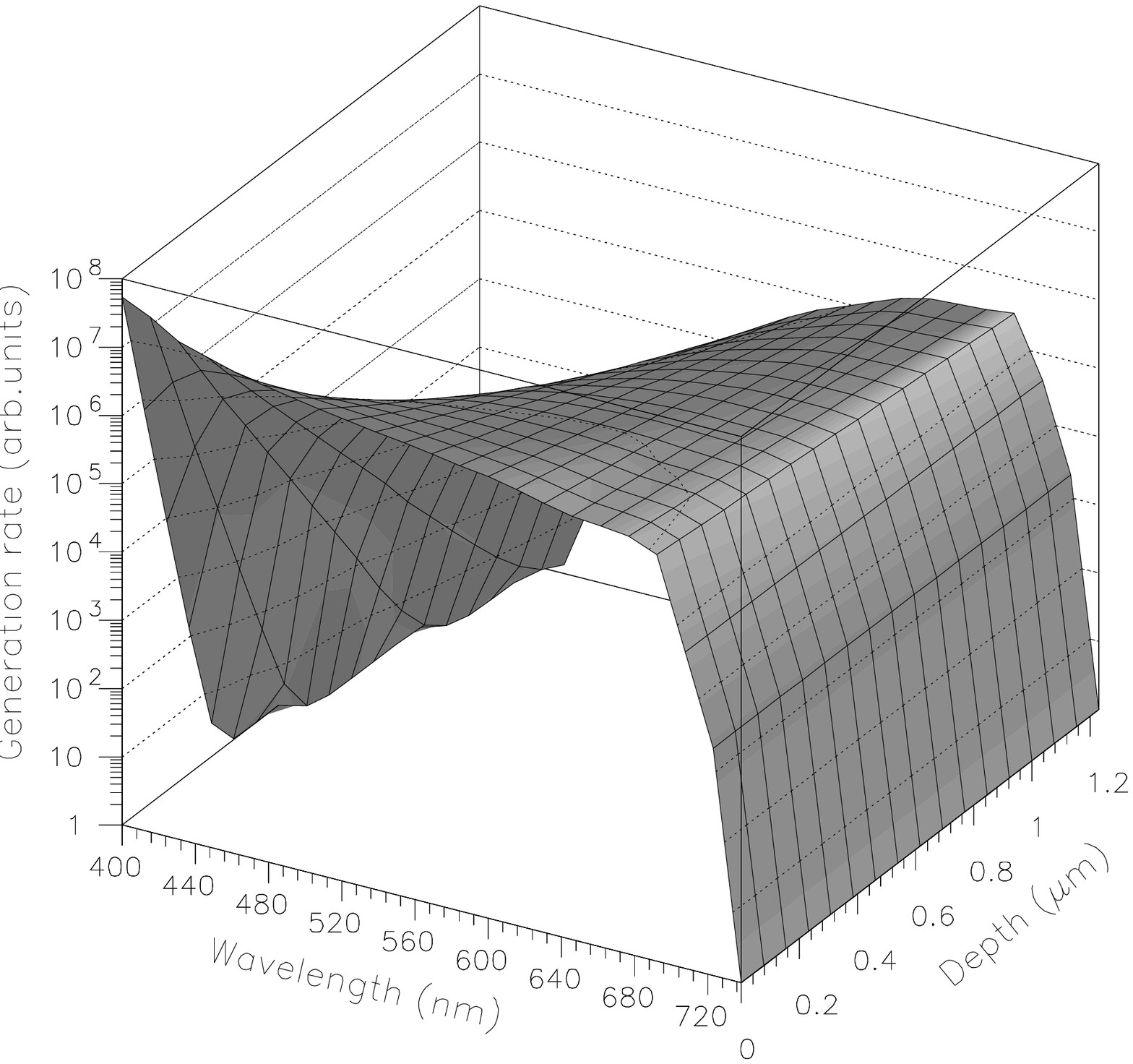}
{Generation rates in an ungraded \pin\ solar cell for wavelengths
above the bulk bandgap\label{fig3dgenhomo}}

\bigeps{htbp}{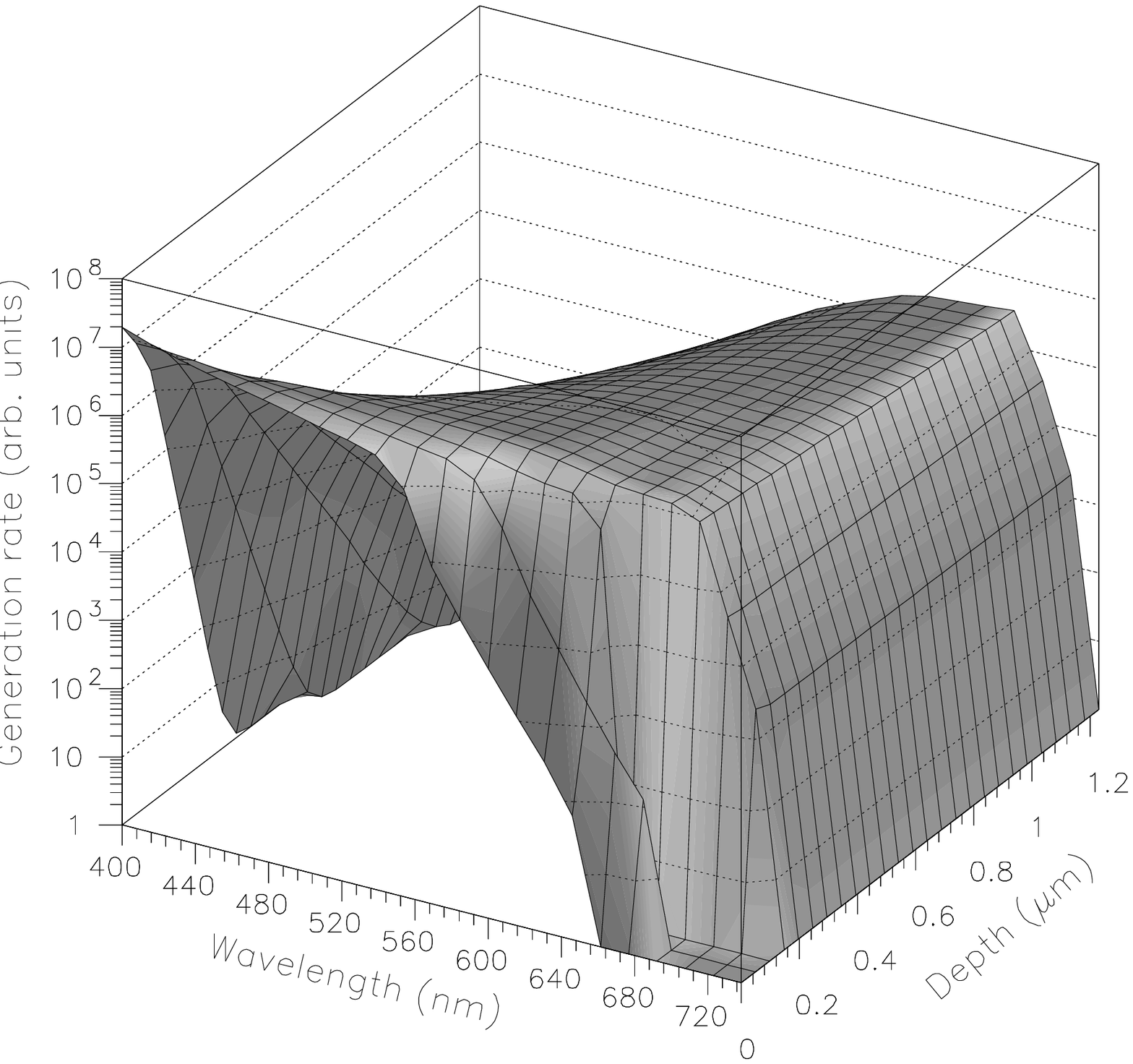}
{Generation rates in a compositionally graded \pin\ solar for 
wavelengths above the bulk bandgap $E_{b2}$\label{fig3dgengrade}}

\subsubsection{Generation Rate}

Figures \ref{fig3dgenhomo} and \ref{fig3dgengrade} show
the generation rates for the structures of figure \ref{figfluxratio}
as a function of depth and position. In the homogeneous cell, the 
generation at each wavelength decreases exponentially with increasing 
depth. The exponential decay however very nearly approximates a constant
for wavelengths within about 50nm of the bandedge.

The graded case is considerably different. Considering
the p layer first, we see that the generation rate at short
wavelengths is decreased because of the lower absorptivity
near the surface. 

At lower photon energies, the generation rate falls to zero as the high
Al fraction material towards the front of the cell ceases to absorb
light. The generation rate at 650nm for example only becomes
significant at a depth of about $0.1\mu m$, which is over two 
thirds of the total p layer width. 


We conclude that the grade improves the generation profile
both by shifting minority carrier generation rate in the p 
towards the i layer, and by increasing the generation rate in the
latter. An increase in generation rate in the n region is also visible
but is not expected to modify the n region $QE$ considerably because
the light levels in this layer remain low. The magnitude of the subsequent 
$QE$ enhancements in all three regions of the solar cell is discussed below.

\subsection{Minority Carrier Transport\label{secmodeltransport}}

In chapter \ref{sectheory} we saw that the $QE$ of graded structures
depends on the diffusion length and diffusivity of the neutral layers.
Both of these parameters are composition dependent functions.

Neither quantity can directly be measured in the QWSC structures. We
are therefore obliged to make a number of assumptions regarding the
functional form of both the diffusion length and the diffusivity.

\subsubsection{n Layer Minority Carrier Transport}

In the structures we are considering, the n layer remains ungraded,
and only constant values of $L_{p}$ and $D_{p}$ are required. Since the
n region contribution in our case is small, we cannot estimate values
of these parameters from $QE$ measurements. We rely instead on the values
reviewed in chapter \ref{secparameters}. Although this approach is  
inaccurate, the effect on the absolute $QE$ is small because of the 
very low n region $QE$ contribution in \pin\ \algaas\ devices.

\subsubsection{p Layer Minority Carrier Diffusion Lengths}

The spatial variation of $L_{n}$ and $D_{n}$ is an important
factor in determining the overall p layer $QE$. A number of 
assumptions must be made to compensate for the lack of direct 
measurement.

Appendix \ref{secapp1} shows that the $QE$ of 
ungraded p layers is directly proportional to $L_{n}$ for
low absorption and low $L_{n}$. Experimental measurements of the
$QE$ of ungraded devices with different values of $X_{b2}$
can therefore be used to build up a picture of the functional 
dependence of $L_{n}$ on aluminium composition. In turn, this
gives the spatial variation of $L_{n}$ in compositionally graded
structures.

We make the further assumption that growth conditions are reproducible
and the resulting table of $L_{n}$ as a function of $X_{b2}$ can be 
generalised to graded structures.

\subsubsection{p Layer Diffusivity\label{secdiffusivityfield}}

Chapter \ref{sectheory} showed that the $QE$ of ungraded
p layers depends on the ratio $S_{n}/D_{n}$. Since the
recombination velocity $S_{n}$ is device dependent and cannot
be measured directly in our structures, we cannot extract the
functional form of $D_{n}$ from $QE$ spectra. We must therefore
make a further assumption allowing us to parametrise the
effect of the diffusivity in terms of a single parameter.

Chapter \ref{sectheory} further showed that $D_{n}$ affects the
$QE$ of graded structures in two ways. The first is similar to
the ungraded case and involves the recombination velocity parameter
$\cal S$. It was also demonstrated however that $\cal S$ has a
weaker effect in graded structures because of the bandgap gradient
effective field ${\cal E}_{f}$.

\bigeps{htbp}{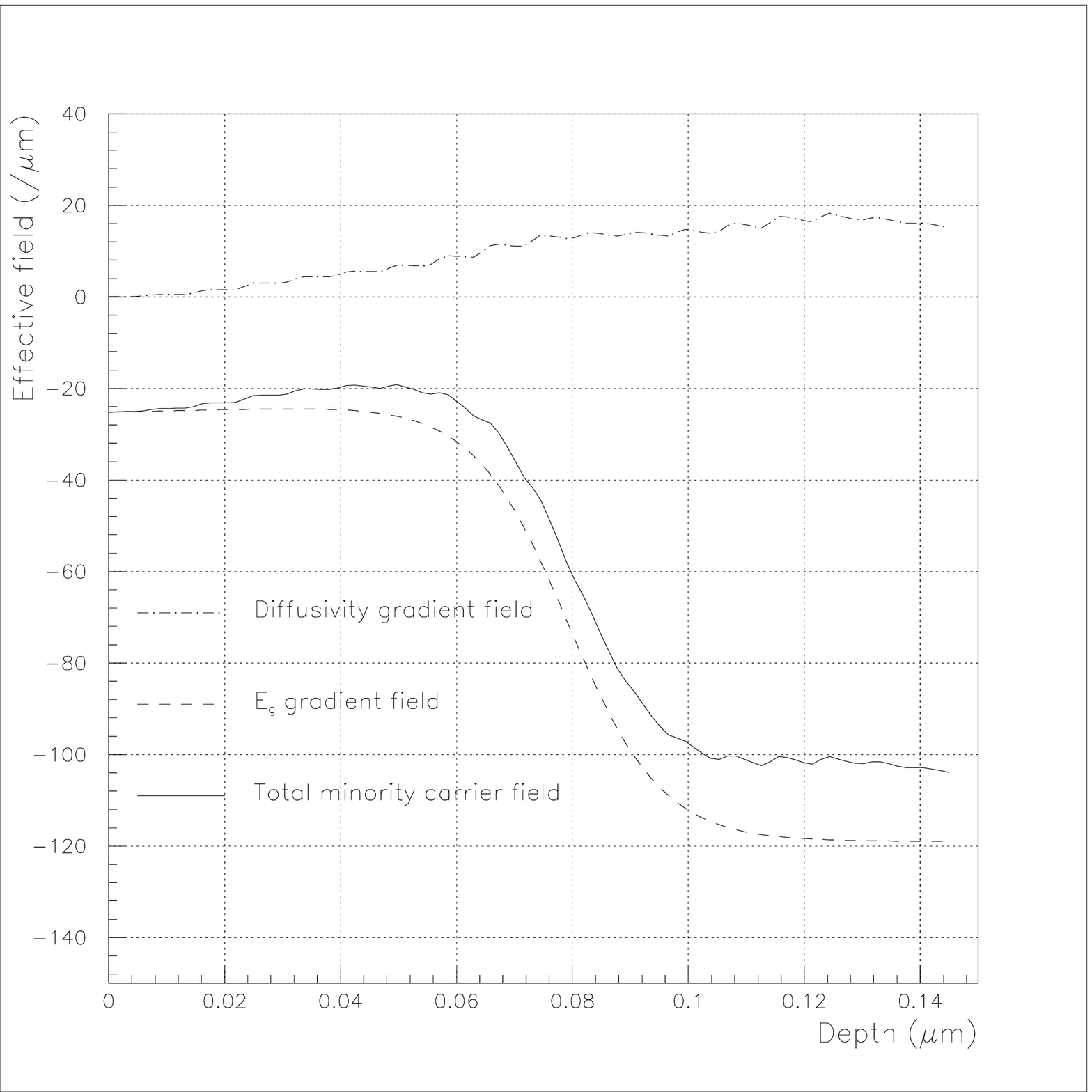}
{\label{figfields} Effective electric field in per micron as a 
function of position for a p layer graded from $X_{b1}=67\%$ to 
$X_{b2}=30\%$ } 

$D_{n}$ also appears as a term $D'_{n}/D_{n}$ which we define
as the diffusivity field ${\cal D}_{f}$ in units of per metre:
\begin{equation}
\label{diffusivityfieldposition}
{\cal D}_{f}=\frac{\nabla_{\!x}D_{n}}{D_{n}}
\end{equation}
This term expresses the current flow caused by minority carriers in
regions of high mobility moving towards regions where minority carriers
are less mobile. Although it is expressed here in terms of the
differential with respect to distance $\nabla_{\!x} D_{n}$, it can be 
re-expressed in terms of the aluminium compositional gradient
\begin{equation}
\label{diffusivityfield}
{\cal D_X}=\frac{\nabla_{\!X}}{D_{n}}
\end{equation}
This form, which we will call the diffusivity field, has the advantage 
of being independent of the device geometry, and is dimensionless. 
The diffusivity in terms of these two forms is
\begin{equation}
\label{positionandaldifffields}
\begin{array}{l l}
D_{n}(x) & =D_{n}(0) \exp\left({\cal D}_{f}x \right) \\
 & =D_{n}(0)\exp \left( {\cal D_{X}} (X(x)-X_{b1}\right) \\
\end{array}
\end{equation}
where $X(x)$ is the aluminium composition at position $x$. The diffusivity
field is related to ${\cal D}_{f}$ via the relation
\begin{equation}
{\cal D_{X}}={\cal D}_{f}\left( \frac{x_{w_{p}}}{X_{b2}-X_{b1}}\right)
\end{equation}
The diffusivity field defined this way has a value of about -4 if
we use the Hamaker parametrisation of chapter \ref{secparameters},
for p type \algaas\ doped at about $\rm 2\times 10^{18} cm^{-3}$.

Since comparisons between samples with different geometries are more
clearly apparent with this form, we shall use it in the modelling
section. We continue to use the device dependent form ${\cal D}_{f}$ 
here, however, because it has the same units as the effective field
${\cal E}_{f}$ and is therefore better suited to this discussion.
In terms of ${\cal D}_{f}$, then, the total field ${\cal E}_{t}$ acting
on the electron minority carriers is
\begin{equation}
\label{totalfield}
{\cal E}_{t}={\cal E}_{f}+{\cal D}_{f}
\end{equation}

According to experimental data presented in reference 
\cite{walukiewicz92} (reviewed in chapter \ref{secparameters}),
$D_{n}$ decreases rapidly with increasing aluminium
fraction, before levelling off for indirect material. Figure       
\ref{figfields} shows the resulting fields ${\cal D}_{f}$, 
${\cal E}_{f}$ and ${\cal E}_{t}$ where ${\cal E}_{f}$ is calculated
as in chapter \ref{secparameters} and ${\cal D}_{f}$ is found from
the experimental data of reference \cite{walukiewicz92}.

Since the diffusivity is higher in \algaas\ with a low aluminium 
fraction, ${\cal D}_{f}$ is positive and, unfortunately, opposes the
drift field ${\cal E}_{f}$ due to the bandgap gradient. The figure
shows that a significant decrease in ${\cal E}_{t}$ may be expected.

The functional form of $D_{n}$ therefore may have important consequences
for the overall effect of a compositional grade on minority carrier
transport efficiency. Since we cannot experimentally determine the
compositional dependence of ${\cal D}_{f}$, we must find a parametrisation
which can reduce this poorly defined function to a single fitting 
parameter.

An exponential parametrisation for $D_{n}$ suggested by
Hamaker \cite{hamaker85} was reviewed in chapter \ref{secparameters}.
The Hamaker parametrisation reproduces the 
functional form of the experimental data given in reference
\cite{walukiewicz92}. For the example shown in figure \ref{figfields}, 
the Hamaker value is  ${\cal D}_{f}\sim 10\mu m^{-1}$, which agrees
qualitatively with the published experimental data. 
Because of material variability, however, we do not expect the Hamaker
parametrisation to yield accurate values of ${\cal D}_{f}$ for
our material. 

In the general case of an exponential parametrisation for $D_{n}$,
${\cal D}_{f}$ is a position independent constant for linear compositional 
gradients. Moreover, it is independent of the magnitude of $D_{n}$. This 
means that the magnitude of the diffusivity influences the $QE$ of a 
graded structure only through the surface boundary parameter $\cal S$. 
Since, as we have mentioned ealier in this section, $\cal S$ has a 
weak effect in graded structures, we are left with a single 
principal fitting parameter which is the diffusivity field $D_{f}$.

We will therefore assume an exponential form for $D_{n}$ along the lines
suggested by Hamaker but will treat the diffusivity field 
${\cal D}_{f}$ as a position independent fitting parameter.

\subsubsection{Summary of the Transport Model}

Modelling of the $QE$ from the neutral regions is therefore subject
to a number of assumptions which we will review here.

The first is that the n region contribution is small. We only
treat ungraded n layers, and rely on published
data for transport parameters in this case. 

For ungraded p layers, we have seen that parameters are well
known except for $L_{d}$, $D_{n}$ and $S_{n}$. Because the diffusion
length in our structures is smaller than the p layer width $x_{p}$,
$L_{n}$ can be estimated from the $QE$ close to the bulk
bandedge at energy $E_{b2}$.

The parameters $S_{n}$ and $D_{n}$ in ungraded structures can be
separated as a single surface recombination parameter $\cal S$ 
(see chapter \ref{sectheory}) which is used as the sole fitting 
parameter in the short wavelength regime.

For compositionally graded structures, we make the assumption 
that the epitaxial material is of reproducible
quality. The values of $L_{n}$ derived from modelling the $QE$ of
ungraded structures with different values of $X_{b2}$ are
used to build up a table of the compositional dependence of $L_{n}$.
Under the aforementioned assumption, this provides the position
dependence of $L_{n}$ for the grandad structures.

We finally make the assumption that the functional dependence of
$D_{n}$ on aluminium fraction is exponential. The diffusivity field
${\cal D}_{f}$ is therefore a constant, which is independent of the
magnitude of the parametrisation $D_{n}(x)$. Since chapter 
\ref{sectheory} has shown that the $QE$ of graded structures is
only weakly dependent on the surface recombination parameter
$\cal S$, we are left with a single dominant fitting parameter which 
is ${\cal D}_{f}$.

The number of assumptions involved in this modelling prevents
us from expecting quantitative predictions of the $QE$ of graded
structures. We expect, however, to qualitatively describe the
dependence $QE$ enhancement factors due on the one hand to better
light distribution and on the other, to improved minority carrier
transport in graded solar cell structures.

\subsection{Minority Carrier Concentration}

\bigeps{htbp}{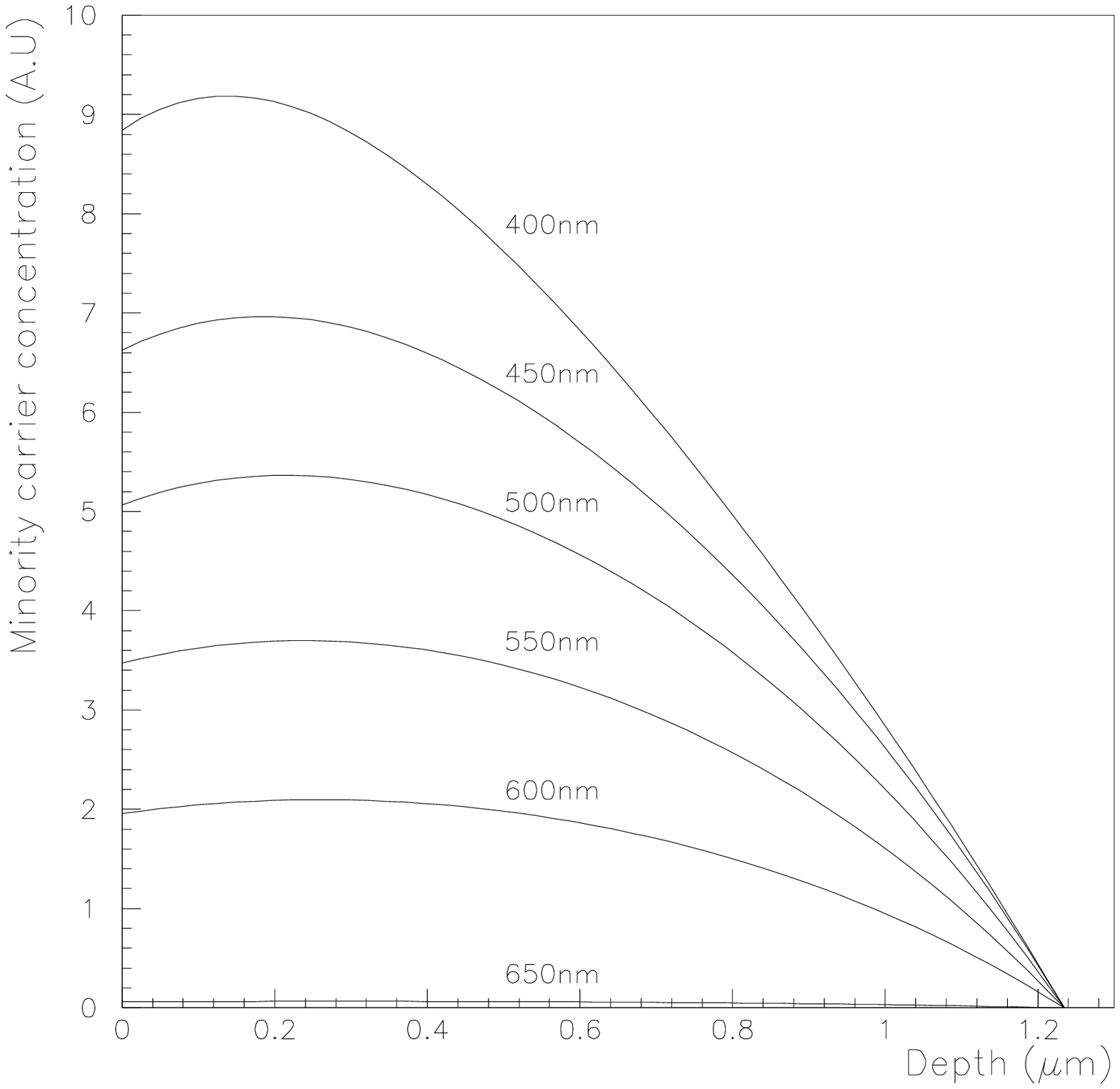}
{Minority carrier concentration in an ungraded p layer for a 
range of wavelengths.\label{fighomn}}

\bigeps{htbp}{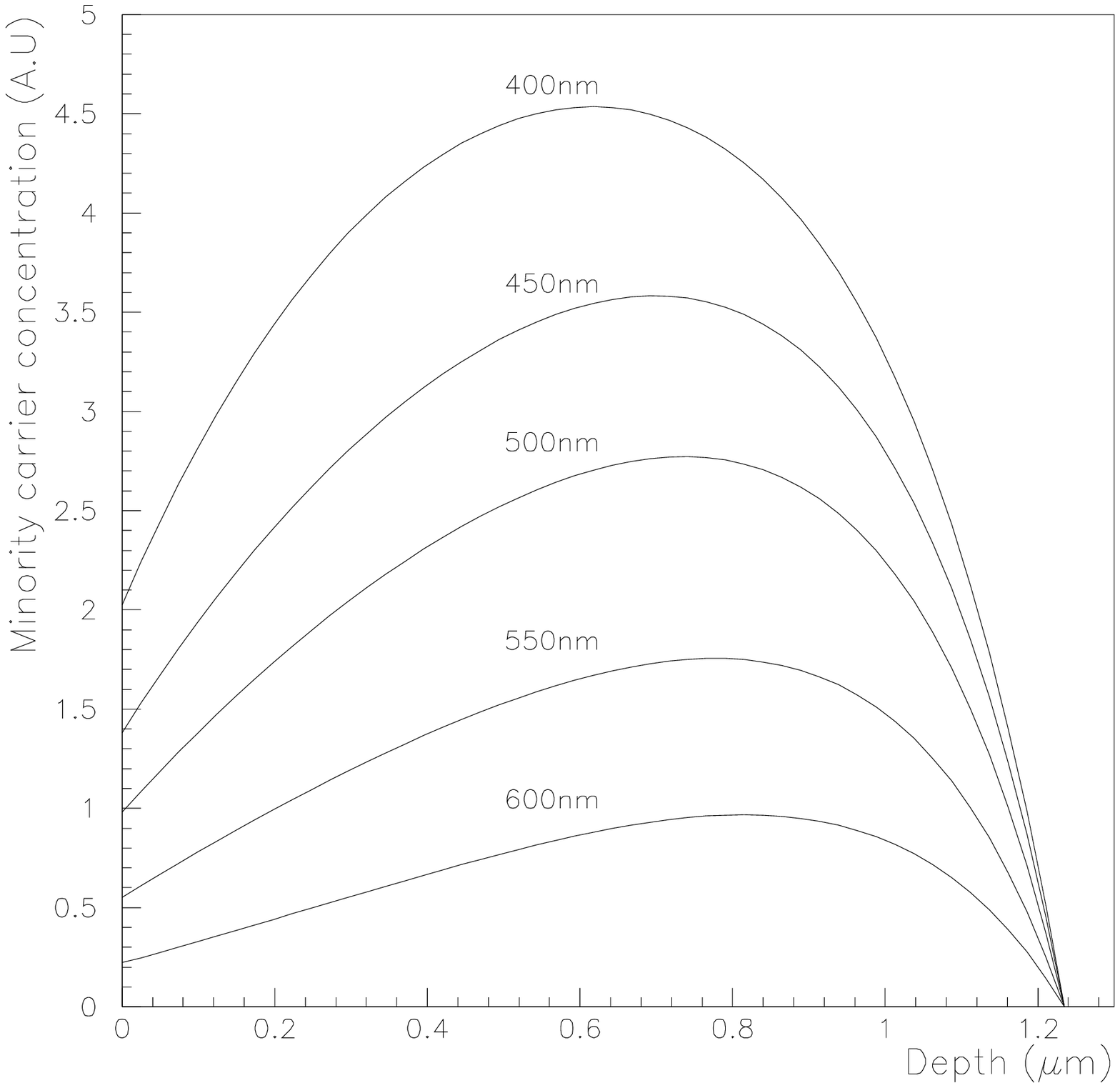}
{Minority carrier concentration in a p layer graded from 30\% to 40\%
Al for a range of wavelengths.
\label{figlowgraden}}

\bigeps{htbp}{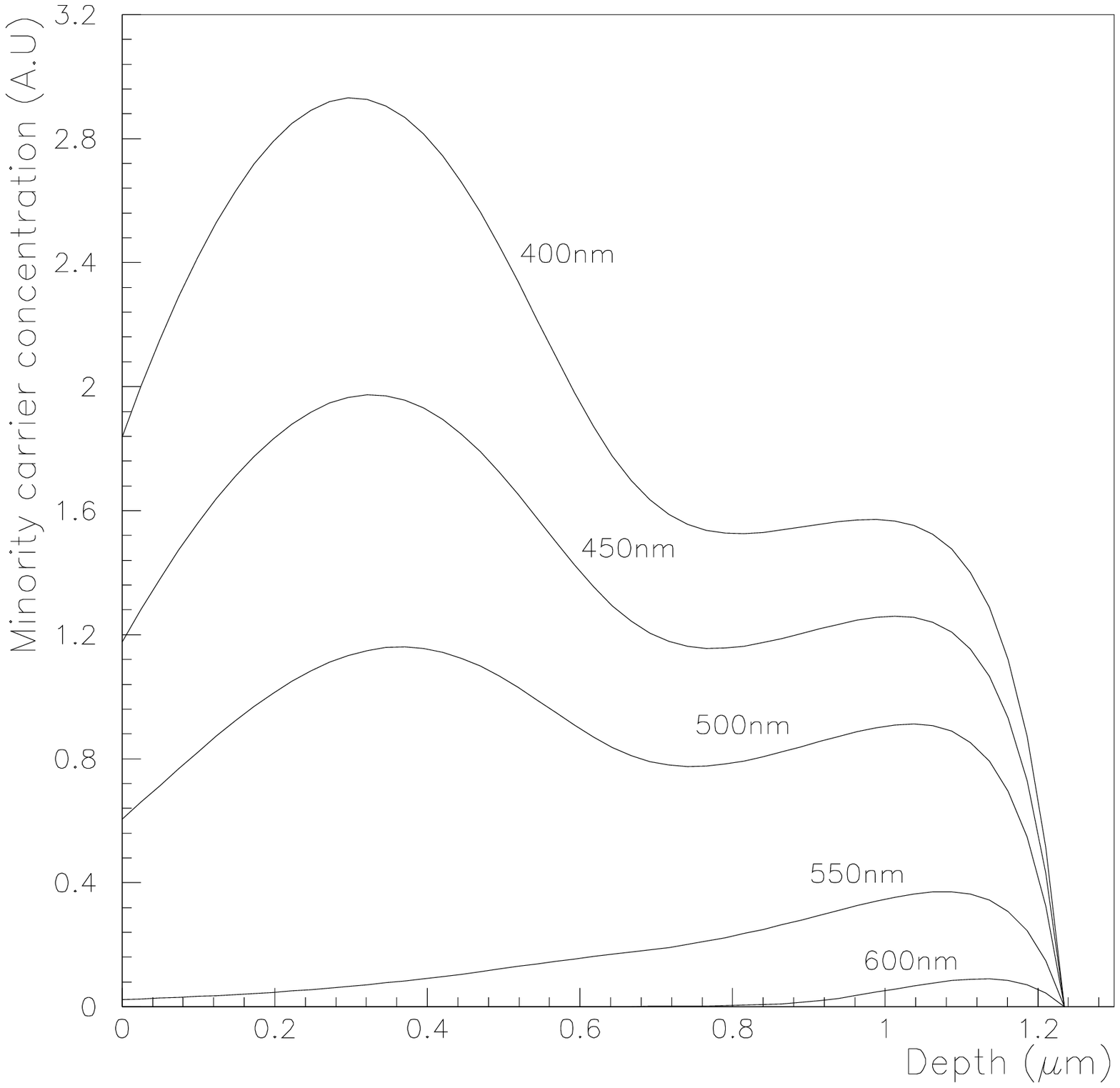}
{Minority carrier concentration in a p layer graded from 30\% to 67\%
Al for a range of wavelengths.
\label{fighighgraden}}

Figures \ref{fighomn} and \ref{figlowgraden} show 
the minority carrier concentrations for an ungraded p layer 
and a cell graded from 30\% to 40\%. The concentrations are
given in arbitrary units but are subject to the same normalisation
factor.

We see that the minority
carrier concentration in the graded case decreases in magnitude by
a factor of about two, and is shifted towards the depletion 
region. The decrease in concentration is due partly to decreased 
absorption and partly to the modifications in minority carrier
transport efficiency caused by the grade. The effective field
in particular sweeps carriers to the junction. Increased bulk
recombination due to the higher aluminium fraction also plays
a role. 

Figure \ref{fighighgraden} shows the case of a p layer graded
from 30\% to 67\%. In this case, the two separate field regions
are clearly reflected in the minority carrier concentration.
The lower direct transition region near the depletion layer
has a lower minority carrier concentration in this case because
the higher field sweeps carriers towards the depletion layer
more efficiently than the lower field in the indirect region
near the surface.

The overall minority carrier concentration is again lower than
the previous two cases we have looked at for the same reasons.
The change is greater in this case because of the increased
compositional gradient.

\subsection{QE and Short Circuit Current\label{secqex}}

The quantum efficiency for the ungraded and graded cells
described in the previous sections is shown in figure
\ref{figqehg}. The breakdown of percentage increase in
p, i, n and total $QE$ is given in figure \ref{figqehgratio}.
We note that the p layer contribution is increased at short 
wavelengths but decreased near the bulk band edge.

The improvement at very short wavelengths is the result
of increased carrier collection efficiency in the p layer.
This is due to improved minority carrier transport and
the shifting of the minority carrier generation profile
towards the i region. The deterioration near the band 
edge however is due to reduced light absorption in the
p layer at low photon energies. 

\bigeps{htbp}{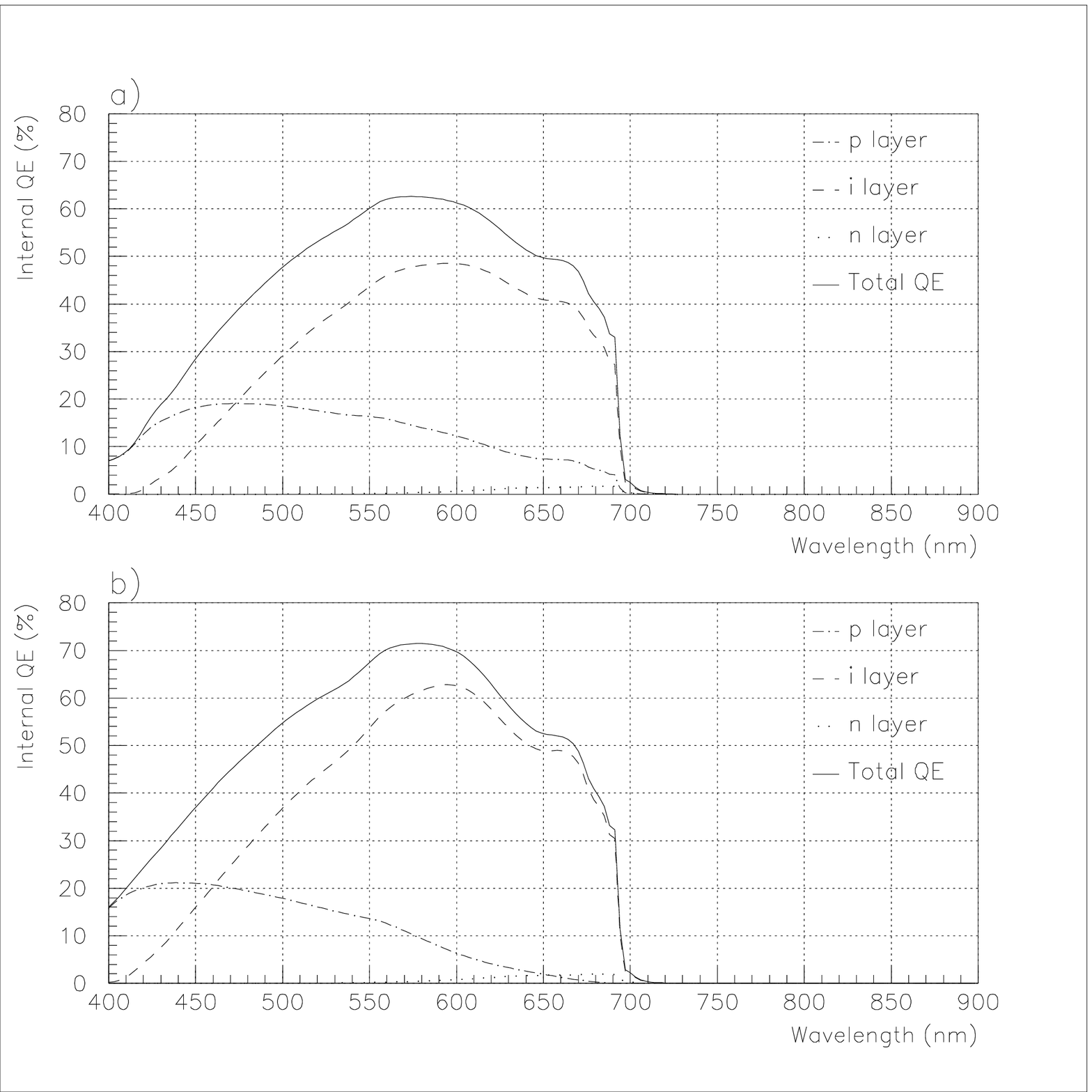}
{$QE$ of a) a 30\% \pin\ and b) a \pin\ with a p layer graded from 
$x_{b1}=67\%$ to $x_{b2}=30\%$ \label{figqehg}}

\bigeps{htbp}{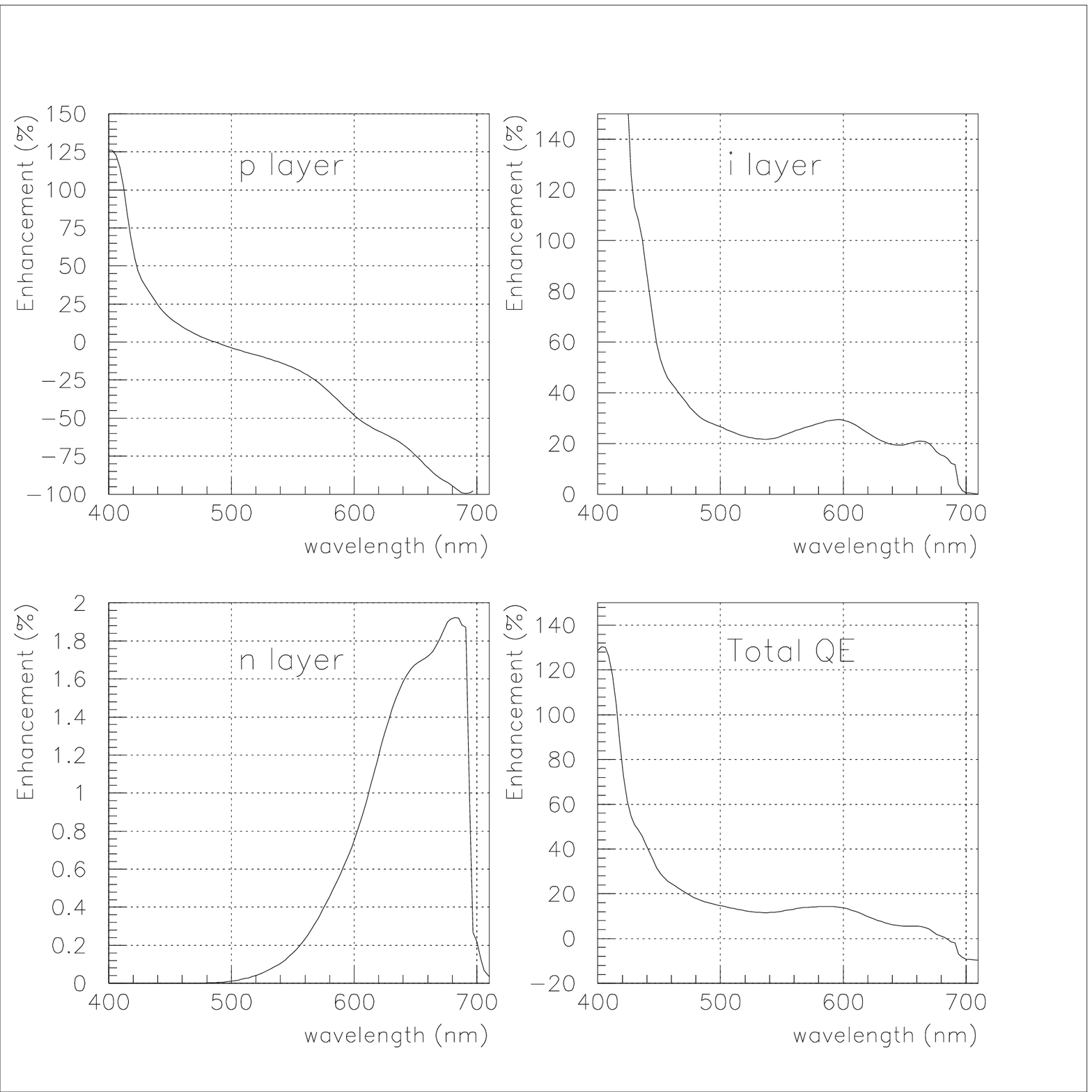}
{Percentage increase in $QE$ for p, i, n layers and total enhancement
for a graded \pin\ device relative to an ungraded control.
\label{figqehgratio}}

Figure \ref{figexampleqex}a shows a prediction of short circuit
current as a function of the topmost aluminium fraction in the p
layer. Such calculations are subject to large uncertainty however,
particularly given the large distribution of possible diffusivity
gradients. The calculation of figure \ref{figexampleqex}a uses
a gradient which is about half that of the published data of
chapter \ref{secparameters} which as we shall see in chapter 
\ref{secqeresults} is consistent with experimental $QE$ spectra.
Although the model cannot indicate the magnitude of the diffusivity,
this deviation of a factor of two in gradient is well within
the margins expected.

With the uncertainties due to ${\cal D}_{f}$ in mind,
we note that \jsc\ initially rises sharply as the
effects of surface recombination become less important. For
higher grades, the gain in i region contribution is increasingly
balanced by a reduction of the p layer current. In this example
the $QE$ reaches a clear optimum. Relatively small changes to the
diffusivity can have significant effects on the behaviour of the $QE$
at high aluminium fractions. 

The common feature of such theoretical studies is the increase
of $QE$ up to $X_{b1}\sim 50\% \pm 5\%$. Thereafter,
more optimistic parametrisations of $\cal D$ than that used in
the figure show a $QE$ which continues to increase monotonically as 
a function of $X_{b1}$, albeit at a slower rate. 

We conclude that a successful grade in a cell with $X_{b2}\sim 30\%$
requires a grade of approximately $X_{b1}\geq 40\%$.
A higher grade is not expected to significantly affect the overall
short circuit current, and may in fact improve it, depending on the
depth dependence of the diffusivity. This is partly due to the fact
that two enhancement mechanisms are present, and that even devices
with disappointing transport parameter profiles will benifit from 
enhanced i region $QE$ because light intensity $QE$ enhancement
mechanism is independent of transport and is greater for higher
grades.

\bigeps{htbp}{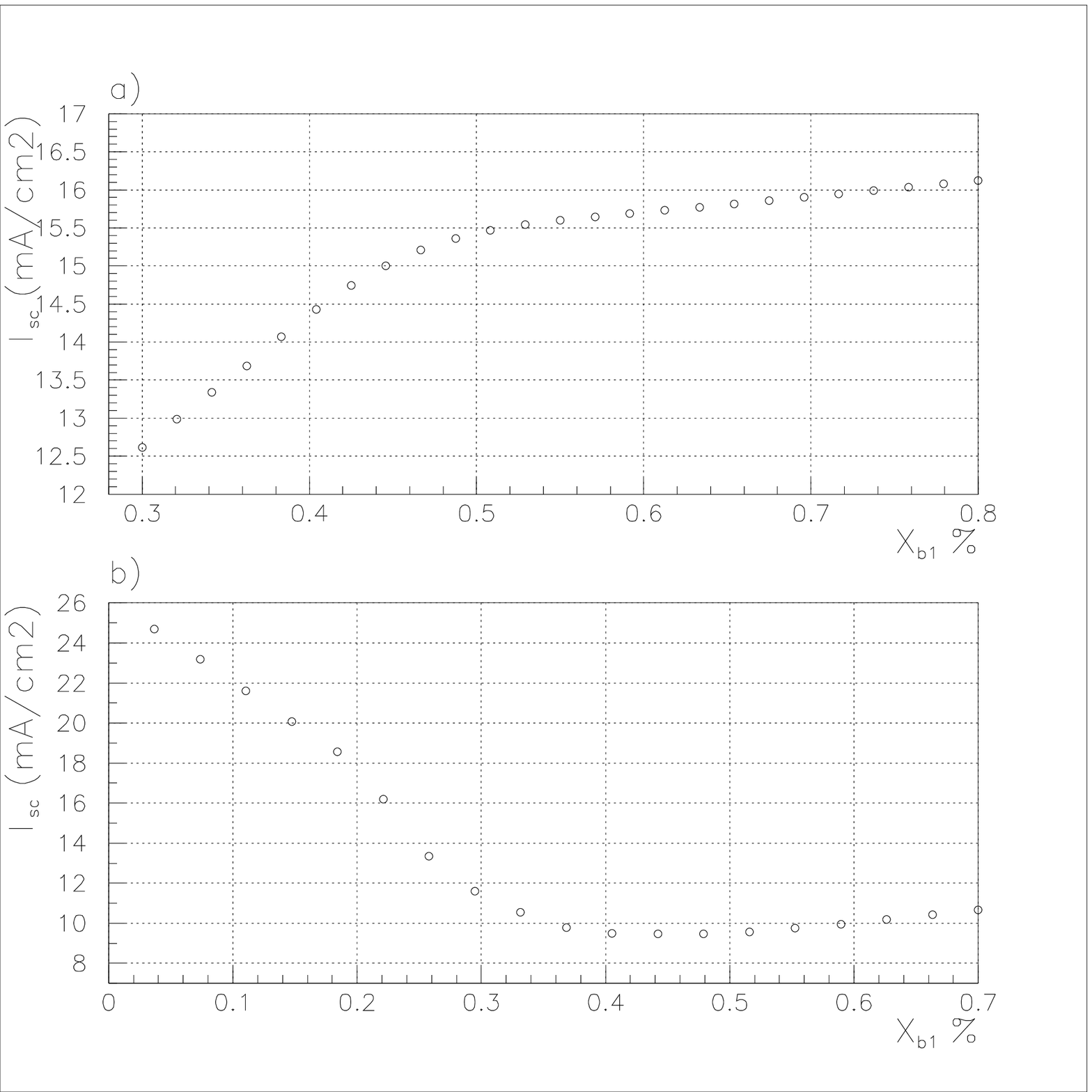}
{\jsc\ of a graded \pin\ cell as a function of the higher bandgap 
$X_{b1}$ \label{figexampleqex}}

The choice of the higher aluminium fraction is not therefore
critical from a modelling point of view. It is possible to
favour one enhancement mechanism over another without seriously
affecting the overall short circuit current.

In real devices, however, this must be offset against the
technological problems involved in growing and processing
\algaas\ layers which are graded to high aluminium fractions
over the short distances involved.

We note finally that in good material, a grade may in
fact reduce the overall \jsc. Figure \ref{figexampleqex} b)
shows the extreme example of a series of \jsc\ predictions
starting with a good \gaas\ cell. In this case the $QE$ of
the cell is rapidly degraded as the efficient \gaas\ material
at the front of the cell is replaced with \algaas\ with increasing
aluminium fraction.

\subsection{Windows\label{secwindows}}

Figure \ref{fighomowindowremoval} shows the $QE$ of an ungraded
p layer with and without a window for a recombination velocity
of $500m/s$, which is frequently observed in our samples.

If we assume the recombination
velocity is unchanged, we see a significant increase in $QE$ because 
less light is lost before the incident spectrum reaches the p layer. 
However, removal of the window layer is expected to increase surface 
recombination. The lowest curve on the graph shows the worst case 
scenario of a recombination which is
large enough to force the minority carrier concentration 
to negligible levels at the front of the cell. We note that 
increased light transmission in this case cannot compensate the 
loss of carriers at the surface.

The behaviour in graded structures however is different.
Appendix \ref{secapp1} shows that the effect of surface recombination
is reduced by the effective field. In good material with
appropriate contacts (see chapter \ref{secexperiment}), the
surface recombination velocity is expected to be very small compared to
the effective field term.

Figure \ref{figgradewindowremoval} shows calculations analogous to those
of figure \ref{fighomowindowremoval} but for a structure graded from
30\% to 67\% aluminium. In this case, both calculations assuming an
unchanged surface recombination and a worst case infinite surface
recombination after removal of the window predict an enhanced $QE$.

Although the predicted gain in $QE$ is not large, we conclude that
window removal may be beneficial because the graded p layer fulfills
the function which the window is designed to perform. 

\bigeps{htbp}
{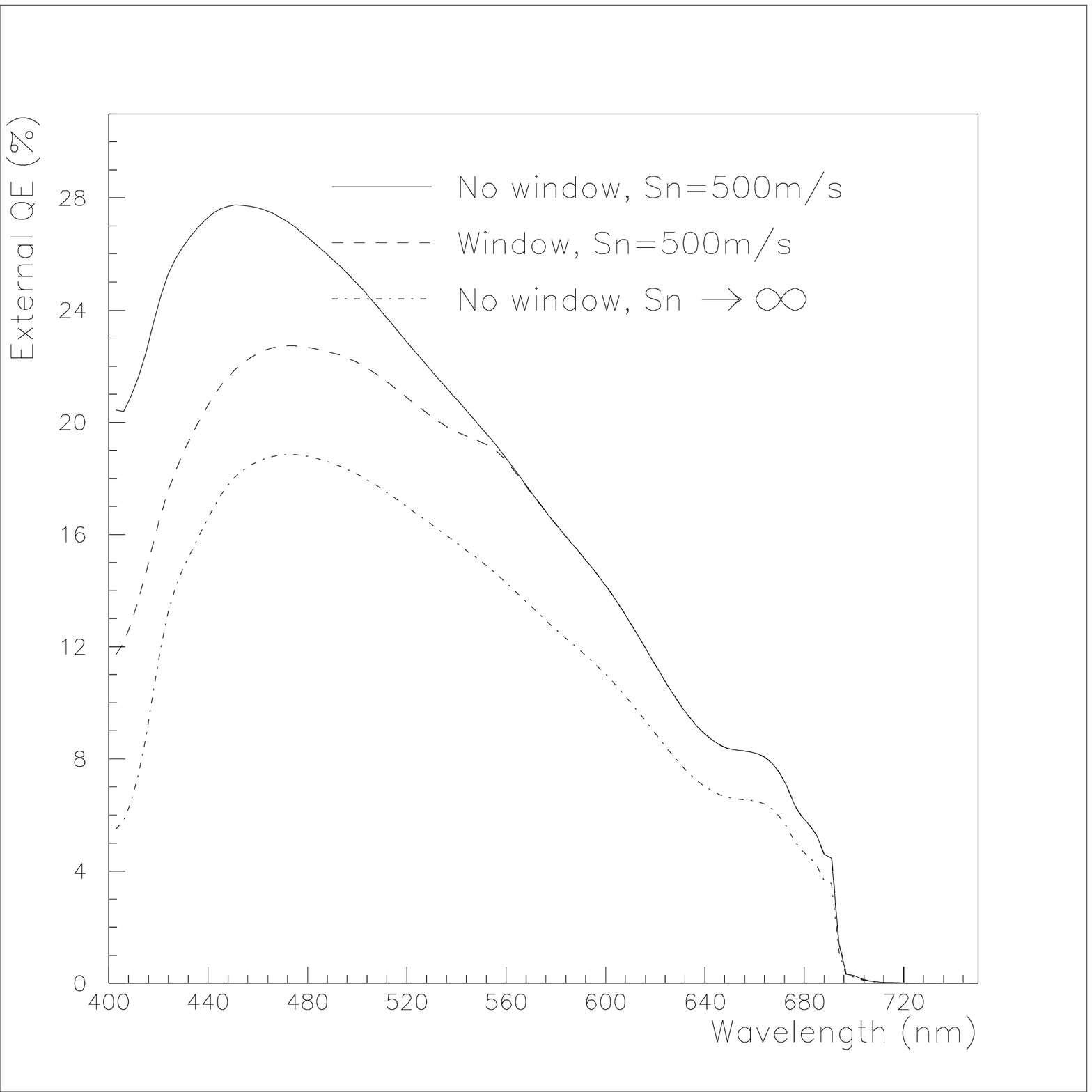}
{Influence of window layer removal on the $QE$ of an ungraded p layer,
including surface recombination effects\label{fighomowindowremoval}}
\bigeps{htbp}
{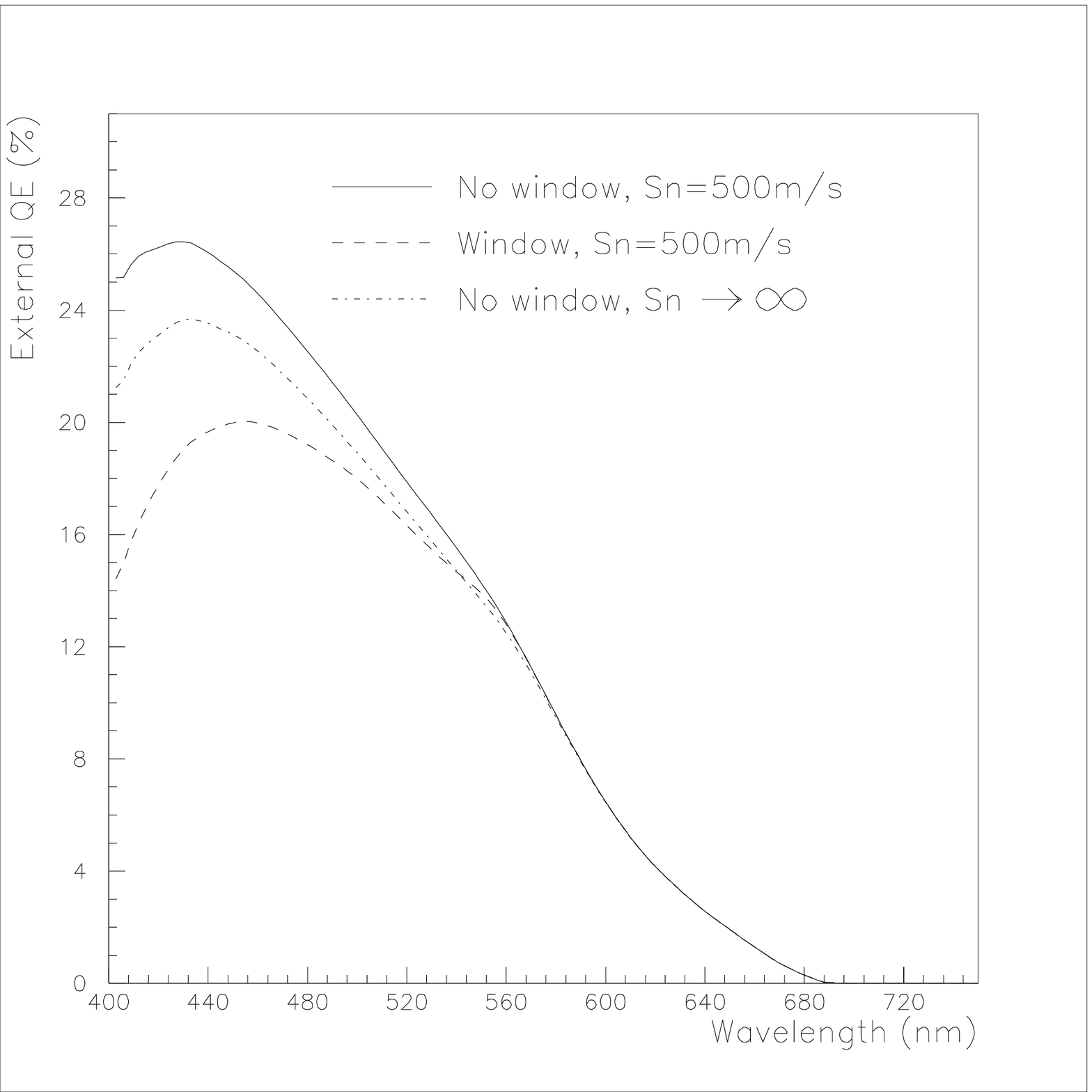}
{Influence of window layer removal on the $QE$ of a graded
p layer $QE$, showing weak dependence on surface recombination velocity
\label{figgradewindowremoval}}

\section{Further QE Enhancement Below the Barrier Bandgap}

The quantum wells in a standard QWSC design typically absorb
less than half the incident light. The remaining fraction is
absorbed in the substrate and does not contribute to the
photocurrent.

The quantum well photocurrent can be increased by ensuring
that the light is reflected back through the cell after the
first pass. This is achieved by removing the \gaas\ substrate
and coating the back of the n layer with a metallic back surface
mirror, as described in chapter \ref{secexperiment}

This technique has consequences for all optical functions
associated with the cell, which are examined in the following
sections.

The modelling results presented assume a measured wavelength
dependent front surface reflectivity and a back surface 
reflectivity of 95\%.

\subsection{Generation, Flux and QE in Mirrored Cells}

Figure \ref{figmirrorflux} shows how the addition of
a back mirror alters the light intensity as a function of position
in a QWSC. The light levels at long wavelengths are significantly
increased. The main overall consequence is a more uniform light
intensity throughout the solar cell. 
The position and wavelength dependent Fabry-Perot oscillations
in light intensity are due to interference phenomena.

\bigeps{tp}{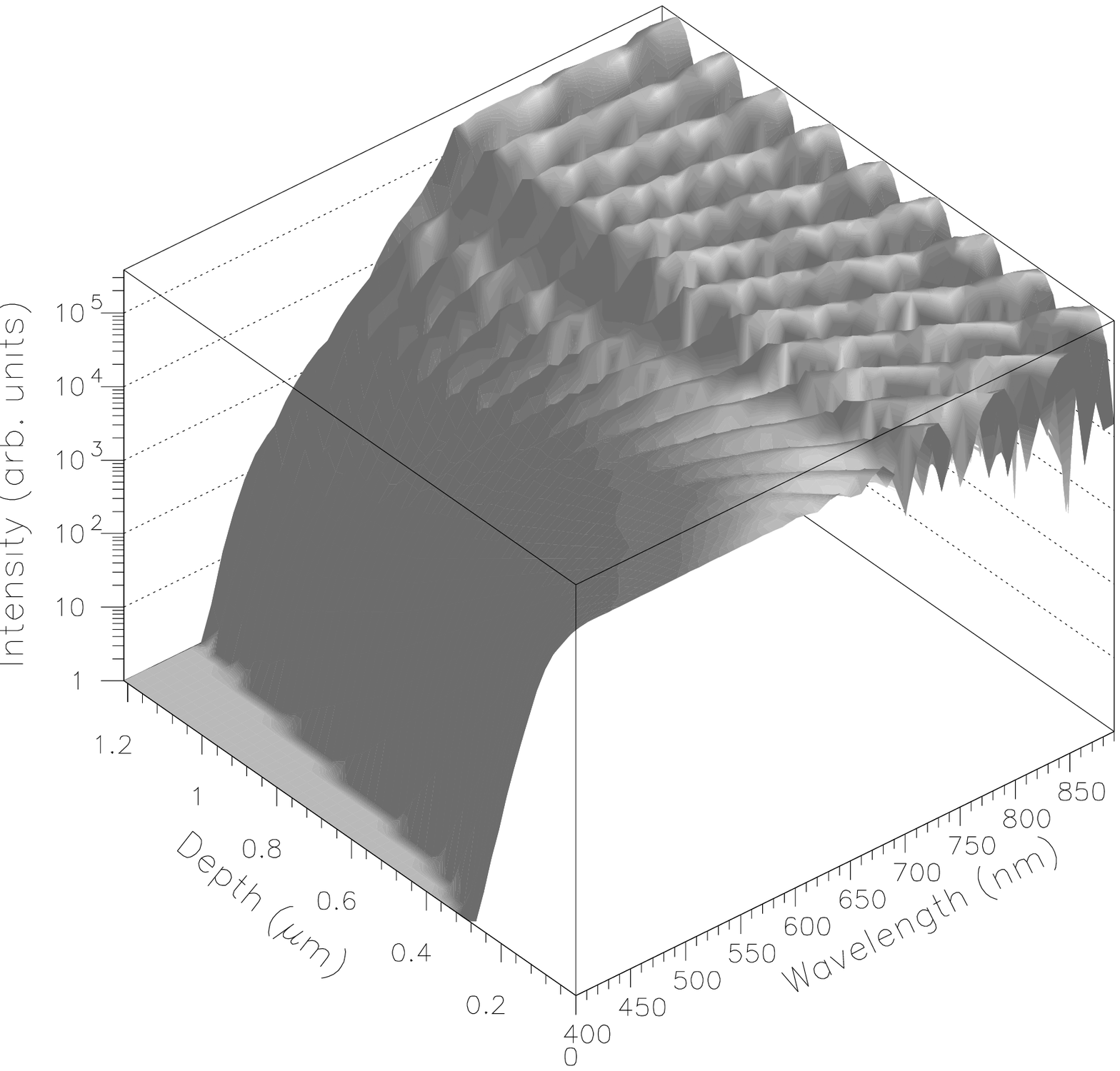}
{Light flux in a mirror backed QWSC
\label{figmirrorflux}}

Figure \ref{figqwscgeneration} shows the generation rate of a 30 well 
graded QWSC as a function of position and wavelength. The extended quantum
well generation is clearly visible for wavelengths between 700nm and
870nm. The detail of the generation in individual wells is not visible 
since, as we saw in section \ref{sectheory}, the model considers the 
i region as a weighted average of quantum well and bulk material.

Figure \ref{figmirrorgeneration} gives the generation rate for an 
identical cell with a back surface mirror, again as a function of 
wavelength and position. Interference effects impose a strong position 
dependent oscillation in the quantum well generation which is absent 
in the non mirrored device.

The magnitude of the generation enhancement is visible in figure
\ref{figmirrorgenratio}a, which shows the generation rates in both cases,
integrated over wavelengths from 400nm to 900nm. Significant
enhnancements are seen for the i and n regions. In the design
used for this example, insufficient light makes a second pass
through the light to significantly affect the generation rate
in the p layer.

Figure \ref{figmirrorgenratio}b shows the enhancment in the i layer, 
expressed as a percentage of the generation rate in the cell without a 
mirror. Since the $QE$ of the i layer is proportional to the i 
layer generation rate, an enhancement of approximately 30\% in $QE$ 
is expected for this example.

\bigeps{htbp}{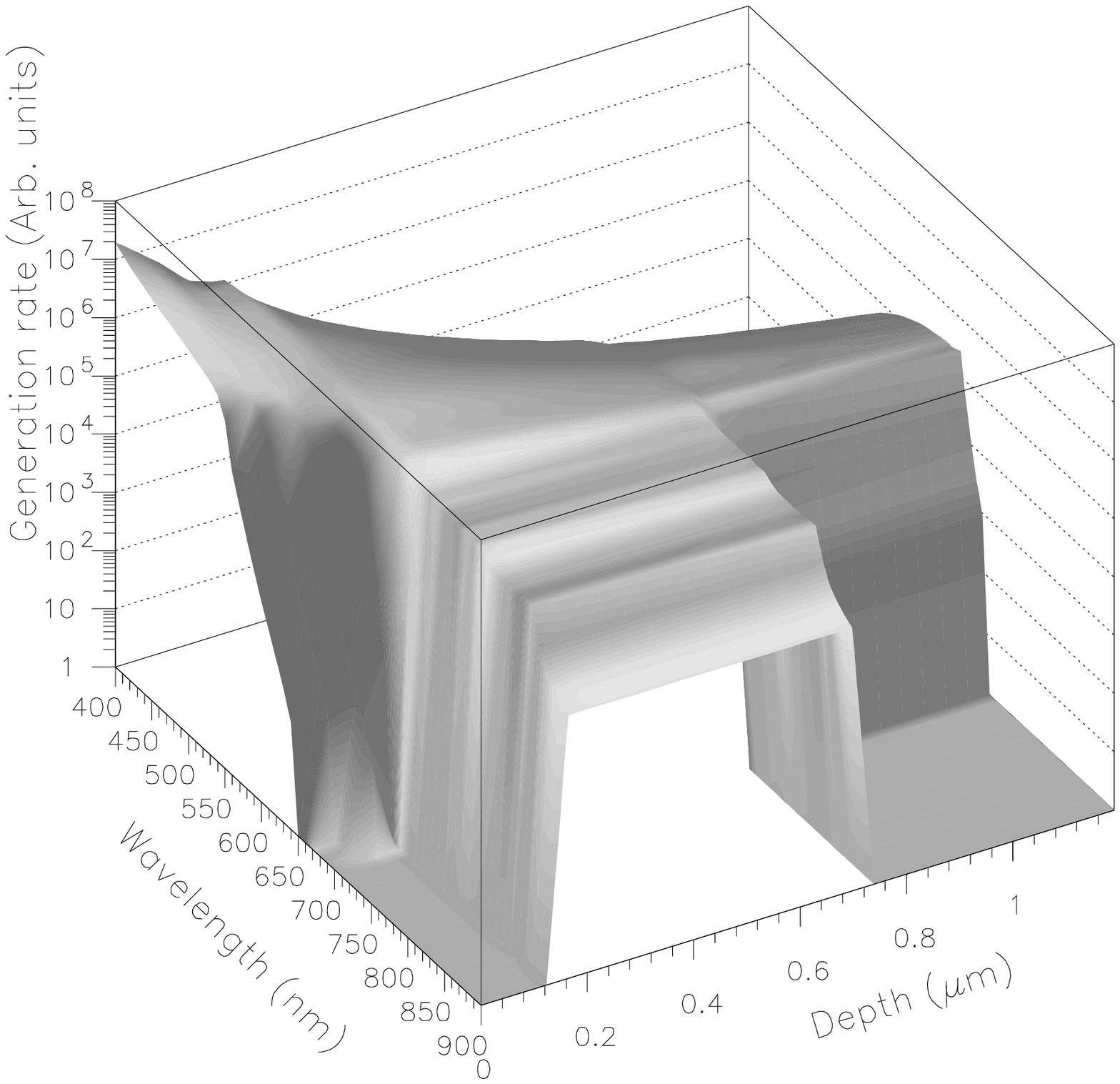}
{Generation rate in a 30 quantum well QWSC
\label{figqwscgeneration}}

\bigeps{htbp}{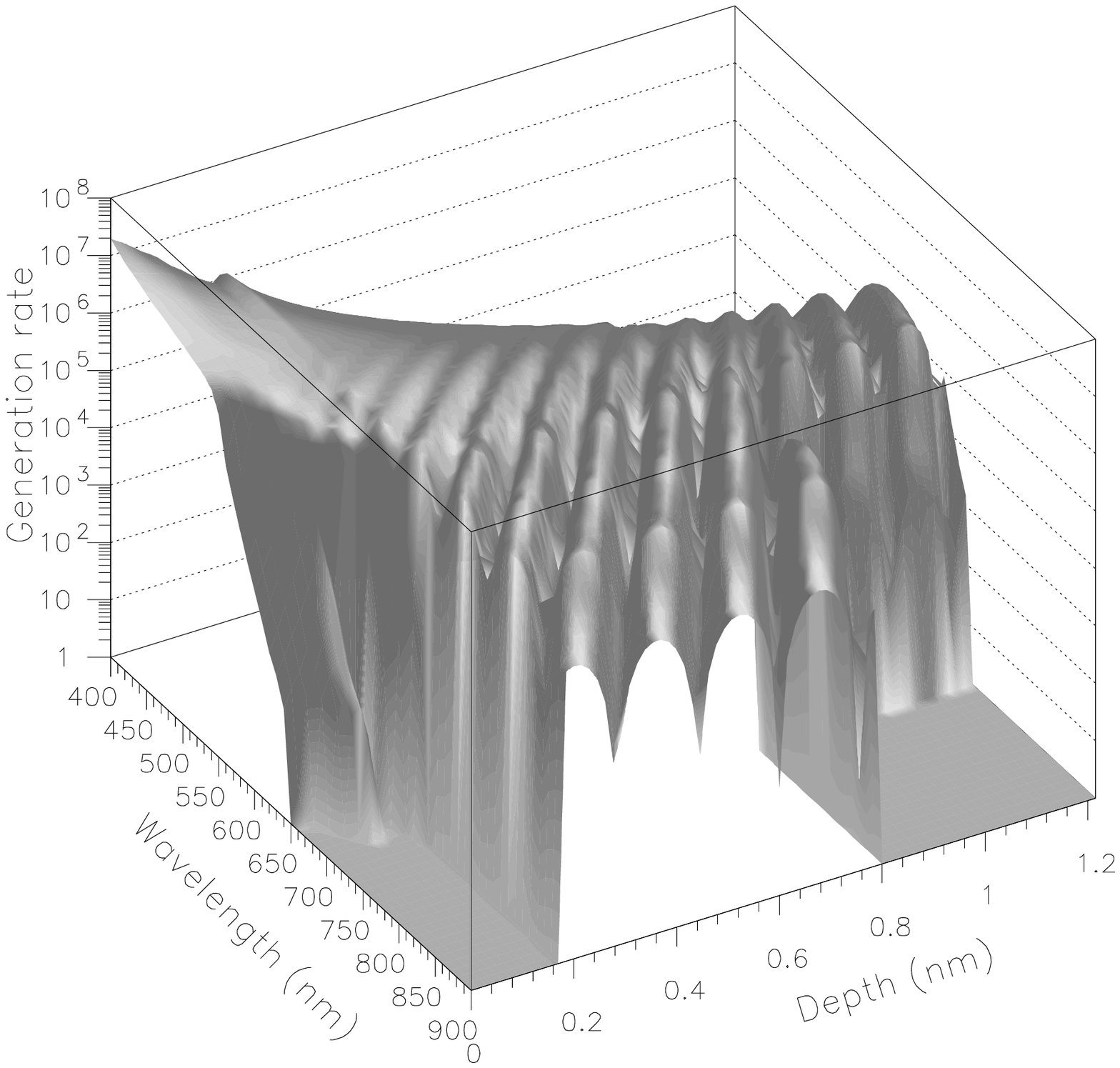}
{Generation rate in a mirror backed QWSC
\label{figmirrorgeneration}}

%
%

\bigeps{htbp}{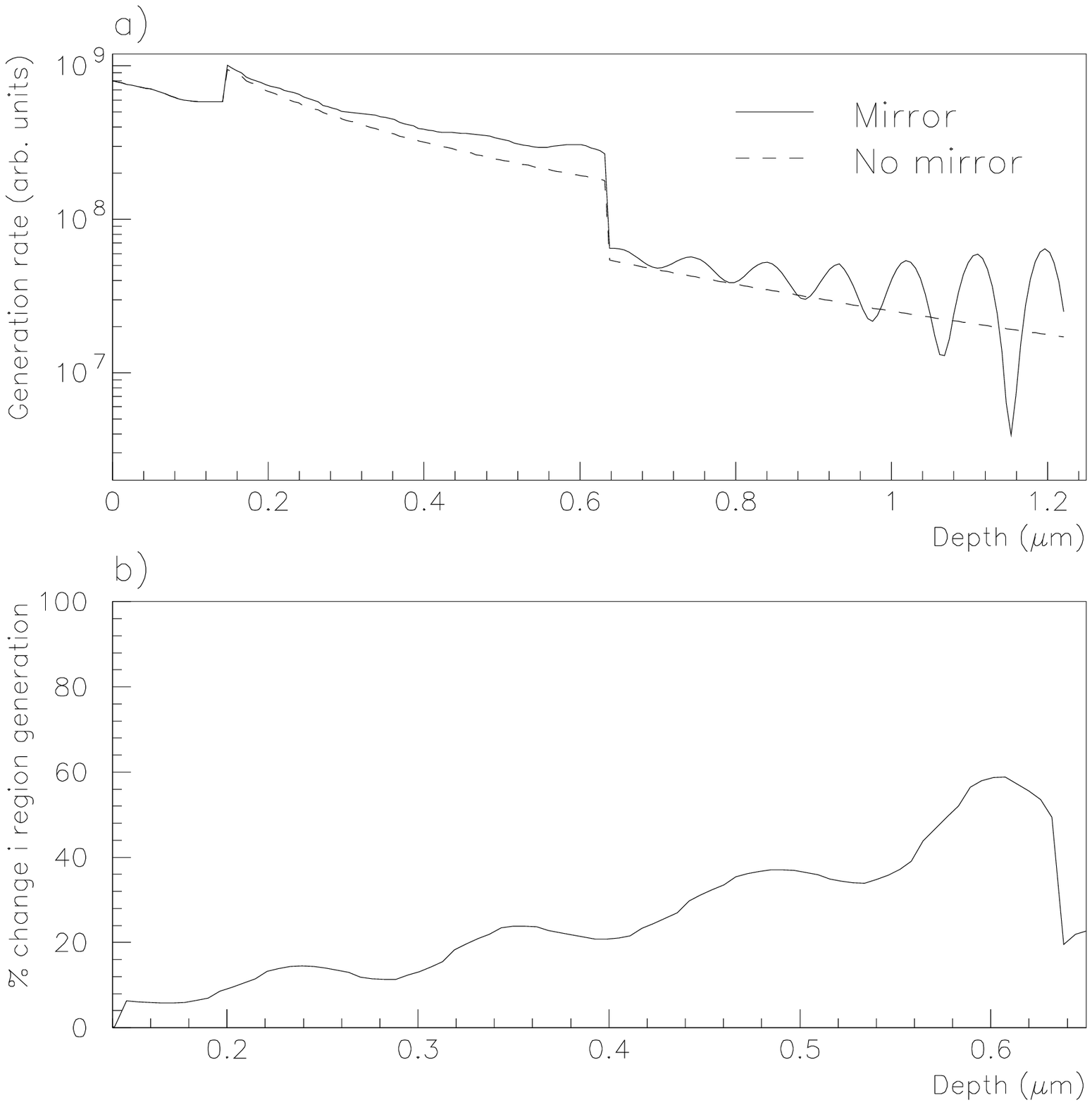}
{Generation rate enhancement a) in all layers of a graded QWSC with
and without a back mirror and b) percentage increase in the intrinsic region.
\label{figmirrorgenratio}}
\bigeps{htbp}
{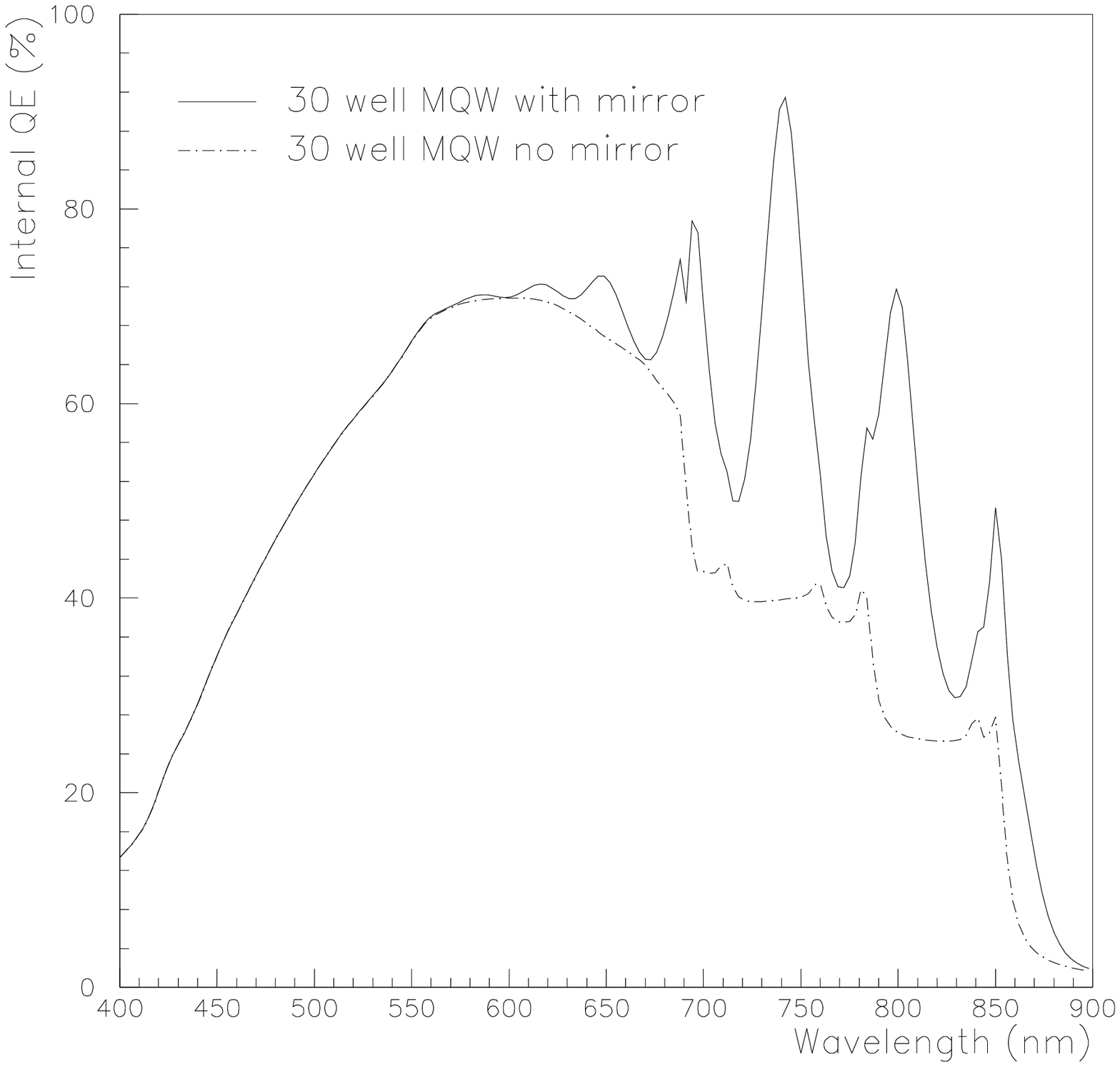}
{Quantum efficiency of a thirty well QWSC with a back mirror compared
to an indentical cell without a mirror.
\label{figmirrorqe}}

\bigeps{htbp}
{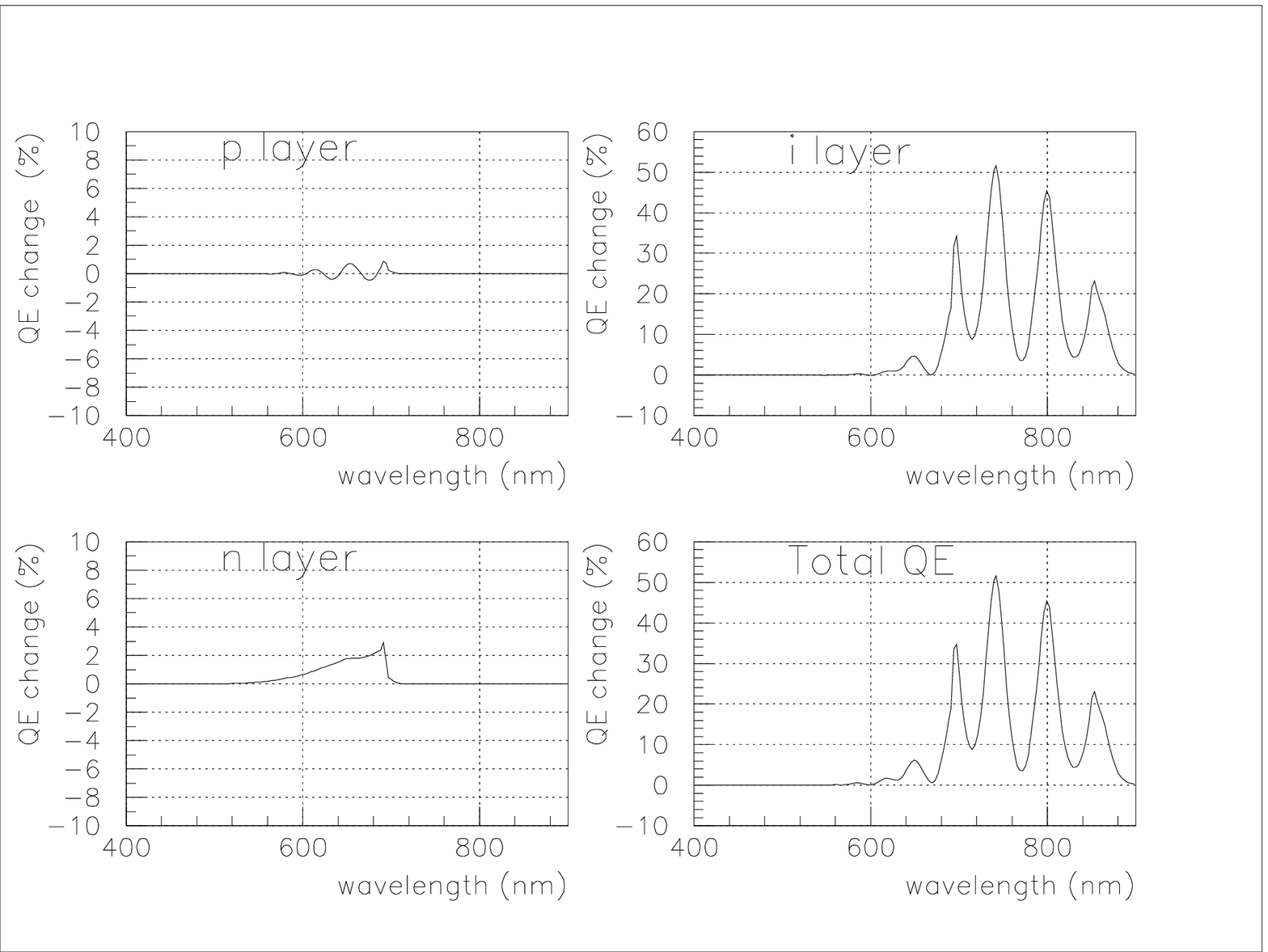}
{Absolute $QE$ increase in a thirty well QWSC with a back mirror compared
to an indentical cell without a mirror.
\label{figmirrorqeincrease}}

Finally, figure \ref{figmirrorqe} shows the $QE$ enhancement 
expected in a 30 well ungraded QWSC with a back mirror, assuming
95\% specular reflection from the back mirror and an internal front
surface reflectivity given by the mean 70nm SiN AR coat. The breakdown
in $QE$ enhancement layer by layer is shown in figure 
\ref{figmirrorqeincrease}. Contributions to the $QE$ increase from
the neutral p and n regions are expected to be negligible.

Enhancement for photon energies above $E_{b2}$ is minimal
since most of the light at these energies is absorbed
during the first light pass.

The wavelength at which the Fabry-Perot oscillations in the
$QE$ spectrum appear are critically determined by the bulk
bandgap $E_{b2}$. This provides a very useful characterisation
tool for determnining the exact value of $E_{b2}$ in QWSC samples

The overall predicted i region enhancement in the quantum
well absorption range is of the order of 25\% for such a device.
This enhancement ratio decreases to about 20\% for fifty well samples
because of the greater light absorption on the first pass.

\section{Conclusions}

In the preceding sections we have reviewed previous work on
$QE$ and \jsc\ enhancement in ungraded cells and explained
the limitations of the optimisation in ungraded QWSCs.

We have separately considered further enhancements above the
barrier bandgap for graded QWSCs and below the barrier bandgap
for any QWSC cell.

Above the barrier bandgap we have seen that the spatial variation
of optical and transport parameters increases the $QE$ 
by reducing minority carrier generation near the surface of
the solar cell and by increasing light transmission to the 
highly efficient i region.

The compositional grade results in improved
carrier collection in a graded p layer because of the presence of
an effective field. The consequent improvement in $QE$ due to this 
mechanism is difficult to predict quantitatively because of the added
complexity introduced by the spatial variation of materials parameters,
and the assumptions made necessary by the lack of knowledge
regarding the position dependence of transport parameters in
particular. We have presented a modelling scheme which uses
previously derived theoretical results in conjunction with published and
measured experimental data to permit modelling of the $QE$ of graded 
layers with a minimum set of free parameters.

The model in this case can only be used as a qualitative
guide to suggest optimal parameters for graded p layers, but
may be improved by comparison with experimental data.

Below the barrier bandgap the \jsc\ from the quantum wells can
be greatly enhanced by coating the back of the QWSC with a mirror.
The large fraction of light which is not absorbed upon a first
pass is reflected back and forth in the cell, increasing light
absorption in the quantum wells and hence the \jsc.

%% file: chap6.tex

\chapter{Device fabrication and characterisation\label{secexperiment}}

The QWSC design is based on a MQW photodiode detector described
by Whitehead \cite{whitehead90}.
Each successive generation of QWSC design passes through design, growth,
processing and characterisation stages. 

\section{Sample Growth\label{secgrowth}}

Successive growth runs are determined by a combination of the
characterisation of earlier samples and optimisations carried
out using the short circuit current model. Sample growth was
carried out using two epitaxial growth techniques, which are 
briefly described below.

\subsection{Metalorganic Vapour Phase Epitaxy (MOVPE)\label{movpe}}

MOVPE growth was carried out by John Roberts at the EPSRC III-V
Semiconductor Facility at Sheffield University using a Quantax 
MOVPE reactor. Details are given in reference \cite{jroberts94}.

MOVPE growth consists in passing reagent chemicals in vapour form
over a heated substrate. The precursor gases dissociate at the
surface of the substrate, depositing atoms at the rate of about
one monolayer per second.

Substrate temperatures in our case varied
between 560C and 600C. For the \algaas\ system, the reagents
were trimethyl gallium (TMG, $\rm Ga(CH_{3})_{3}$), trimethyl aluminium
(TMA, $\rm Al(CH_{3})_{3}$) and arsine ($\rm AsH_{3}$). 

The p-type dopant in early studies was zinc in the form of dimethyl
zinc (DMZ, $\rm Zn(CH_{3})_{2}$). This was later switched to carbon doping
in the form of $\rm CCl_{4}$ to reduce p-type dopant diffusion. 
n-type doping was achieved using disilane ($\rm Si_{2}H_{6}$).
These reagents are transported to the growth chamber in a hydrogen
carrier gas at atmospheric pressure.

The MOVPE growth technique has a number of attractive
features. The most interesting from the perspective of solar cell
production is the high productivity, given that commercial MOVPE reactors
are currently capable of growth on many wafers at the same time.
Moreover, the gas sources used in growth can be replenished without
shutting down the reactor. MOVPE growth is therefore significantly
cheaper than some other epitaxial growth methods.

Compositional control is, in principle, lower in MOVPE than in
MBE and suffers from a lack of in-situ measurement techniques
such as reflection high energy electron diffraction. Accordingly,
growth mechanisms in MOVPE are less well understood than in MBE in
particular. Nevertheless, near monolayer control is routinely achieved
in good MOVPE material.

However, most of the MOVPE material we have characterised suffers
from electrical problems due to high background doping levels.
As we saw in chapter 
\ref{secsolarcells}, a high background doping adversely affects the 
open circuit voltage of the QWSC. Early material showed tolerable
net dopant densities. We shall see in subsequent sections that
a lower silicon background doping density, for example, may in
fact lead to a higher background doping density due to the
loss of a compensation effect for the carbon background.

\subsection{Molecular Beam Epitaxy (MBE)\label{secmbe}}

Molecular beam epitaxy was performed by Christine Roberts in the
III-V Interdisciplinary Centre for Semiconductor Materials in
Imperial College. 

MBE growth consists in firing collimated beams of molecules
at a substrate heated to between 480C and 630C, which is situated in an 
ultra high vacuum growth chamber. 
Collimated molecular beams are generated in heated crucibles,
which contain the precursors in solid or liquid form. Evaporation produces 
the molecular beam. The molecules dissociate at the surface of the 
substrate, depositing the metal atoms. 

Compositional control is achieved by changing the temperature of 
the solid sources. Very sharp compositional changes of the 
order of a monolayer are possible simply by placing a shutter in front of
any given source. The sample is generally rotated during growth to ensure 
uniform growth across the wafer.

Dopants in this study are Be for p-type and Si for n-type
\algaas. As we shall see in subsequent sections, the Be dopant is
prone to significant diffusion at the doping levels and growth 
temperatures used in this study.

This growth method is relatively well understood, since an array of
characterisation tools may be placed in the growth chamber. A review
of these characterisation tools has been written by Foxon
\cite{foxon90}.

The method is complementary to the MOVPE technique. It is inherently
more costly, but can achieve higher purity due partly to the
use of UHV techniques, cryogenic paneling, the wide range of
in-situ characterisation techniques available, and the relatively
well understood growth mechanisms. 

Although the background doping levels in MBE material were generally
low, decreasing material quality with increasing aluminium fraction
has consistently been observed.

\section{Processing\label{secprocessing}}

Processing was carried out by Malcolm Pate at the III-V Semiconductor
facility at Sheffield. This last stage before obtaining finished devices
is critical, since the devices are exposed to potentially damaging
environments.

The most common device configuration is the mesa
photodiode. The early
studies described in reference \cite{paxman92} used both anti-reflection
coated (AR) and non AR coated (NAR) versions of this structure
in order to save processing time on poor material. This
study concentrates on the AR coated devices.

The principal stages of the processing routine are tried and tested,
and involve relatively few processing steps.
Reference \cite{pate91} gives a summary. The principle innovation
is the addition of a back surface mirror. We now outline the main
steps involved in turning a piece of QWSC wafer into the finished
device.

\subsection{Device Geometry\label{secpdandladevices}}

The photodiode devices are $1000 \mu m$ circular mesa structures
defined by photolithography followed by an etch.
Gold ring contacts are then laid down on the mesa, defining the
600 $\mu m$ diameter optical window.

NAR devices retain the protective  $\sim 40nm$ thick 
\gaas\ capping layer which is grown on the front surface of the cell
to prevent \algaas\ oxidation and surface damage in handling. 
AR devices are subjected to two further 
steps. The first of these is the removal of the GaAs capping layer 
with a selective etch which stops short at the \algaas\ window,
which acts as an etch-stop. A reliable etch stop requires an 
aluminium fraction of at least 30\% Al.

Following the removal of the cap, a SiN AR coat of thickness $\sim 70$\AA\
is deposited onto the optical window. This step involves a brief exposure 
of the unprotected device to air, followed by device exposure to a 
silane/nitrogen plasma deposition machine at about 300C. SiN is deposited
at a rate of approximately 50nm per minute.

Where possible, a piece of unprocessed QWSC wafer is AR coated
at the same time as the devices, since these are too small
to allow accurate measurement of the reflectivity. Such pieces
of wafer were received for about half the samples processed.

Design and processing of large area devices follows similar
steps and is described in \cite{guido95}.

\subsection{Mirror Backed Devices\label{secmirrordevices}}

The addition of a back mirror to the QWSC devices was first tried
with mesa photodiode devices. The initial steps are identical to
those listed in the previous section. In addition to these, however,
the light absorbing GaAs substrate is removed only under the optical window 
with a selective etch which uses the n layer as an etch-stop.

This leaves a $\sim 300 \mu m$ hole of diameter $\sim 600 \mu m$. The
gold metal constituting the mirror is then deposited in
this hole, in direct contact with the back surface of the n layer.

\subsection{Effects of Processing\label{secprocessingeffects}}

Exposure of QWSC wafers to processing leads to unavoidable
consequences in device performance. Effects which are directly
related to processing include the wavelength optimisation of the
AR coat and the electrical characteristics of the solar cells.
The former may be improved by tuning the thickness of the AR
coat to suit a particular cell, whereas the electrical characteristics
such as series and parallel resistance may be optimised by the
choice of appropriate alloys and mask sets at the contacting
stage.

Other processing effects are accidental and relate to the
internal quantum efficiency of the devices. The removal
of the cap has been known to cause problems in some MBE
material for reasons as yet unknown, leading to a poor
window surface. Malcolm Pate has reported difficulties with
the etch stop in some cases. This may
be due to poor material quality at the surface, or an
unexpectedly low aluminium mole fraction at the front surface.

Another source of damage is oxidation following the removal of
the protective cap and subsequent exposure of the topmost \algaas\
layers of the cell to air. Since these \algaas\ layers are particularly 
susceptible to oxidation, this exposure must be as brief as possible 
to avoid oxygen diffusion through the thin \algaas\ window to the 
p layer beneath.

A general decrease in device quality is generally seen as the
number of processing steps is increased. This is particularly
visible in mirrored samples which show variable device quality.
Surface contamination is frequently observed in these samples. 

On the other hand, we shall see
in subsequent sections that there is evidence to suggest that the
AR coating stage has a passivating effect, leading to improved
electrical characteristics in forward bias for initially poor
devices. High quality samples on the other hand are generally 
degraded by the AR coating process.

We conclude that the processing stage unfortunately introduces
a variability in the characteristics of the finished QWSC devices.
This variability is difficult to ascribe to any given cause, even
with the limited number of steps involved.


\section{Characterisation\label{seccharacterisation}}

Experimental data were obtained using equipment designed
and built by Mark Paxman and described in reference \cite{paxman92}. 
The two routine characterisation tools are photocurrent (PC) spectroscopy,
monochromatic current-voltage (MIV), and  measurements of
the anti-reflection coating (AR measurements).

Other characterisation techniques carried out by
collaborators include transmission electron microscopy (TEM) and
secondary ion mass spectroscopy (SIMS).

\subsection{Photoconductivity Experiment\label{secphotocurrentexp}}

QE, MIV and AR measurements were all carried out on the photoconductivity 
rig which has been extensively covered in references
\cite{paxman92}, \cite{barnes94} and \cite{zachariou96}.

Illumination is provided by a tungsten-halogen lamp. The light is passed 
through a grating monochromator which is capabable of
wavelength resolution of the order of $0.3nm$ (reference \cite{paxman92}).
Two power supply units (PSUs) enable the device to be driven to reverse
or forward bias. 

A pinhole placed at the exit slit of the monochromator transmits light
to a collimator. The image of the pinhole is then projected on to the
surface of the device. The diameter of the pinhole must be made smaller 
than that of the optical window in order to ensure that no light is lost
by reflection from the contacts. The light power incident on the
device is usually of the order of 100nW.

The light beam is modulated by a chopper wheel at $\sim$ 180Hz.
A lock-in amplifier is then used to measure the resulting photocurrent.
The use of lock-in techniques ensures that the DC background 
due to ambient light and the dark current is removed from the 
measurement.

The incident power is measured by a calibrated silicon photodiode 
placed at the same position as the QWSC. This additional calibration
measurement can then be used to compute the QE as a function of
wavelength.

The experiment is controlled by a computer running a Turbo Pascal
program written by Mark Paxman. It should be noted that
measurements at short wavelengths are prone to large errors
because of the low reponse of the Si photodiode in this
wavelength range. The error over most of the wavelength
range has been quoted as $\sim \pm 5\%$ in reference \cite{paxman92}.

Measurements performed on this rig are AC and are designed to
eliminate DC signals due to the dark currents of both the
solar cell devices and the detector. The resulting measurement
gives the photocurrent, as opposed to the DC light current
which includes a dark current component.

\subsection{Monochromatic current-voltage characteristics
\label{secmivexp}}

Monochromatic IV (MIV) measurements consist of measuring
photocurrent as a function of applied bias at a fixed
wavelength. The measurement is normalised to 100\% at the
voltage independent saturation level reached in reverse bias.
This technique provides information on the quality of both growth
and processing. 

Measurements are normally carried out above the
bulk band-edge in order to prevent distortion of
the MIV with voltage because of the modification of quantum
well absorption with Stark shift. 
Below the bulk band-edge, full $QE$ measurements (see section 
\ref{secQEexp} below) as a function of bias must be taken in 
order to identify a wavelength range in which the $QE$ is relatively 
insensitive to the Stark shift.

Typical MIV curves for good and moderate devices are shown 
in figure \ref{figexamplemivs}. The "square" curve, showing a  
MIV which starts to fall off at zero bias was measured
for a 30 well MBE grown ungraded QWSC with $X_{b2}\simeq 20\%$. 
The second curve showing a MIV which starts to decrease at -4V 
represents data for an identical sample grown in the same series
with $X_{b2}\simeq 30\%$.

\bigeps{tbp}{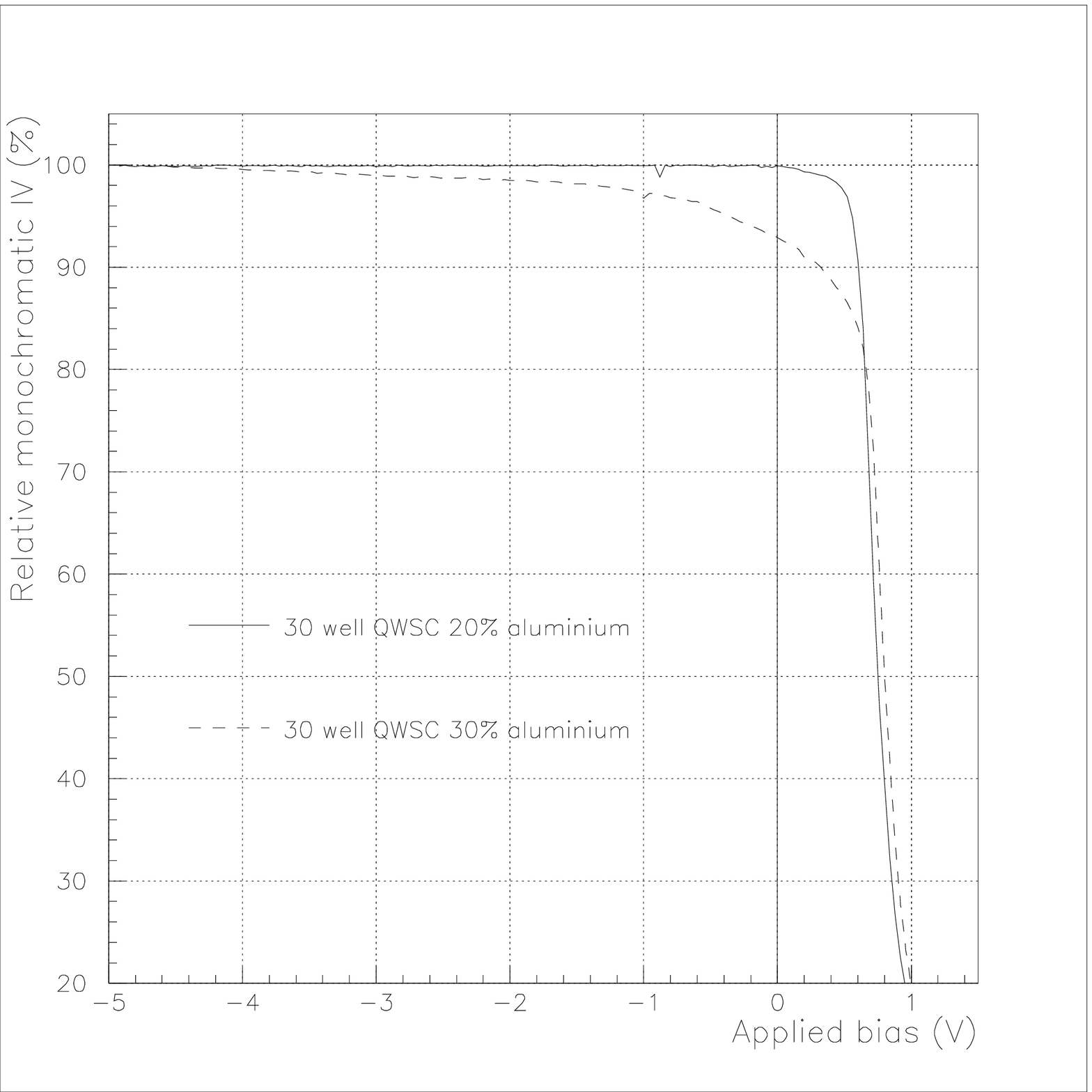}
{Typical MIV curves for good and moderate 
solar cell devices.\label{figexamplemivs}}

Good devices such as the 20\% sample maintain a constant 
photocurrent into forward
bias. At a moderate forward bias, the signal 
falls abruptly. A cut-off occurring in forward bias may be due
to the failure of the experimental technique, since the lock-in
amplifier begins to detect the increasingly large AC variation
in the dark current caused by the chopped light. Details of these
problems have been given by Paxman \cite{paxman92} and Barnes 
\cite{barnes94}.
In reverse bias, the MIV remains constant until the breakdown voltage,
not shown here because the MIV measurement does not measure the dark
current. At this point, the dark current increases very rapidly. 
Detection of the photocurrent against this background becomes difficult, 
and the resulting data noisy.

Devices with significant background doping levels such as the 30\% device 
show a photocurrent which falls off at voltages well below the
$\rm V_{oc}$. This current drop is due not
to the increasing influence of the dark current, but to the decrease 
in field across the intrinsic region, and subsequent failure of the 
\pin\ structure to operate as a solar cell.

With increasing reverse bias, the PC of such cells 
reaches a maximum. We saw in section \ref{secniproblem} 
that the voltage at which the drop in photocurrent occurs can be used to 
estimate the background doping. The method however becomes inaccurate 
for devices with doping levels in excess of $\rm \sim 10^{16}cm^{-3}$
and in samples which do not conform to the abrupt \pin\ doping profile
assumed in the model.



\subsubsection{Device quality}

Many devices of lower quality depart significantly from the picture
described. This may be due either to growth or processing problems.

Samples with doping profiles which do not conform to the sharp
\pin\ doping profile assumed in the previous discussion will
behave in a more complex manner. If for example the voltage is
varied from negative to positive bias, the p and n side
depletion widths will vary in a manner which reflects the
doping profiles on both sides. An estimate of this profile
may only be obtained if the one is much greater than the other.

In practice, this method cannot be exploited further because
such variations may also result from device history.
Difficulties in processing in particular may lead to electrical 
problems. This may be manifest as a very noisy MIV measurement 
or as the absence of a plateau in reverse bias allowing the curve 
to be normalised to 100\%.

We conclude that the MIV measurement is useful in identifying
promising devices and furthermore indicates whether the devices
conform to the idealised structure assumed by the modelling
programs.

\subsection{Quantum efficiency measurements\label{secQEexp}}

Measurements quoted in chapter \ref{secmivar}
are always taken at voltages for which the i region of the cell 
is fully depleted, in order to maintain the field across the i region
and fulfill the assumptions in the model. Although the PC spectrum
in the wells moves with applied bias due to the Stark shift, this
dependence is quoted by reference \cite{jennynint1096} as being 
of the order of meV for typical biasses of up to a 
few volts  and shifts the quantum well absorption edge by only a 
few nm for the moderate biasses used in the measurements.

The MIV technique of the previous subsection provides a useful
guide to the voltage at which the QE measurement should be taken.

\subsection{Reflectivity measurement\label{secreflectivityexp}}

The last routine characterisation tool is the reflectivity
experiment. A slight modification of the experimental setup
described above is required. As mentioned in section
\ref{secprocessing}, pieces of wafer are generally AR coated
at the same time as the photodiode devices. Direct reflectivity
measurements are possible on large area devices.

The reflectivity measurement consists of placing the AR coated sample
at the position normally occupied by a device. Light reflected from
the surface is collected and measured by the Si photodiode at as low
an angle as possible. The sample is then removed, and the
incident power measured directly by the Si photodiode, the
optical path length being kept constant.
The reflectivity is then given by the ratio of the two signals.
Beam angles are typically below 5 degrees. Measurements
were repeated on different dates and a maximum error of about
5\% in reflectivity at long wavelengths was found. 

A good knowledge of the reflectivity is required
for accurate modelling of QE spectra. For some photodiode devices
however, pieces of AR coated wafer were unavailable. In
this case, we are forced to use an average of previous
measurements for the AR coating method used. A significant
error is introduced in this case, which may be qualitatively
reduced by examining the devices under a light. Typical AR
coats are a characteristic blue colour.

Reflectivity measurements give additional indications of
processing or growth problems. This is typified by a large
variation in reflectivity over the AR coated sample.

\subsection{External characterisation}

In addition to the experiments described above, more sophisticated
characterisation techniques were applied to a selection of structures
by collaborators.

\subsubsection{Transmission electron miscroscopy (TEM)}

TEM measurements were carried out by Professor Grunbaum
in the Materials Department of Oxford
University.

Transmission electron microscopy consists of firing a collimated
electron beam through a thinned cross section of a sample. 
The electron beam is diffracted by the crystal lattice, and the
diffraction pattern recorded on a fluorescent screen or on
photographic film. Details of this technique are given in the
review articles by Holt \cite{holt90} and Hutchinson 
\cite{hutchinson90}.

Resolution of the order of 0.15nm is possible on state of the art 
microscopes. Compositional contrast is possible because of the
different electron beam scattering factors for \algaas\ with different
values of $x$. TEM micrographs can therefore give unprecedented direct
measurement of the physical dimensions of QWSC samples. 
Measurements were carried out on selected samples to verify the
dimensions of the quantum wells in particular, and of the larger
scale p layer thickness.

\subsubsection{Secondary ion mass spectroscopy (SIMS)}

SIMS measurements were carried out by Federico Caccavale of Padua
University. 

The technique is described in detail by Clegg \cite{clegg90}.
This measurement involves firing 
a collimated beam of ions at the sample. The high energy
incident beam removes atoms from the first few monolayers
of the sample. The flux of these sputtered atoms is then
measured by a mass spectrometer.

The incident beam progressively digs a pit in the surface of the
sample, providing compositional information as a function of
depth. The technique must be tuned by choice of incident atomic
beam and mass spectrometer setting to detect the spatial variation
of different elements. The focus in the two measurements carried
out for this project was the variation of doping density and
aluminium composition with depth.

Depth resolution is dependent on factors including the
type of atom used in the incident beam and the beam energy.
Use of oblique incident atomic beams and non-normal incidence
to prevent mixing at the surface of the sample can achieve
resolutions of the order of a few nm.

\section{Conclusion}

In this chapter we have described growth, fabrication and 
characterisation of QWSC devices. We have seen
that the behaviour of the finished devices may depart 
significantly from the theoretical picture given in
previous chapters. Both growth and processing may contribute
to this deviation.

The most readily identifiable problem is that of background
doping. This is revealed by MIV measurements, which are also
used to suggest an appropriate bias at which to measure
the QE.

Although QE measurements may be taken at zero bias in
good material, it is frequently necessary to force the
device into reverse bias to ensure that the approximations
in the model are consistent with the experiment. 

Reflectivity measurements are essential to a correct analysis
of the $QE$ of a device. In some cases such measurements had to
be extrapolated from an average over previous AR coating runs.

Apart from these routine experimental tools, we have reviewed
the SIMS and TEM characterisation techniques which were carried
out by collaborators. These additional characterisation methods
may be used to verify or adjust the geometry and composition
of QWSC samples.

%% file: chap7.tex
\chapter{Preliminary Characterisation\label{secmivar}}

In this chapter we present results of the MIV and reflectivity
techniques reviewed in chapter \ref{secexperiment}.
These are the preliminary characterisation tools
used to make an estimate of the device quality. The
reflectivity furthermore is required for subsequent analysis
of quantum efficiency data.

The samples in this chapter are referred to by their growth
name, and relevant design characteristics are given in the
text. 


\section{Monochromatic IV Measurements\label{secmivmeasurement}}

A good device is characterised by a 
MIV curve which remains constant into forward bias, before
showing an abrupt cut-off for reasons given in chapter
\ref{secexperiment}. If this cut-off occurs at forward
biasses greater than about 0.6V, attribution is difficult
because of experimental considerations.

If the cut-off occurs at reverse bias and the acceptor, 
net background and donor doping levels $N_{A}$, $N_{BG}$ and 
$N_{D}$ layers are abrupt, the drop in photocurrent
can be related to the net background doping level $N_{BG}$,
as described in chapter \ref{secsolarcells}. 

\begin{table}
\begin{center}
\begin{tabular}{|c||c|c|c|c|c|c|} \hline
$V_{s}$ (V)& 0 & -2 & -4 & -6 & -8 & -10 \\  \hline
$N_{BG}$ ($10^{16}cm^{-3}$) & 1.0 & 2.0 & 3.0 & 4.0 & 5.8 & 5.9 \\ \hline
\end{tabular}
\figcap{Predicted value of net background doping $N_{BG}$ for a range
of MIV cut-off voltages $V_{S}$, for a standard 30 well QWSC sample with
$X_{b2}=30\%$\label{tblnbgpredictions}}
\end{center}
\end{table}

Calculations of $N_{BG}$ for a range of cut-off voltages are shown in
table \ref{tblnbgpredictions}. The variation of predicted
$N_{BG}$ is only weakly dependent on the cut-off voltage
$V_{S}$. The technique can therefore only provide
a rough estimate of $N_{BG}$, because the values
of $V_{S}$ generally vary only within a few volts.
A further limitation is introduced by the poorly understood 
electrical influences on the MIV characteristic, and by the fact 
that the abrupt doping assumption is frequently erroneous, as
shown by SIMS measurements in chapter \ref{secqeresults}.
The position dependence of the p layer doping in particular
has been investigated by SIMS in both MOVPE and MBE samples.
This position dependence introduces a further voltage dependence
which  makes the precise attribution of the MIV voltage shoulder
difficult. Furthermore, the shoulder is generally imperfectly
defined in poor samples, and may be subject to errors of
several volts.

Jenny Nelson has developed a model, in collaboration with
Alex Zachariou, which estimates the background doping by
modelling the $QE$ at zero volts, and using $N_{BG}$ as the
main fitting parameter. Varying $N_{BG}$ in this model
alters the effective p, i and n layer thicknesses.
The p layer in this case is modelled as two undepleted
layers with doping levels $N_{A}$ and $N_{BG}$ respectively.
This approach allows a more quantitative estimation of
$N_{BG}$, but nevertheless still assumes square doping 
profiles.

The model presented in this thesis does not use this approach.
The MIV data in the following discussion is principally used as
an indication of material quality. In particular, we will use
it to qualitatively assess those samples in which $N_{BG}$ is large
enough to degrade device operation. We further use the MIV
data as indicative of the reverse bias at which to take $QE$
measurements.

MIV data for QWSC devices grown for this project is generally
poorer than the corresponding data for the \pin\ controls.
We now examine this trend as a function of growth method
and describe the effect of changing growth conditions on
this problem.

In the final part of this section we look at the effect of the
AR coating process on the MIV characteristics of the samples,
and compare these results with data from a passivation study
carried out at Eindhoven University of Technology as a result
of these observations.

\subsection{Trends in QWSC and \pin\ Device MIV Data}

\subsubsection{MOVPE Devices\label{secmovpedevices}}

\smalleps{tp}{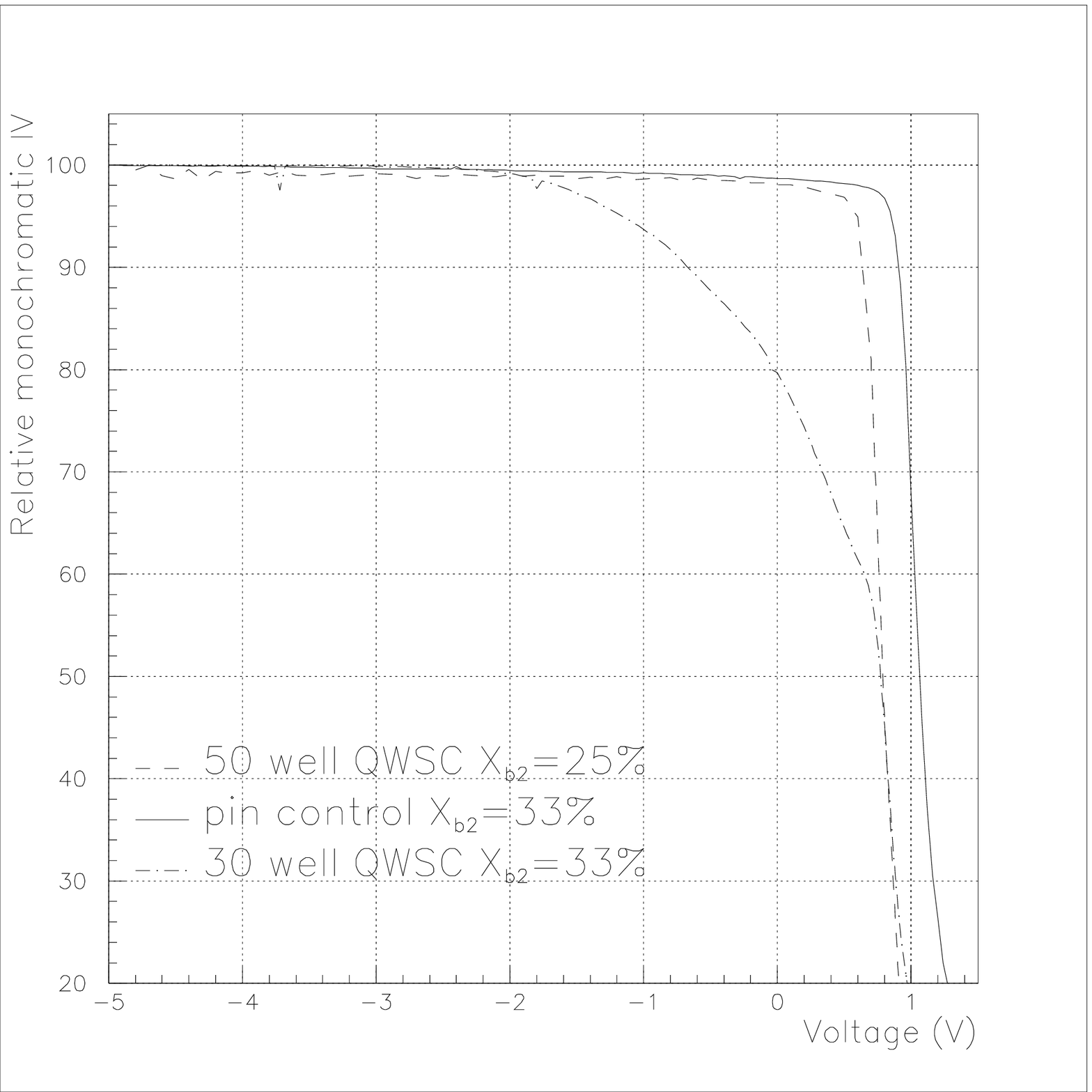}
{MIV data for an ungraded MOVPE grown 30 well QWSC and \pin\ 
control pair with $X_{b2}=33\%$ aluminium before AR coating,
showing deterioration of MIV in QWSC devices\label{figpinqwmovpemiv}}

Figure \ref{figpinqwmovpemiv} shows MIV data for the MOVPE grown 30 
well QWSC and \pin\ control pair QT468a and QT468b. The
nominal aluminium composition is $X_{b2}=33\%$. Neither device in 
this example is AR coated in order to minimize the variability
introduced by processing. Also shown is data for a 50 well QWSC
grown previously on a different MOVPE reactor.

The MIV data for the 30 well QWSC device shows a voltage 
independent plateau in reverse bias. A cut-off is seen at about
-2V. Assuming that this current decrease is due to a loss of
field in the i region as explained in chapter \ref{secsolarcells},
we can estimate the net background doping as 
$N_{BG}\simeq \rm 2\times 10^{16}cm^{-3}$. The net background doping
for our MOVPE material is p type residual carbon \cite{jroberts}.

The curve for the \pin\
device is constant up to the point at which the measurement becomes
unreliable for reasons given in chapter \ref{secexperiment}. The
MIV cut-off in this case cannot therefore reliably be attributed to 
a loss of field in the i region. The cut-off 
point at about 0.8V however allows us to estimate the upper 
limit of the net background doping as $N_{BG}\leq \rm 5\times 
10^{15} cm^{-3}$.

The 50 well device MV370 of figure \ref{figpinqwmovpemiv} was grown 
on the Sheffield MR100 machine, which was subsequently removed
from the \algaas\ growth program. The high quality MIV of this
50 well sample shows that QWSC devices with a large number of
wells and a low net background doping can be grown by MOVPE.
Data for other high quality MOVPE grown QWSC samples of this 
type have been published \cite{paxman93}. The data in this case
again only allows us to set an upper limit estimate on the 
net background doping, which is $N_{BG}\leq \rm 2\times 10^{15}cm^{-3}$.

John Roberts \cite{jroberts} has suggested that the systematic
difference in MIV seen between \pin\ and MQW samples may be
caused by a lack of self compensation between oxygen and carbon
impurities in samples with multiple quantum wells due to reduced
oxygen impurity levels in the quantum wells. This
leads to a higher than expected p type carbon background.
Moreover, the difference in net background doping levels
on either side of a quantum well barrier interface is expected
to lead to poor interface quality and high interface recombination
velocities.

John Roberts has further suggested that \algaas\ material will 
have poor minority carrier transport properties despite the low net 
background doping levels, because of the high impurity levels. We shall see
in subsequent sections that the \algaas\ minority carrier transport
properties decrease with increasing aluminium content,
even though the net background doping levels observed in \pin\
structures such as that of figure \ref{figpinqwmovpemiv} appear low.

\smalleps{htbp}{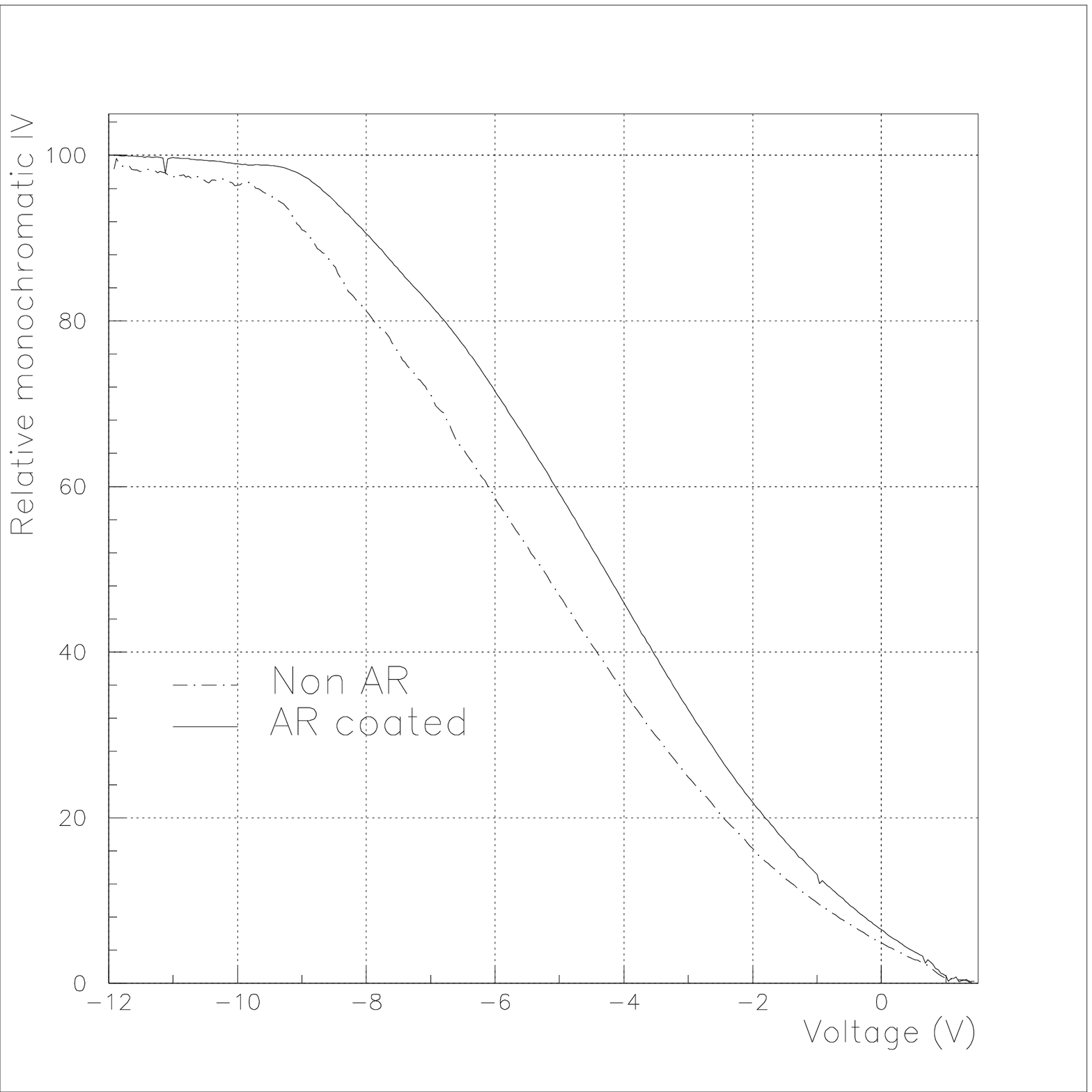}
{MIV measurements for optimised device QT794 for NAR
and AR coated devices\label{fig794mivs}}

Very poor MIV characteristics have been recorded for some MOVPE
samples. Figure \ref{fig794mivs} shows the MIV data for optimised
sample QT794a. This 50 well sample has no window, and a thin p layer
nominally graded from $X_{b1}=40\%$ to $X_{b2}=30\%$. The MIV data 
in this case begins to drop off at -9.5V, suggesting 
$N_{BG}\simeq 3\times 10^{16} \rm cm^{-3}$. SIMS data taken by
Federico Caccavale were aimed at verifing sample geometry, composition 
and intentional doping profiles. The p type carbon background 
doping sensitivity was $\rm \sim 10^{17}cm^{-3}$ and was unfortunately too 
insensitive to verify this value of $N_{BG}$.

Further discussions
concerning this sample in the $QE$ modelling section of this
chapter show that the minority carrier efficiency in this sample
is severely reduced, as suggested by John Roberts.

\subsubsection{MBE Devices\label{secmbemivs}}

QWSC and \pin\ samples grown by MBE show a similar trend to MOVPE
samples discussed in the previous section. Figure \ref{figqwscival}
shows the MIV data for three ungraded 30 well QWSC samples grown and
processed in the same run. The nominal composition of these samples was
$X_{b2}=20\%$ for U2027, $X_{b2}=30\%$ for U2029 and $X_{b2}=40\%$ for U2031.
Similar data for their respective controls U2028, U2030 and U2032 which 
have matching composition but no wells is shown in figure
\ref{figpinival}. 

\smalleps{hbp}{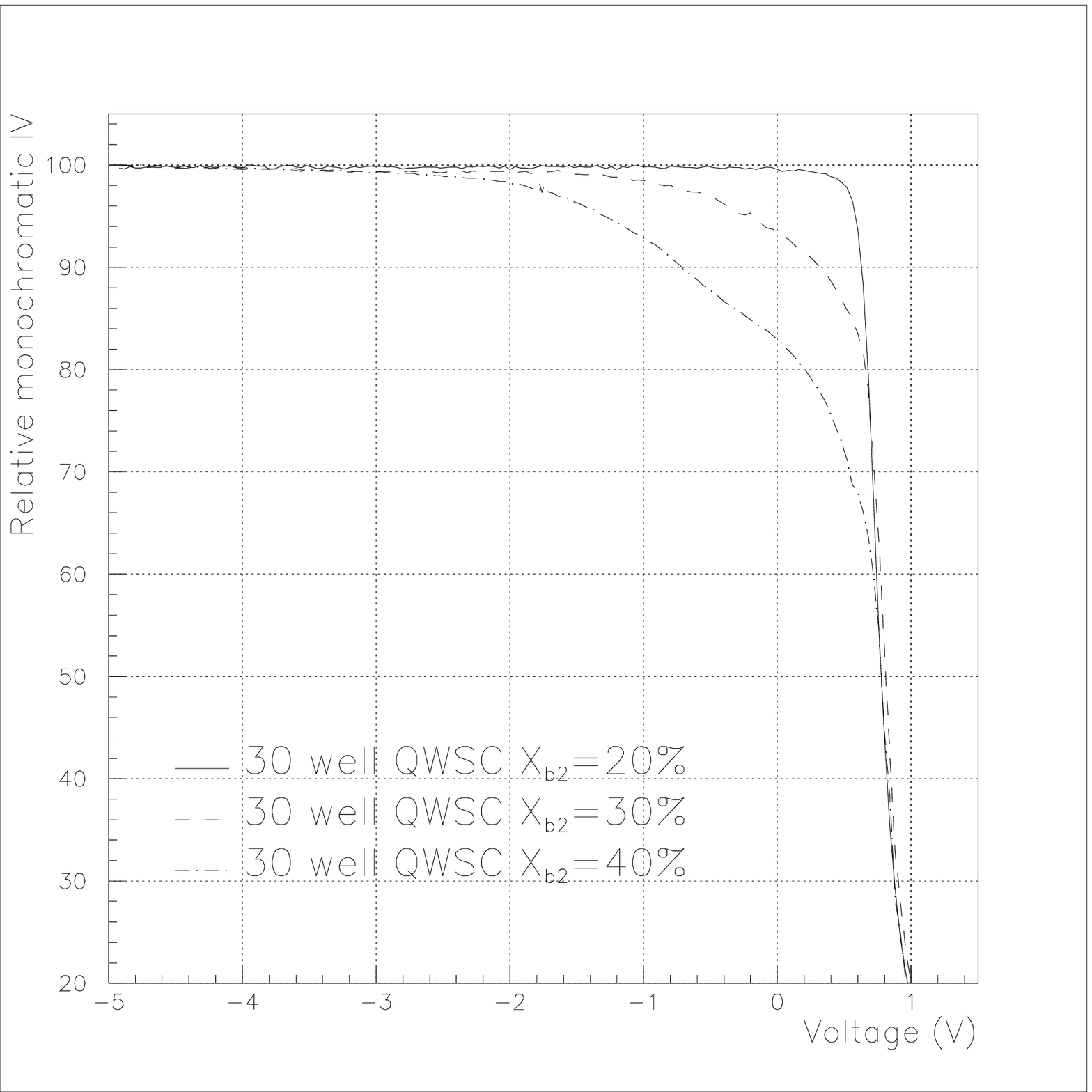}
{MIV data for a series of 30 well QWSC ungraded devices grown by MBE
with nominal aluminium fractions $X_{B2}$=20\%, 30\% and 40\% 
\label{figqwscival}}
\smalleps{htp}{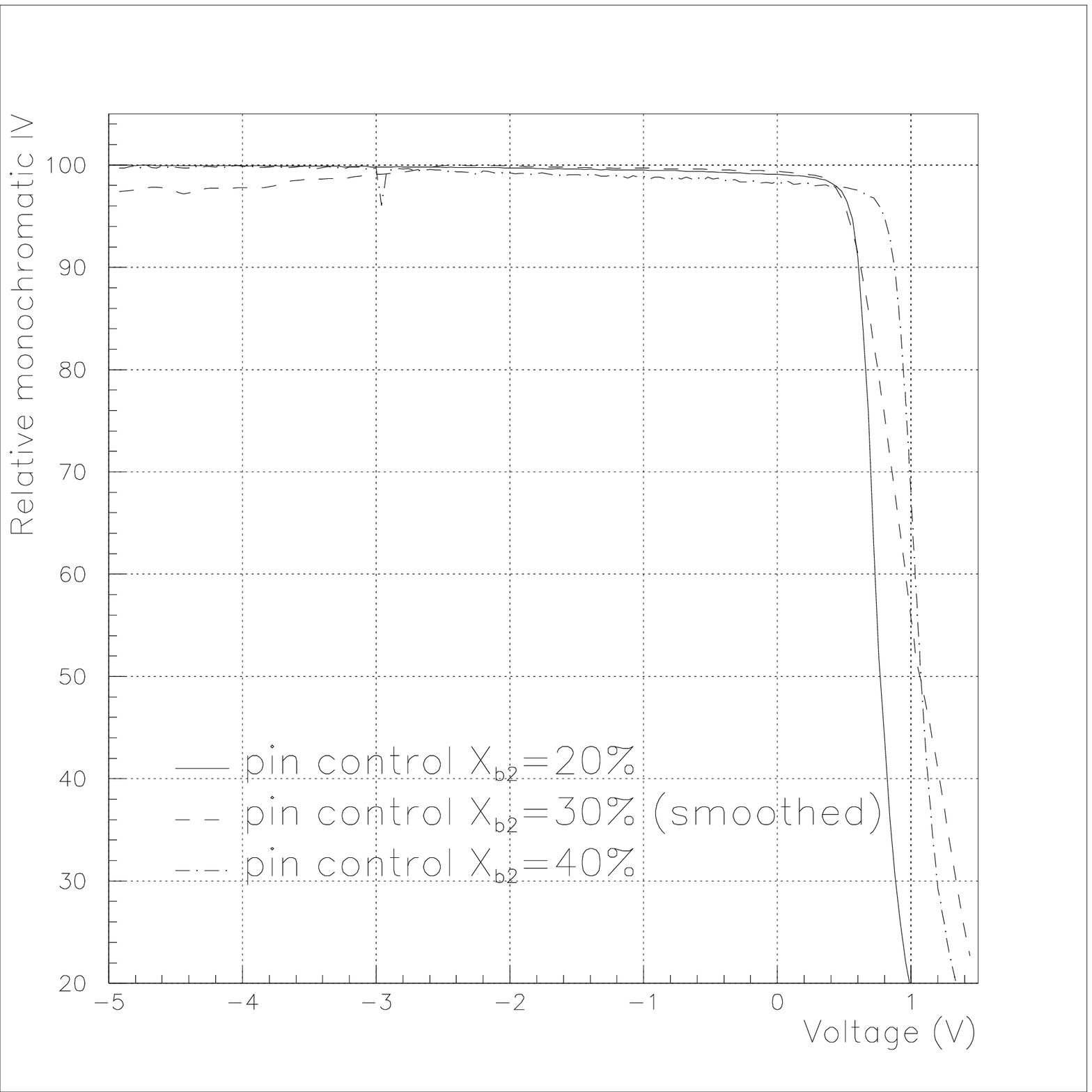}
{MIV data for a series of \pin\ ungraded devices grown by MBE
with nominal aluminium fractions $X_{B2}$=20\%, 30\% and 40\% 
\label{figpinival}}. 

The data for the 20\% \pin\ U2028 has 
been smoothed to eliminate noise spikes.
We note that the MIV data at 20\% are nearly ideal in both
QWSC and \pin\ cases with estimated limits on the net background
doping of $N_{BG}\leq \rm 7\times 10^{15}cm^{-3}$. 
The same limit is found for the 30\% and 40\% aluminium \pin\ samples.

The QWSC MIV data however become increasingly
poor as the aluminium content is increased, whereas the \pin\ control
MIV data remain good. Estimates of the net background doping yield values
of $N_{BG}\simeq \rm 2\times 10^{16}cm^{-3}$ for the 30\% and 40\% 
QWSC structures. Although the two curves are significantly different,
the MIV cut-off method of determining $N_{BG}$ is not sufficiently
sensitive to differentiate between the net background doping levels in
the two samples. A model such as Jenny Nelson's model which was mentioned
in the introduction to this section would however be more sensitive
to $N_{BG}$ in this case.

\smalleps{tp}{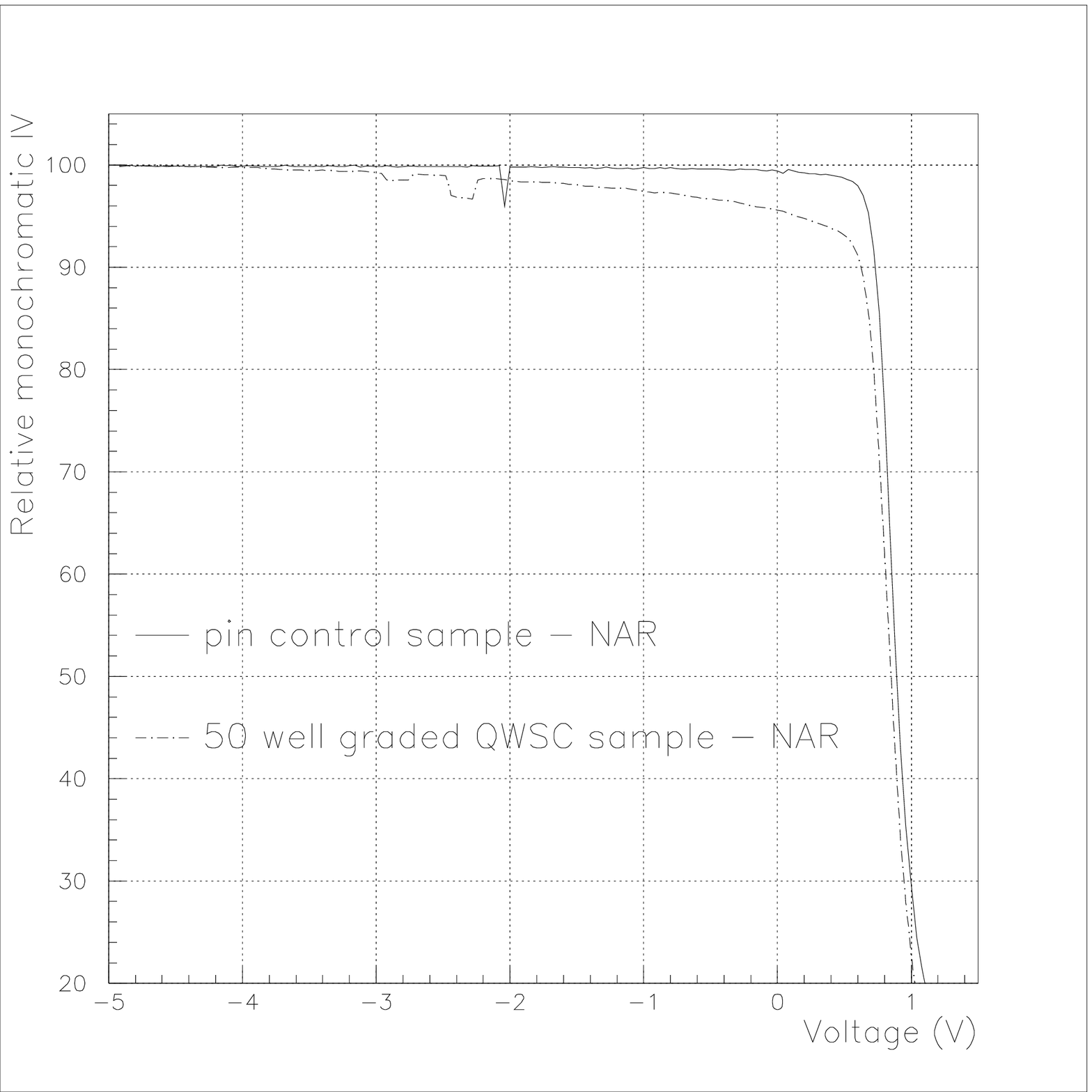}
{MIV data for the 50 well graded sample U6041 and its \pin\
control U6045, indicating low net background doping \label{figgoodmbemivs}}

Finally, figure \ref{figgoodmbemivs} shows MIV data for NAR devices 
processed from \pin\ and 50 well QWSC structures grown in the latter
part of this project. Although the nominal lower aluminium fraction
$X_{b2}$ for these samples was $30\%$, modelling in both cases
indicates a much lower $X_{b2}\simeq 24\%$. Similar behaviour
is again apparent, with the \pin\ sample performing better
than the QWSC.

In this case, however, a phenomenological difference with
the 30 well samples of figure \ref{figqwscival} can be seen.
Instead of a steady decrease from the cut-off voltage
seen in reverse bias for the 30 well samples, the current for the
50 well device of figure \ref{figgoodmbemivs} decreases slowly from
about -3V, but is still at about 95\% of the reverse bias maximum
at zero bias. Furthermore, SIMS data  presented
in chapter \ref{secqeresults} indicates diffusion of the Be p type 
dopant from the p layer into the i layer. This leads to an 
exponentially decreasing Be doping profile in the i layer. 
This means that the depletion width in the p layer will begin to 
decrease more rapidly with an applied bias varying in the positive 
direction, as the depletion edge scans through regions of the p
layer with exponentially decreasing doping. 

We therefore ascribe the slow decrease in MIV from 100\% at
-3V to about 95\% at 0V to the inhomogeneity in the doping
profile, and the fact that it does not match our assumptions
regarding square doping profiles.

The voltage cut-off at about 0.6V is then too low to allow us
to reliably suggest an absolute value of $N_{BG}$. However,
we can again set an upper limit which is $N_{BG}\leq 2\times 10^{15}$.
This agrees with the upper limit set by the SIMS measurements.
Net background doping levels of this magnitude are sufficiently
low to allow efficient solar cell operation for 50 well QWSC
designs.

Previous studies presented in Mark Paxman's thesis 
\cite{paxman92} include measurements of MBE material grown by
Tom Foxon at Phillips Research Laboratories in 1990.
This is sample G951, which was a 50 well QWSC with
an aluminium fraction of $X_{b2}=33\%$, and was shown
to have a net background doping level of
$\rm N_{BG}\simeq 3\times 10^{14}cm^{-3}$.

As in the case of MOVPE material, we conclude that the
background doping problem is not unsurmountable.

\subsection{Growth Optimisation Studies}

\subsubsection{MOVPE Ga Precursor Optimisation}

An attempt was made to reduce the background impurity
concentration in MOVPE samples in collaboration with
John Roberts of the III-V Semiconductor Growth Facility
in Sheffield. This involved comparing device quality
before and after an improved TMG source was used for
the MOVPE growth.

\smalleps{tp}{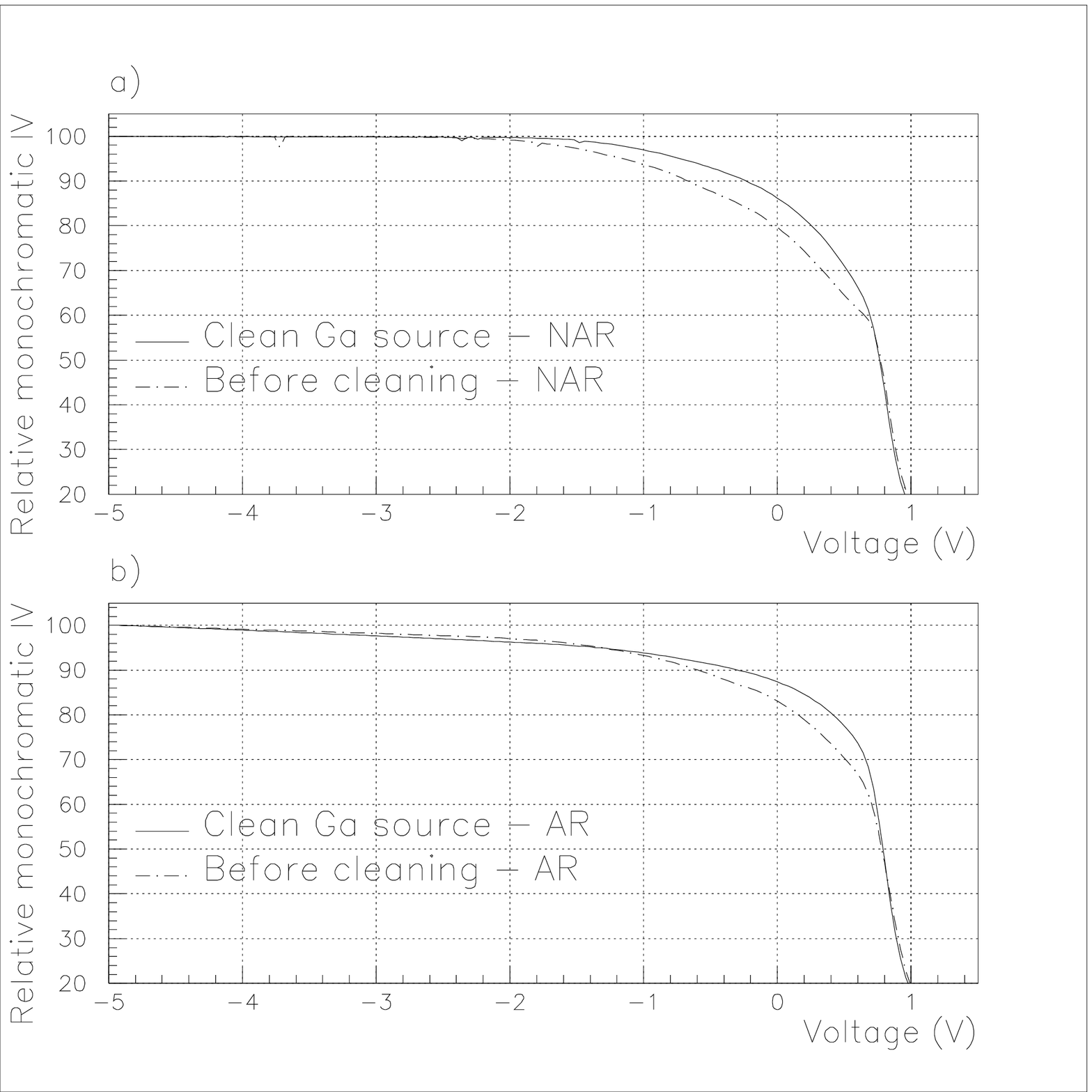}
{MIV measurements for the identical pair of QWSC devices QT468a
and QT528. The latter was grown with a cleaner Ga source and shows
an improved MIV characteristic.\label{figcleanga}}

Figure \ref{figcleanga} shows MIV measurements for QT468a and
the repeat structure QT528. The latter is identical to QT468a
but was grown with the cleaner TMG source. 
The MIV data for both NAR and AR devices is better for the material
grown with the cleaner TMG source. This supports the principle that 
poor MIV data is indicative of high impurity levels in the growth
precursors.

We note however that the voltage at which the MIV drops is very 
similar in both cases. The simple theory outlined in chapter 
\ref{secsolarcells} would therefore predict the same net background
doping level in both cases of $N_{BG}\simeq \rm 2\times 10^{16}cm^{-3}$.
This further indicates that the main source of background impurities
is the TMA source.

\subsubsection{MOVPE Growth Temperature Optimisation}

MOVPE growth temperature may significantly affect growth
morphology and impurity incorporation during
growth. In particular, carbon incorporation decreases at
lower temperatures. This is offset by an increased defect
density \cite{jroberts}.

A series of identical single quantum well (SQW) samples
were grown at three different growth temperatures in order
to identify an optimum growth temperature. The temperatures
were 560C for QT574b, 580C for QT874a and 600C for QT874c. 
The MIV characteristics for these samples
are given in figure \ref{figmovpetemp}. We again consider NAR 
devices in order to minimize the device variability introduced by the
AR coating process. The p layer in all samples was grown at
700C.

\smalleps{tp}{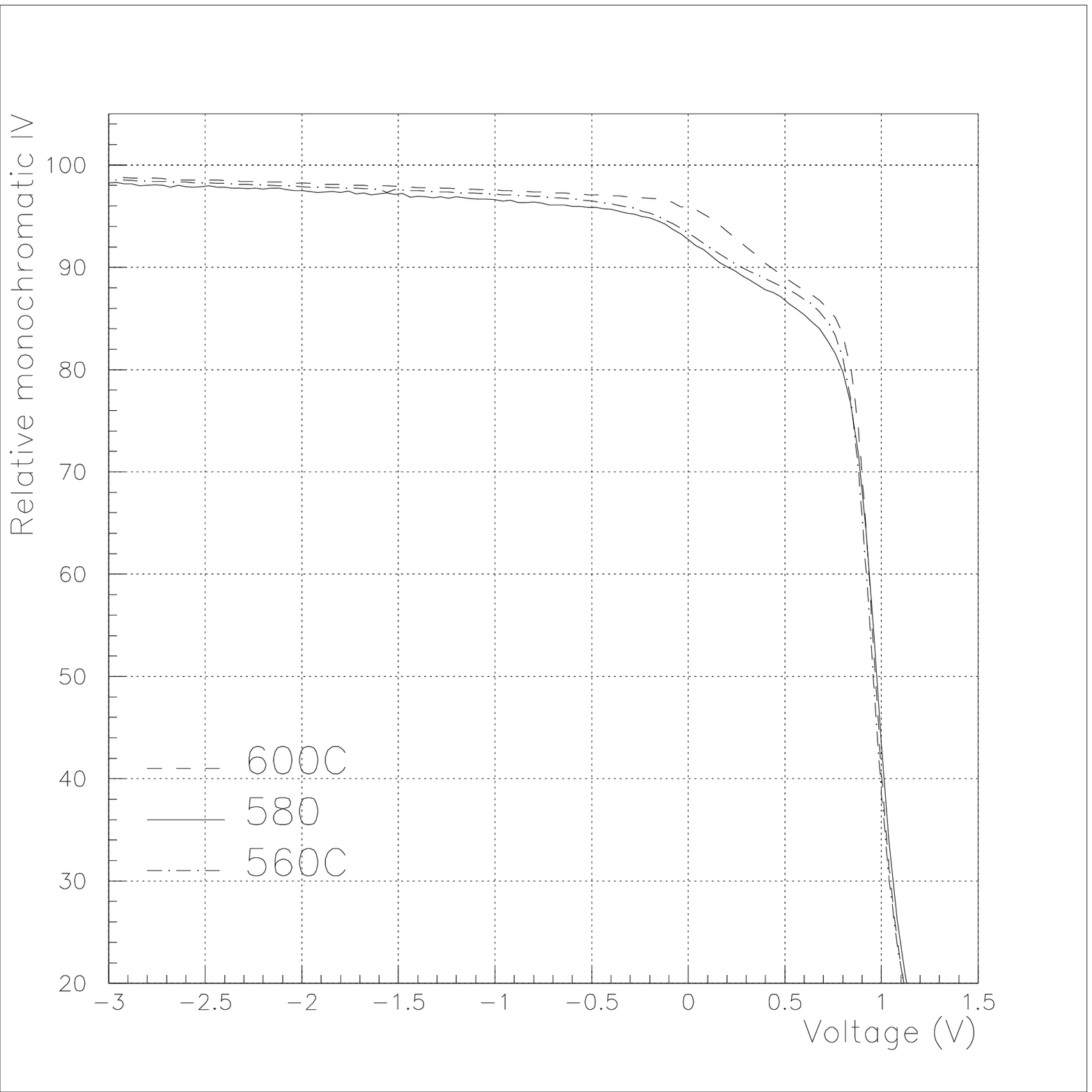}
{MIV data for identical MOVPE single quantum well samples grown at
560C, 580C and 600C, showing little variation.\label{figmovpetemp}}

Although the data for 600C appears higher
than the other two at 0V, the difference is only of the order
of 3\%, which is comparable or smaller than the experimental
uncertainty. The MIV measurement in this case cannot distinguish 
between the three samples. Clearer conclusions can be 
drawn from PL spectra taken by J.S. Roberts \cite{jroberts}. The
relative strengths of PL signals measured from the wafers relative
to the 560C sample were 3.5 for 580C and 1.35 for 600C.
A temperature of 580C was therefore proposed as the 
optimum by John Roberts. He suggests that this temperature
represents a good compromise between increased background carbon
incorporation at higher growth temperatures and greater defect densities
at lower temperatures.

\subsection{Passivation effects}

\subsubsection{Comparison of MIV Data of AR Coated and As-Grown Devices}

Figure \ref{figarmivdown} shows MIV data for the \pin\ sample
QT468b for samples with and without an AR coat. The initially
good MIV characteristic is degraded by the AR coat deposition.
The decrease in MIV in reverse bias however shows a constant
gradient. Linear behaviour of this type may be due to a change
in the effective doping profile brought about by the AR coating
process. Other influences such as electrical problems related to
the extra processing step cannot however be ruled out. 

Figure \ref{figarmivup} shows similar data for
QT640a, which is a QWSC sample with the same dimensions as QT468b 
but which has a graded p layer. The initially poor MIV
characteristic is improved by the AR coating process. The
improvement however does not significantly change the voltage
point at which the MIV starts to fall.
The net background for both NAR and AR devices is therefore 
estimated at $\rm N_{BG}\sim 2\times 10^{16}cm^{-3}$.

The degradation of good samples and improvement of poor ones
upon the deposition of the AR coat has frequently been observed
in the devices processed for this project.

\smalleps{htbp}{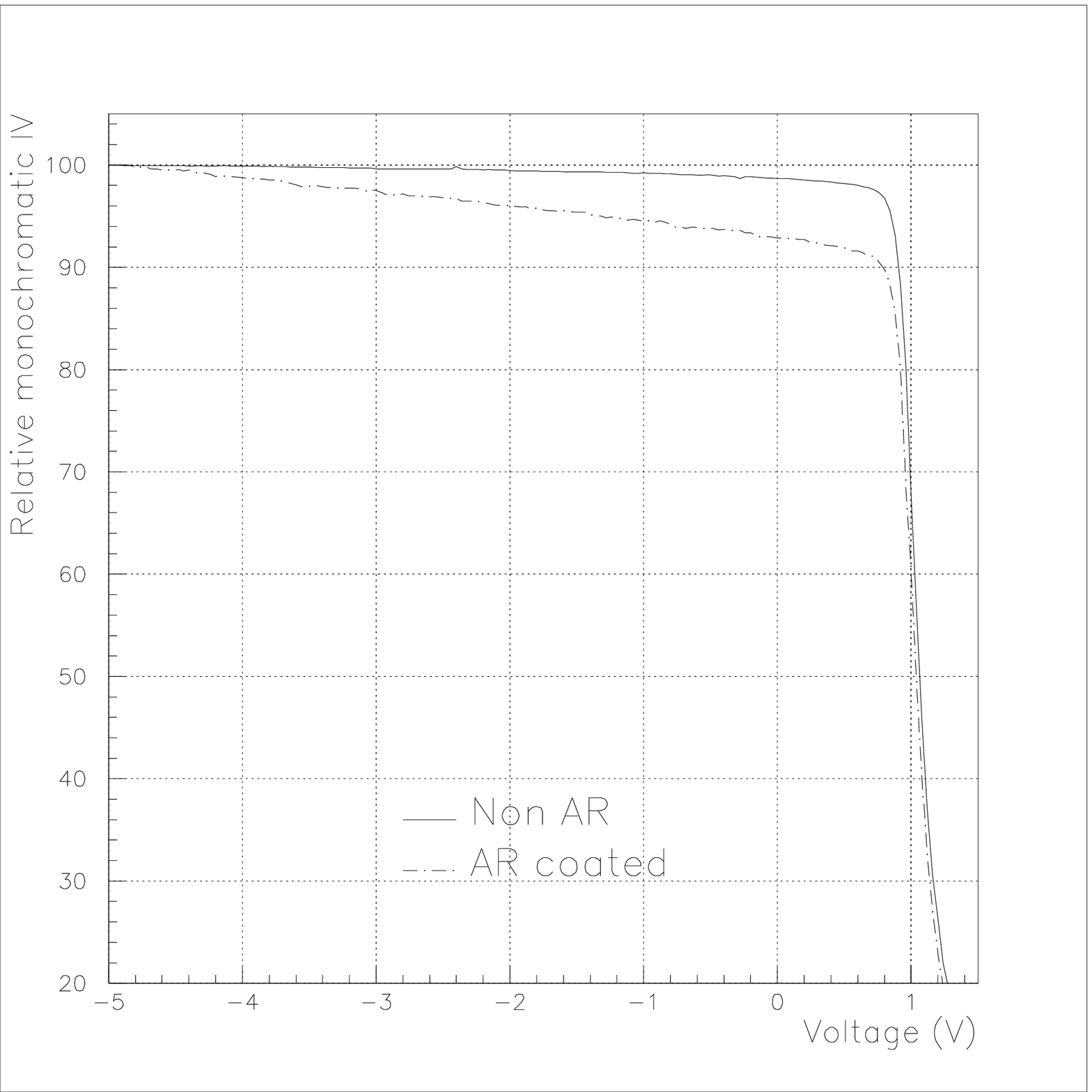}
{Degradation of MIV for a good \pin\ device QT468b\label{figarmivdown}}
\smalleps{htbp}{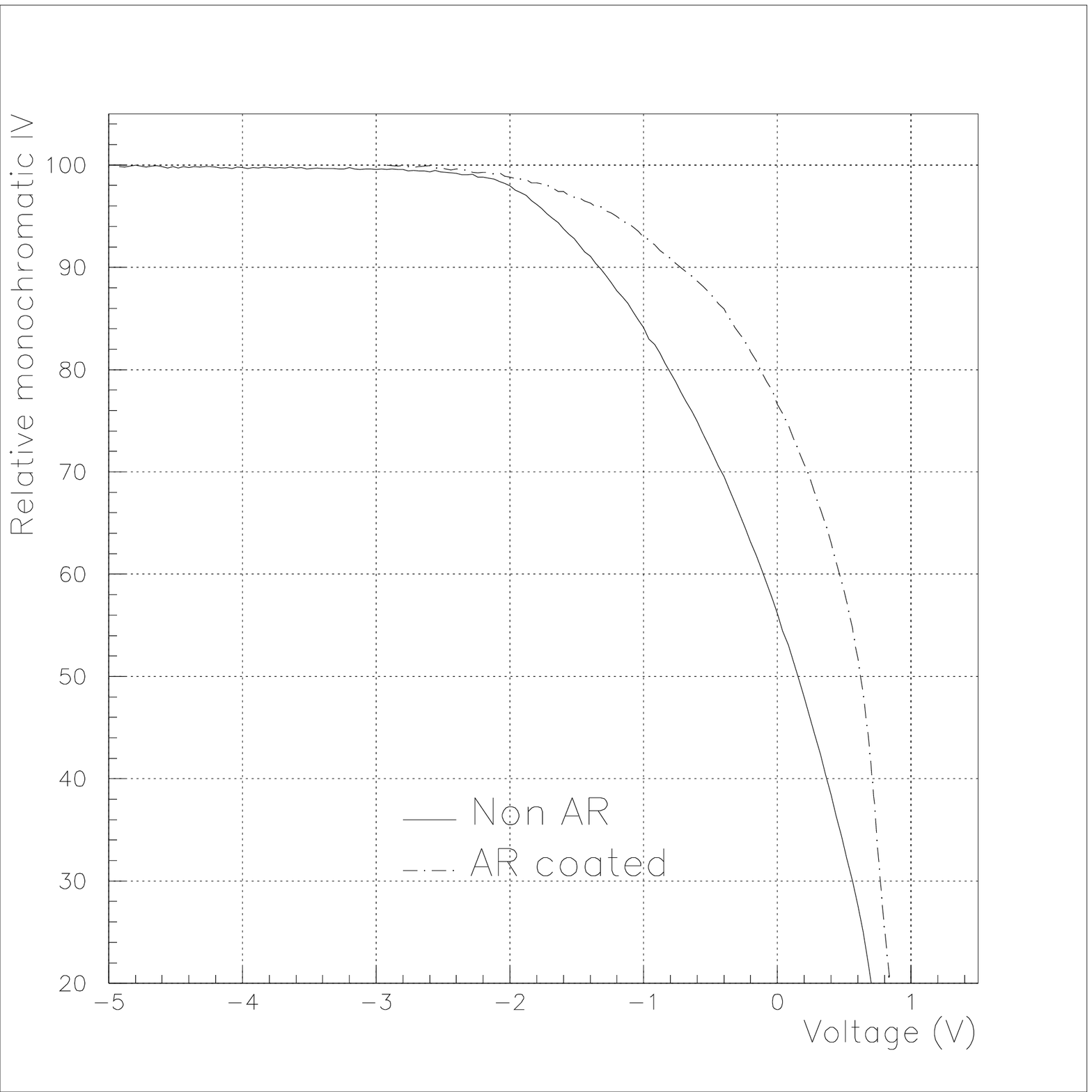}
{Improvement in MIV for poor QWSC device QT640a\label{figarmivup}}

Such behaviour was also observed in samples grown at Eindhoven
Technical University. Peter Ragay {\em et al.} \cite{ragay93}
of the III-V Solar Cell group at Eindhoven have carried out 
passivation studies which show similar behaviour in QWSC samples.

MIV data for the 50 well QWSC sample W276 is given
in figure \ref{figpass}. The wafer was grown by MBE at Eindhoven
Technical University. The two curves in the graph show the device
characteristic before and after exposure to a hydrogen passivating
plasma. Hydrogen ions diffuse through the crystal lattice and
bond to impurities and defects. These may include materials defects,
minority carrier traps, and ionised donor and acceptor
impurities in the crystal. The mechanisms are reviewed for GaAs
in \cite{murray90} and for \algaas\ in \cite{pavesi92}.

Although the MIV data for the passivated device is much
improved, electrical characterisation by Guido Haarpainter
indicated that the treatment significantly increased the series
resistance of the device. This is ascribed to the degradation of
the intentional p and n layer doping due to unintentional
passivation of the donor and acceptor dopant atoms.

The analogies between passivated and AR devices suggest that the
hydrogen present in the silane plasma may have a passivating effect
on AR devices. The exposure time of one to two minutes is however
very much smaller than the typical times of 30 minutes quoted by 
Ragay \cite{ragay93}, and does not benefit from the driving
field which is usually employed to assist ionic diffusion.
This however must be set against the fact that
the samples we are considering are very thin. The i layer in the
samples quoted above is approximately $0.15\mu m$ from the surface
of the sample, which is smaller than the thicknesses of several
microns usually studied in the literature. Furthermore, our samples 
show a weaker change in MIV than the Eindhoven samples.
Confirmation of this hypothesis would require further 
characterisation.

Passivation effects in indium phosphide solar cells have been
reported by Chatterjee and Ringel \cite{chatterjee96}. 
This study observed an increased
$V_{oc}$ in passivated $p^{+}$ -- $n$ samples. Deep level transient 
spectroscopy confirmed that deep levels were passivated.
Consequences for the series resistance and overall efficiency
of the cells are not given in this reference.

We conclude that passivation may be of use in improving the
dark currents of solar cells by passivating deep levels, and
may also decrease the $N_{BG}$.
The expected degradation of the intentional doping however 
presents a significant problem. It is expected to increase the
series resistance and hence reduce the efficiency of the solar
cell.
 
\smalleps{htbp}{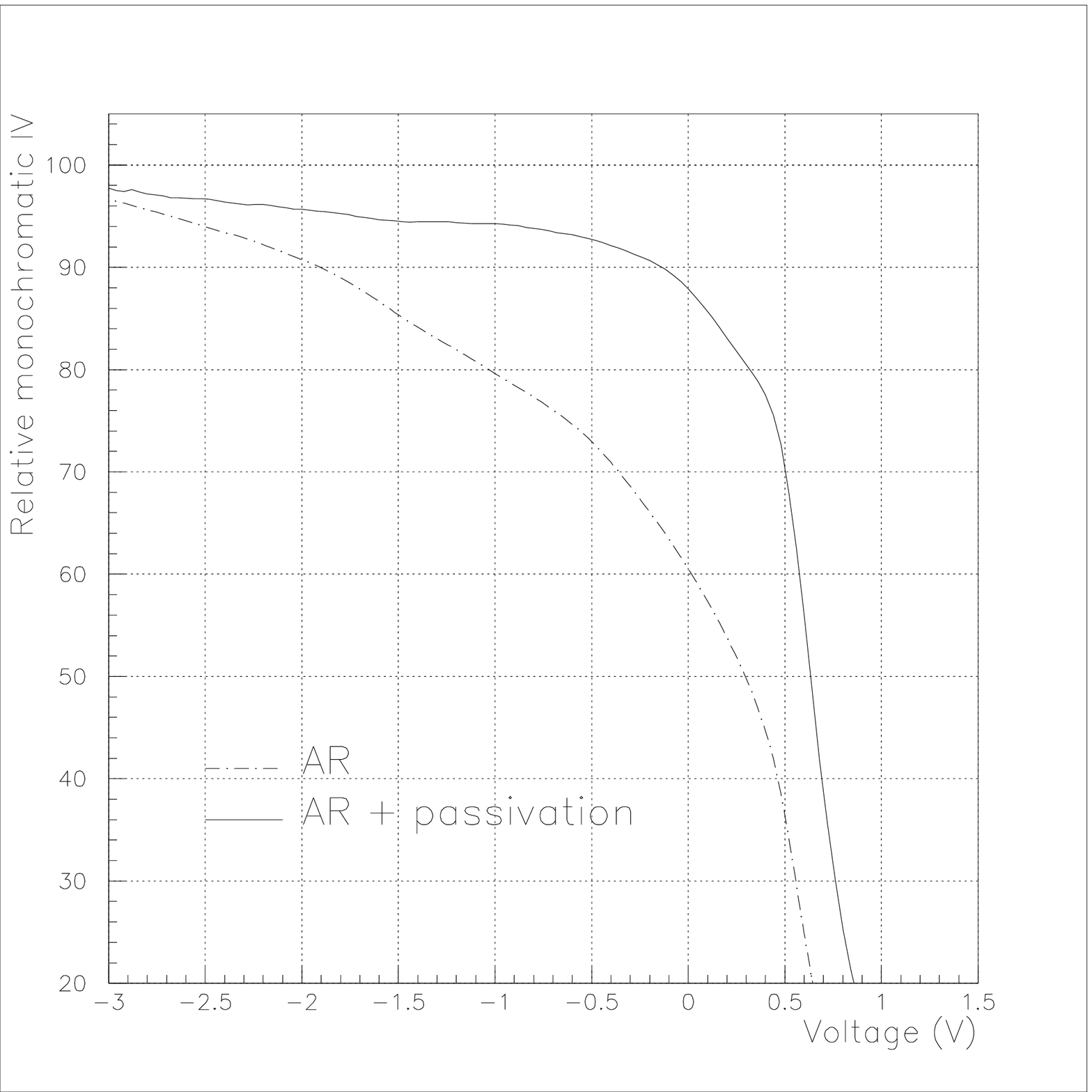}
{MIV data for the 50 well QWSC sample W276 before and after
exposure to a passivating hydrogen plasma\label{figpass}}

\subsection{Conclusions}

High net background doping levels
have proved to be a significant problem in both MOVPE and MBE
material with values of $N_{BG}$ qualitatively estimated as
greater than $\rm 10^{16}cm^{-3}$ observed in both systems.
The systematic difference between QWSC and \pin\ devices has
been ascribed to improved compensation properties in \pin\
structures. The background doping problem is more systematic
in MOVPE grown material.

Efforts to reduce the background doping levels in MOVPE material
have so far met with little success and the high levels of $N_{BG}$
estimated for this growth method for QWSC devices continue to 
pose problems.

Previous MOVPE and more recent MBE material has shown that both systems
are capable of net background doping densities of the order of
$\rm 10^{15}cm^{-3}$.

The AR coating process has been shown to significantly alter
the MIV characteristics of solar cell devices. In general,
the AR coat degrades the MIV data of good devices but improves that
of poor devices. Samples which fall into the
latter category show similarities with a passivated sample
from Eindhoven University of Technology. This is likely to be due
to unintentional passivation of intentional and unintentional
impurities in our samples by the hydrogen in the silane plasma 
used for the AR coating processing step. It is difficult
to make use of this effect in practice since the passivation
process unavoidably alters the intentional
doping levels. This has been indirectly observed as an increased
series resistance in the passivated Eindhoven device which
degrades solar cell efficiency.


\subsection{Reflectivity Measurements\label{secreflectivities}}

\smalleps{tp}{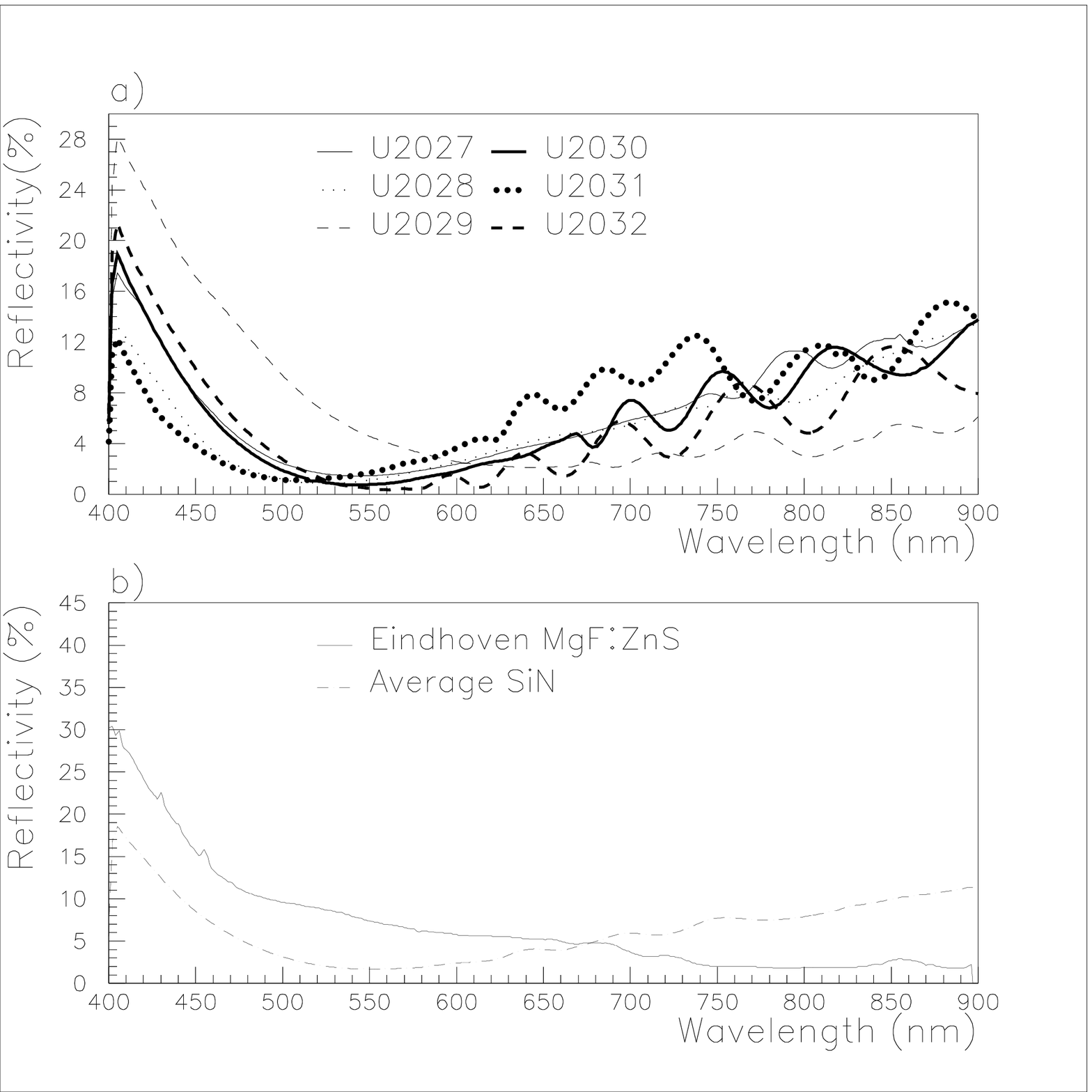}
{Reflectivities for QWSC and \pin\ devices for the U2027-U2032
MBE series\label{fig2000reflectivities}}
\smalleps{htbp}{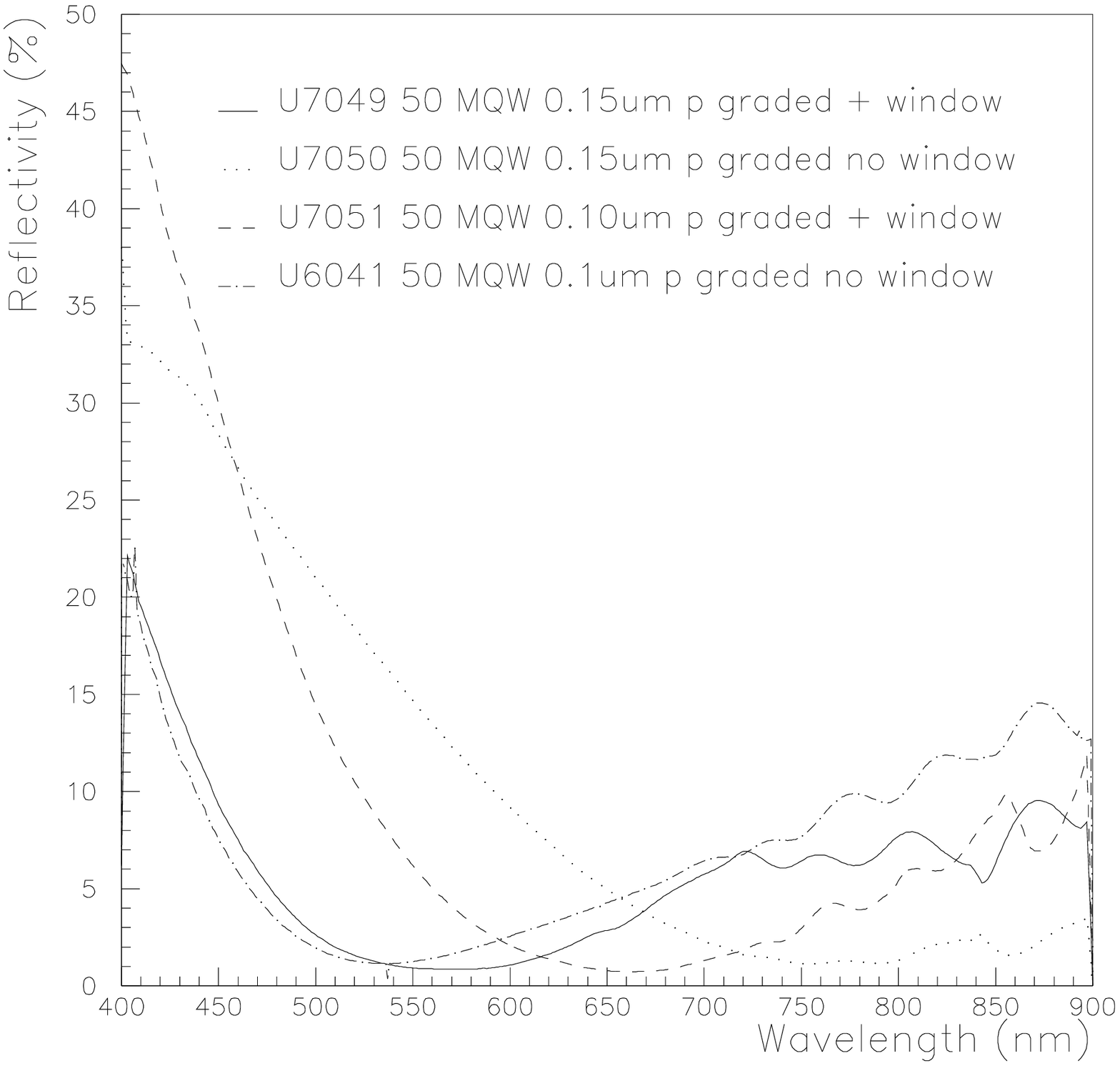}
{Reflectivities of a variety of samples with and without windows,
showing that the removal of the window does not noticeably affect
the reflectivity\label{figwinreflectivities}}

We now briefly consider spread and reproducibility of
reflectivity measurements. These are important in
determining the internal $QE$ of devices, and cannot
always be measured directly. 

Figure \ref{fig2000reflectivities}a shows the spread in
reflectivity for a range of samples processed in the
same AR coating run. Of the six samples shown in the figure,
only sample U2029 deviates significantly. 

These data form the basis of the average SiN reflectivity curve which was
discussed in chapter \ref{secexperiment}. This mean reflectivity is
shown figure \ref{fig2000reflectivities}b, and is used to model the
reflectivity of those samples for which a dedicated reflectivity
measurement sample is not provided.

Also shown in figure \ref{fig2000reflectivities}a is a standard
reflectivity for the Eindhoven MgF:ZnS double layer AR
coat deposited at Eindhoven. We note that it is optimised
at longer wavelengths, but is small over a wider wavelength
range than the SiN single layer coat. 

The curves in figure \ref{figwinreflectivities} however
show reflectivity data for samples grown towards the end of
the project, just before the AR coating kit was upgraded.
A large spread is seen for the samples AR coated at this
time. This is consistent with discussions with Malcolm
Pate of the III-V Semiconductor Facility at Sheffield
University regarding the problems experienced with the
original deposition equipment prior to the upgrade.

It has been suggested \cite{paxman93} that the thin window may be
used in conjunction with the SiN layer to produce a
double layer AR coat. The optimum suggested window thickness
was 30nm for the 70nm SiN AR coat.

The data of \ref{figwinreflectivities}
includes two samples with and without windows, but with
a nominally indentical SiN AR coat. The window thickness
in this case was 20nm instead of 30nm. Both samples were coated 
in the same processing run. We note that the data for the large area
device processed from the U6041 wafer, which has no window, is
very similar to the mean reflectivity of figure 
\ref{fig2000reflectivities}.

The reflectivity of the fifty well sample U7049 is again very
similar in magnitude and wavelength dependence, despite the 
fact that it has a window. The two other reflectivity curves shown 
in figure \ref{figwinreflectivities} are very high at short wavelength, 
indicating that the SiN coat is thicker than expected. The 
reflectivity for these samples is effectively optimised 
at longer wavelengths.

We note furthermore than if we were to normalize the reflectivities
of these four samples to the same wavelength, the windowless sample
U7050 would have the lowest reflectivity.

The window thickness in this example is 50\% different from the
optimum. The large differences observed in these data however lead
us to conclude that the reflectivity of our samples is primarily
determined by factors other than the presence or absence of
a window. The reflectivity overall is reasonably reproducible,
with the exception of samples processed towards of the
project, just before the AR coating kit was overhauled by
Malcolm Pate.

\subsection{Conclusions}

In this chapter we have considered results and implications of
the MIV and reflectivity characterisation techniques.
The first section demonstrated the use of the MIV measurement to 
estimate sample quality in MOVPE and MBE samples. The technique
has been shown to qualitatively demonstrate the magnitude of the
background doping $N_{BG}$. Quantitative analysis however
requires further analysis of the kind developed by Jenny Nelson
and Alexander Zachariou \cite{zachariou96}.

We established that QWSC structures generally suffer from higher 
levels of $N_{BG}$ than their \pin\ controls. This has been assigned
to the loss of a compensation mechanism in QWSC structures by discussion
with John Roberts.
We further saw that previously grown QWSC samples have demonstrated
that this problem is not insurmountable. Some recently grown
MBE samples have confirmed this, although they have not reached
the standard laid down by a sample (G951) grown at Phillips
Research Laboratories in 1991. Overall, we conclude that
MOVPE material is more variable than MBE material, and is
subject to higher values of $N_{BG}$ and a wider departure
from nominal design parameters.

Consequences of the AR coating process for device performance
have been considered. An interesting obervation is that the
MIV characteristics of poor NAR samples tend to improve in
corresponding AR devices. This has shown interesting correlation
with work by Peter Ragay \cite{ragay93} on hydrogen passivation.

Reflectivity measurements were considered and shown to be reasonably
consistent. Samples processed towards the end of the project however
showed more variability. An average reflectivity was presented in
order to enable modelling of samples for which the reflectivity
was not directly measurable.

%% file: chap8.tex
\chapter{Experimental and Modelled $\bf QE$ Spectra\label{secqeresults}}

The first section of this chapter gives an overview of the techniques
used to model selected types of devices and outlines the information
which may be extracted from the modelling, together with limitations
in each case.

We then present experimental and modelled $QE$ spectra for a set
of samples with unusual i region compositional profiles. We then
examine a set of ungraded samples grown by MOVPE and MBE, in order
to estimate the dependence of the minority carrier transport on
aluminium fraction. A set of graded devices are characterised, 
using the information derived from the previous sections. 
The final section presents $QE$ spectra for QWSC devices equipped 
with back surface mirrors.

The samples are referred to by their growth
name, and relevant design characteristics are given in the
text. Supplementary information is included in the graphical output
of the modelling program.

\section{Principles of Data Fitting\label{secfitprinciples}}

In this section we present experimental $QE$ data and the corresponding
modelling results for a range of ungraded samples grown by MBE.
We start by reviewing theoretical results from chapter \ref{sectheory}
relevant to modelling experimental data, and establishing the modelling
techniques which will be used throughout this section.

\subsection{Sample geometry}

Sample descriptions are generally accurate, because of the
epitaxial methods used in sample growth. Compositions however
frequently need to be adjusted. This is usually possible
only if a sharp cut-off in $QE$ is present, for example in
\pin\ devices. Aluminium fractions in QWSC samples can be
estimated by fitting the shape of the $QE$ above the base
aluminium fraction $X_{b2}$ and by looking at the position
of the excitonic peaks in the quantum well $QE$.

Adjustments are occasionally necessary to the window thickness,
and are justified by the presence or absence of a characteristic dip 
in the $QE$ for photon energies greater than the window bandgap.

An additional problem in MBE material is dopant diffusion.
The sample U6041 which will be described in some detail below
was characterised by SIMS. This revealed 
that the effective p thickness was approximately 40nm larger
than expected, which for this sample represents an increase
of nearly 50\%. This casts some doubt over modelling
other MBE samples with nominal layer thicknesses. In practice,
modelling has shown that the extra thickness is not
important in most MBE samples because they have p layers
which are thicker than U6041. Furthermore, these samples
do not show the symptoms of this problem. These are
an overestimated i region $QE$ at wavelengths of about 600nm, 
or an overestimation of the quantum well $QE$ resulting 
from unintentional doping of wells near the p layer.

\subsection{Samples with efficient minority carrier transport}

Chapter \ref{sectheory} shows
that the $QE$ is dominated by different parameters in different
wavelength ranges. In particular, the $QE$ at photon wavelengths
corresponding to the band edge energy $E_{b2}$ is dominated by
the i region contribution, and by the absorption and diffusion
lengths in the p and n electrostatically neutral layers.

Section \ref{secgradediregion} will show that the i region $QE$ of
even quite complex samples may be modelled accurately, as long
as the i region geometry and composition are close
to specifications. We will further establish that the model is
useful as a diagnostic tool. It can be used to adjust
compositions, in the cases where the latter
determine a sharp cut-off in $QE$ at a certain wavelength.

In the most general case, fitting of the model to the experimental
$QE$ is subject to three main parameters, which are the recombination
velocity parameter for the front of the p layer
${\cal S}=S_{n}/D_{n}$, and the p and n diffusion
lengths $L_{n}$ and $L_{p}$. The recombination velocity at the back
of the n layer is not a significant parameter. This is because
the n layer is negligible at short wavelengths, and small overall.
Furthermore, the n layer thickness of $\sim 0.6\mu m$ is 
much greater than $L_{p}$.

Appendix \ref{secapp1} shows that it is impossible to draw any
clear analytical conclusions without making assumptions about the 
relative magnitudes of one or the other of these parameters. Experience 
with modelling however shows that different parameters dominate 
in different wavelength ranges. The diffusion lengths $L_{n}$ and 
$L_{p}$ are most significant near the band edge energy $E_{b2}$, 
whereas $\cal S$ is important at short wavelengths. Fitting the 
theory to the data in this case is an iterative process
involving repeated adjustment of the three parameters 
until a best fit is found for all wavelength ranges. Samples 
falling into this category are generally those with good material
quality and thin i layers, such as the low aluminium fraction
30 well samples. 

We will present results from samples grown at Eindhoven Technical University
which also conform to this picture. In this case, however, the $QE$ at short
wavelengths is good, suggesting a very low value of $\cal S$.
The fitting then reduces to a two parameter fit, with the short
wavelength $QE$ being exclusively determined by $L_{n}$, and $L_{p}$
being estimated subsequently by fitting the band edge $QE$.

\subsection{Samples with simplified modelling}

More exact modelling is possible in samples conforming to
the simplified analytical expressions for the $QE$ derived
in Appendix \ref{secapp1}.

Samples with small n region contributions in particular allow the 
$QE$ near the band edge to be fitted mainly in terms of $L_{n}$. A 
further coincidental simplification arises from the fact that the
n region contribution is generally small in samples where
$L_{n}$ is less than the effective p layer thickness
$x_{w_{p}}$. This is due to the fact that both p and n
contributions to the $QE$ are decreased in poor material.
In this case, results from chapter \ref{sectheory} and
Appendix \ref{secapp1} have shown that the $QE$ for photon
energies near $E_{g2}$ is very weakly dependent on $\cal S$.
The overall $QE$ in this case can essentially be modelled
independently at short and long wavelengths, in terms of
$\cal S$ and $L_{n}$ only. Samples falling into this category 
include those with base aluminium fractions $X_{b2}$ greater than about
$30\%$ and samples with thick i regions.

\subsection{Samples with high surface recombination parameters}

At short wavelengths,
the $QE$ generally depends on both the recombination parameter
$\cal S$ and the electron minority diffusion length $L_{n}$. As we saw
above, the discussion need only consider the value
of $\cal S$ in the p layer. As mentioned in chapter 
\ref{secoptimisation} we cannot separately estimate $S_{n}$ 
and $D_{n}$.

A consequence of the analytical model of chapter \ref{sectheory}
is that values of $\cal S$ comparable to the term $1/L_{n}$ 
are required to affect the $QE$ significantly at short 
wavelengths. Since diffusion lengths in the majority of our 
samples are only a fraction of a micron, it follows that
surface recombination only becomes significant if
${\cal S} > \rm 10^{7}m^{-1}$. Values quoted in chapter 
\ref{secparameters} however set an upper limit on
$\cal S$ of about $10^{6}m^{-1}$.
The larger values required to fit some samples indicate
either unusually low mobilities together
with high recombination velocities, or with a p layer
width which is greater than the specifications. More 
fundamentally, this behaviour may be due to a failure of the 
recombination velocity to adequately represent 
minority carrier transport at the front of the cell.

The recombination velocity
represents the high defect concentration near the surface
as a sheet of charge, which acts as a carrier sink. In reality, these 
defects are distributed over a finite depth, and may be caused
not only by crystal defects such as dangling bonds,
but by surface oxidation which may extend some distance
into the sample. If this damage is extended over a depth which is 
not negligible compared to the thickness of the p layer, the latter 
will be effectively shortened, and very high surface recombination
velocities required to reproduce experimental $QE$ data.

The following discussion attempts to model samples with
recombination velocities and diffusivities which are
consistent with published data. For samples with thin p layers
and low diffusion lengths, this in practice means values
of $\cal S$ which are too low to have a significant effect.

When such an approach is not possible,
we shall assume that the surface recombination picture does
not accurately describe the minority transport at the front
of the cell, and substitute a recombination parameter which
is large enough to saturate this mechanism. In practice, this
means values such that ${\cal S} > 1/L_{n}$, which are
effectively infinite insofar as they are large enough to
reduce the minority carrier concentration at the front of
the cell to negligible levels.

\subsection{Graded samples\label{secmodellinggrades}}

\smalleps{t}{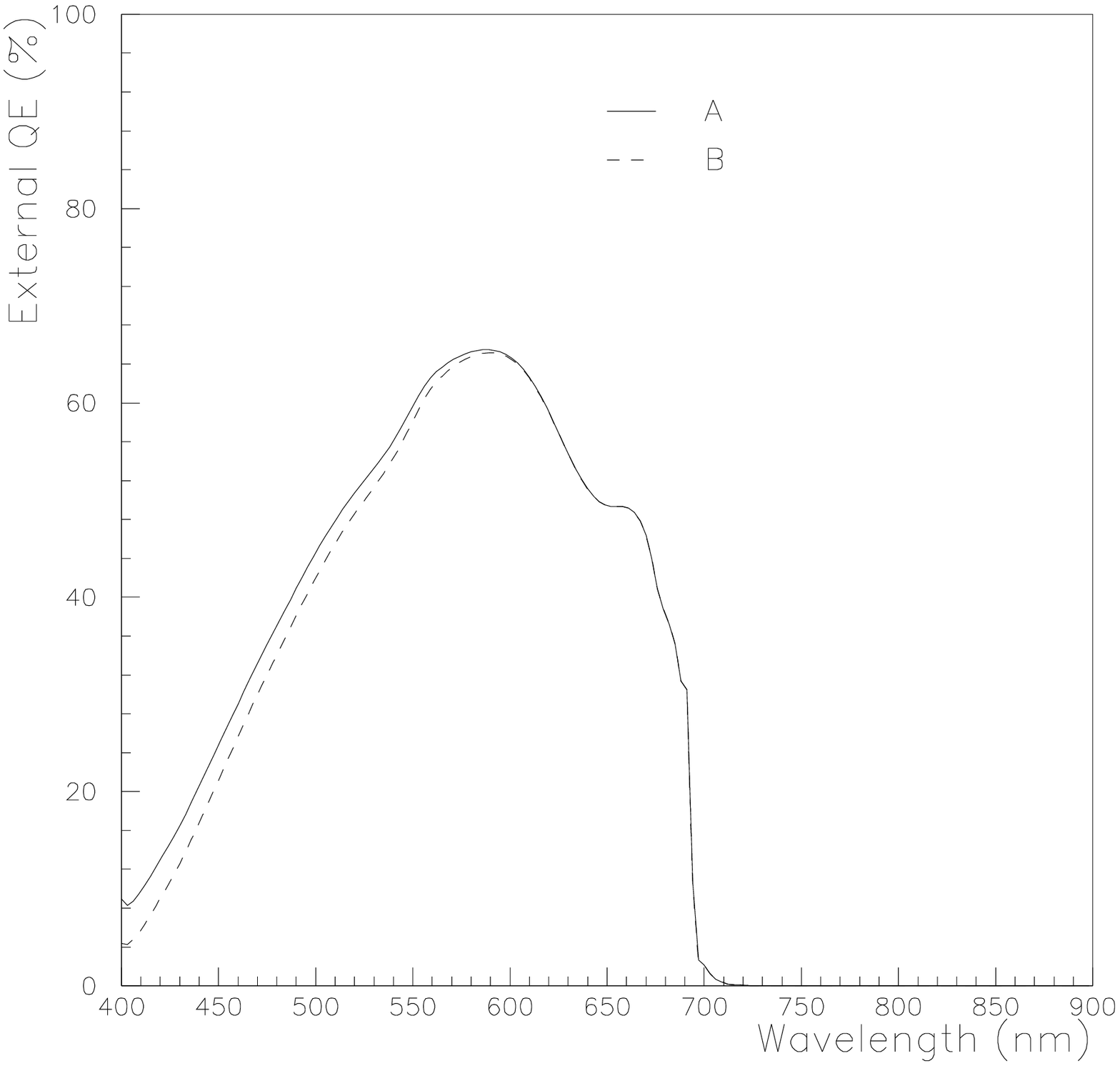}
{$QE$ for two linear parametrisations of $L_{n}$ as a function of
composition. For
curve A), $L_{n}$ is kept constant at $0.08\mu m$. Curve B 
represents the worst case scenario of $L_{n}(0)=0.08\mu m$ and 
$L_{n}(x_{w_{p}})=0.01\mu m$\label{figlnvariationingrades}}

A number of important differences appear when modelling
graded samples. The first is that the p region $QE$
for wavelengths corresponding to energy $E_{b2}$ is generally
small, particularly for grades with a large compositional gradient.
The $QE$ near the band edge in graded samples is therefore principally
determined by the i and n region contributions. Since we will not
consider samples with compositionally graded n layers, this simplifies
the fit near band edge $E_{b2}$.

Fitting the overall $QE$ of graded p samples is necessarily limited
by the assumptions we are forced to make regarding minority
carrier transport in compositionally graded material. In particular,
the assumption that the estimated dependence of $L_{n}$ on aluminium
fraction and position can be extrapolated from the values used 
to model samples with different $X_{b2}$ is clearly liable to 
lead to large errors. Figure \ref{figlnvariationingrades} shows two
$QE$ calculations. Curve A has no diffusion length gradient, and
curve B has a higher than expected $L_{n}$ gradient. The calculation
is for a value of ${\cal D_{X}}=-4$. We note that even in this extreme
case, the $QE$ is relatively insensitive to the variation in $L_{n}$.
The $QE$ in this case is therefore determined primarily by
the diffusivity field $\cal D_{X}$  defined in section 
\ref{secdiffusivityfield}.

The theoretical considerations in chapter \ref{sectheory}
and empirical tests using the model have shown that the
recombination parameter $\cal S$ has little effect on the
$QE$ of the p layer at short wavelengths.
Fitting the model to experimental $QE$ data for graded samples is
greatly simplified, since $\cal D_{X}$ is the determining
parameter at short wavelengths.

Although the diffusivity field $\cal D_{X}$ is in principle 
determined solely by the diffusivity $D_{n}$, it will in 
practice be influenced by other parameters. This is a
consequence of reducing the modelling to a two parameter fit
which is dominated by $\cal D_{X}$. Inaccuracies in the assumptions
we have made regarding the diffusion length, for example, are
subsumed into the modelled value of $\cal D_{X}$.

Modelling graded structures therefore follows similar guidelines
to modelling the $QE$ of ungraded structures, with the difference
that the $L_{n}$ is eliminated as a parameter and the band edge
response is determined by the i and n region contributions.
The short wavelength $QE$ will be modelled in terms of $\cal D_{X}$
and to a lesser extent the recombination parameter $\cal S$.
This simplified approach is made necessary by the lack of detailed
knowledge of the position dependence of the minority carrier
transport parameters.

\subsection{Modelling of the quantum well $QE$}

The quantum well $QE$ has been modelled in previous work by
Jenny Nelson, which has been reviewed in \cite{paxman93}.
For photon energies corresponding to electronic levels near 
the top of the well, the model for this wavelength range 
underestimates the $QE$ to some extent because of 
superlattice states which are not included, and because
of the change from a two dimensional to a three dimension
density of states, which the model assumes is abrupt.

With the exception of this correction, the model has been shown 
to reliably reproduce the quantum well $QE$, including excitonic 
features, and is sufficiently accurate for our purposes.
The model is expected to be accurate for the first quantum well 
continuum, which is deep enough in energy to approximate the 
theoretical model accurately.

The parameters for the quantum well $QE$ are kept fixed, since
they are well understood and have been reviewed extensively
\cite{paxman93}.

\subsection{Mirror backed samples}

Modelling of mirror backed samples introduces two new parameters.
The first is the phase change $\psi_{b}$ experienced by light
internally reflected from the back surface mirror. This is
a variable quantity which is poorly known \cite{whitehead94}.

The parameter $\psi_{b}$ can however be accurately determined
from the position of the wavelength dependent Fabry-Perot 
oscillations which are observed in mirrored samples.

In principle, no other parameters are required, since the phase
change $\psi_{f}$ upon internal reflection at the front 
surface semiconductor-air interface is 
0 radians \cite{macleod69}, the back surface reflectivity is well 
known from the reflectivity of the relevant metal, and the front 
reflectivity is generally measured.

In practice, we shall see that some samples show clear signs of
non specular reflections from the back surface, and show an
enhancement in $QE$ which is greater than that expected for
light reflected normally from the back surface. This case is
further characterised by lower Fabry-Perot amplitudes than
expected. In some samples, they may be absent altogether.

Finally, although the front surface reflectivity is measured,
the model frequently predicts larger Fabry-Perot peaks than
are in practice observed. The modelling in this case can be
improved by the introduction of a front reflectivity factor
$\cal F$. This reduces the modelled amount of light reflected
back into the cell upon the third pass, and is a factor of
the front surface reflectivity $R_{f}$ of section 
\ref{sectheorygenerationdefs}. A value of $\cal F$ smaller
than 1 reduces the amount of light reflected 
back into the cell from the front surface. This
reduces the amplitude of the Fabry Perot peaks, but has little
or no effect on the overall enhancement factor.
The origin of this factor is not clear, but may be due to
the fact that the light is not incident normally, or to an
unintentional roughnening at the back surface mirror.
Furthermore, we saw in section \ref{sectheorygenerationdefs}
that the internal reflections occurring at the interfaces between
the quantum wells and the barriers are not included
in the model. A brief calculation showed however that the effect is
expected to be small. More discussion
of this point is given in section \ref{secmirrorqe}.
Finally, the uncertainty in reflectivity also
plays a role, since measured reflectivities were not available
for the mirrored samples.

\smalleps{ht}{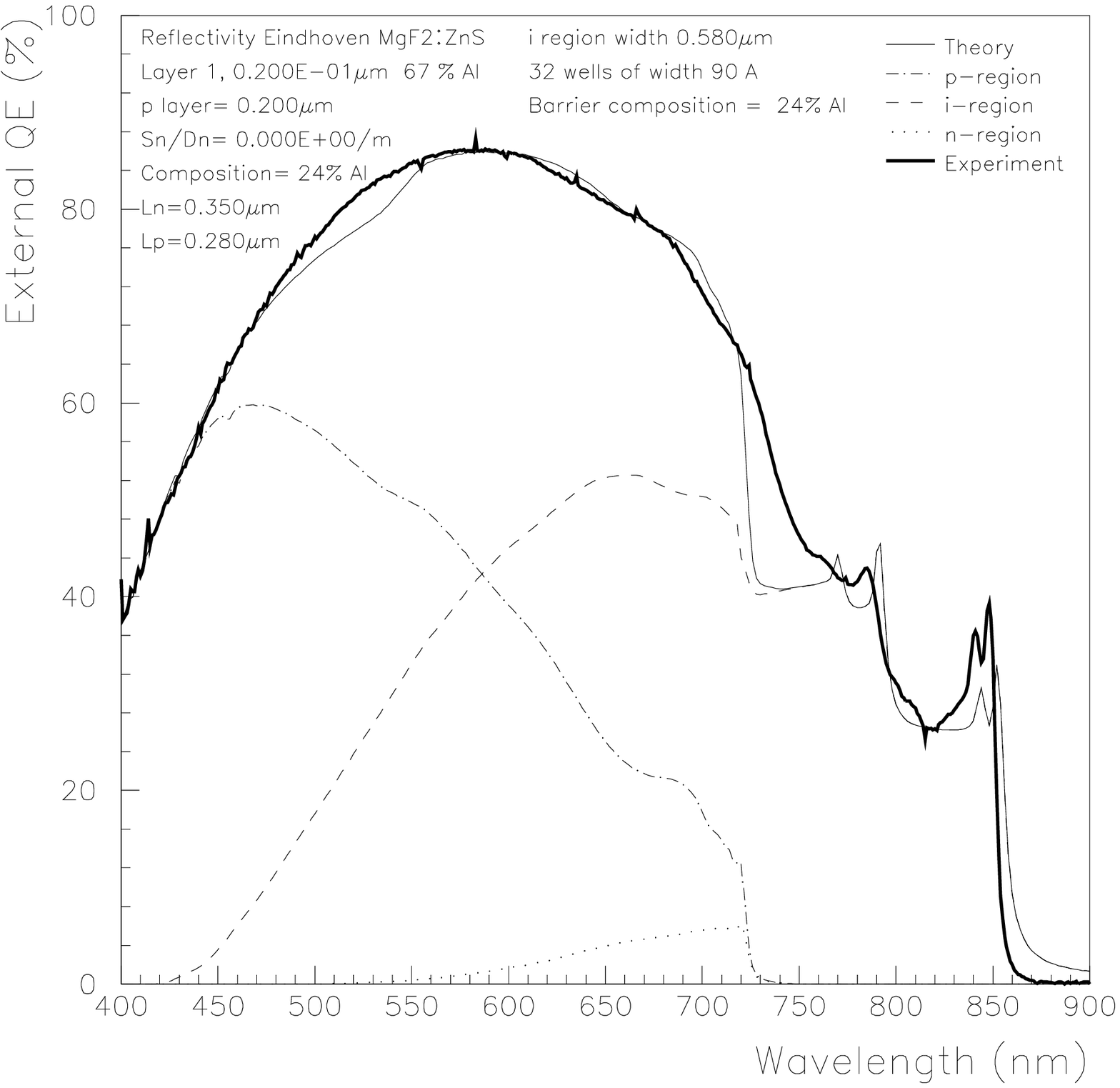}
{$QE$ of 32 well QWSC Eindhoven structure W405\label{figw405}}

\section{Graded i Region Samples\label{secgradediregion}}

A series of samples grown by MBE in Eindhoven University
of Technology was studied. All structures have identical, ungraded
window, p and n layers but have different i region specifications.
These samples were initially grown to compare the efficiency
of QWSC structures, \pin\ controls and \pin\ heterostructures
with graded i layers, as reported by Ragay \cite{ragay93}.
These samples are also useful because they allow us to
test the accuracy of the $QE$ model in reproducing experimental
data for devices with complex i region designs. 

MIV data for these samples were uniform and reproducible,
indicating good device quality and a low value of 
$\rm N_{BG}\leq 5\times 10^{15}cm^{-3}$.

The samples
consist of a QWSC device, its \pin\ control, a \gaas\ cell
and two samples with linearly graded sections in the i layer.
Although the nominal p layer thickness was $0.5\mu m$, Peter
Ragay has confirmed that the real thickness is $0.2\mu m$ for
all devices. Furthermore, the nominal window layer of $0.04 \mu m$
was estimated as at most $0.02 \mu m$ by the modelling, on the
basis of the high $QE$ value at short wavengths. A thicker window
would produce a characteristic sharp decrease in $QE$ for energies
above the window bandgap. The $QE$ measurements for these Eindhoven 
samples were taken by  Ernest Tsui.

\smalleps{htbp}{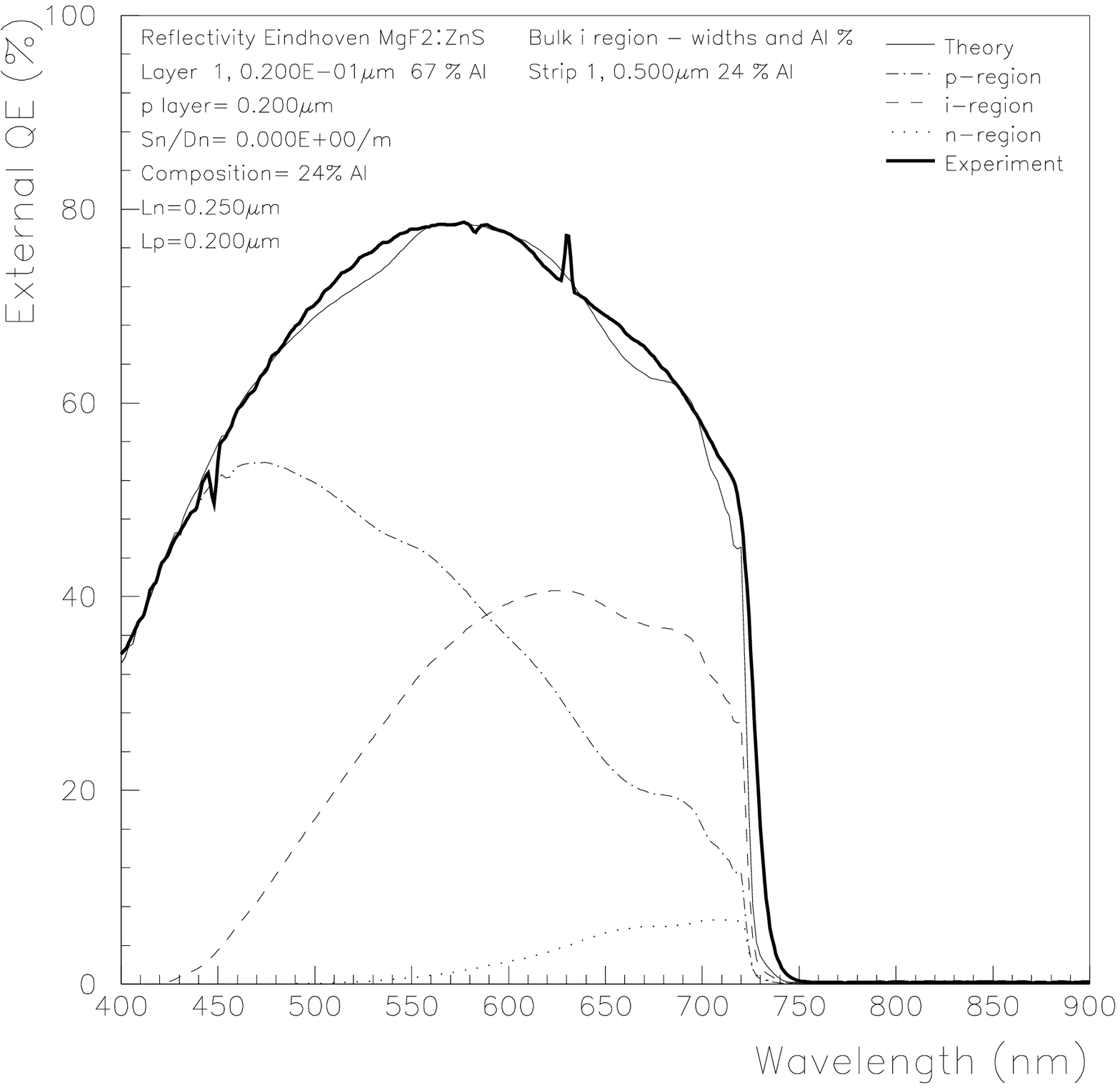}
{$QE$ of \pin\ Eindhoven sample W406\label{figw406}}

Figures \ref{figw405} and \ref{figw406} show the experimental data
and best fit from the model for a QWSC structure W405 and its
\algaas\ control \pin\ W406. The i layer of W405 contains 32
wells of width $\rm 90 \AA $ separated by $\rm 60 \AA $ barriers.
The i layer of the control W406 consists of $0.5\mu m$ of \algaas\ with
a nominal aluminium fraction of $X_{b2}=20\%$. The model however
indicates that the true aluminium content is about 
$X_{b2}\simeq 24\%$ because of the sharp cut-off in $QE$ observed
at about 730nm.

We first consider the $QE$ for photon energies above the neutral
layer bandgap $E_{b2}$. The interface recombination parameter 
$\cal S$ is estimated as too small to  have any effect on the 
$QE$ at short wavelengths, and is set to zero.

In this case of negligible recombination parameter $\cal S$, the $QE$
of the neutral layers depends only on $L_{n}$, $L_{p}$ and
the reflectivity. The modelling assumes a standard reflectivity
supplied by Peter Ragay, and assumes that the layer specifications
are correct, with the exception of the window layer thickness as
mentioned above.

Modelled diffusion lengths are of the order or
greater than the neutral layer widths at about $0.3\mu m$. The
hole diffusion length $L_{p}$ is slightly smaller than the
electron diffusion length $L_{n}$ as expected. These are close to
Hamaker's values, which are $L_{n}\simeq 0.39\mu m$ and 
$L_{p}=0.20\mu m$ respectively.

We conclude that the minority carrier transport in the
neutral layers of these devices is efficient, and the MBE 
material of high quality. 

\smalleps{htbp}{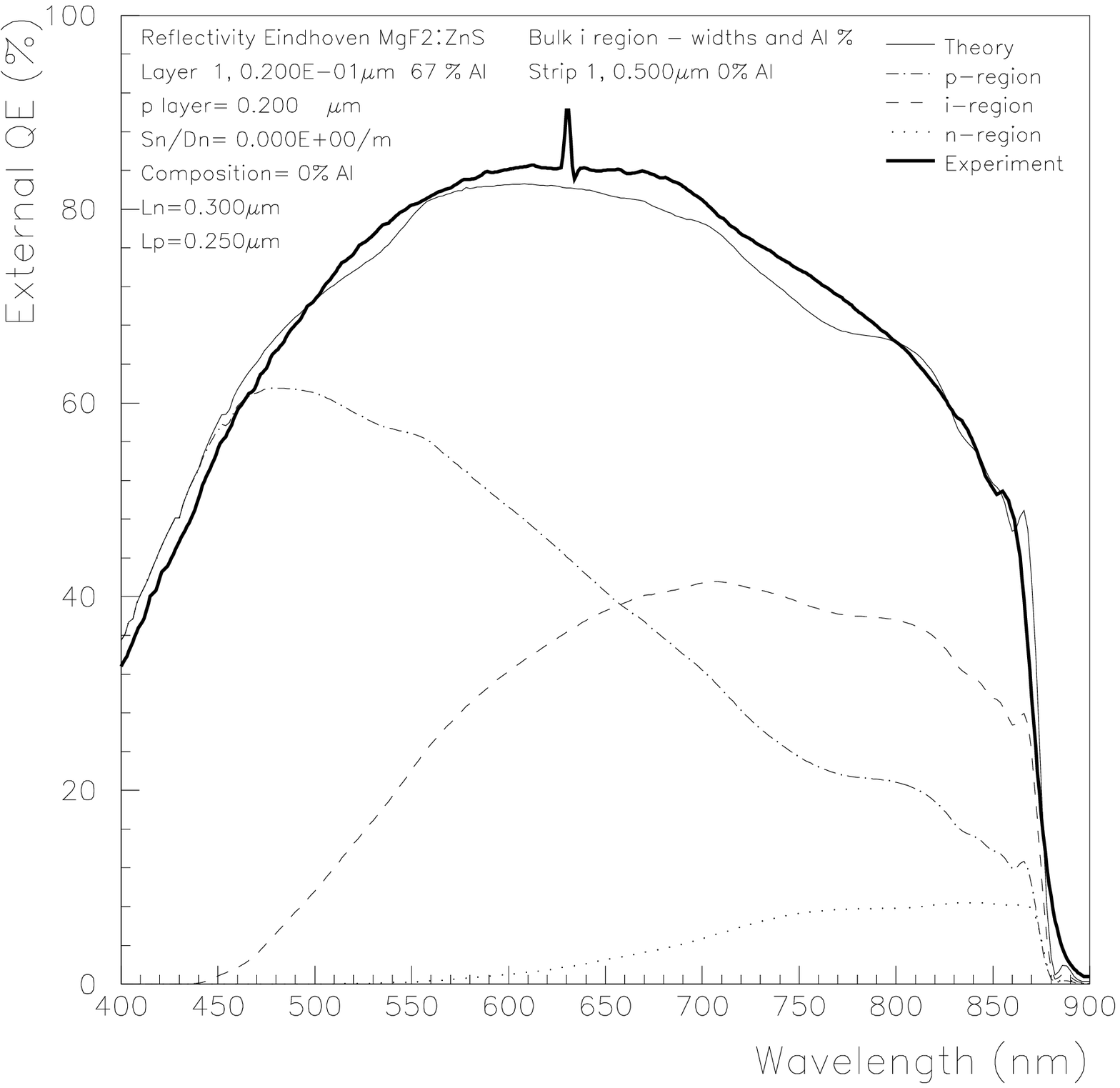}
{$QE$ of \gaas\ structure W407\label{figw407}}

Below bandgap $E_{b2}$, the quantum well $QE$ in sample W405 is
accurately reproduced. The main source of error in this case is
the reflectivity. The accuracy of the fit in this wavelength
range suggests that the Eindhoven AR coating process is 
reproducible. Furthermore, it serves to confirm the assumption
of carrier collection efficiencies close to 100\% in the
space charge region.

Figure \ref{figw407} shows similar data for sample W407.
The device specification is identical to W406 except that
the \algaas\ is replaced by \gaas\, leading to an ungraded
\gaas\ cell. The modelling in this case indicates minority
carrier transport which does not differ significantly from
the previous two samples. In general, significantly higher
diffusion lengths are expected in \gaas.

\begin{figure*}
\unitlength 0.70mm
\linethickness{0.4pt}
\begin{picture}(165.00,92.67)
\put(43.00,75.00){\line(1,0){2.00}}
\put(45.00,75.00){\line(0,-1){28.00}}
\put(45.00,47.00){\line(1,0){20.00}}
\multiput(65.00,47.00)(0.14,-0.12){184}{\line(1,0){0.14}}
\multiput(90.00,25.00)(0.14,0.12){184}{\line(1,0){0.14}}
\put(115.00,47.00){\line(1,0){50.00}}
\put(165.00,47.00){\line(0,1){0.00}}
\put(28.00,86.67){\vector(0,1){0.2}}
\put(28.00,12.00){\line(0,1){74.67}}
\put(28.00,92.67){\makebox(0,0)[cc]{Al \%}}
\put(28.00,75.00){\line(1,0){2.33}}
\put(28.00,47.00){\line(1,0){2.00}}
\put(28.00,25.00){\line(1,0){2.00}}
\put(26.00,75.00){\makebox(0,0)[rc]{67}}
\put(26.00,47.00){\makebox(0,0)[rc]{$X_{b2}=24$}}
\put(26.00,25.00){\makebox(0,0)[rc]{0}}
\put(44.00,79.00){\makebox(0,0)[cc]{Window}}
\put(55.33,51.00){\makebox(0,0)[cc]{p layer}}
\put(90.00,51.00){\makebox(0,0)[cc]{i layer}}
\put(140.00,51.00){\makebox(0,0)[cc]{n layer}}
\put(165.00,7.67){\makebox(0,0)[rc]{Depth}}
\put(165.00,14.00){\vector(1,0){0.2}}
\put(25.67,14.00){\line(1,0){139.33}}
\end{picture}
\caption{Schematic compositional profile of graded i region sample 
W408\label{figw408bands}}
\end{figure*}
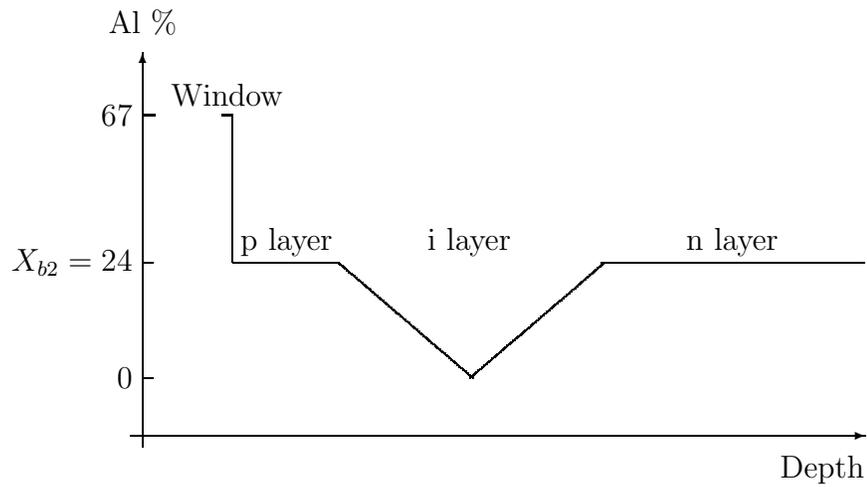

\smalleps{htbp}{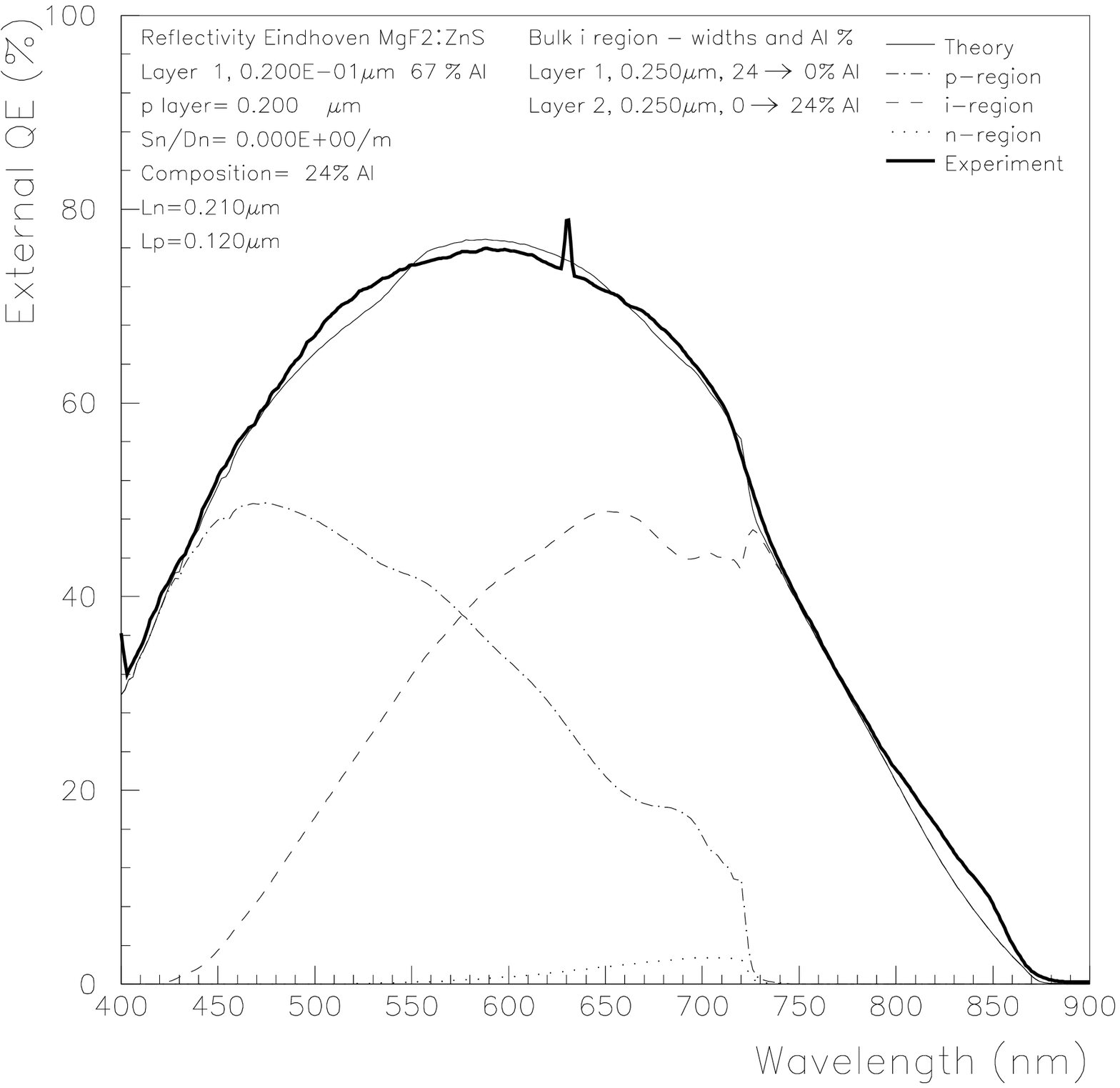}
{$QE$ of graded i region sample W408\label{figw408}}

The schematic of figure \ref{figw408bands} shows compositional profile
of graded i region sample W408. In the depletion
approximation, the $QE$ below the
bandgap $X_{b2}$ is dependent only on the reflectivity and the
integrated absorption. Figure \ref{figw408} shows the experimental
and modelled $QE$ for this sample. We note that the short wavelength
response for photon energies above $E_{b2}$ requires slightly smaller
diffusion lengths than samples W405-W407. The aluminium fraction
$X_{b2}$ is again estimated at about 24\%.

The model predicts the i region contribution remarkably accurately,
confirming the reliability of the absorption coefficient parametrisation
reviewed in chapter \ref{secparameters} for aluminium fractions between
0 and, in this case, 24\%.

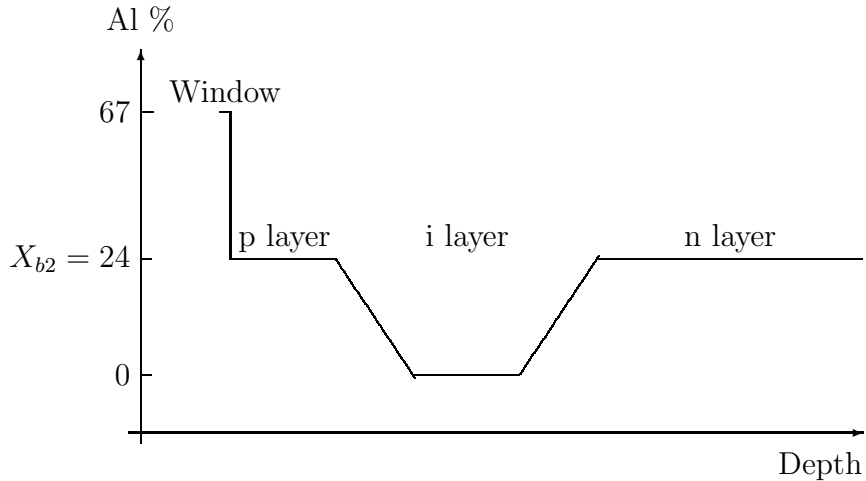
\begin{figure*}
\unitlength 0.70mm
\linethickness{0.4pt}
\begin{picture}(165.00,92.67)
\put(43.00,75.00){\line(1,0){2.00}}
\put(45.00,75.00){\line(0,-1){28.00}}
\put(45.00,47.00){\line(1,0){20.00}}
\put(115.00,47.00){\line(1,0){50.00}}
\put(165.00,47.00){\line(0,1){0.00}}
\put(28.00,86.67){\vector(0,1){0.2}}
\put(28.00,12.00){\line(0,1){74.67}}
\put(28.00,92.67){\makebox(0,0)[cc]{Al \%}}
\put(28.00,75.00){\line(1,0){2.33}}
\put(28.00,47.00){\line(1,0){2.00}}
\put(28.00,25.00){\line(1,0){2.00}}
\put(26.00,75.00){\makebox(0,0)[rc]{67}}
\put(26.00,47.00){\makebox(0,0)[rc]{$X_{b2}=24$}}
\put(26.00,25.00){\makebox(0,0)[rc]{0}}
\put(44.00,79.00){\makebox(0,0)[cc]{Window}}
\put(55.33,51.00){\makebox(0,0)[cc]{p layer}}
\put(90.00,51.00){\makebox(0,0)[cc]{i layer}}
\put(140.00,51.00){\makebox(0,0)[cc]{n layer}}
\put(165.00,7.67){\makebox(0,0)[rc]{Depth}}
\put(165.00,14.00){\vector(1,0){0.2}}
\put(25.67,14.00){\line(1,0){139.33}}
\multiput(65.00,47.08)(0.12,-0.18){126}{\line(0,-1){0.18}}
\put(80.00,25.00){\line(1,0){20.00}}
\multiput(100.00,25.00)(0.12,0.18){126}{\line(0,1){0.18}}
\end{picture}
\caption{Schematic compositional profile of graded i region sample 
W409\label{figw409bands}}
\end{figure*}

Figure \ref{figw409bands} shows the schematic compositional profile of
the last Eindhoven sample W409. This sample is similar to W408 except
that the i region grades are sharper, and a $0.2\mu m$ \gaas\ layer
is grown in the centre of the i region. Modelled minority carrier
transport in the neutral layers is similar to W408. The i region
$QE$, for photon energies below $E_{b2}$, is again satisfactorily
reproduced.

\smalleps{htbp}{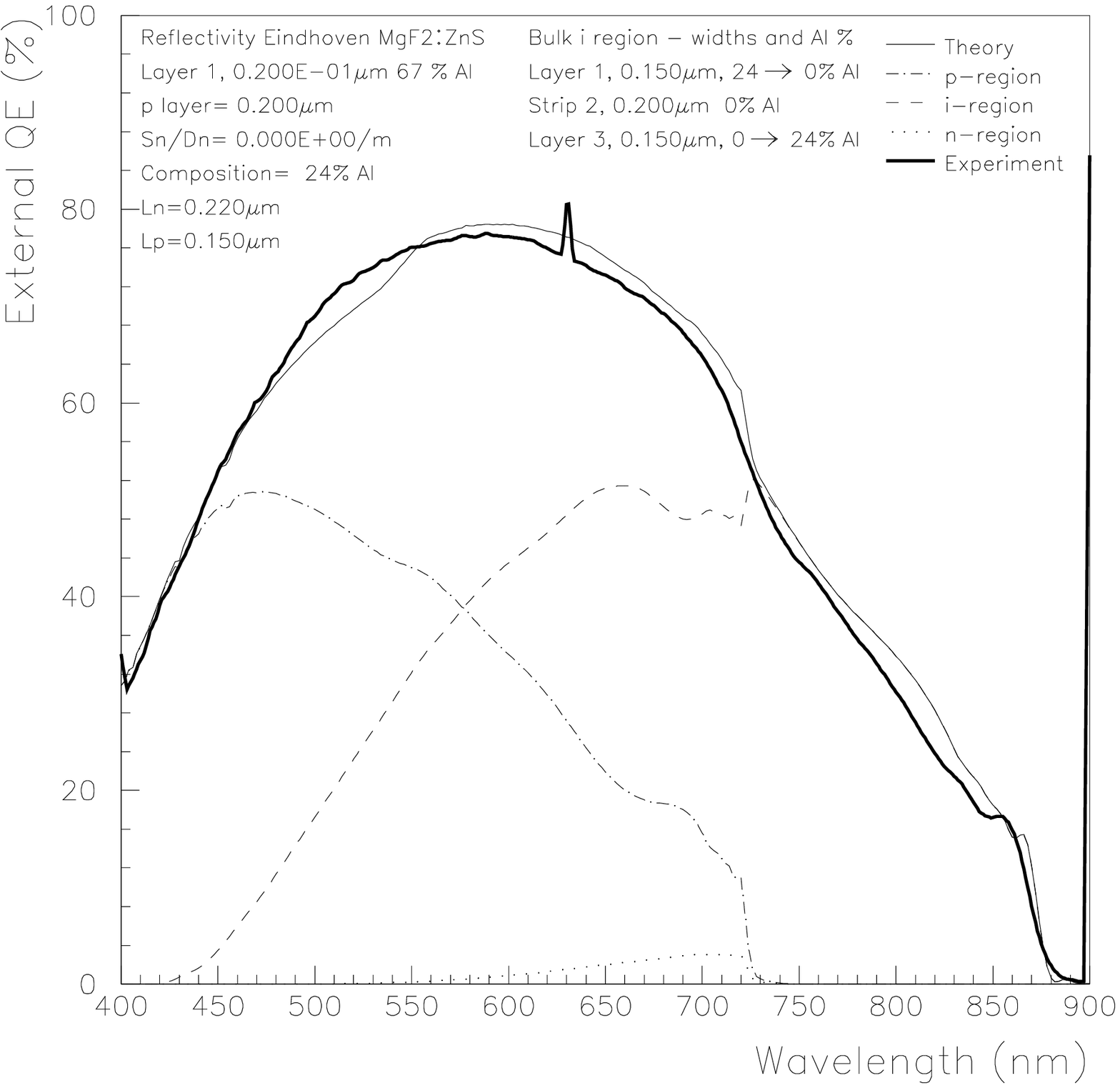}
{$QE$ of graded i region sample W409\label{figw409}}

\begin{figure*}[htbp]
\unitlength 1.00mm
\linethickness{0.4pt}
\begin{picture}(113.33,91.00)
\multiput(30.00,50.00)(0.12,0.24){42}{\line(0,1){0.24}}
\put(35.00,60.00){\line(1,0){15.00}}
\multiput(50.00,60.00)(0.12,-0.23){42}{\line(0,-1){0.23}}
\put(55.00,50.33){\line(0,-1){20.33}}
\put(55.00,30.00){\line(1,0){9.00}}
\put(64.00,30.00){\line(0,-1){5.00}}
\put(64.00,25.00){\line(1,0){2.00}}
\put(66.00,25.00){\line(0,1){5.00}}
\put(66.00,30.00){\line(1,0){9.00}}
\put(75.00,30.00){\line(0,1){20.00}}
\multiput(75.00,50.00)(0.12,0.24){42}{\line(0,1){0.24}}
\put(80.00,60.00){\line(1,0){15.00}}
\multiput(95.00,60.00)(0.12,-0.23){42}{\line(0,-1){0.23}}
\put(100.00,50.33){\line(0,-1){20.33}}
\put(100.00,30.00){\line(1,0){13.33}}
\put(104.67,34.33){\makebox(0,0)[lc]{GaAs substrate}}
\put(65.00,16.33){\makebox(0,0)[cc]{InGaAs well}}
\put(20.00,85.33){\vector(0,1){0.2}}
\put(20.00,30.00){\line(0,1){55.33}}
\put(20.00,91.00){\makebox(0,0)[cc]{Al \%}}
\put(17.67,30.00){\line(1,0){4.66}}
\put(20.00,13.67){\vector(0,-1){0.2}}
\put(20.00,30.00){\line(0,-1){16.33}}
\put(20.00,7.33){\makebox(0,0)[cc]{In \%}}
\put(20.00,60.00){\line(1,0){2.00}}
\put(22.00,50.00){\line(-1,0){2.00}}
\put(20.00,25.00){\line(1,0){1.67}}
\put(13.67,30.00){\makebox(0,0)[rc]{0}}
\put(13.67,25.00){\makebox(0,0)[rc]{20}}
\put(13.33,50.00){\makebox(0,0)[rc]{$X_{b1}=25$}}
\put(13.33,60.00){\makebox(0,0)[rc]{$X_{b2}=81$}}
\put(45.67,13.33){\line(0,1){2.00}}
\put(45.67,18.67){\line(0,1){2.00}}
\put(45.67,24.00){\line(0,1){2.00}}
\put(45.67,29.33){\line(0,1){2.00}}
\put(45.67,34.67){\line(0,1){2.00}}
\put(45.67,40.00){\line(0,1){2.00}}
\put(45.67,45.33){\line(0,1){2.00}}
\put(45.67,50.67){\line(0,1){2.00}}
\put(45.67,56.00){\line(0,1){2.00}}
\put(45.67,61.33){\line(0,1){2.00}}
\put(45.67,66.67){\line(0,1){2.00}}
\put(45.67,72.00){\line(0,1){2.00}}
\put(45.67,77.33){\line(0,1){2.00}}
\put(45.67,82.67){\line(0,1){2.00}}
\put(84.67,14.33){\line(0,1){2.00}}
\put(84.67,19.67){\line(0,1){2.00}}
\put(84.67,25.00){\line(0,1){2.00}}
\put(84.67,30.33){\line(0,1){2.00}}
\put(84.67,35.67){\line(0,1){2.00}}
\put(84.67,41.00){\line(0,1){2.00}}
\put(84.67,46.33){\line(0,1){2.00}}
\put(84.67,51.67){\line(0,1){2.00}}
\put(84.67,57.00){\line(0,1){2.00}}
\put(84.67,62.33){\line(0,1){2.00}}
\put(84.67,67.67){\line(0,1){2.00}}
\put(84.67,73.00){\line(0,1){2.00}}
\put(84.67,78.33){\line(0,1){2.00}}
\put(84.67,83.67){\line(0,1){2.00}}
\put(43.33,64.33){\makebox(0,0)[rc]{p layer}}
\put(87.00,64.33){\makebox(0,0)[lc]{n layer}}
\put(65.00,64.33){\makebox(0,0)[cc]{i layer}}
\put(91.00,8.67){\vector(1,0){0.2}}
\put(41.00,8.67){\line(1,0){50.00}}
\put(65.00,3.00){\makebox(0,0)[cc]{Depth}}
\end{picture}
\caption{Schematic compositional profile of GRINCH laser structure 
NRC1038\label{fig1038bands}}
\end{figure*}
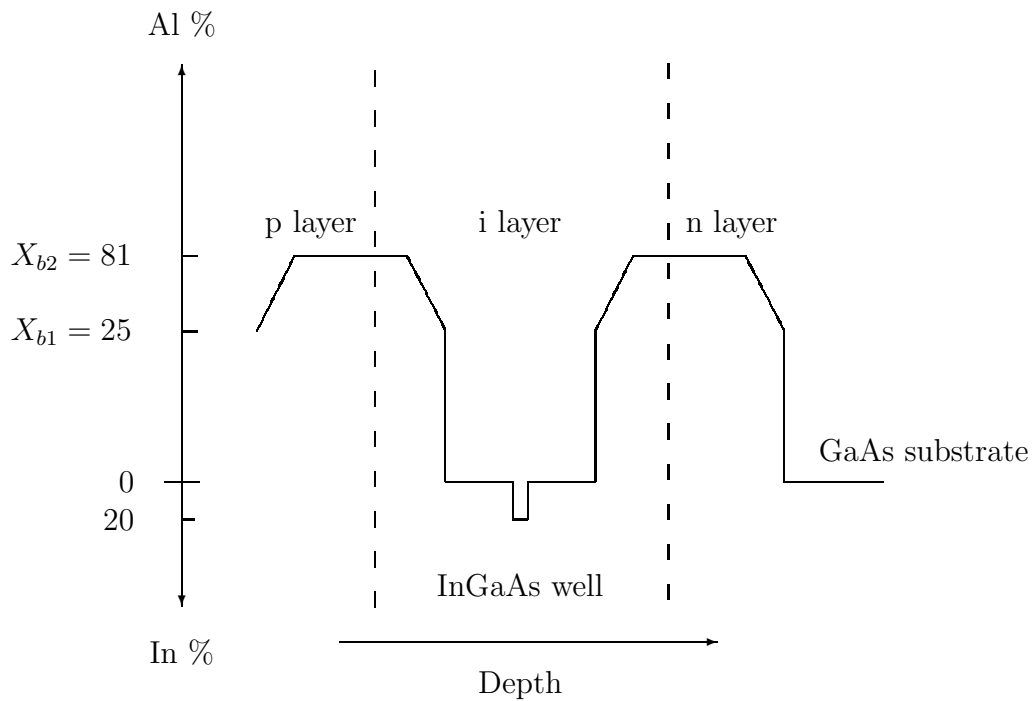

\smalleps{htbp}{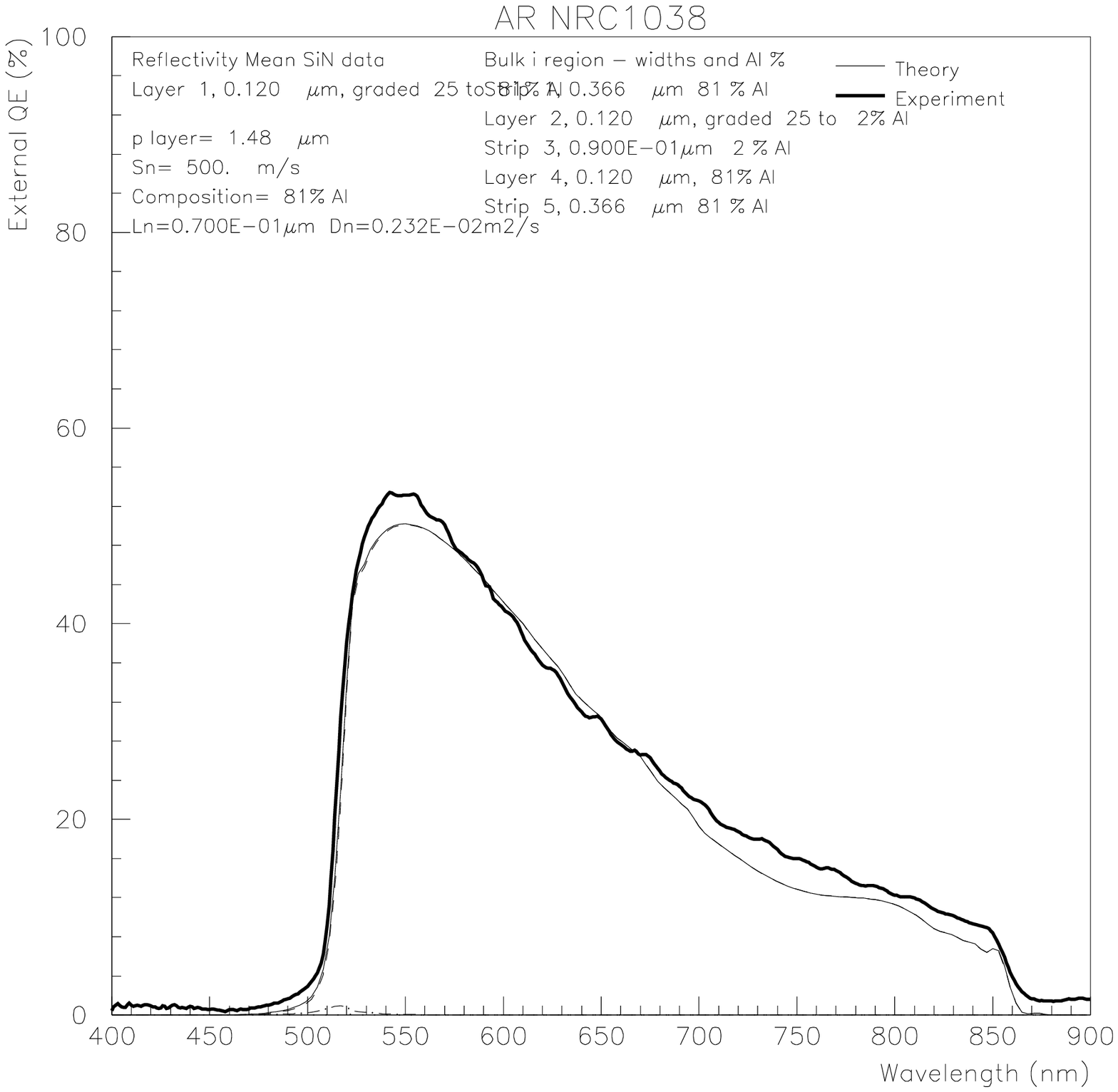}
{$QE$ of GRINCH structure NRC1038\label{fig1038}}

Figure \ref{fig1038bands} shows a schematic compositional profile
of the GRINCH laser structure NRC1038. This structure has been
investigated by colleagues as a promising structure for a high
$V_{oc}$ QWSC structure since studies have shown that the dark
current in such structures is very low. The measured and modelled
$QE$ data for this structure are given in figure \ref{fig1038}.

The p layer of this structure is very thick. It consists of a
heavily doped graded layer starting at $X_{b1}=25\%$ at the
surface and rising to a nominal $X_{b2}=70\%$ at a depth of 120nm.
This is followed by $1.484\mu m$ of \algaas\ of nominal
composition $X_{b2}$.
Only a negligible amount
of photons with energy greater than the higher p layer bandgap
$X_{b2}$ reach the intrinsic region. The p layer $QE$ is 
negligible because of the very high aluminium fraction in
this layer. The associated low minority electron diffusion 
length in this case prevents photogenerated minority electrons
from reaching the depletion layer. A sharp cutoff is therefore
observed in the $QE$ at energy $E_{b2}$ corresponding to a photon 
wavelength of about 500nm. The main p layer composition can be estimated
from this cutoff, and is of the order of $X_{b2}\simeq 81\%$.

The i region consists of seven different sections,
as illustrated in the schematic. These include both ungraded
sections with composition $X_{b2}$ and grades from $X_{b2}$ to
$X_{b1}$. The centre of the i layer is \gaas .
The $\rm In_{0.2}Ga_{0.8}As$ well at the centre of the i layer 
is modelled as an equivalent thickness of \gaas\ since the model is
not equipped with an InGaAs quantum well absorption routine.
The aluminium fraction $X_{b1}\simeq 25\%$ is assumed accurate,
since we have no sharp threshold from which to make a reliable estimate.
The $QE$ cutoff at the nominal \gaas\ band edge occurs at a slightly
higher photon wavelength than expected, and corresponds to \algaas\
with a composition of 2\% Al.

Overall, the $QE$ of this GRINCH laser structure is well reproduced
by the model. Since the $QE$ of the i layer is critically determined
by the amount of light transmitted through the p layer, this fit
confirms that the absorption of the graded section at the front of
the p layer has been accurately modelled, with the proviso that
the lower aluminium fraction $X_{b1}$ is assumed to be 25\% as
specified in the growth menu.

\subsection{Conclusions}

The series of Eindhoven samples presents a number of design features
which have allowed us to test the reliability of the theoretical model
for complex structures grown on a different MBE system. In particular,
the graded i region samples allow us to test the accuracy of the
depletion approximation and of the integrated absorption. The
$QE$ in both long and short wavelength ranges has been accurately
modelled with typical values of the transport parameters.

These structures also demonstrate high diffusion lengths for both 
electrons and holes, albeit at relatively low aluminium fractions.
Finally, the interface recombination parameter $\cal S$ is negligibly
small for these samples, despite the rather large compositional change
from $\sim 20\%$ to $\sim 67\%$ between the p layer and the window.

The GRINCH laser structure NRC1038 has been used to test the model
in the case of a complex structure where the composition changes
rapidly and where the $QE$ is essentially determined solely by the
integral of the absorption coefficient over the cell.

\section{Ungraded QWSC and \pin\ samples\label{sechomoqe}}

\subsection{MOVPE samples}

\smalleps{htbp}{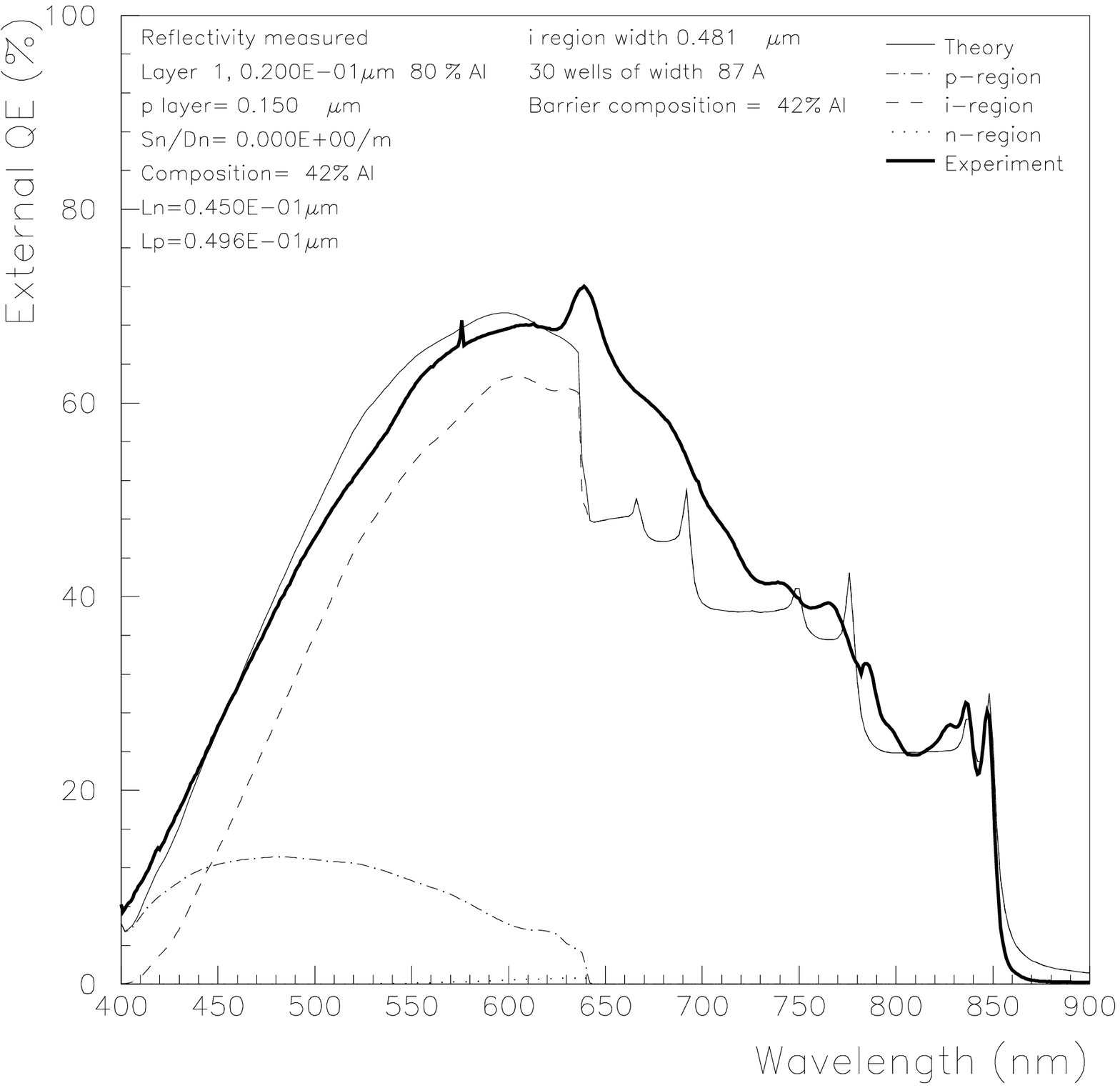}
{$QE$ modelling and data for nominally 30\% aluminium 30 well
sample QT468a\label{figar468a}}
\smalleps{htbp}{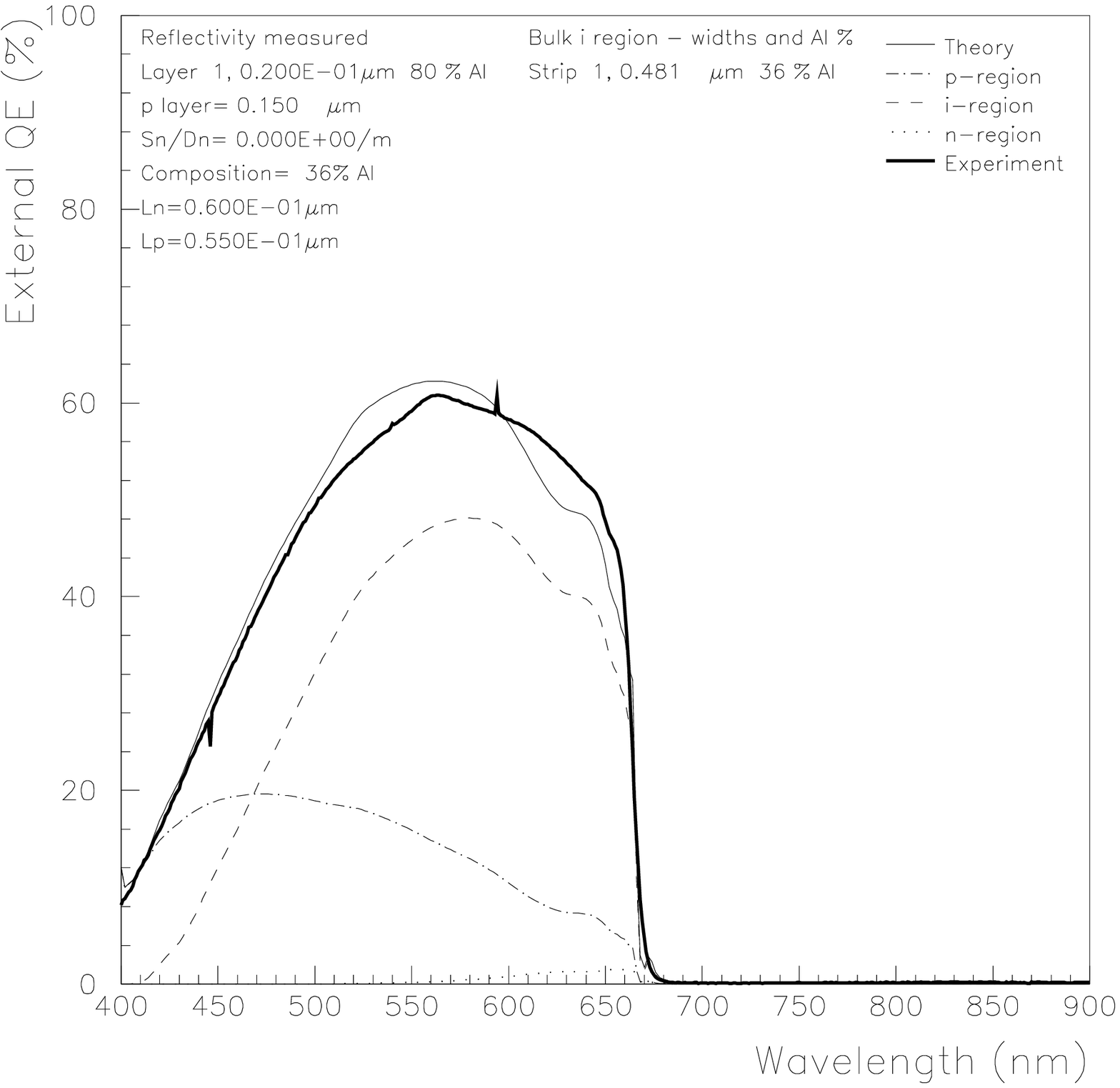}
{$QE$ modelling and data for QT468b, the \pin\ control to QT468a
\label{figar468b}}
\smalleps{htbp}{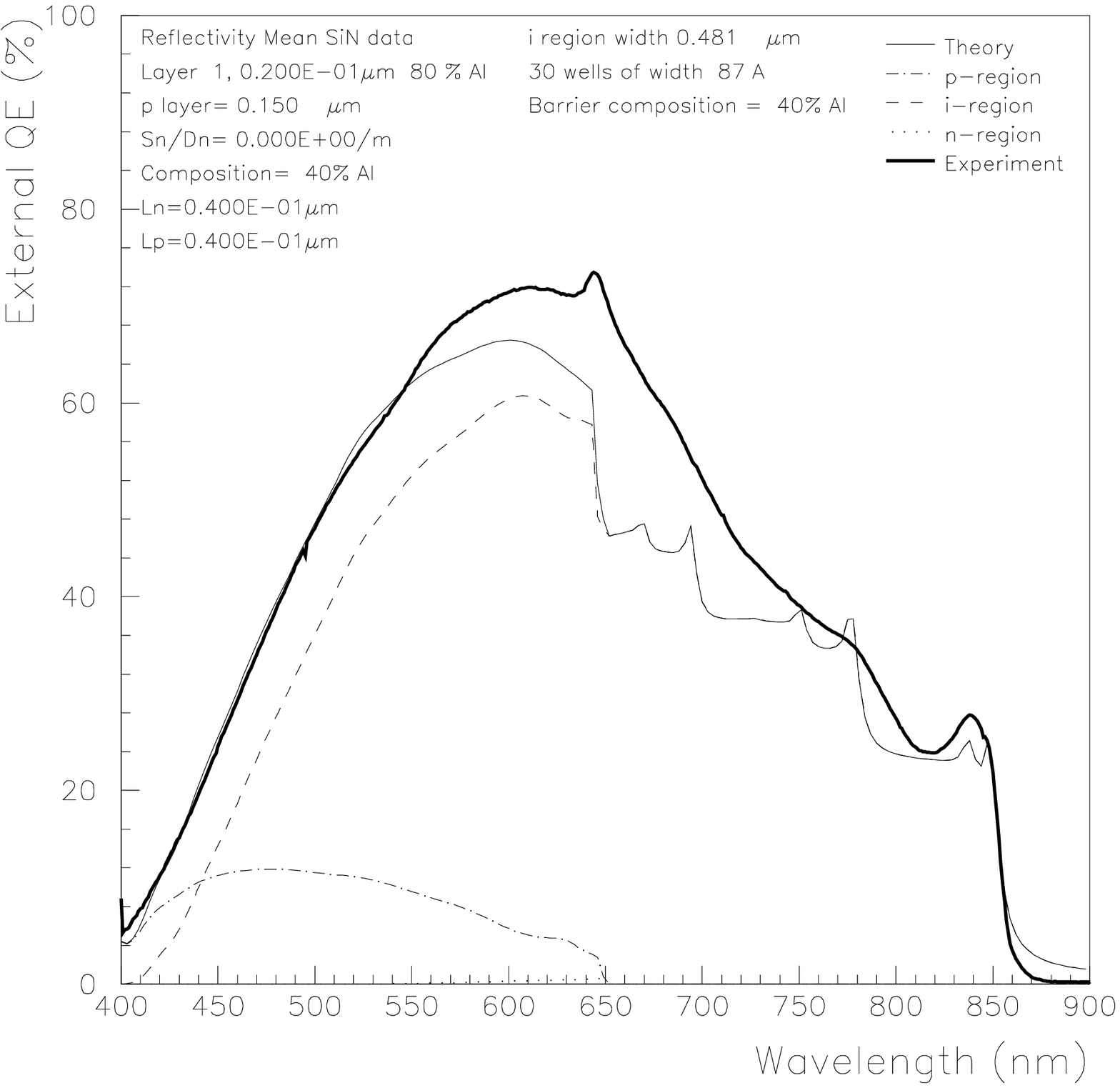}
{$QE$ modelling and data QT528, which is a repeat of QT468a with
a cleaner Ga source\label{figar528}}

All MOVPE samples were grown with nominal aluminium
fractions of about $X_{b2}=30\%$. They conform to
the same basic design, which consists of an 80\% aluminium
window layer of thickness $0.02\mu m$, a $0.15\mu m$ p
layer, carbon doped at $1.34\times 10^{18} \rm cm^{-3}$. The
n layer consists of $0.6\mu m$ Si doped at 
$6 \times 10^{17} \rm cm^{-3}$. 

No direct reflectivity measurements were available for the
samples presented in this section. We therefore use the
mean SiN data mentioned earlier in section 
\ref{secreflectivities}.

The composition of samples frequently departs
from the specifications by a few percent.
Experimental $QE$ and modelling for the 30 well
QWSC QT468a and its control sample QT468b are shown in
figures \ref{figar468a} and \ref{figar468b}. The
nominal aluminium fraction in both cases is $X_{b2}=33\%$.
Modelling however suggests the rather high value 42\% for 
QT468a and 36\% for QT468b. The value of 42\% for QT468a is,
as we shall see in section \ref{secmirrorqe} 
supported by modelling the Fabry-Perot
peaks of mirrored samples.

The fitting for QT468a suggests a p layer diffusion length of
$L_{n}=0.075\mu m$. The n layer $QE$ is negligible. The
modelling uses a negligible surface recombination parameter,
which is set to 0.

Sample QT468b shows a shortfall in $QE$ near the band edge.
This is not ascribed to a large n layer $QE$, given
that a value of $L_{p}$ much larger than $L_{n}$ would be
required to raise the theoretical $QE$ in this wavelength
range to the experimental level. The shortfall may be due
either to an increased effective i region thickness or to
the error introduced by the use of the mean SiN reflectivity. Finally,
it may be due to the uncertainty in the absorption coefficient
which was mentioned in chapter \ref{secparameters}. The
error in modelling however is not serious, in view of the
$\pm 5\%$ error on the experimental data.

Finally, figure \ref{figar528} shows modelled and measured
QE for sample QT528, which is a repeat of QT468a but with
a cleaner Ga source. The composition is modelled as $X_{b2}=40\%$.
The modelled diffusion length is significantly shorter than
QT468a, at $L_{n}=0.04\mu m$. For material with these very
low diffusion lengths, minority carrier loss at the front
of the cell is dominated by bulk recombination. The modelling
becomes insensitive to the surface recombination parameter
$\cal S$, which cannot therefore be ascertained with any
degree of confidence. It is therefore set to zero.

The model significantly
underestimates the experimental $QE$ at wavelengths between
about 550nm and 650nm. This may again be due to the assumption
of a mean SiN reflectivity. Possible departures from sample
specification might also account for the shortfall in
theoretical $QE$ but cannot be established whithout further
characterisation, for example SIMS.

The modelling however suggests that minority carrier
transport is much lower than that observed in the
Eindhoven samples. The diffusion lengths are slightly
lower than the values suggested by Hamaker, and
vary significantl between growth runs.
The surface recombination parameter in all three samples
was negligible.

\section{MBE samples\label{sechomombeqe}}

The samples U2027 to U2032 and U4036 mentioned in section 
\ref{secmbemivs} were modelled with a view
to establishing how the minority carrier transport
depends on the base aluminium fraction $X_{b2}$ for
MBE material.
The samples all consist of
a 67\% aluminium window of width $0.02 \mu m$, a $0.15\mu m$
p layer Be doped at $\rm 2\times 10^{18} cm^{-3}$, a $0.481 \mu m$
i layer, and finally a $0.6 \mu m$ n layer Si doped at 
$\rm 6\times 10^{17} cm^{-3}$.

The samples took the form of ungraded QWSC devices and their
\pin\ controls. All samples were grown in the same growth run with
the exception of the 20\% \pin\ sample U2028 which was initially
grown with a \gaas\ i region instead of the intended \algaas.
This device was repeated in a subsequent growth run as sample U4036.

Experimental $QE$ and modelling data for the AR coated devices
processed from these samples are presented in graphs
\ref{figar2027} to \ref{figar2032}.

\smalleps{htbp}{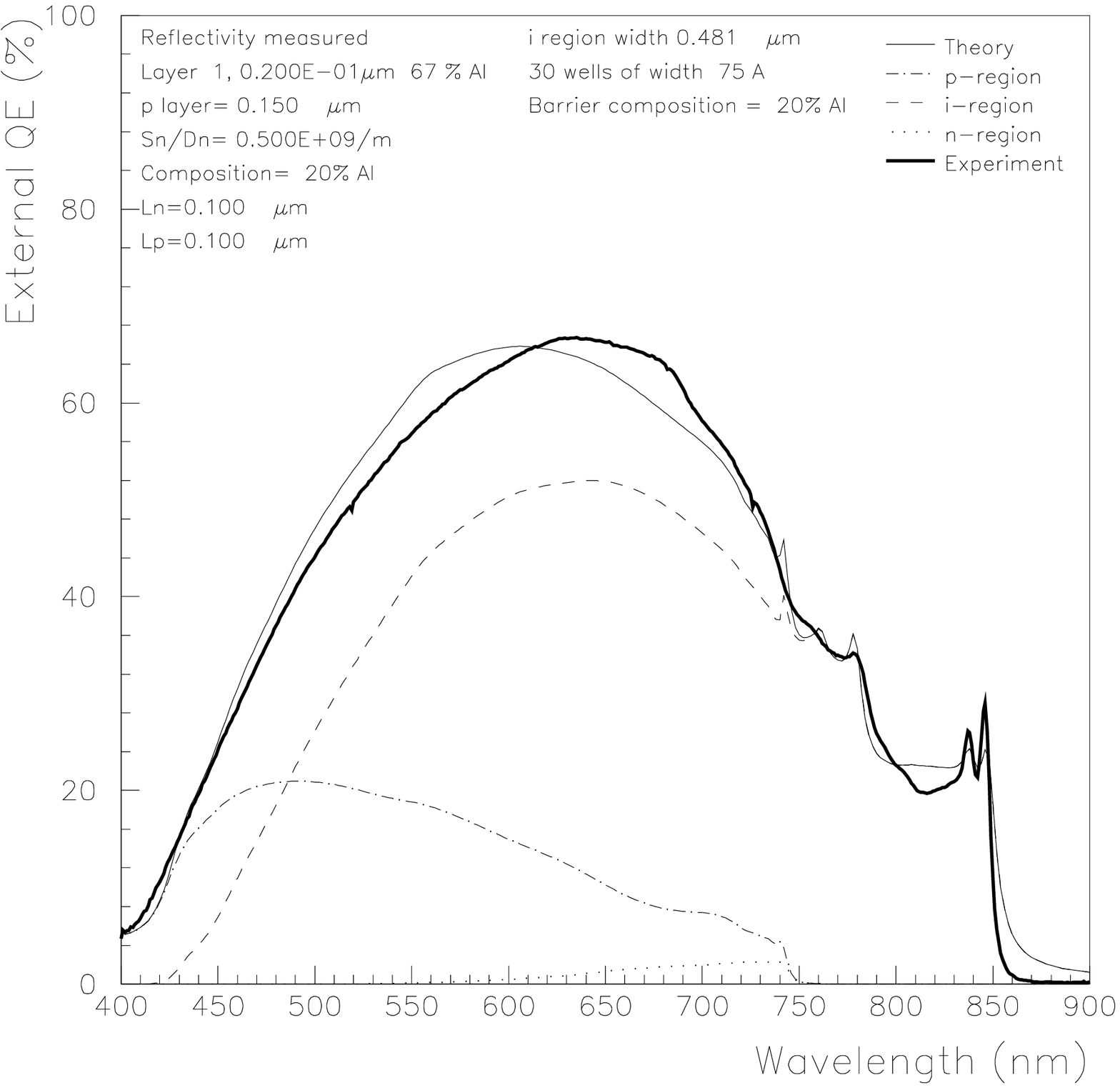}
{$QE$ modelling and data for nominally 20\% aluminium 30 well
sample U2027\label{figar2027}}
\smalleps{htbp}{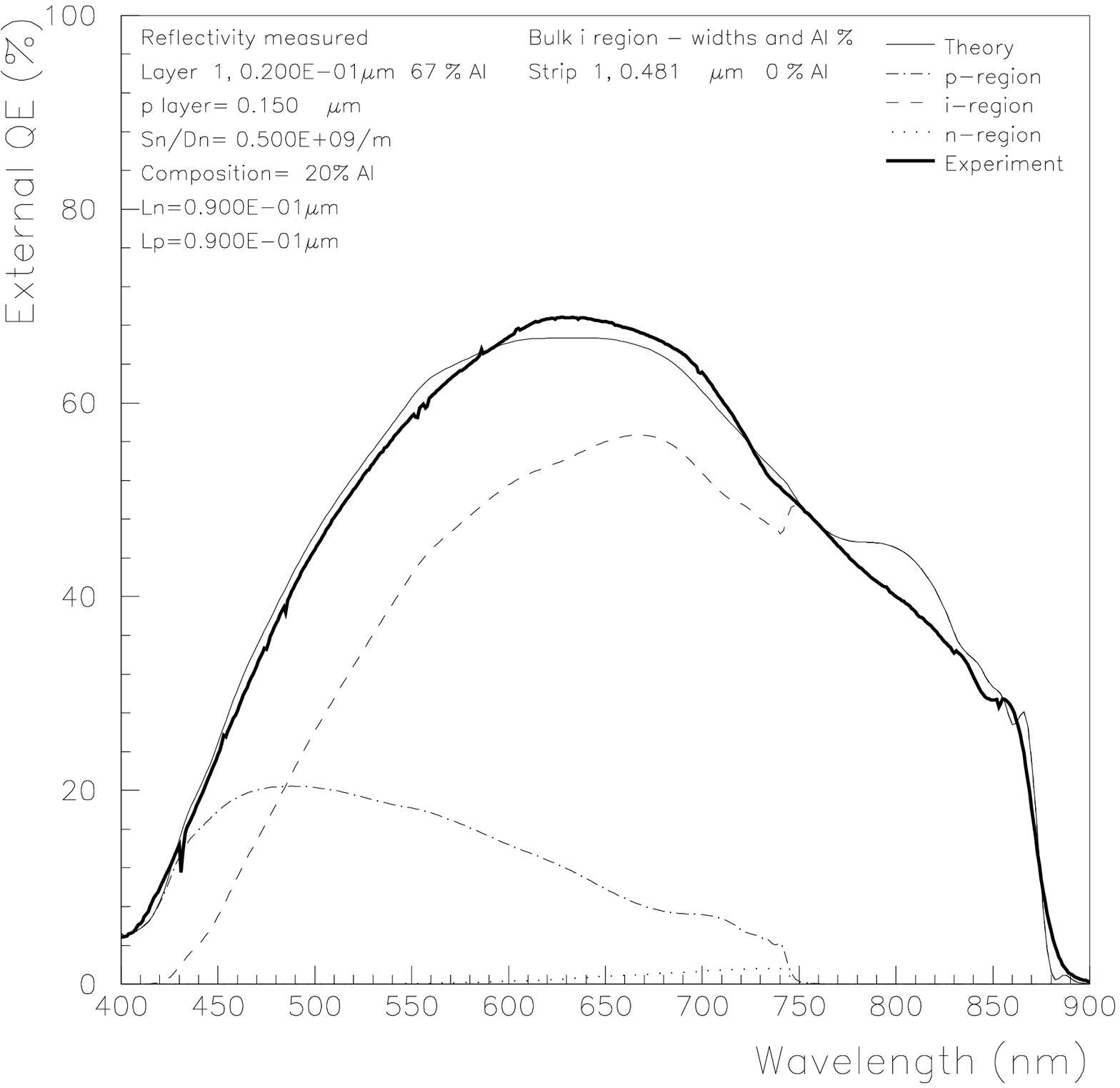}
{$QE$ modelling and data for U2028, which consists of 30\%
p and n layers with a GaAs i region\label{figar2028}}
\smalleps{htbp}{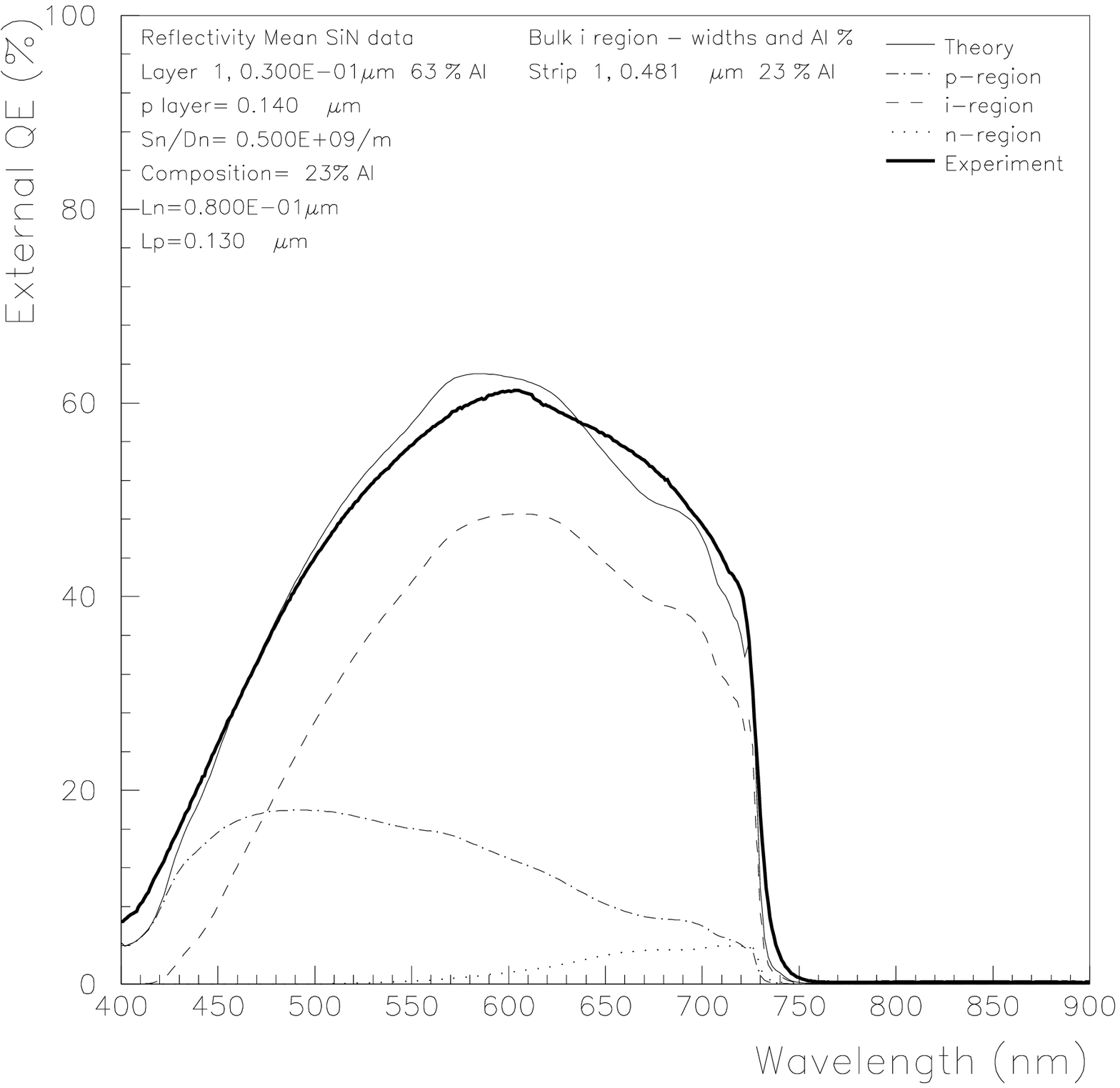}
{$QE$ modelling and data U4036, which is a \pin\ repeat of U2028
with a base aluminium fraction of $X_{b2}=23\%$ \label{figar4036}}

\subsection{20\% samples}

Samples U2027 and U2028 are accurately modelled with aluminium
fractions of $X_{b2}=20\%$. U4036 however has a composition close 
to $23\%$. 

Although the modelling inaccuracies for these samples are of the 
order of the experimental error, the fitting procedure for these samples
presents some problems concerning choice of parameters.
The modelled $QE$ peaks at a shorter wavelength than the
data. This could in principle be modelled more closely by increasing 
$L_{p}$ to boost the n layer near the band edge, at about 750nm, and
reducing $L_{n}$. The short wavelength reponse is poor, and sets a 
limit on the maximum value of $L_{n}$. Since, in general, holes are
less mobile than electrons, the electron diffusion length should
be greater than the hole diffusion length. The Hamaker 
parametrisation for example suggests that $L_{n}$ is about twice
as large as $L_{p}$ for all aluminium fractions. 

We cannot in principle rule out that $L_{p}$ may be
greater than $L_{n}$, for example due to particular growth
or geometrical considerations. In the absence of more detailed
information, however, we choose to set a limit on the maximum 
value of $L_{p}$, which is that it should be less than or 
equal to $L_{n}$

The short wavelength response then poses a problem because a
value of $L_{n}$ high enough to fit the band edge causes the
model to overestimate the $QE$ at short wavelengths, even though
we use a quasi-infinite recombination parameter.

The fits presented in figures \ref{figar2027} to \ref{figar4036}
are the result of a compromise between these two constraints.
The resulting diffusion lengths of $L_{p}$ and 
$L_{n}\sim 0.10 \mu m$ are smaller than those given by the Hamakers 
parametrisation, which predicts $L_{n}\sim 0.39 \mu m$ and 
$L_{p}\sim 0.20 \mu m$.

Finally, the quantum wells in the QWSC sample are modelled as 
$\rm 70 \AA$, which is significantly smaller than the nominal
$\rm 87 \AA$. Furthermore, we note that the model overestimates
the $QE$ contribution in the first quantum well continuum,
at about 800nm. The misfit in this wavelength range is not
serious, however, and is of the order of the experimental
error of $\pm 5\%$ in $QE$.

This may be interpreted as indicating a loss mechanism in
the quantum wells, possibly due to the increased 
recombination at the interface between quantum wells 
and the barriers. Alternatively, dopant diffusion, combined
with a diffusion length which is closer to the Hamaker value
would reduce the $QE$ at short wavelengths and increase it near the
band edge $E_{b2}$. More characterisation is required to
establish whether or not this is the case.

\subsection{30\% samples}

\smalleps{htbp}{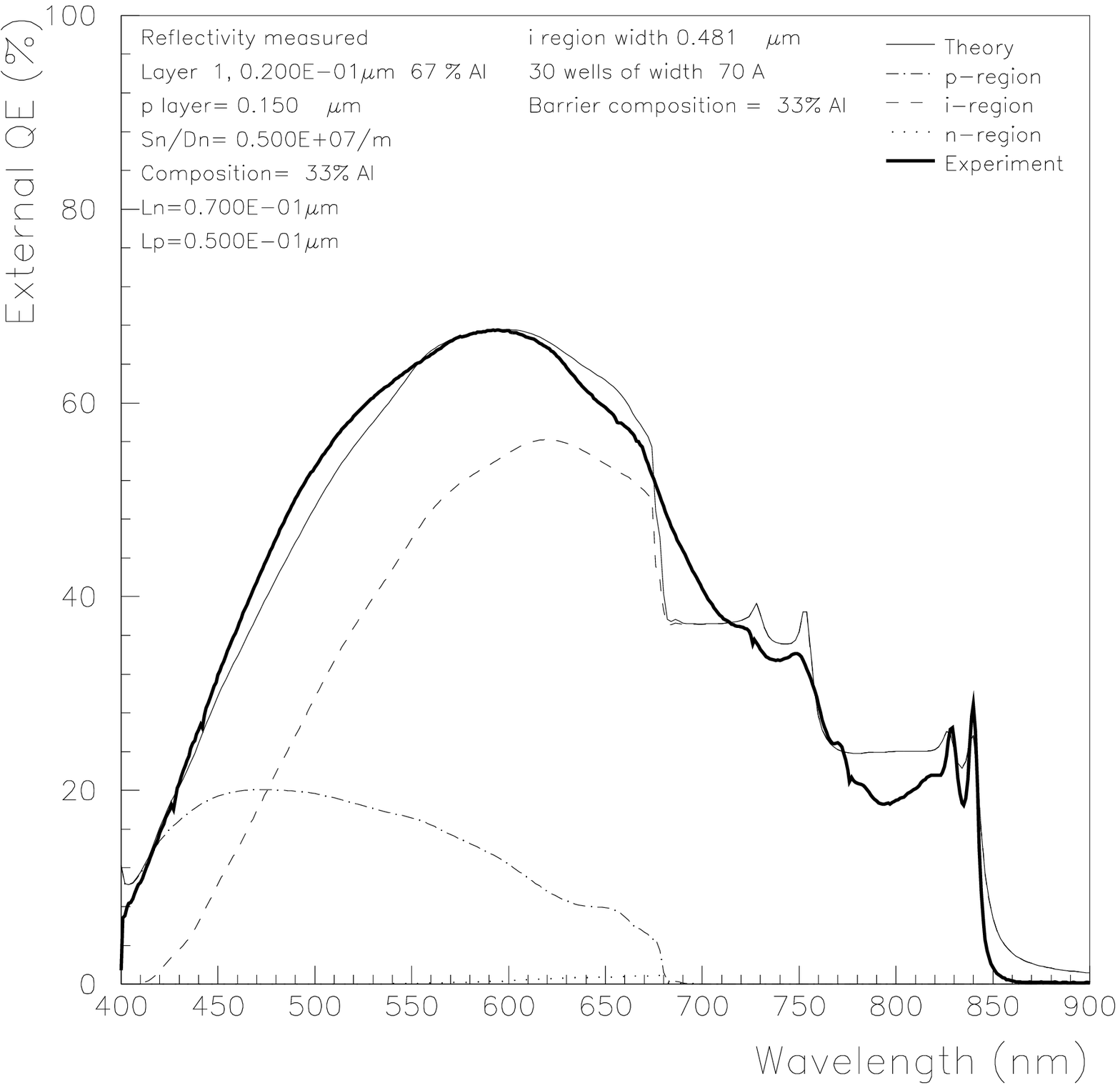}
{$QE$ modelling and data for nominally 30\% aluminium 30 well
sample U2029\label{figar2029}}

\smalleps{htbp}{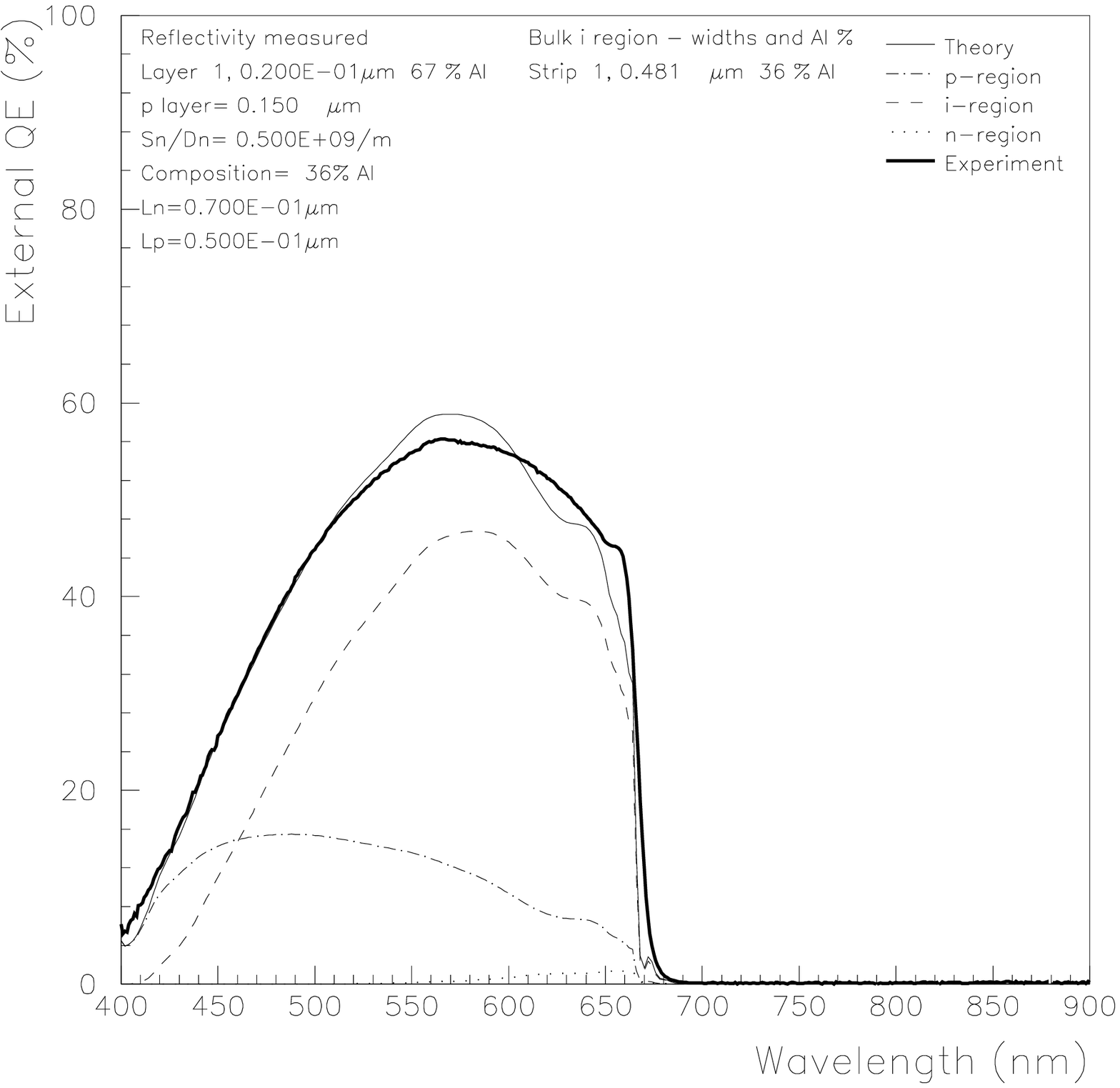}
{$QE$ modelling and data U2030, the \pin\ control to U2029
\label{figar2030}}

Figures \ref{figar2029} and \ref{figar2030} show $QE$ data and
modelling for U2029 and U2030. The true aluminium fractions  are
estimated at $X_{b2}=33\%$ and $X_{b2}=36\%$ respectively. 
The n layer contribution of the QWSC device appears negligible,
and no reliable estimate of $L_{p}$ is possible. The recombination
parameter $\cal S$ is modelled as small but finite. 
$L_{n}$ is then the single significant fitting parameter, and is
modelled as about $L_{n}\sim 0.07\mu m$. This compares
favourably with the Hamaker value of $0.053 \mu m$.

The quantum well $QE$ is again overestimated by the model. This
may be ascribed to the same reasons as were given for U2027 in the
previous section. 

For the \pin\ samples U2030, the n layer $QE$ is again negligible.
The surface recombination parameter in this case appears
effectively infinite, although the predicted diffusion length
is identical to the QWSC sample. Problems in the fit are similar
to those seen for sample U4036, but are again comparable to the
experimental error.

The estimate of $L_{n}$ in this case is more reliable than
in the case of the $20\%$ samples because the fitting procedure
is simpler and the resulting match between theory and experiment
better.

\subsection{40\% samples}

The $40\%$ samples U2031 and U2032 are mutually more consistent,
and are shown in figures \ref{figar2031} and \ref{figar2032}.
Both have a modelled aluminium fraction of $47\%$ instead of
the nominal $40\%$. The n region $QE$ is again vanishingly
small. $\cal S$ in both cases appears too low to model,
and is set to zero. 

The diffusion lengths in this case are modelled as 
$L_{n}\sim 0.040\mu m$ for the QWSC and $L_{n}\sim 0.055\mu m$
for the \pin\ sample. Hamaker in this case suggests $0.02\mu m$.

The quantum wells in this case fit very closely in the 
first continuum. This however contradicts the possible
explanation for an overestimated well $QE$ in the $20\%$ and
$30\%$, since the interface recombination in a $40\%$ sample
should in principle be higher.

\smalleps{htbp}{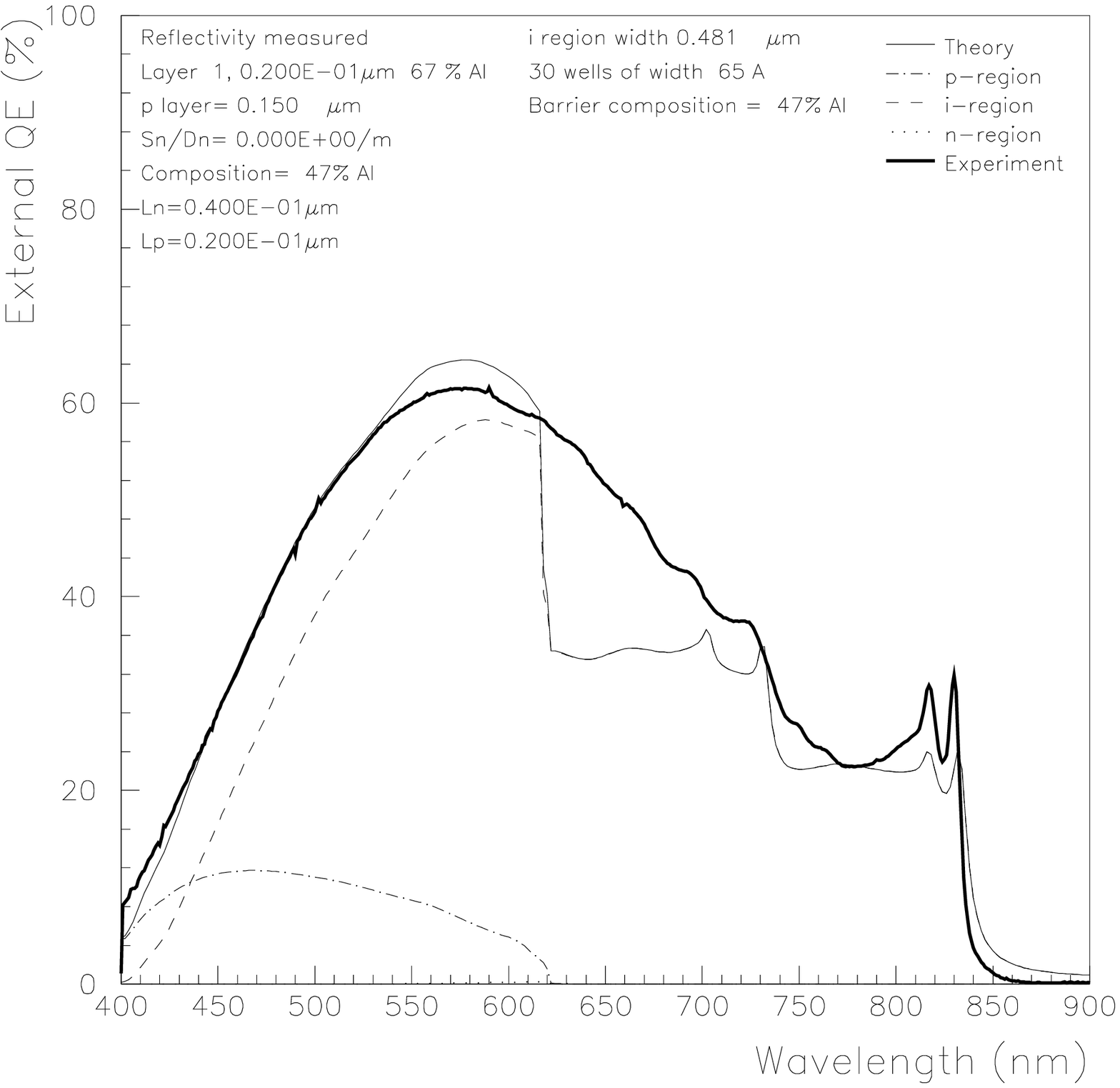}
{$QE$ modelling and data for nominally 40\% aluminium 30 well
sample U2031\label{figar2031}}
\smalleps{htbp}{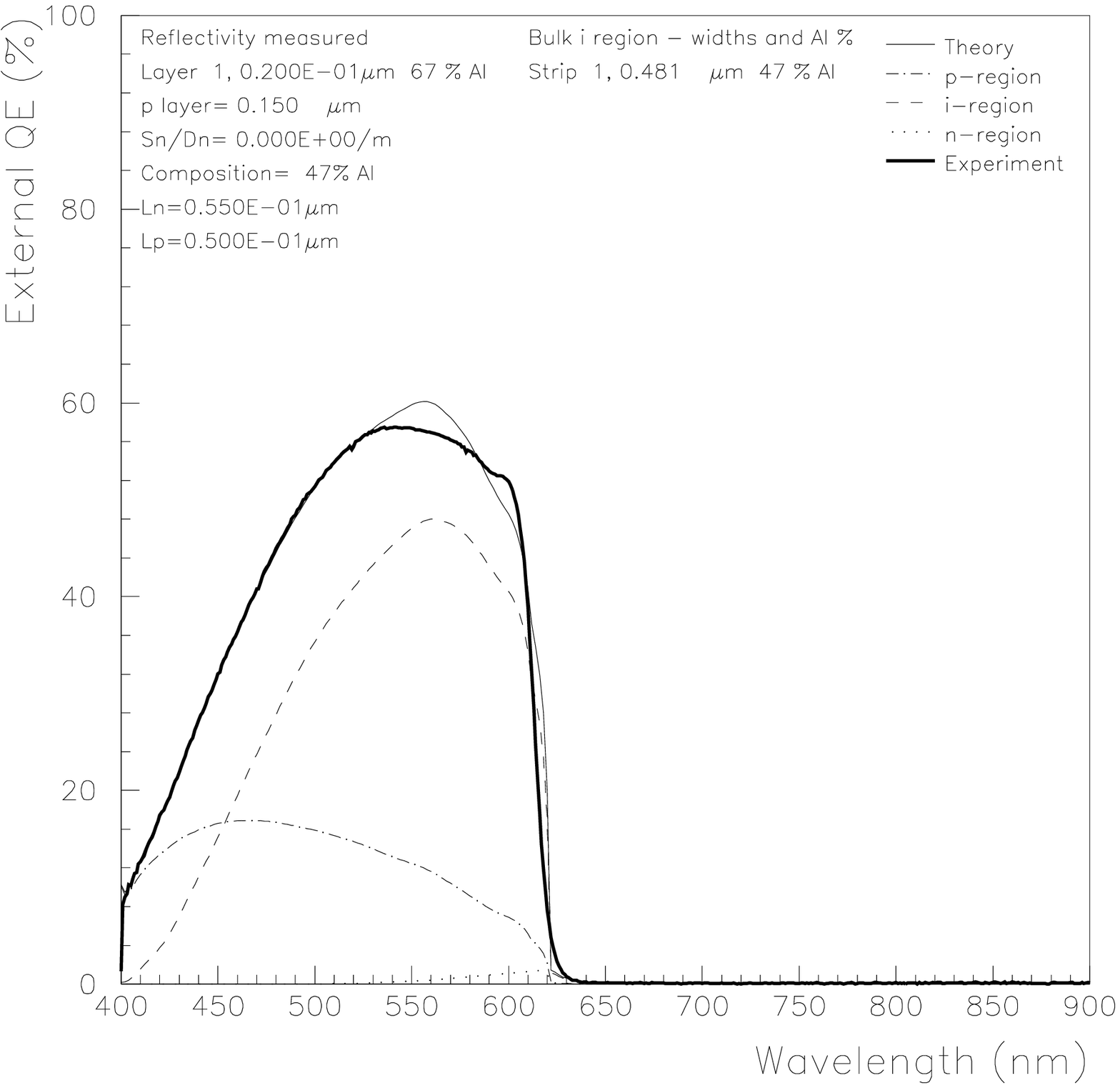}
{$QE$ modelling and data U2032, the \pin\ control to U2031
\label{figar2032}}

\subsection{Diffusion length tabulation \label{seclntabulation}}

\smalleps{htbp}{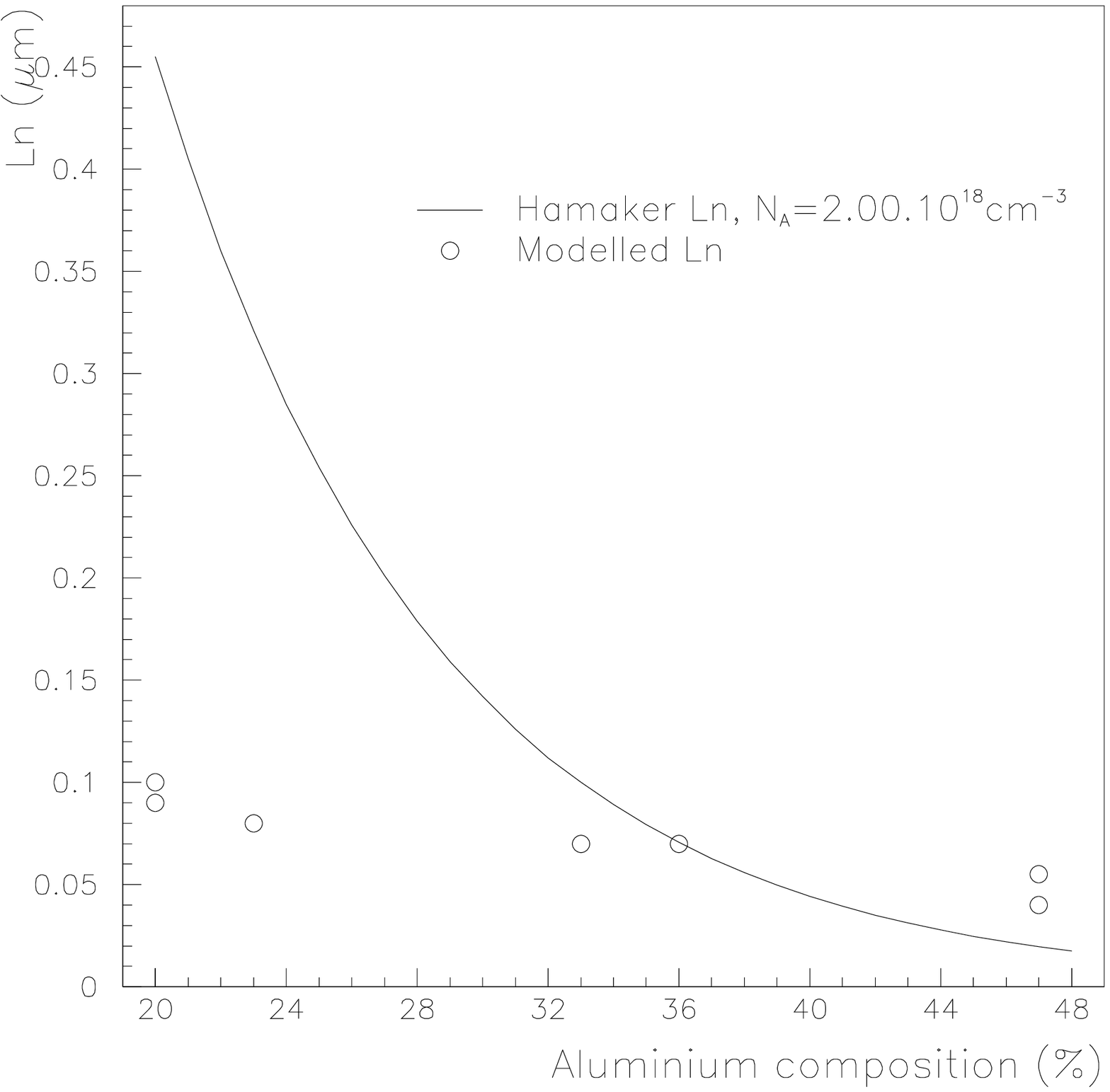}
{Modelled minority electron diffusion length $L_{n}$ 
for \pin\ and QWSC samples, compared with Hamaker's 
parametrisation for the MBE doping level doping levels
\label{figlntable}}

Figure \ref{figlntable} shows the modelled value of 
$L_{n}$ as a function of aluminium fraction $X_{b2}$
for QWSC and \pin\ MBE grown ungraded structures of the previous
section. Also shown is the Hamaker parametrisation \cite{hamaker85} 
for the relevant doping level.

The modelled values show a weaker dependence on aluminium
fraction than the Hamaker parametrisation. The modelled values
decline slowly with increasing $X_{b2}$. These values
define the piecewise linear parametrisation which is used
to define the position dependence of $L_{n}$ in the graded
structures of the section \ref{secgradeqe}.

We conclude however that the transport parameters derived from the
$QE$ spectra systematically indicate poorer minority carrier transport
efficiency than has been reported in the highest quality \algaas.
Bachrach\cite{bachrach72}, for example, has reported diffusion
lengths of the order of 2.3$\mu m$ in p doped \algaas\ with a composition
of $x=36\%$, although at a lower doping level of 
$N_{A}\sim 10^{17}\rm cm^{-3}$. The review by Ahrenkiel
\cite{ahrenkiel92} shows that diffusion lengths greater than those
given by the Hamaker parametrisation are possible.

\subsection{Conclusion}

In this section we have modelled a selection of
samples from the two available growth systems. The
resulting p layer diffusion length has in all cases proved
smaller than the p layer width.  Although the MOVPE
material seems to have more variable minority carrier
transport properties, the values obtained are close
to those of the MBE material. A small mismatch between
theory and experiment has been observed in a number of
samples. This may be due the uncertainty in the absorption
coefficient mentioned in chapter \ref{secparameters},
although other influences such as the reflectivity,
depletion widths and i region width cannot be ruled out.

The resulting tabulation of the electron minority carrier
diffusion length $L_{n}$ is more weakly dependent
on the aluminium fraction than suggested by the parametrisation
due to Hamaker \cite{hamaker85}. The results further indicate
that the minority carrier transport efficiency in both MOVPE and
MBE material is relatively poor compared to values in the
literature.

The surface recombination parameter $\cal S$ is low in most of
the samples we have studied. Notable exceptions are the MBE
grown samples with low aluminium composition and one \pin\ sample
from the same growth system. These samples used a quasi-infinite
$\cal S$. Supplementary characterisation is required to establish
whether this is due to innacurate modelling
of the minority carrier transport near the surface of the cell,
or to geometrical and compositional factors.

\section{Compositionally Graded p layer samples\label{secgradeqe}}

In this section we present data and theoretical modelling of the $QE$
of graded p layer samples grown by MOVPE and MBE. We will quote
estimates of the short circuit current density $J_{sc}$ calculated
from the experimental $QE$ data for an illumination of AM1.5. These
figures do not include contact shading.

In the case of MBE samples, we will use the
diffusion length parametrisation based on the modelling of
ungraded samples. In the case of MOVPE samples, we do not
have as wide a range of samples with different aluminium
fractions. The previous section however has demonstrated that the 
diffusion length in both growth systems is similar. We shall therefore use
the same table of $L_{n}$ as a function of aluminium fraction for
MOVPE material.

\subsection {MOVPE Samples\label{secgrademovpeqe}}

\smalleps{hp}{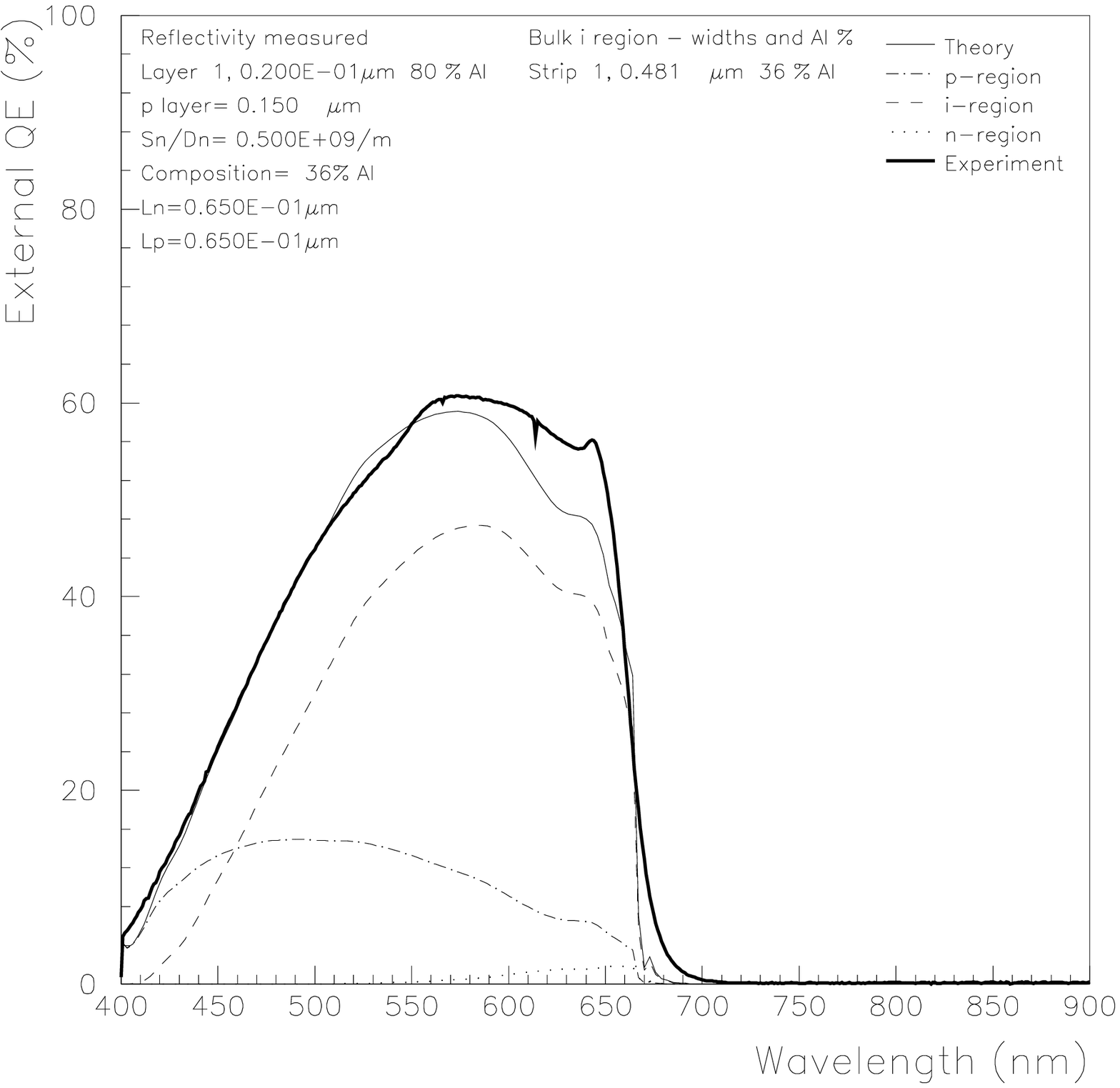}
{Experimental and modelled $QE$ of ungraded MOVPE \pin\ 
QT601a \label{figar601a}}
\smalleps{hp}{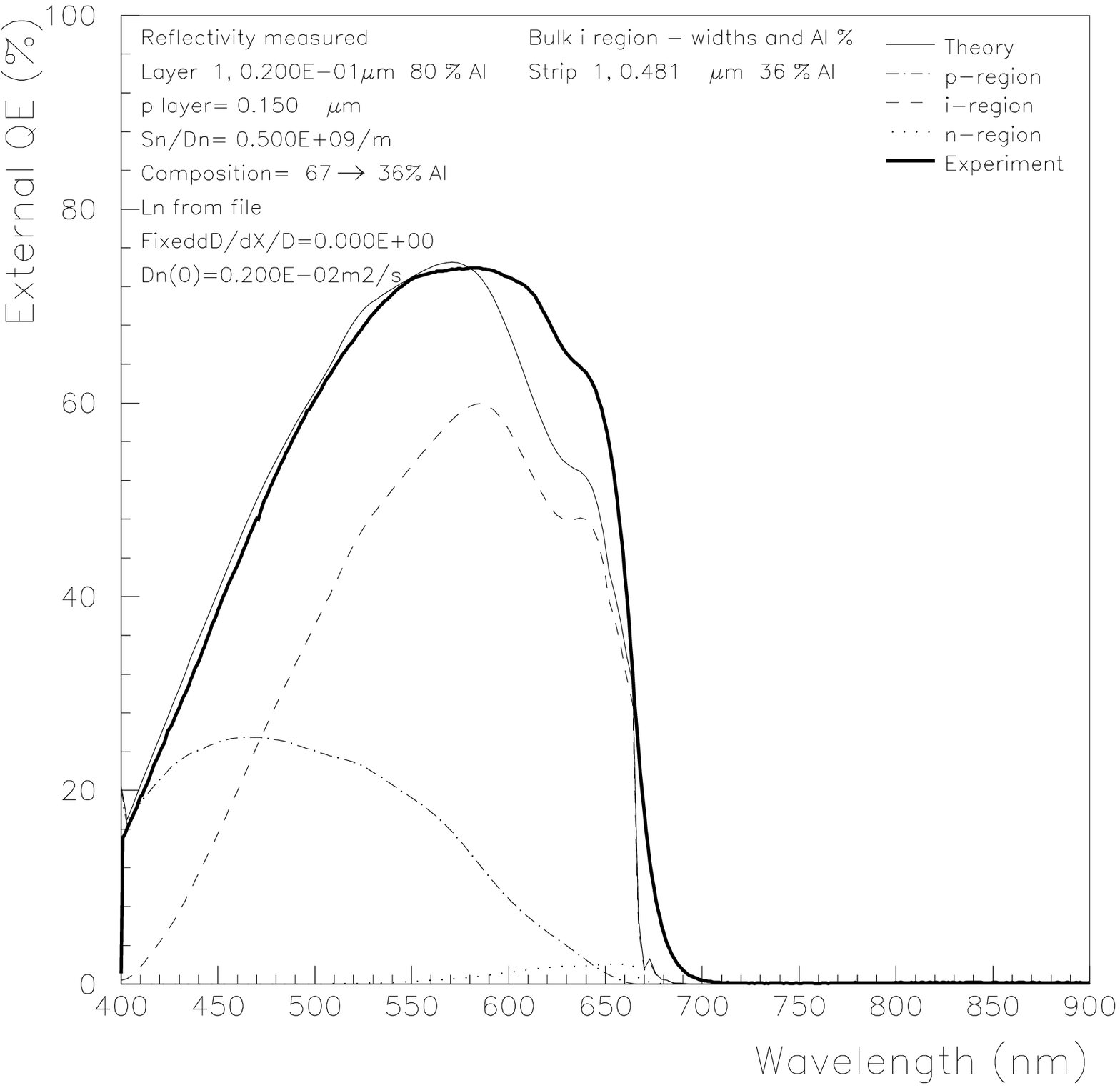}
{Experimental and modelled $QE$ of graded MOVPE \pin\ 
QT601c \label{figar601c}}

The graded p design was initially tested by graded
\pin\ device QT601c and its control structure QT601a, which
is identical but has no grade. The graded p in sample QT601c
is $0.15\mu m$ thick and is graded from $X_{b1}=67\%$ to 
$X_{b2}=33\%$, and has a window of thickness $0.02 \mu m$ with
a composition of 80\% aluminium. 
John Roberts however has indicated that growth problems
cast some doubt over the real value of the aluminium fraction $X_{b1}$ 
at the top of the grade. The $QE$ data and modelling for these samples
is shown in graphs \ref{figar601a} and \ref{figar601c}.

The fit for both samples requires a quasi-infinite recombination
parameter and $L_{n}$ and $L_{p}\sim 0.06\mu m$. The fitting problems
here are again reminiscent of samples such as the 20\% ungraded
devices of the previous section, but are more significant.
A larger value of $L_{p}$ would again improve the $QE$ fit, 
but cannot be justified without more information.

The $QE$ of graded sample QT601c suffers from a similar problem near
the band edge. The short wavelength $QE$ however is accurately
modelled with ${\cal D_{X}}=-1.5$, versus the Hamaker value of 
-4.2. The grade in this case enhances the short circuit current 
in AM1.5 from $J_{sc}=8.06 \rm mA/cm^{-2}$ for QT601a to 
$J_{sc}=10.35\rm mA/cm^{-2}$. This represents an overall increase
in projected $J_{sc}$ of 28\%.

A set of three samples was subsequently grown to investigate
the performance of a grade in MOVPE QWSC designs. The first
of these is QT640a, which is similar to QT468a but for a grade 
in the p layer with nominal composition $X_{b1}=67\%$ and 
$X_{b2}=33\%$. Sample QT640b is identical to QT640a and has 
the same grading scheme, but has a wider i region of $0.78\mu m$
containing 50 wells. Finally, QT641 is a \pin\ control to QT640b
but does not contain a grade. This last \pin\ sample
was grown to compare material quality between the two growth
runs.

\smalleps{hp}{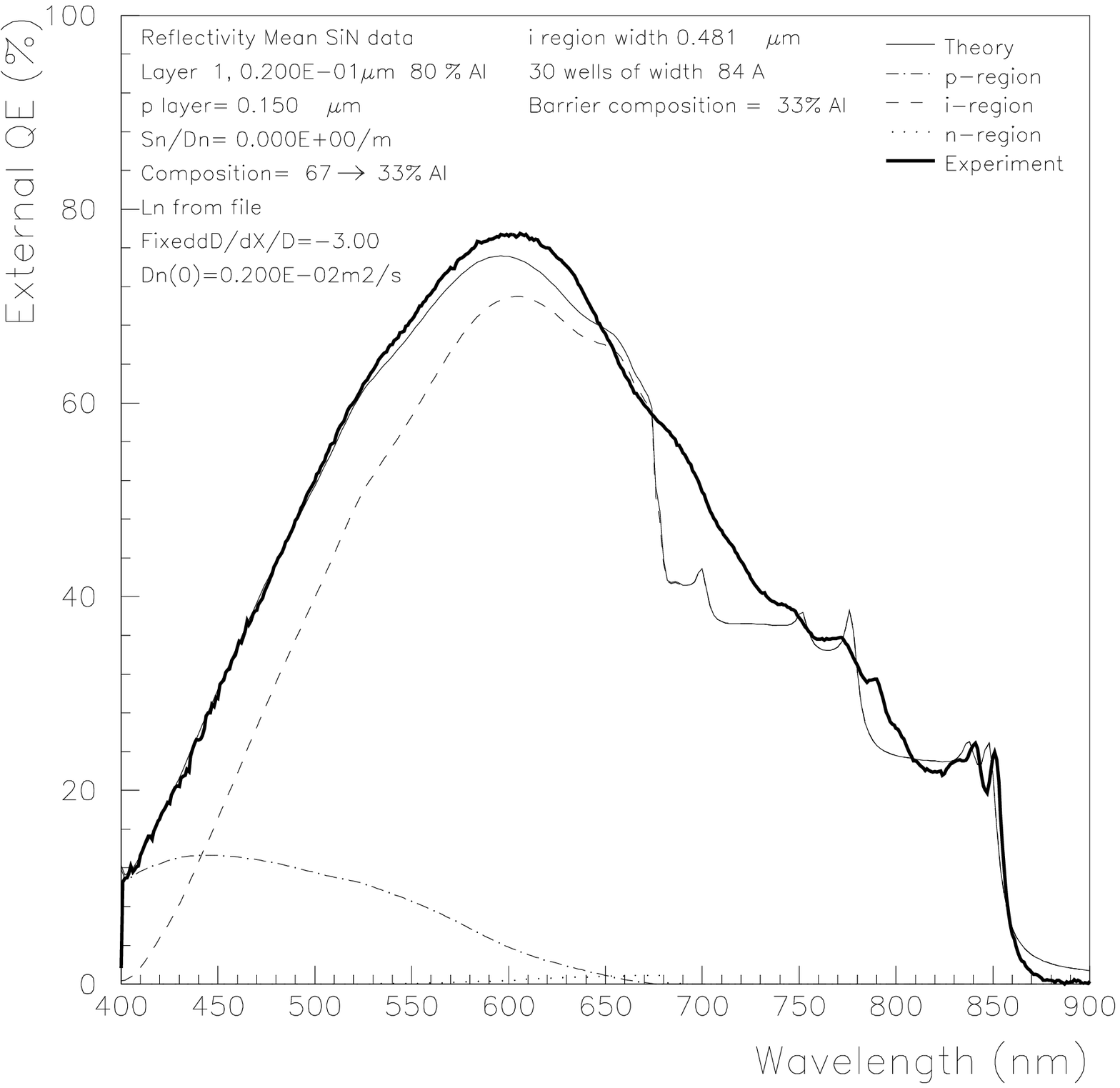}
{$QE$ of 30 well QWSC QT640a, with p layer 
grading ranging from 30\% to 67\% \label{figar640a}}
\smalleps{hp}{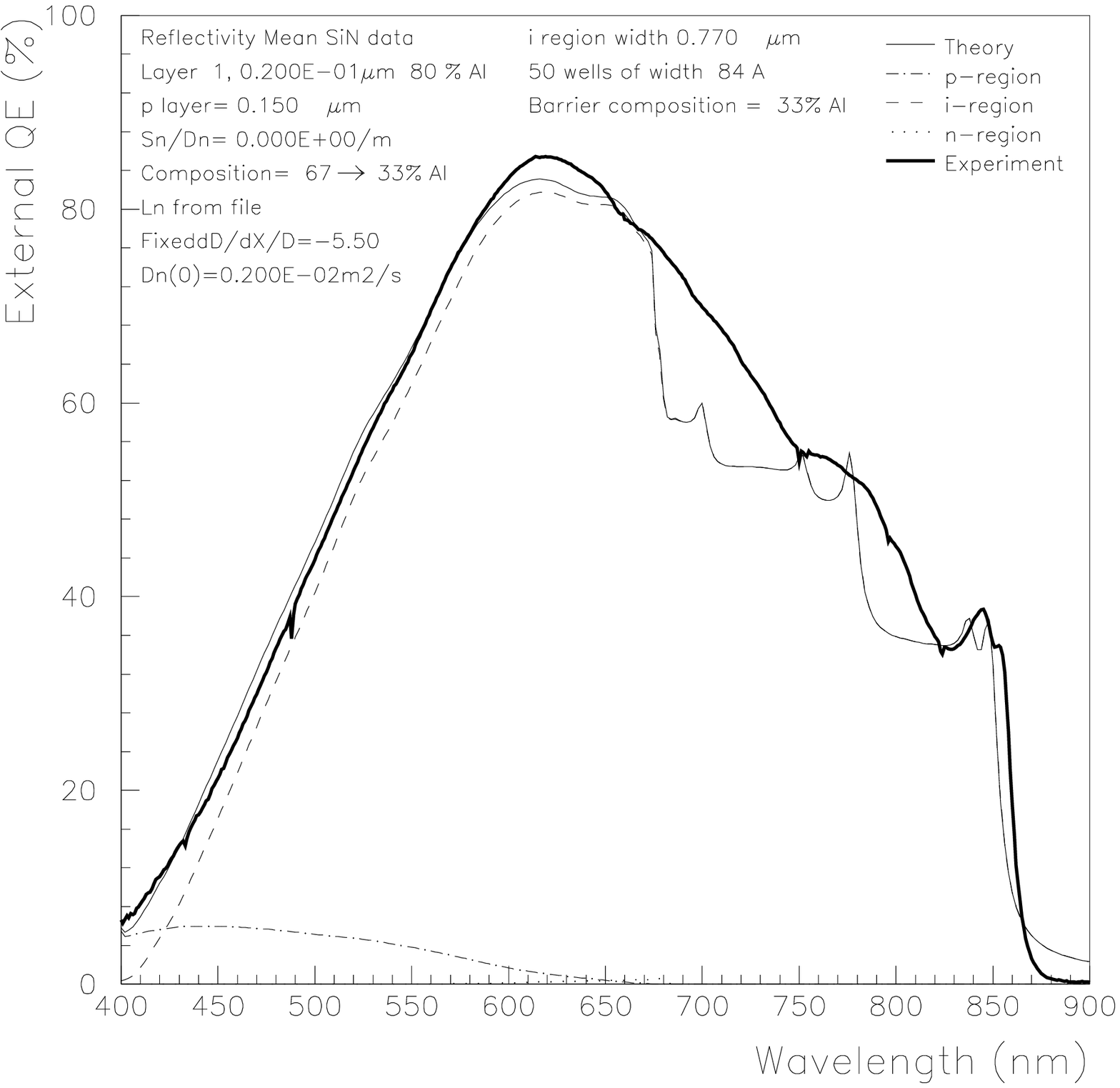}
{$QE$ of 50 well QWSC QT640b with p layer 
grading ranging from 30\% to 67\%\label{figar640b}}
\smalleps{htbp}{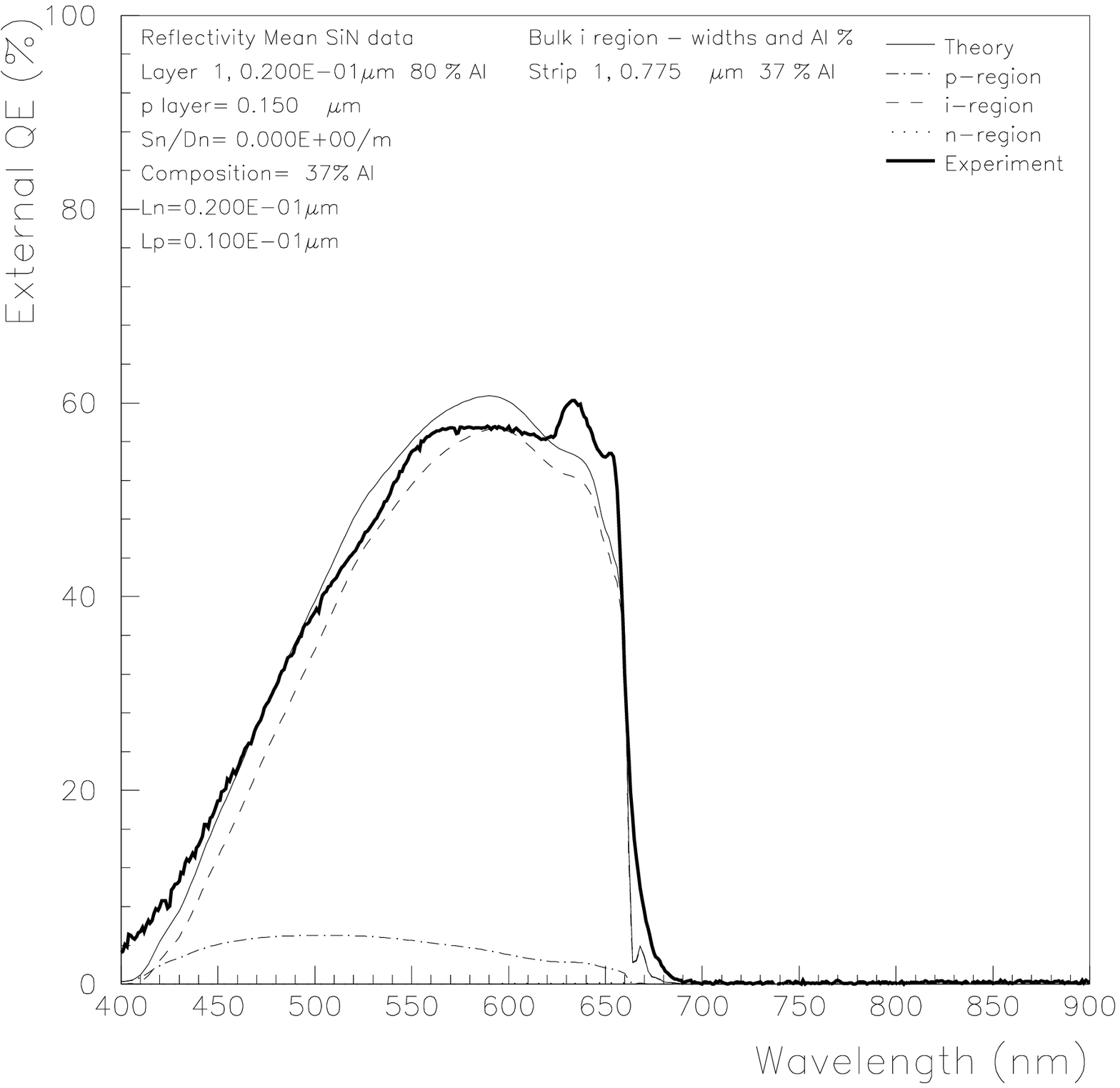}
{$QE$ of QT641, the \pin\ control to 50 well 
QWSC graded 30-67\%\label{figar641}}

The experimental $QE$ and modelling for QT641 is shown in
figure \ref{figar641}. The experimental $QE$ is very poor
at short wavelengths, indicating poor minority carrier
transport. The modelled p layer $QE$ is dominated by
bulk recombination losses, and the parameter $\cal S$ is not
significant in the fitting. It is therefore set to zero.
Finally, the n layer $QE$ is negligible. 

The two QWSC samples bear out the conclusion that
minority carrier transport efficiency for these samples
was lower than for the previously grown samples QT601a and 
QT601c. Although the grades in both QWSC samples are of
identical design to that of QT601c, they perform 
more poorly. 

The modelling and data for these samples is
shown in figures \ref{figar640a} and \ref{figar640b}.
The modelling assumes a mean SiN reflectivity.
The diffusivity field  for QT640a was ${\cal D_{X}}=-3$.
The p layer $QE$ of sample QT640b in particular is seriously 
degraded, and is modelled with ${\cal D_{X}}=-5.5$. The rather
large value of $\cal D_{X}$ for QT640b is expected to be at least
partly due to the assumption that the $L_{n}$ tabulation based
on the MBE samples of the previous section is over-optimistic
for these samples.

\smalleps{ht}{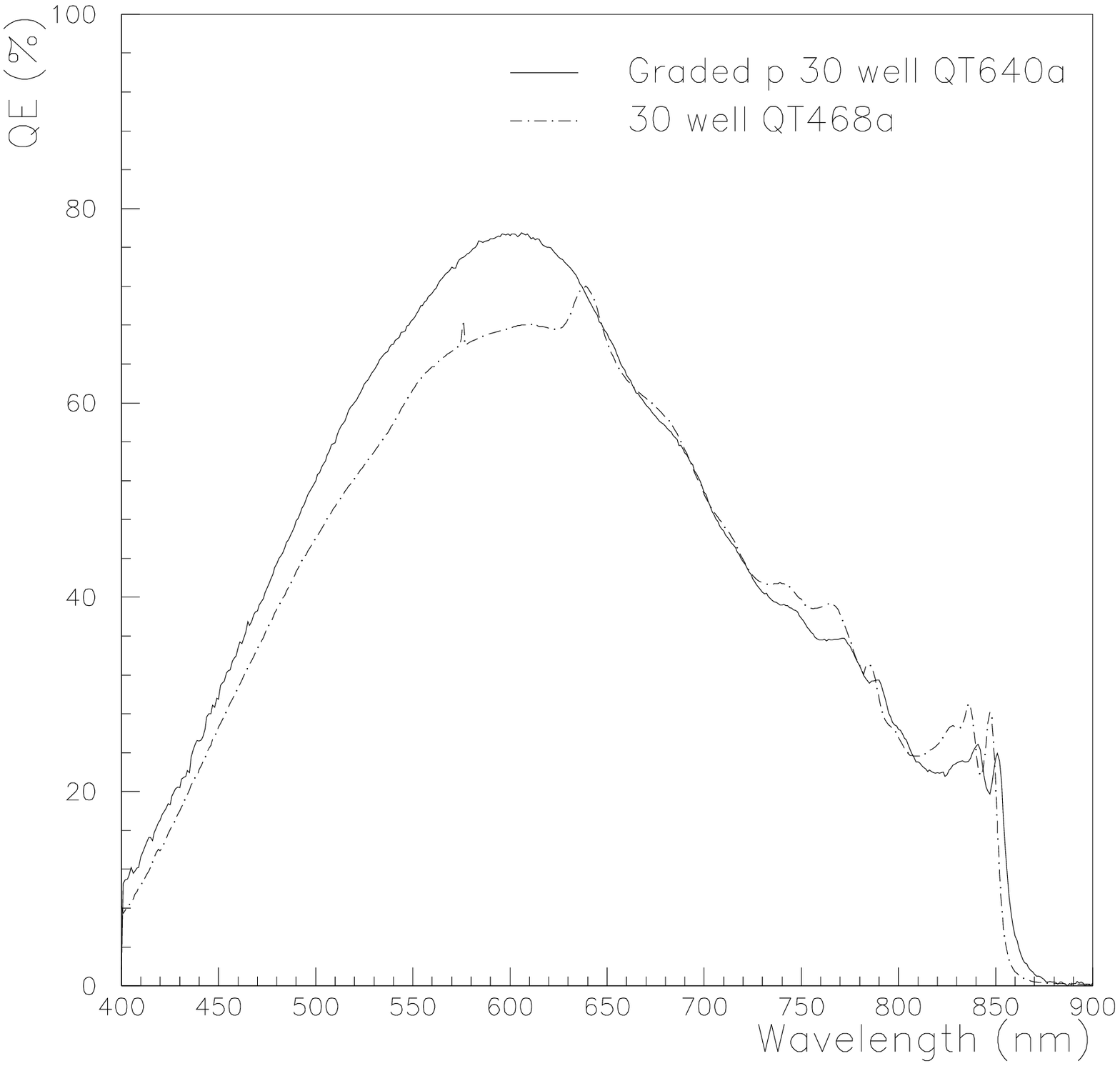}
{Comparison of thirty well sample QT468a and similar
structure QT640a which has a graded p layer \label{fig468a640a}}

Figure \ref{fig468a640a} compares the $QE$ of QT640a with
QT468a, which has the same design but for the absence of
the grade. The $QE$ enhancement at short wavelengths is
too small to be significant. We can state however that the
grade has at the very least maintained the p layer $QE$
in poorer material. This does not take into account the
different reflectivities in the two samples however, which
was better in the ungraded sample.

At intermediate wavelengths, a significant $QE$ enhancement is
observed because of increased light transmission to the i
layer.

Comparison of short circuit current densities gives
$J_{sc}=14.94\rm mA/cm^{-2}$ and $J_{sc}=14.02\rm mA/cm^{-2}$
respectively. This represents an enhancement of $7\%$ in
$J_{sc}$ for the graded sample, despite the inferior material
carrier transport in sample QT640a. We conclude that these 
samples illustrate the principle of a grade in QWSC samples, despite 
the fact that the ungraded benchmark sample benefits from superior 
minority transport properties.

\smalleps{htp}{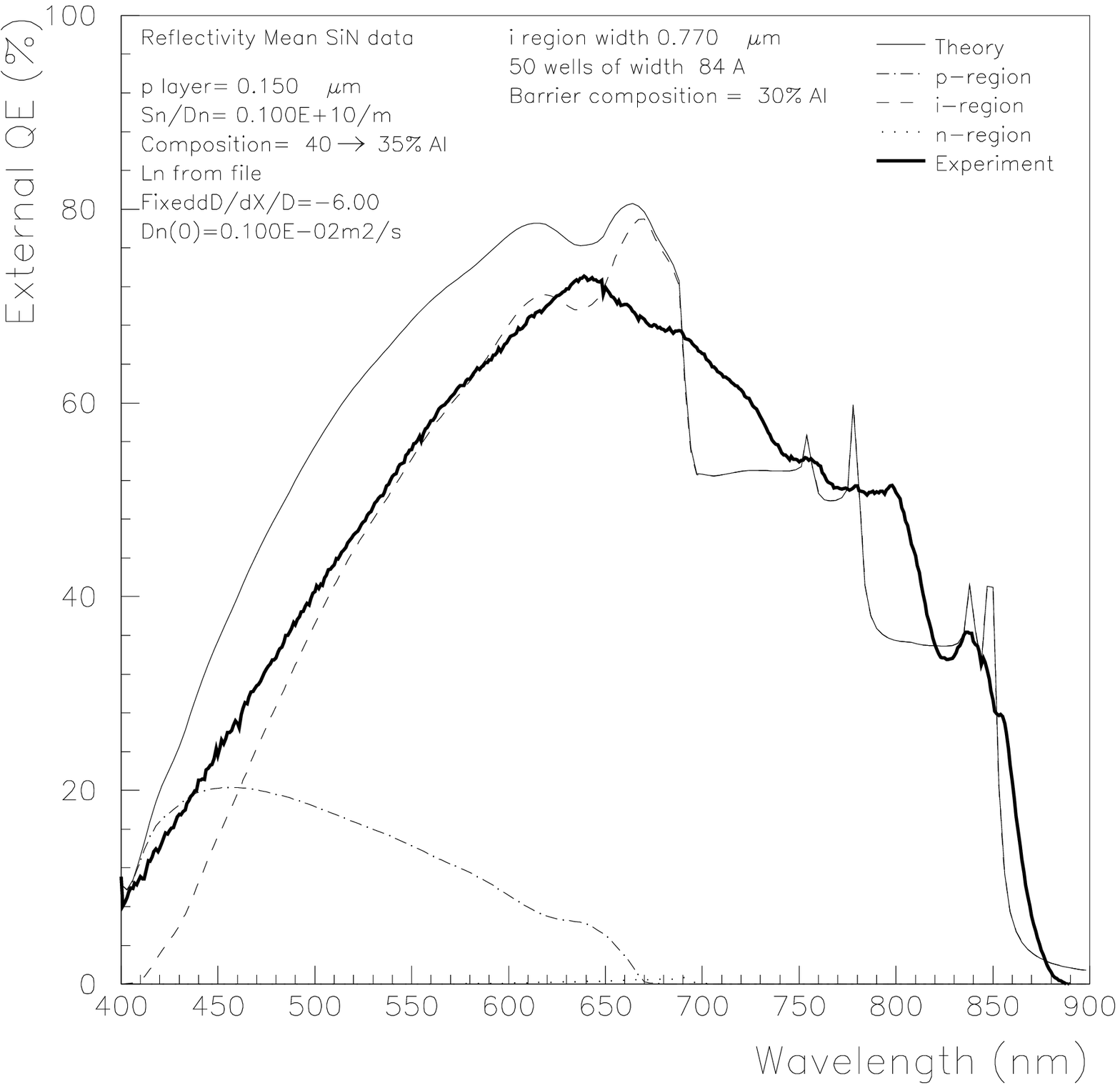}
{MOVPE version QT794a of the optimised 50 well graded QWSC design 
\label{figar794a}}
\smalleps{hp}{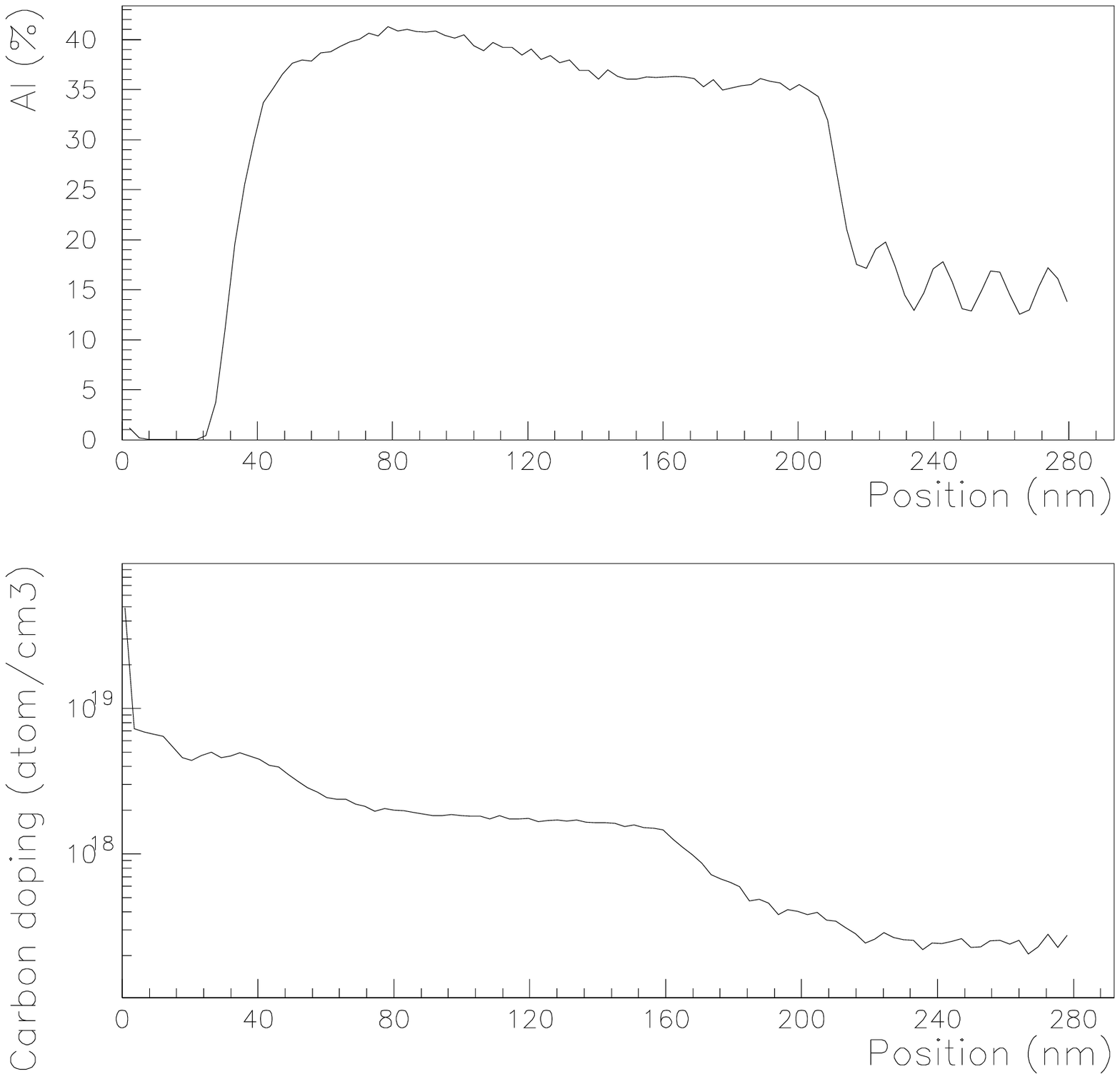}
{SIMS measurements of C doping and aluminium composition in sample
QT794a \label{fig794asims}}

Finally, we present experimental and theoretical data for
sample QT794a which cannot be explained by the model. This sample
has been mentioned in section \ref{secmivmeasurement} and was
shown to have an high $N_{BG}$. 

This sample was designed with a reduced
p layer thickness of $0.1\mu m$. The grading scheme was
aimed at maximising the p layer response by choosing a
moderate grade ranging in composition from $X_{b1}=40\%$ to
$X_{b2}=30\%$. This is at the lower end of the optimised
theoretical grading predictions presented in section
\ref{secqex}, and is intended to ensure good minority 
carrier transport by reducing the amount of aluminium towards 
the front of the p layer. We shall present results in the
MBE section below which demonstrate successful grades over this
compositional range.
Furthermore, the results of section \ref{secwindows} were
exploited and the window in this sample removed.

The results are shown in figure \ref{figar794a}. The modelling
benefits from a measured reflectivity. The MIV data indicated
that the $QE$ measurement had to be performed at at least -10V
in order to ensure full depletion of the device. We note that
the experimental $QE$ between about 500nm and 620nm is
approximately equal to the i region $QE$ alone, suggesting
the the p layer QE is essentially zero in this range.
It is impossible to model the p layer contribution
consistently under this constraint, particularly since the
short wavelength response is not negligible, indicating that
at least some minority carriers from the front of the cell
are collected.

SIMS and TEM measurements were performed 
on this sample and its MBE grown companion U6041, which will be 
presented in the following section. The TEM measurements confirmed 
that the dimensions of this sample were correct. The compositional
SIMS measurements had to be calibrated by inferring the aluminium
content from $QE$ spectra. The resulting aluminium profile is
shown in figure \ref{fig794asims}. The depth calibration
used by Federico Caccavale in the SIMS results was independently
verified by the TEM measurements. The mean sum of the well and
barrier thickness was estimated as ($15.3nm\pm 0.5nm$), which is
experimentally indistinguishable from the nominal value of
$14.7nm$ set down in the sample description.

\subsection{MBE\label{secgradembeqe}}

For MBE material, it was decided to continue the
study of $QE$ as a function of base aluminium fraction
$X_{b2}$ for QWSC graded structures. The samples are, in
the first instance, based on those studied in section 
\ref{sechomombeqe}. They are all Be doped at a level
of $N_{A}=1.34\times 10^{18}\rm cm^{-3}$ in the p layer
and Si doped with $N_{D}=6.00\times 10^{17}\rm cm^{-3}$
for the n layer. The main design characteristics of the 
samples studied here are summarised in table \ref{tblmbegrades}.
No reflectivity measurements could be performed on the
samples U4034-U4035 due to the absence of reflectivity
measurement samples in the processing run.

\begin{table}
\begin{center}
\begin{tabular}{| l || c | c | c | c | c |}
\hline
Sample & U4032 & U4033 & U4034 & U4035 & U6041 \\ \hline
\hline
Window ($\mu m$)  & 0.03 & 0.03 & 0035 & 0.035 & 0 \\ \hline
$X_{b1}$ (\%)     & 40 & 40 & 67 & 67 & 40 \\ \hline
$X_{b2}$ (\%)     & 20 & 20 & 30 & 40 & 30 \\ \hline
$x_{i}$ ($\mu m$) & 0.81 & 0.481 & 0.481 & 0.481 & 0.81 \\ \hline
Wells & 50 & 30 & 30 & 30 & 50 \\ \hline
\end{tabular}
\figcap{Principle design parameters of MBE grown 
graded p QWSC samples\label{tblmbegrades}}
\end{center}
\end{table}

\smalleps{htbp}{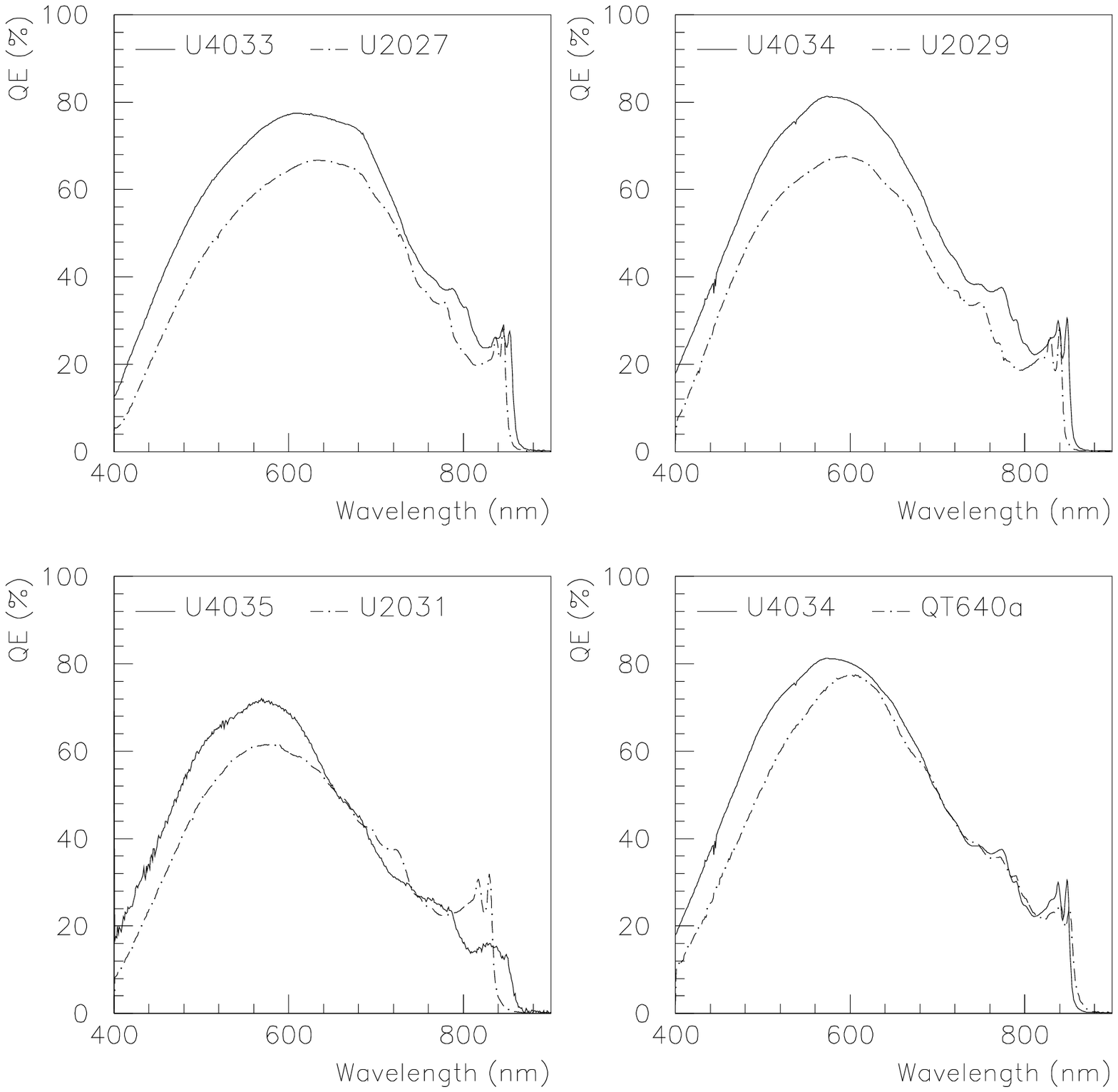}
{Comparisons of $QE$ for QWSC samples with and without grades, for a 
variety of grading schemes. Also shown is comparison of similar graded
QWSC structures grown by MBE and MOVPE\label{figmbegrades}}

The four graphs of figure \ref{figmbegrades} compare the experimental
$QE$ of ungraded QWSC structures U2027, U2029 and U2031 with their
graded counterparts. Enhancements in $QE$ are observed in all cases.
The increases in predicted $J_{sc}$ are 21\%, 24\% and 11\%
for the samples with $X_{b2}=20\%, 30\%$ and $40\%$ respectively.
The last figure is lower because of the degraded $QE$ in the
wells.

Figure \ref{figmbegrades} also shows a comparison between 
the MOVPE sample QT640a reviewed
in the previous section and U4034, which has similar specifications.
Although U4034 has a thicker window than QT640a, it nevertheless has 
a significantly better short wavelength $QE$. Comparison 
of modelling results for
these two samples indicates a p layer $QE$ about three times
better for the MBE sample. This confirms the conclusion reached
in the previous section regarding the poor material quality in the
graded MOVPE samples. 

\smalleps{tp}{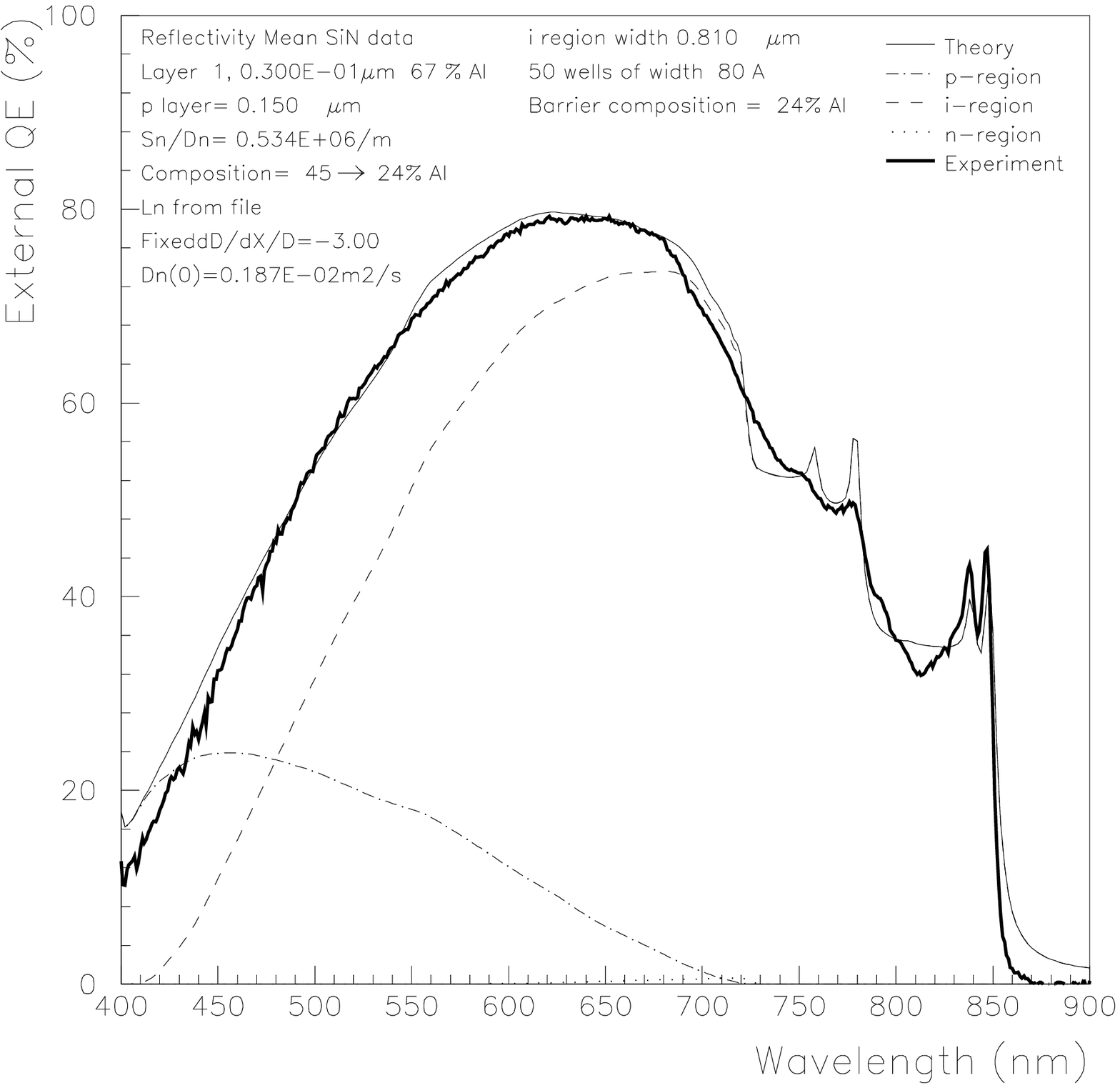}
{MBE 50 well QWSC U4032, graded 20-40\% \label{figar4032}}
\smalleps{htbp}{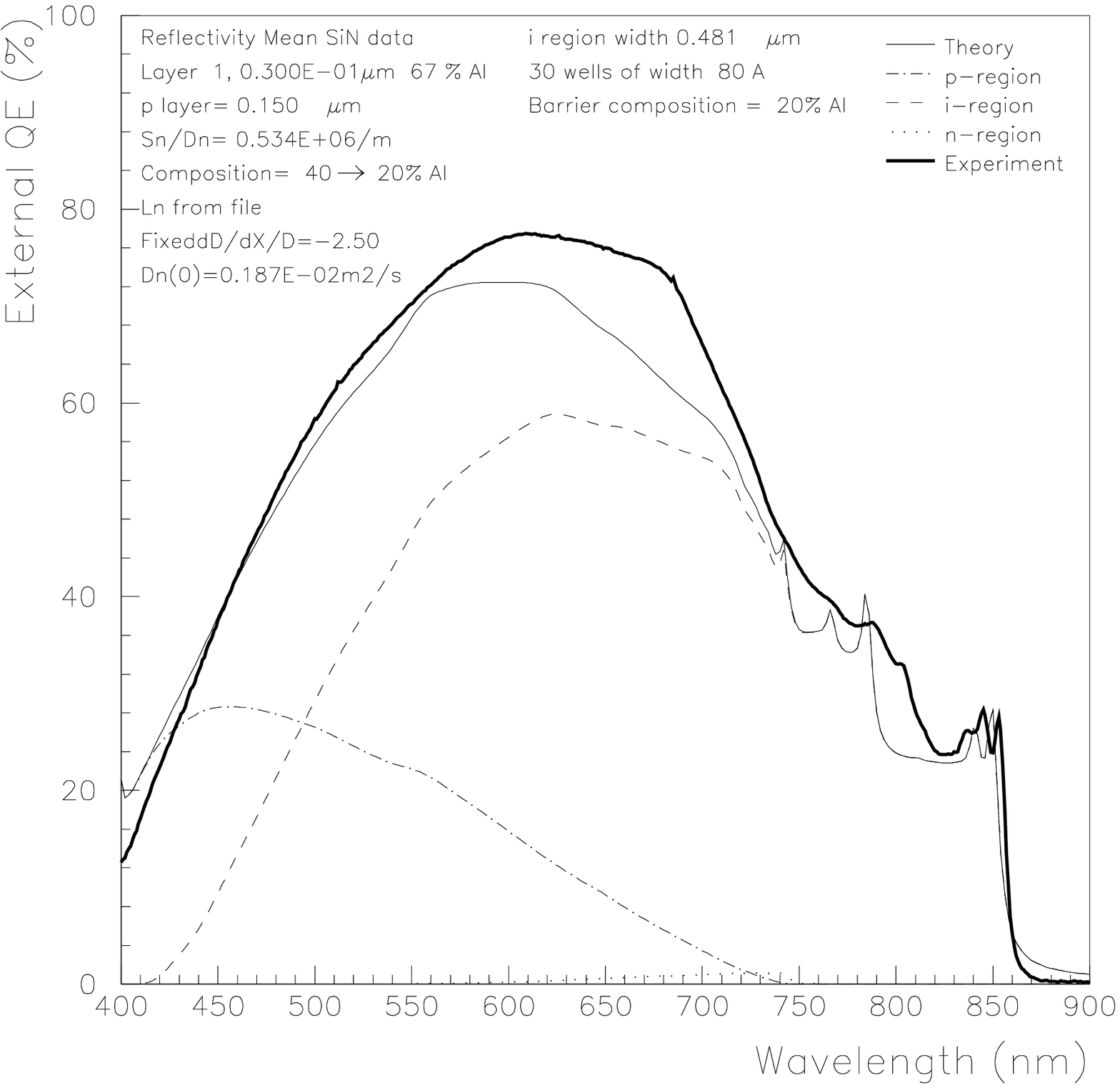}
{MBE 30 well QWSC U4033, graded 20-40\% \label{figar4033}}

\smalleps{hp}{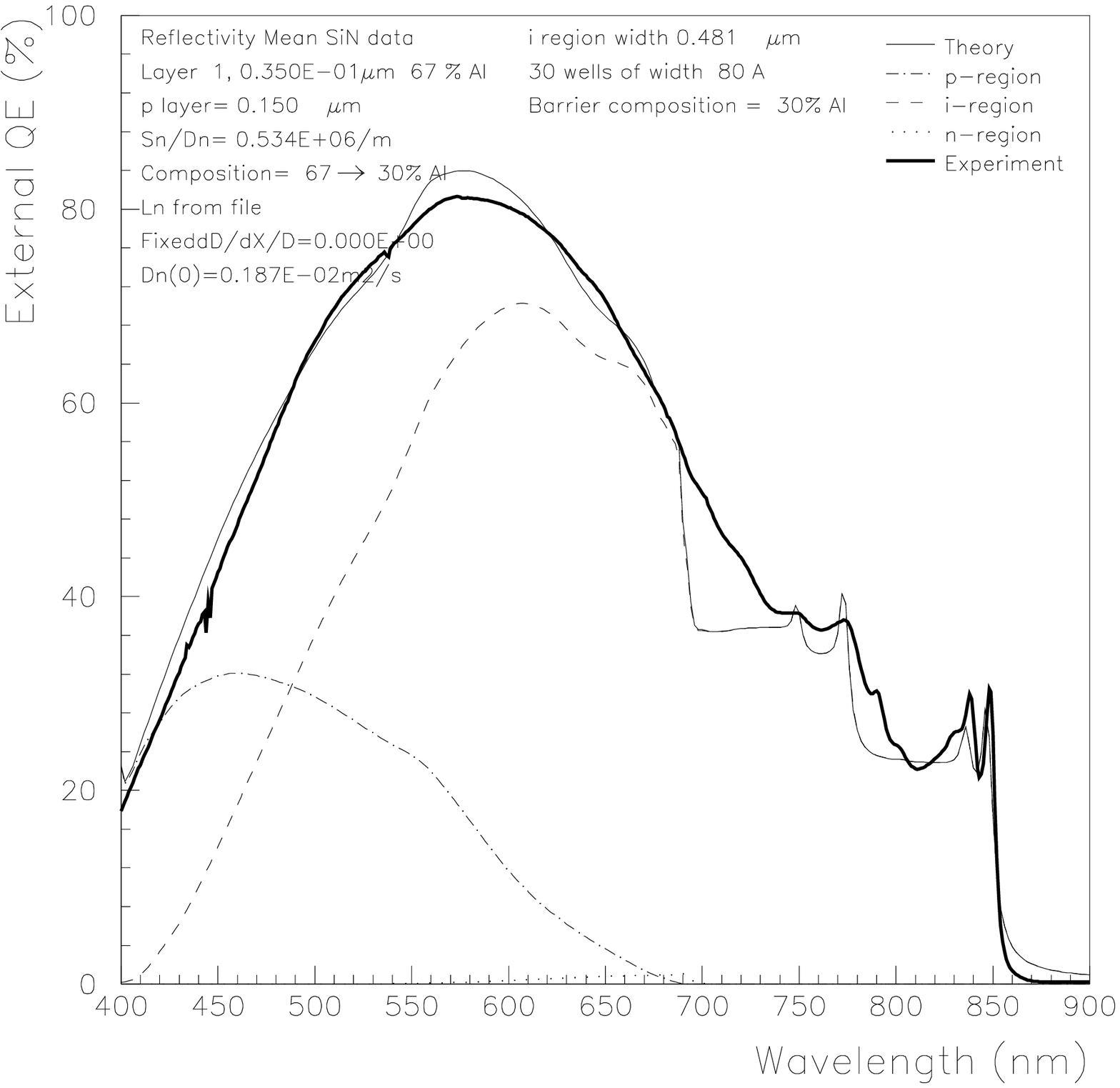}
{MBE 30 well QWSC U4034 graded 30-67\% \label{figar4034}}
\smalleps{htbp}{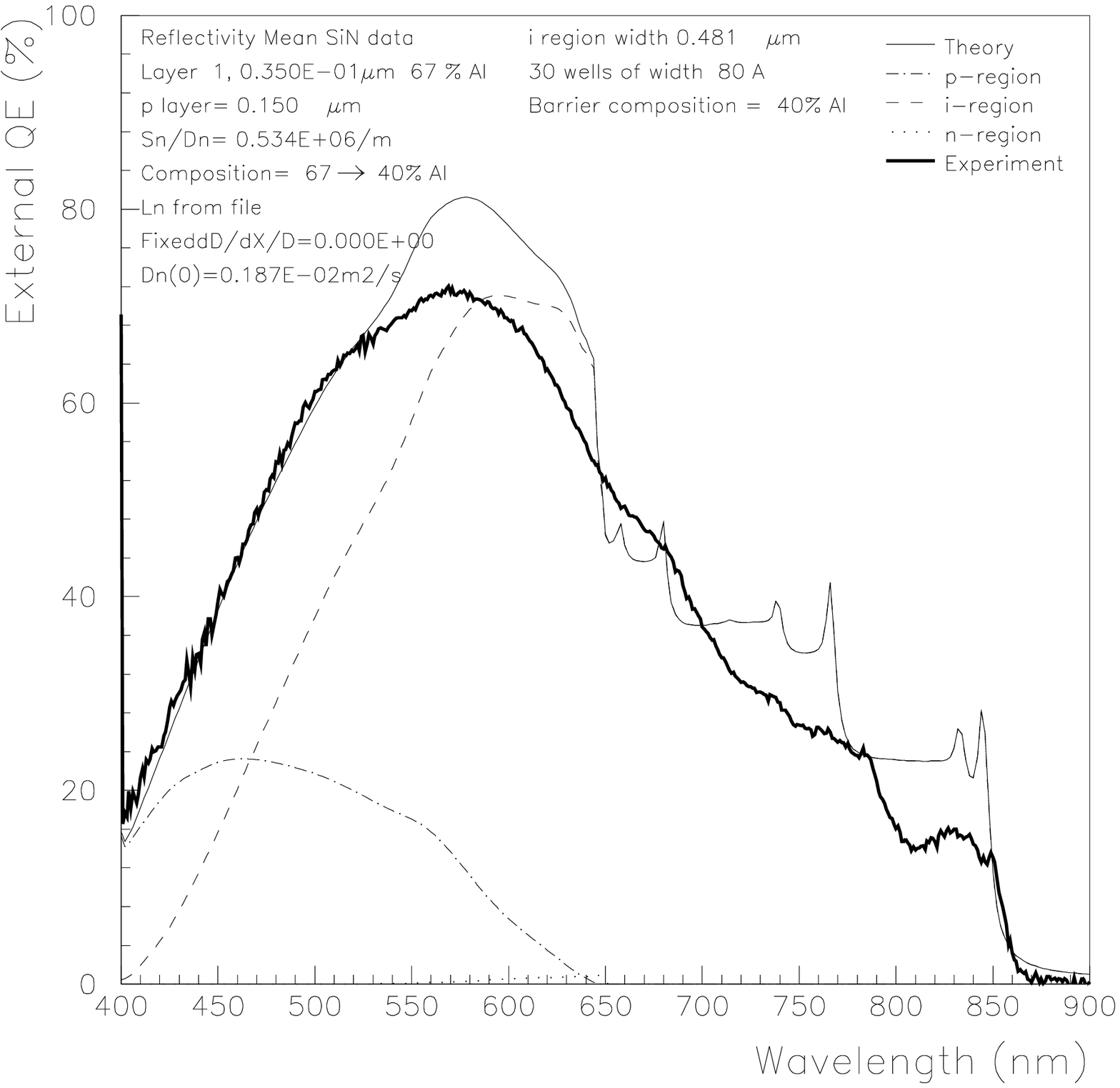}
{MBE 30 well QWSC U4035 graded 40-67\% \label{figar4035}}

Experimental and modelled $QE$ results for these samples are given in graphs
\ref{figar4034} to \ref{figar4035}. Modelling is accurate for U4032
and U4034, but shows significant deviations for U4033 and U4035.

Sample U4032 requires a diffusivity field $\cal D_{X}=$-3 and shows
a slight shortfall in the wells, although this is not significantly
greater than the experimental error.

Sample U4033 presents a serious shortfall in modelled $QE$ near
band edge $E_{b2}$. This has previously been observed in several
MBE and MOVPE samples such as U2027, U4036 and QT601c. As in the
fitting carried out for these samples, the value of $L_{p}$ required
to fit the $QE$ more accurately in this wavelength range is too
great to provide a realistic explanation without direct measurement
of this parameter. We have therefore used the value of $L_{p}$
found to fit previous MBE samples with aluminium fractions of
20\%, which is $L_{p}\sim 0.1\mu m$. The diffusivity field in
this case is $\cal D_{X}$=-2.5.

Sample U4034 (figure \ref{figar4034})
is modelled accurately with a diffusivity field 
$\cal D_{X}=$0. In theory, this indicates a constant diffusivity
in the p layer. In practice, however, this is probably due to the
assumptions we have made regarding the dependence of $L_{n}$ 
on aluminium fraction and a reflectivity which is given by the
average SiN data.

Sample U4035 (figure \ref{figar4035})
is accurately modelled in the short wavelength regime
with $\cal D_{X}=$0. This is ascribed to the same reasons given for
U4034. For this sample, however, there is a clear misfit in i
region contribution. Devices for this sample were generally poor,
and the $QE$ data noisy. The correct attribution of the i region
shortfall requires reprocessing, and could be assisted by
supplementary characterisation such as SIMS measurements. The
short wavelength response however suggests that the p layer
contribution is being accurately modelled.

Finally, figure \ref{fig794a6041} shows optimised sample U6041.
The concepts underlying the design of this sample are identical
to MOVPE grown QT794a and were  outlined in section 
\ref{secgrademovpeqe}. The figure also shows the $QE$ of the
MOVPE device. As in the case of QT640a and U2029, the
MBE device again performs better.

\smalleps{t}{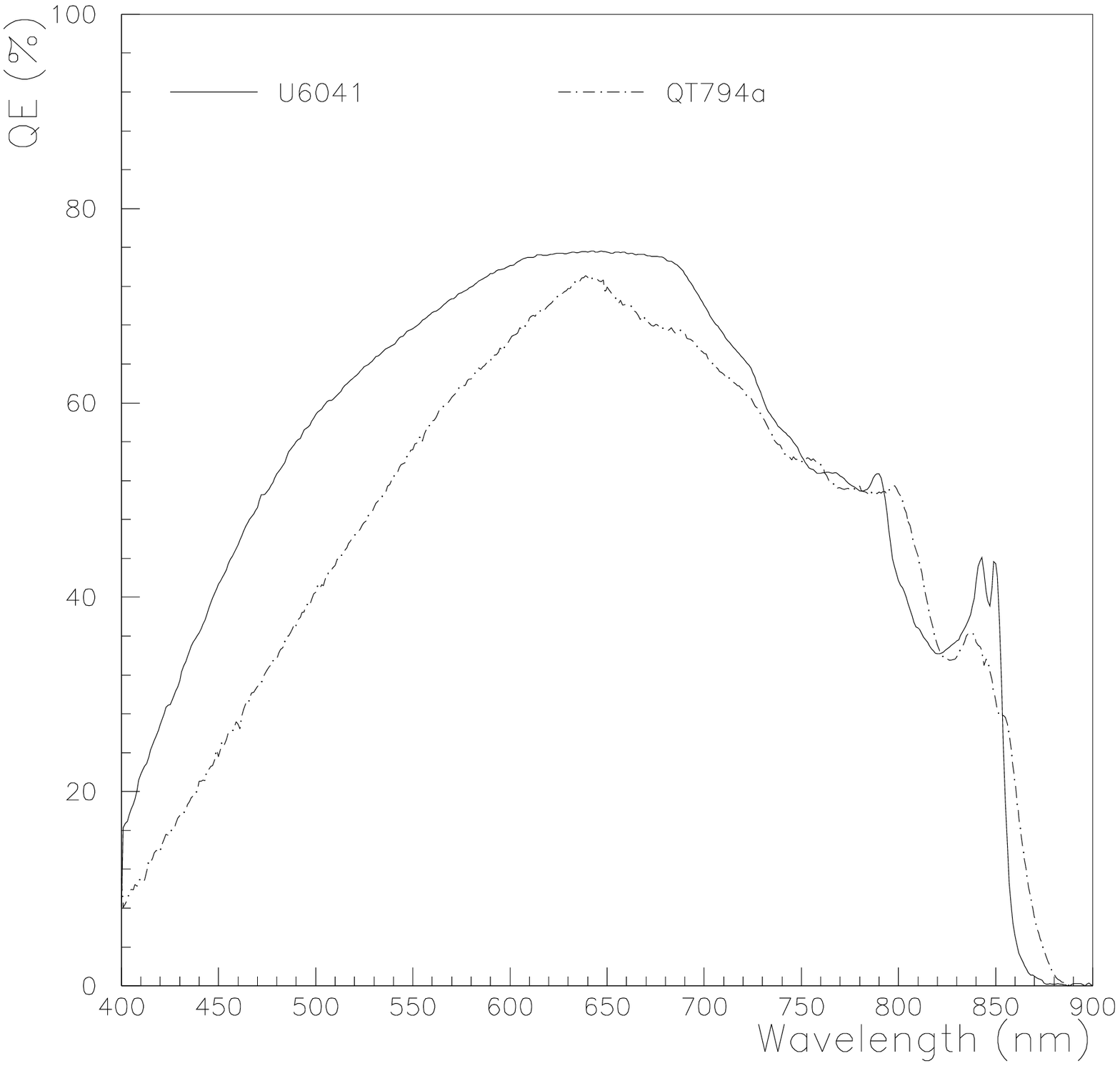}
{Comparison of MOVPE and MBE versions of an optimised 
50 well design QT794a and U6041 with graded p and no
window\label{fig794a6041}}

Initial modelling of the experimental data with nominal
parameter values showed similar problems to those experienced
for QT794a. In particular, the modelled i region $QE$ was
greater than the experimental $QE$ between about 550nm and
600nm. This suggests departure from design characteristics.
Furthermore, the model significantly overestimates the
$QE$ at short wavelengths, indicating an impaired p
layer $QE$.

Extensive characterisation was carrier out on this sample.
Calibration of the SIMS depth scale was verified by TEM,
as for sample QT794a in previous section \ref{secgrademovpeqe}.
TEM data indicated that mean sum thickness of barrier
and well was ($14.1nm\pm 0.5nm$), with the error estimated from the
resolution of the micrograph. The specification was $14.7nm$,
which is not significantly greater.
The total thickness of the multiple quantum well system confirmed 
the depth calibration of the SIMS data presented in figure 
\ref{fig6041sims}.
The graphs show only the most relevant data, which are the aluminium
fraction and Be doping for the first $0.6\mu m$ of the sample,
including the $0.04\mu m$ \gaas\ capping layer. Lower resolution
measurements covering the entire sample were also supplied.

Federico Caccavale has confirmed that the aluminium SIMS signal
in counts per second will be linearly dependent on the aluminium
composition for the range of aluminium fractions in these
samples. Modelling of a mirrored device processed from
the U6041 wafer will be presented in section \ref{secmirrorqe}. 
This has allowed us to fix the lower aluminium
fraction at $X_{b2}=24\%$. The linear relationship between 
SIMS signal and aluminium fraction, together with the value of
$X_{b2}$, enable us calibrate the SIMS mass spectrometer signal. 

The resulting aluminium compositional profile is shown in the lower
part of the figure. We note first that the aluminium
composition peaks at only 29\%. Christine Roberts who grew
this sample has confirmed \cite{croberts96} that grades over thin 
layers are technologically difficult. This is a consequence of the
growth scheme for graded structures, which treats the grade
as a series of ungraded strips grown close together. In
practice, a smooth grade is obtained by ensuring that the
aluminium source is never in equilibrium, and has a constantly
increasing output flux. 

If however the nominally ungraded strips are too thin, the 
aluminium source has too little time to ramp up to the higher 
temperature required for the material with a higher aluminium 
fraction. The result is a lower final aluminium composition
than expected, as seen in this sample. The grading technique
can be improved by characterisation such as the SIMS data 
presented here, which can help in determing an optimum
strip width resulting in both a smooth grade and an
accurate target aluminium fraction $X_{b1}$.

Furthermore, the unexpectedly slow decrease in aluminium
at the front of the p layer has resulted in a
positive bandgap gradient with increasing distance at the
front of the p layer. The origin of this feature is
not clearly understood.

This rounding off of the aluminium profile towards the
top of the sample results in a thin layer with a bandgap
gradient pointing towards the front of the cell. The
resulting field will tend to drive minority carriers
generated towards the front of the cell away from the
depletion region, thereby degrading the short wavelength
$QE$. This unforeseen effect is a result of removing the
window layer. A similar effect in a sample with a window
would not cause this problem, because the step in
the conduction band at the interface between the window
and the p layer acts as a minority carrier reflector.

Finally, we note that the Be dopant profile
extends significantly further than the nominal $0.1\mu m$.
Illegems \cite{illegems77} studied Be diffusion in MBE
\gaas\ and \algaas. He concluded that this dopant was
promising insofar as it allows high doping levels 
($\sim 10^{19} \rm cm^{-3}$) whilst maintaining good
surface morphology in the MBE material. Be diffusion coefficients
are quoted in this work.
Be diffusion has more recently been reported as a problem. 
Lee \cite{lee95} has studied this problem
more recently in \algaas\ as a function of aluminium composition 
and anneal temperature.

Using the correlated TEM and SIMS data, we predict
an effective p layer thickness of $0.14\mu m$.
This fairly conservative value,
is based on the fact that the quantum well signal
fits theory accurately, suggesting that the first wells on the
p side of the i layer are not significantly doped. We
note finally that the residual background doping in this
sample is n type silicon, and were estimated in chapter
\ref{secmivar} as being below $10^{15} \rm cm^{-3}$ by 
both SIMS and the MIV technique.

The resulting $QE$ modelling is shown in figure \ref{figar6041}.
Theory reproduces the experiment closely with a diffusivity field
of -4.5. The surface recombination parameter is set to an effectively
infinite value to account for the drop in aluminium fraction
over the first few nm of the sample. The aluminium composition used
in this case is based on the SIMS scan, and includes the ungraded
$0.04\mu m$ unintentionally doped layer between the p and
i layers. 

The SIMS and TEM data also indicated that the
n layer thickness is $0.72\mu m$ thick instead of the nominal
$0.6\mu m$. 

The dopant diffusion has proved crucial in understanding the $QE$
of this sample. The extension of the p layer and consequential
reduction in i layer thickness explains why modelling using
nominal sample specifications overestimated the i region $QE$.
Similarly, the extra thickness in the p layer and lack of
grade over nearly a third of this layer explains the reduced
p layer $QE$.

The final efficiency of this sample was estimated by
Paul Griffin at ($14.2\% \pm 0.5\%$) under tungsten illumination
corrected to AM1.5 light levels. This is comparable with the efficiency
reached by the previous best \algaas\ sample G951, which was
mentioned in section \ref{secmbemivs}. There is no control
sample to compare the estimated short circuit current enhancement
for this sample.

Despite the departure from device specifications, we have 
shown that the model has adequately reproduced data given corrected 
device specifications. This increases confidence in the ability of 
the model as a diagnostic tool.

\smalleps{tp}{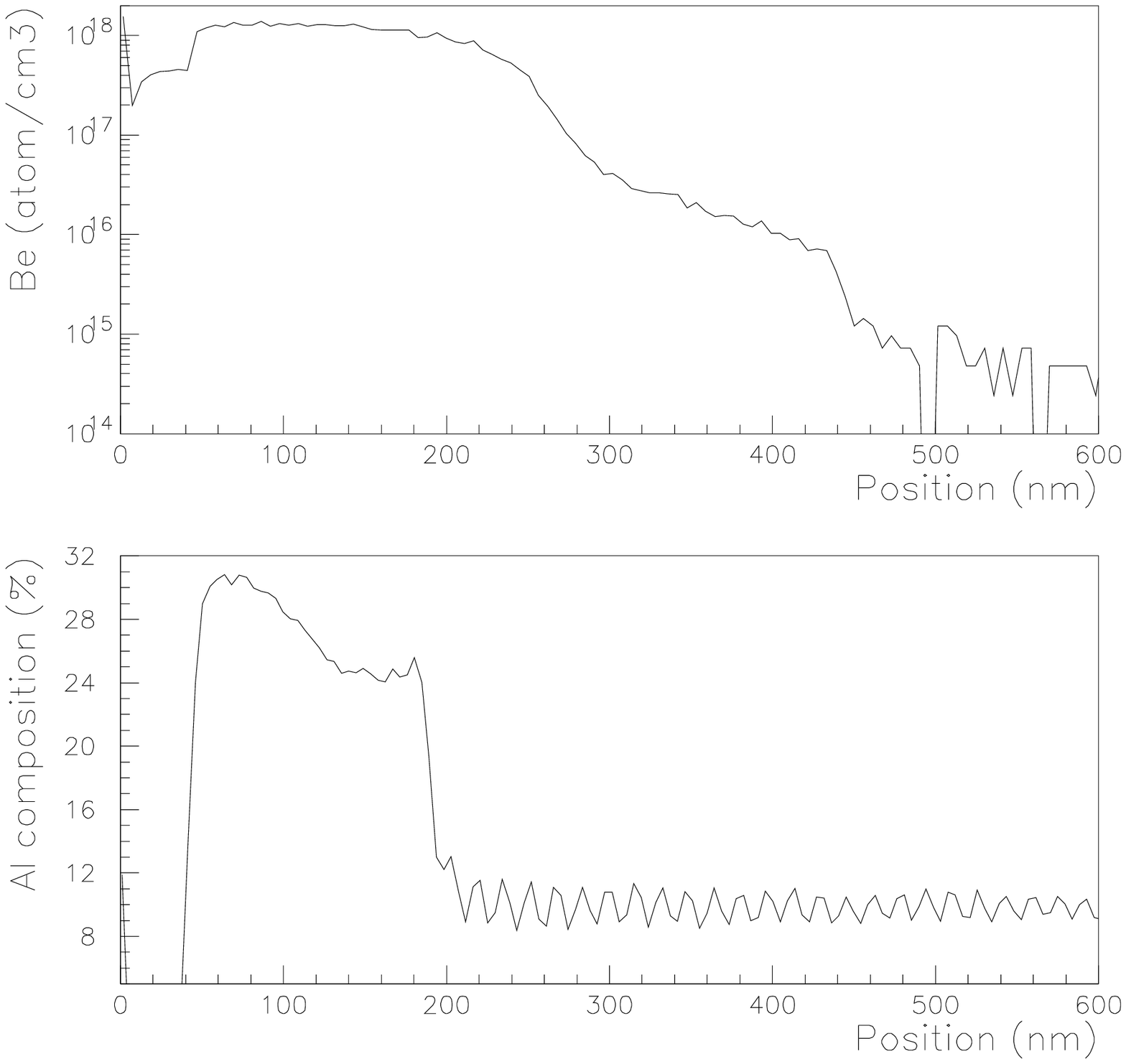}
{SIMS measurements of Be doping and aluminium composition in sample
U6041 \label{fig6041sims}}
\smalleps{htbp}{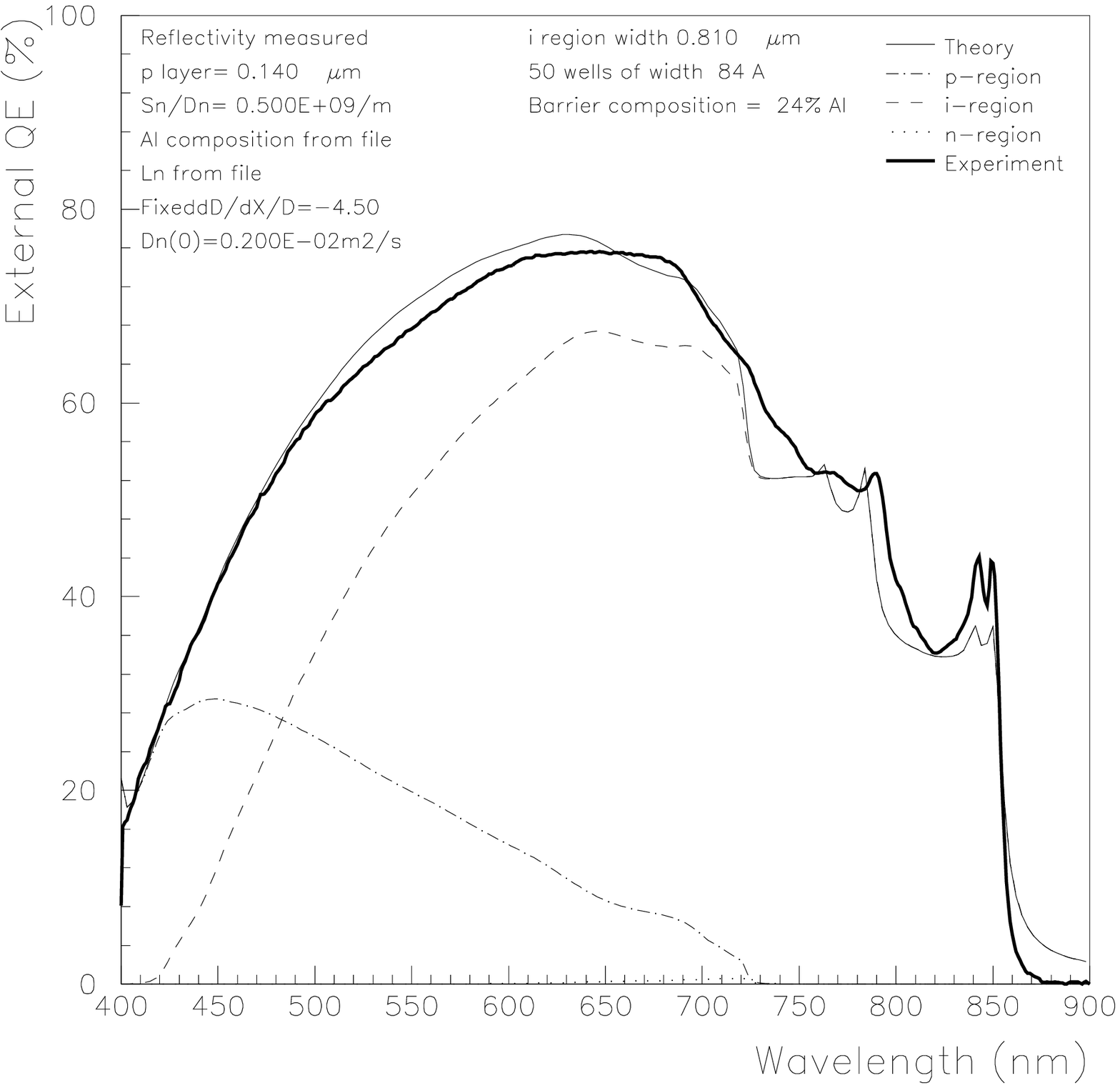}
{MBE 50 well QWSC without a window graded from $X_{b1}29\%$
to $X_{b2}=24\%$.\label{figar6041}}

\subsection{Conclusions}

The study of a wide range of grades has demonstrated their
capacity to increase the $QE$ of \pin\ and QWSC samples
grown both by MBE and by MOVPE.

The case of sample U6041 has served to increase our confidence
in the model both as a diagnostic tool and as a basis for
further optimisations. 

A large amount 
of detailed information about each sample is required. Since
the advanced experimental characterisation cannot be carried
out as a matter of course, we are frequently forced to make
assumptions about composition. The modelling has furthermore
had to use transport data extrapolated from ungraded samples.
Finally, reflectivities are not always available.

As a result, theoretically well defined parameters such as
the the diffusivity field and the recombination
parameter which are used in the modelling are in practice
ill-defined, since modelled values contain systematic errors
due to the assumptions we have mentioned.

Despite these difficulties, the accuracy of the compositionally
graded \algaas\ model has been 
demonstrated for a wide range of samples. In addition, the model
can reproduce changes in $QE$ due to a number of design changes to the 
original structure. These include compositional grades in neutral 
layers and depleted layers from three different growth systems.
Enhancements in $QE$ for samples with graded p layers
have been shown in a wide range of samples.

The diagnostic capabilities of the model have furthermore been
confirmed by comparison with quantitative characterisation
techniques. To our knowledge, this is the first model of compositionally
graded \algaas\ cells to be used to model experimental data.

\section{QE of Mirror Backed Samples\label{secmirrorqe}}

\smalleps{tp}{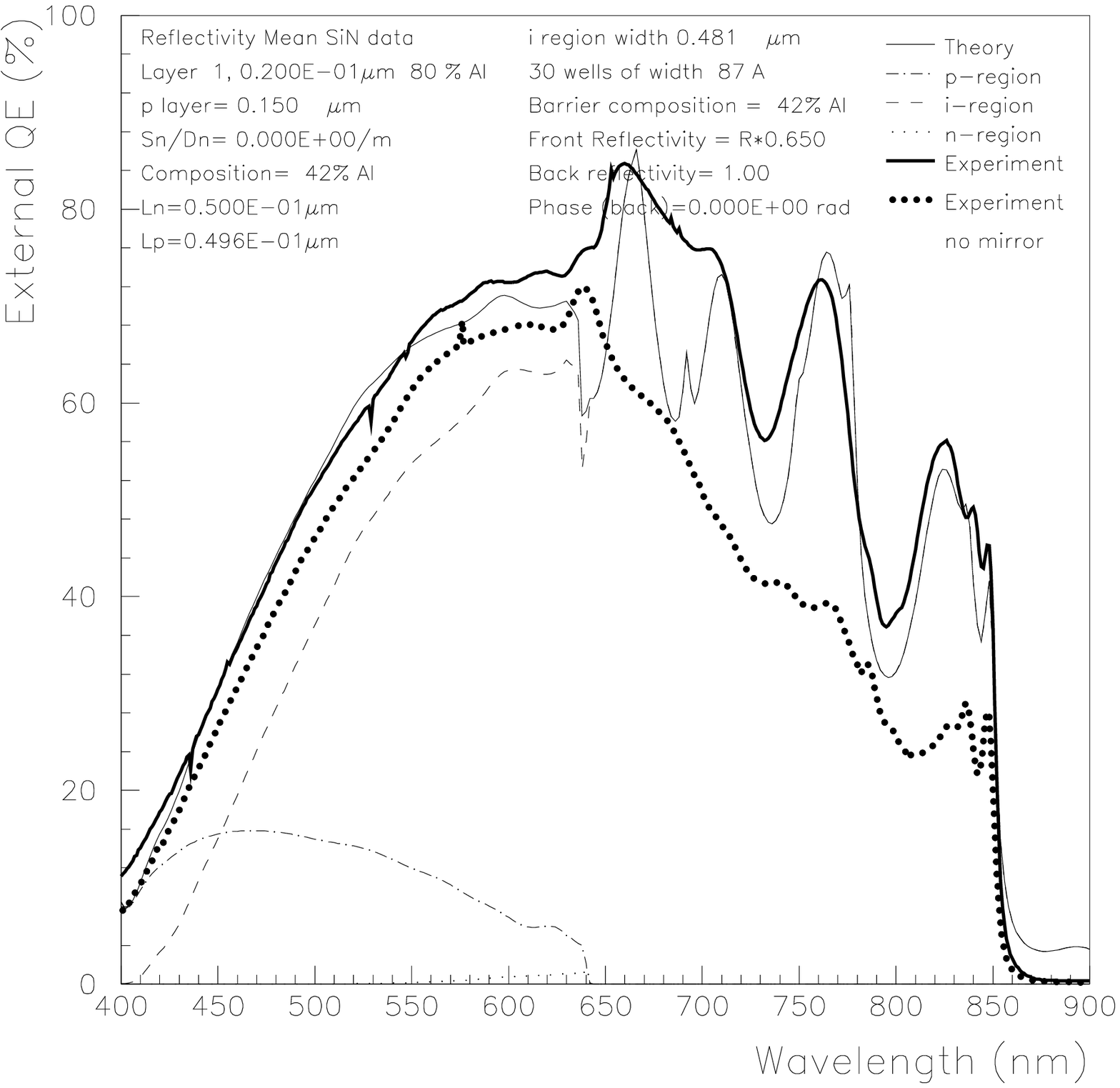}
{$QE$ of mirrored sample QT468a. Also shown for reference is the $QE$
of an AR coated device without a mirror\label{figfp468a}}
\smalleps{htbp}{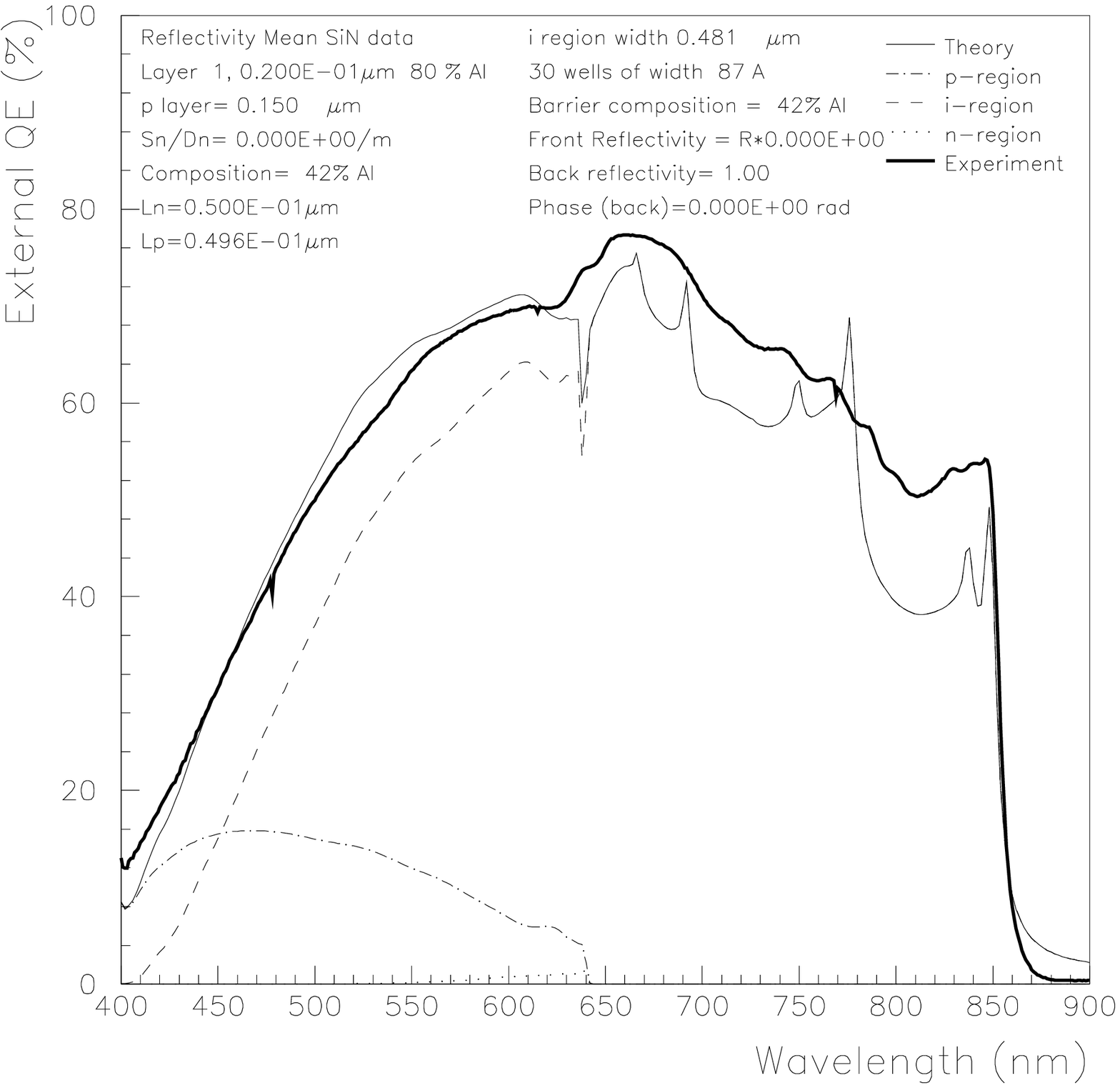}
{$QE$ of a mirrored QT468a device demonstrating non-specular reflection
from the back surface and $QE$ enhancement greater than two passes.
\label{figm468a}}

In this section we investigate the $QE$ enhancement in the
quantum well wavelength range. The technique was tested on
the 30 well sample QT468a. Theoretical modelling and experiment
for this sample is shown in figure \ref{figfp468a}, together
with the $QE$ of an AR coated device without a mirror for
comparison. The increase in predicted $J_{sc}$ for this
device is 27\%. Main fitting parameters are given on the
graphs.

\smalleps{htb}{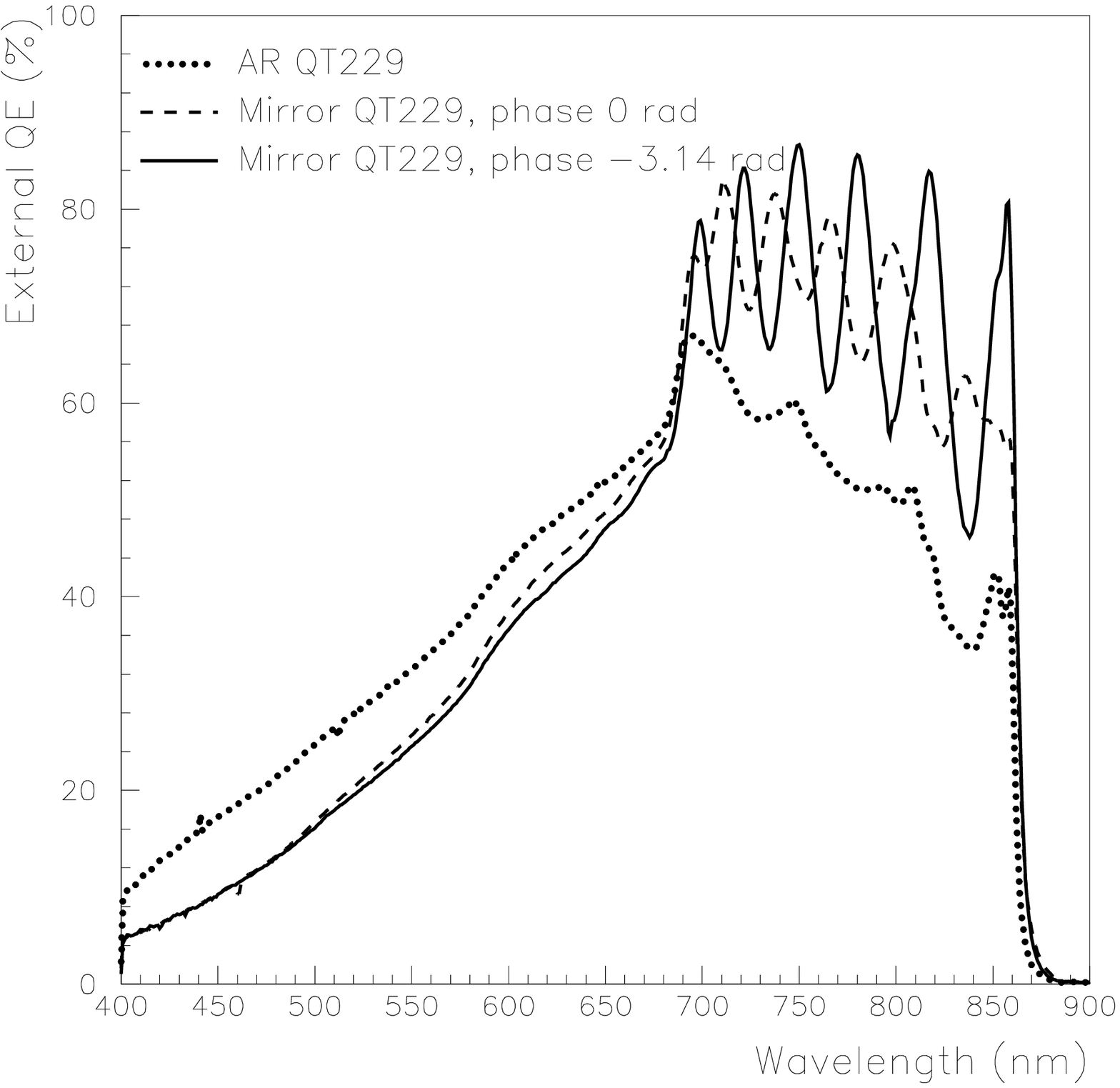}
{Experimental $QE$ enhancement in two mirrored QT229 devices, compared
with a AR device without a mirror \label{figqt229all}}
\smalleps{htbp}{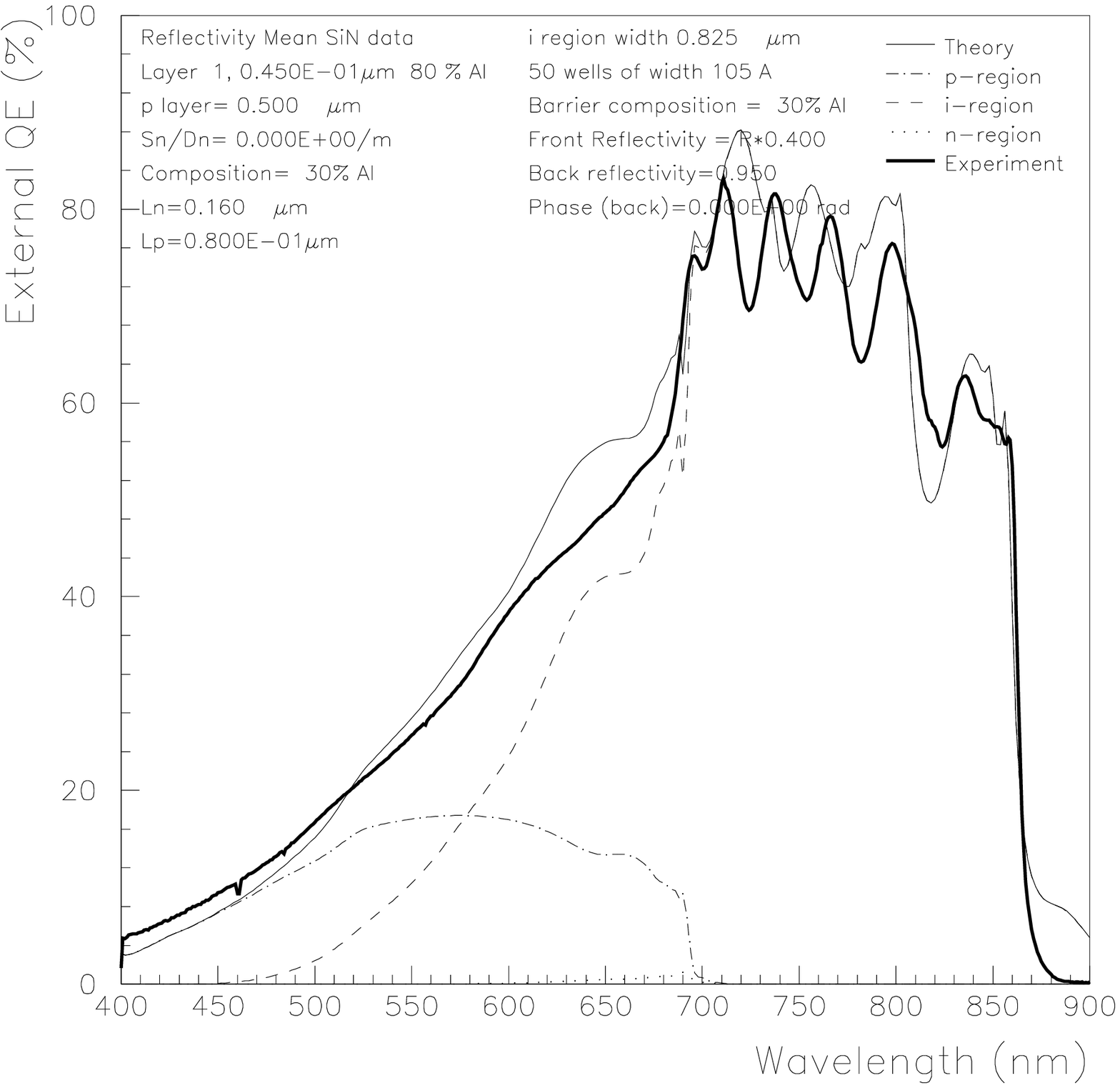}
{$QE$ of a QT229 mirrored device showing a back surface phase
change $\psi_{b}=0$ radian\label{figm229a}}
\smalleps{htbp}{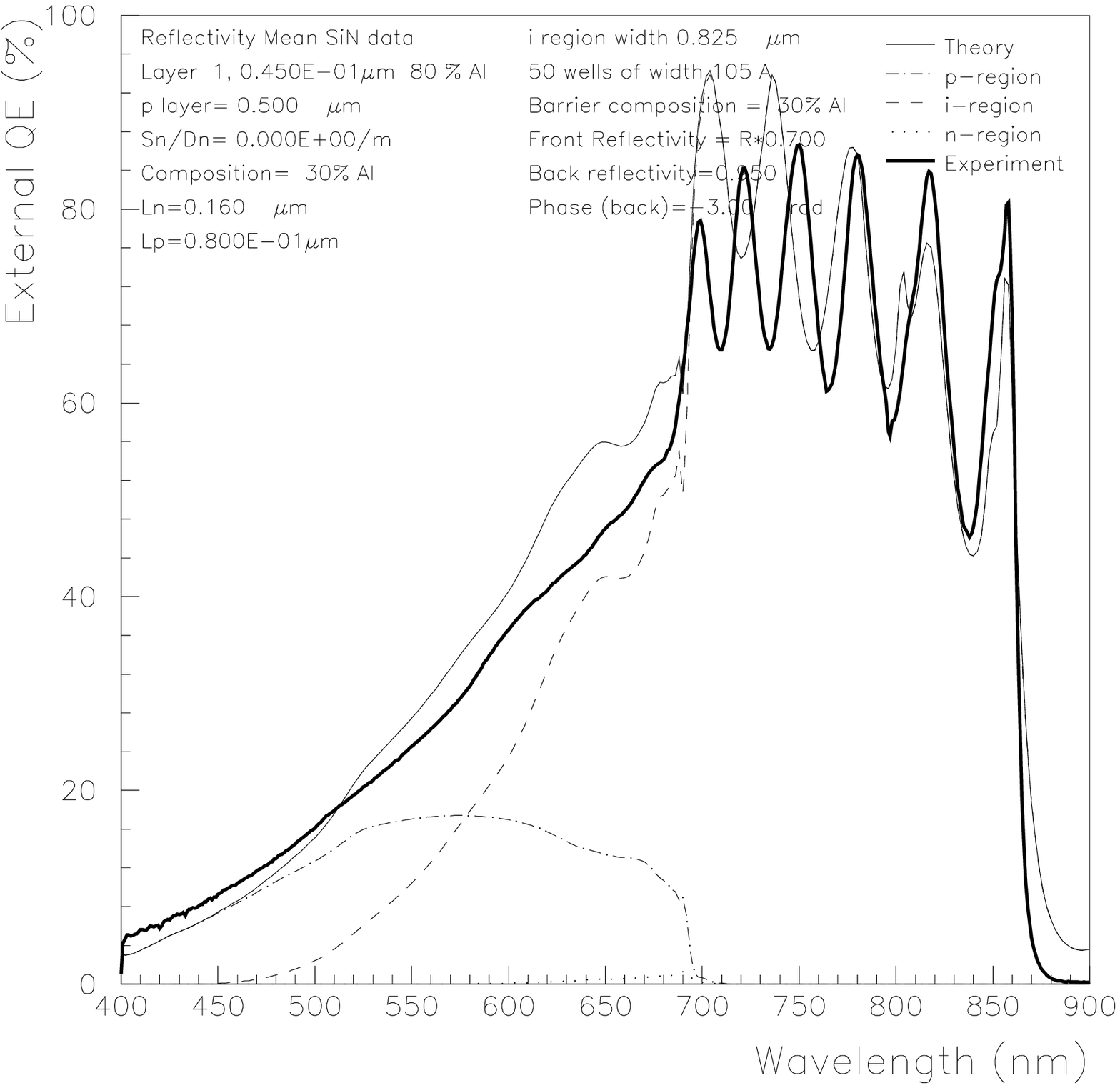}
{$QE$ of a a QT229 mirrored device showing a back surface phase
change $\psi_{b}=-3$ radian\label{figm229b}}

The wavelength dependence of the Fabry-Perot peaks is
accurately modelled. Variations in model accuracy over
the quantum well wavelength range are visible, and are
due to the underlying inaccuracy of the quantum well absorption
model.

Amplitudes of Fabry-Perot peaks in the experiment are variable,
and generally lower than those predicted by the model. The
broadening mechanism responsible for this is not well understood,
but may be partly due to roughness at the surface of the mirror.

Figure \ref{figm468a} illustrates a mirrored device processed from 
the QT468a wafer at the same time as that of figure \ref{figfp468a}.
We note a complete absence of interference peaks in this case.
Moreover, the $QE$ enhancement is more than can be explained by a 
double pass.
We interpret this as indicating that the light is not being specularly 
reflected from the back surface. The randomising effect of reflection
from a rough mirror causes the relative phases of internally reflected
light beams to be randomized. Since the mean path length is also
increased, this results in both increased light trapping in the
cell and the disappearance of the Fabry-Perot peaks.
Both devices processed from the QT468a wafer are modelled with
a zero back surface phase change. 

Figure \ref{figqt229all} shows
experimental data for QT229, which is a 50 well QWSC with a very
thick p layer of nominal thickness $0.5\mu m$. The three
$QE$ curves relate to an AR coated device and two AR coated and
mirrored devices. We note that the positions of the
Fabry-Perot peaks for the two mirrored samples are significantly
out of phase. The modelling for these two samples is given in
graphs \ref{figm229a} and \ref{figm229b}. The modelled phase change
differs by a factor of nearly $\pi$. This result is not unexpected
in the light of discussions with Mark Whitehead \cite{whitehead94}
which indicated that the phase change is device dependent at metal-
semiconductor interfaces.

We note that the frequency of the theoretical
peaks in wavelength is smaller than that of the experiment. This
suggests that the model is underestimating the width of the optical
cavity defined by the mirror and the front surface. The accurate
fits obtained in other samples support the conclusion that the cell
is wider than expected. 

\smalleps{ht}{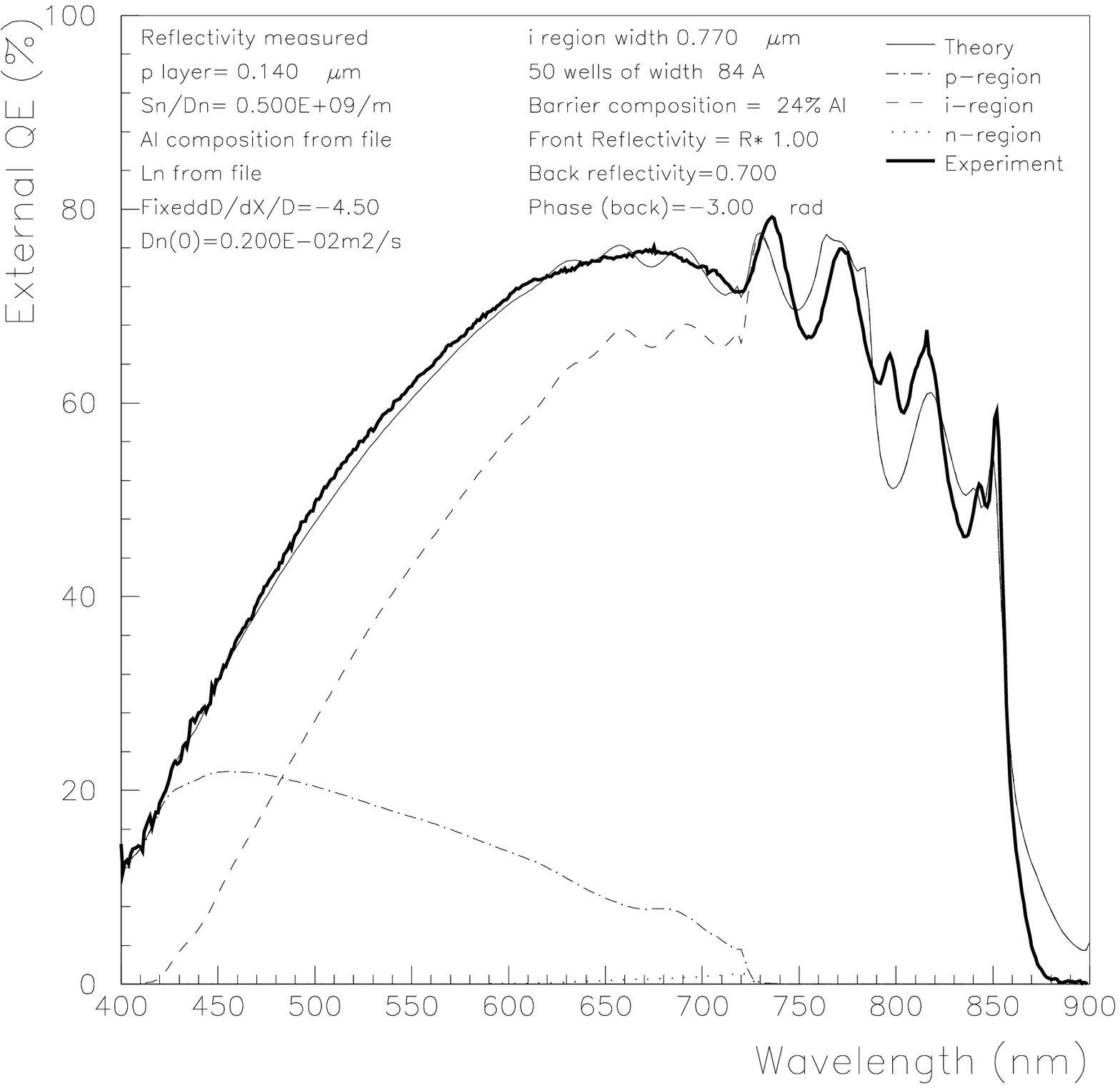}
{$QE$ of a mirrored device processed from the U6041 wafer
\label{figm6041}}

Experimental and theoretical $QE$ for a mirrored device processed
from the U6041 wafer is shown in figure \ref{figm6041}. Only
two devices of this type were processed, one of which was not 
successful.

Despite this, an accurate
fit is obtained using the same parameters as in section 
\ref{secgradembeqe} together with a back surface phase change
of -2.8 radians and a front reflectivity factor ${\cal F}=0.4$,
such as was estimated for the QT229 modelling of figure
\ref{figm229a}.

No measured reflectivity was available for this sample. We have
assumed a poor reflectivity, basing this on the fact that the
mirrored device has a significantly poorer response at short
wavelength than the device without a mirror discussed in
section \ref{secgradembeqe}. This was clearly visible on
the devices, which showed a green tint characteristic of
high reflectivity at short wavelengths. This device was processed
at the time as the sample whose reflectivity was used in the
modelling.

As we explained earlier, this sample was studied by SIMS and TEM.
The wavelength dependence of the Fabry-Perot oscillations is
reproduced.

This sample again illustrates the diagnostic capacilities
of the model, by showing that accurate modelling of the experimental
data is possible if the structure is of good quality and the
composition and geometry is well known.

\section{Optimisations \label{secqeresultsoptimisation}}

\subsubsection{Optimisation of Available Material}

The model can now be used to optimise the QWSC design we
have been considering. Two approaches to this exist. The first
of these is to identify design changes made separately to
samples that have been characterised, and to combine these
in a single optimised sample. 

This relatively conservative
approach is illustrated in figure \ref{figsafeoptimisation}.
This design, which we will call sample A, is based 
on a combination of the Eindhoven AR coat, 
an 80\% window layer, and the grade demonstrated in sample
U2029 together with the back surface mirror seen in
50 well QWSC devices processed from QT229 and U6041. Also 
shown is a prediction for the correponding control without 
a grade or a mirror.

The modelled short circuit currents in this case are
$22.3 \rm mA cm^{-2}$ for A and and $17.0\rm mA cm^{-2}$
for the control. This corresponds
to an enhancement of 31\% with the new design modifications
when compared to the control shown in the figure.

The second and less reliable approach is an extrapolation to
design parameters which have not been observed. Figure
\ref{figdangeroptimisation} shows a similar comparison between
an optimised sample B and its control. The design in this case has an
Eindhoven AR coat, no window, and a thinner $0.1\mu m$ p layer.
In the case of the optimised sample, this p layer is again
graded between $X_{b1}=67\%$ and $X_{b2}=30\%$.

The modelled short circuit current for these predictions
are $J_{sc}=23.6 \rm mA cm^{-2}$ for the optimised design
and $J_{sc}=19.7 \rm mA cm^{-2}$ for the control, or an
enhancement of $20\%$ overall compared to the control without
a grade or mirror.

These results indicate that the more conservative approach
can nearly match the current output of the more adventurous
design. Both samples compare favourably with the previous
best MBE sample G951 which was mentioned in sections \ref{secmbemivs}
and \ref{secgradembeqe}, and which has a theoretical
$J_{sc}=18.2 \rm mA cm^{-2}$, compared to the
experimental value of $(17.5 \pm 0.5)\rm mA cm^{-2}$ 
quoted in Barnham {\em et al.} \cite{barnham94}. This 
estimate is an independent measurement carried
out at the Frauenhofer institute in Germany.

A prediction of the efficiency requires a model of the dark current
of the cell. This is the subject of ongoing work by collaborators.
However, we can make a rough estimate by assuming that the new samples
have the same ${\cal F\!F}=78\%$ and $V_{oc}=1.07V$ which were observed
experimentally in G951. This sets a lower limit to the projected efficiency
of optimisation A since the higher current levels in this cell will 
increase the $V_{oc}$, and the similarity between G951 and design
A should produce the same $\cal F\! F$. With these assumptions, 
the final estimated efficiency for this sample is then $18.6\%$.

The corresponding efficiency for sample B under the same assumptions
is 19.7\%. These results compare well with a good 
conventional ungraded \pn\ cell reported by Amano {\em et al.} 
\cite{amano85}, which had an aluminium
composition of $30\%$ and was 14.6\% efficient under AM1.5
illumination. This cell was the result of extensive growth
optimisation and had large diffusion lengths $L_{p}$ and $L_{n}$,
a high $V_{oc}=1.28V$, a fill factor $F\! F=70.7\% $ and a short
circuit current $J_{sc}=16.2mA/cm^{-2}$.


\smalleps{hpt}{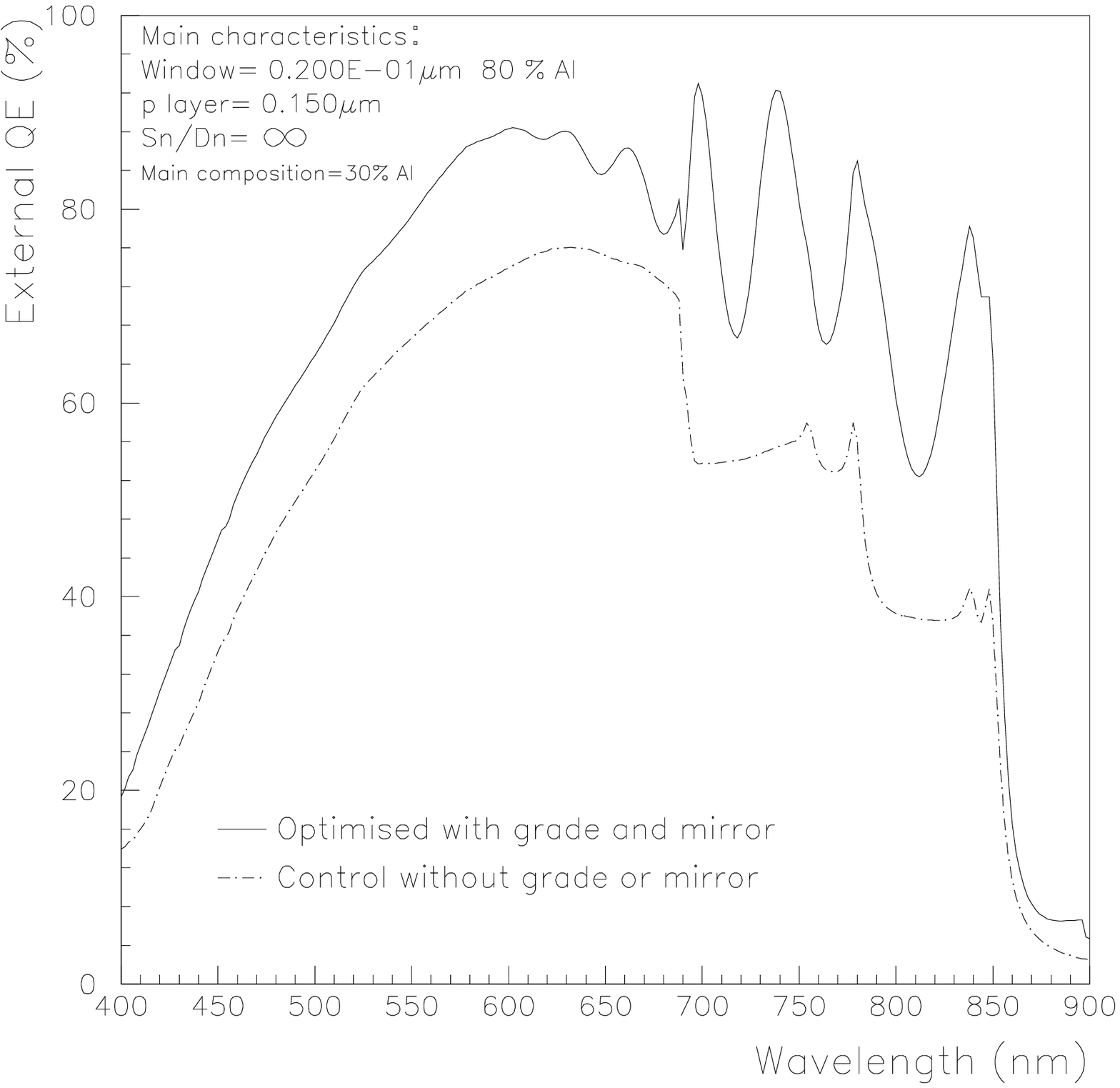}
{Theoretical prediction for optimised sample A combining the
Eindhoven reflectivity, a $0.15 \mu m$ grade with $X_{b1}=67\%$
and $X_{b2}=30\%$, a $0.02\mu m$ window with 80\% aluminium,
a high surface recombination and a back surface mirror.
\label{figsafeoptimisation}}
\smalleps{hpt}{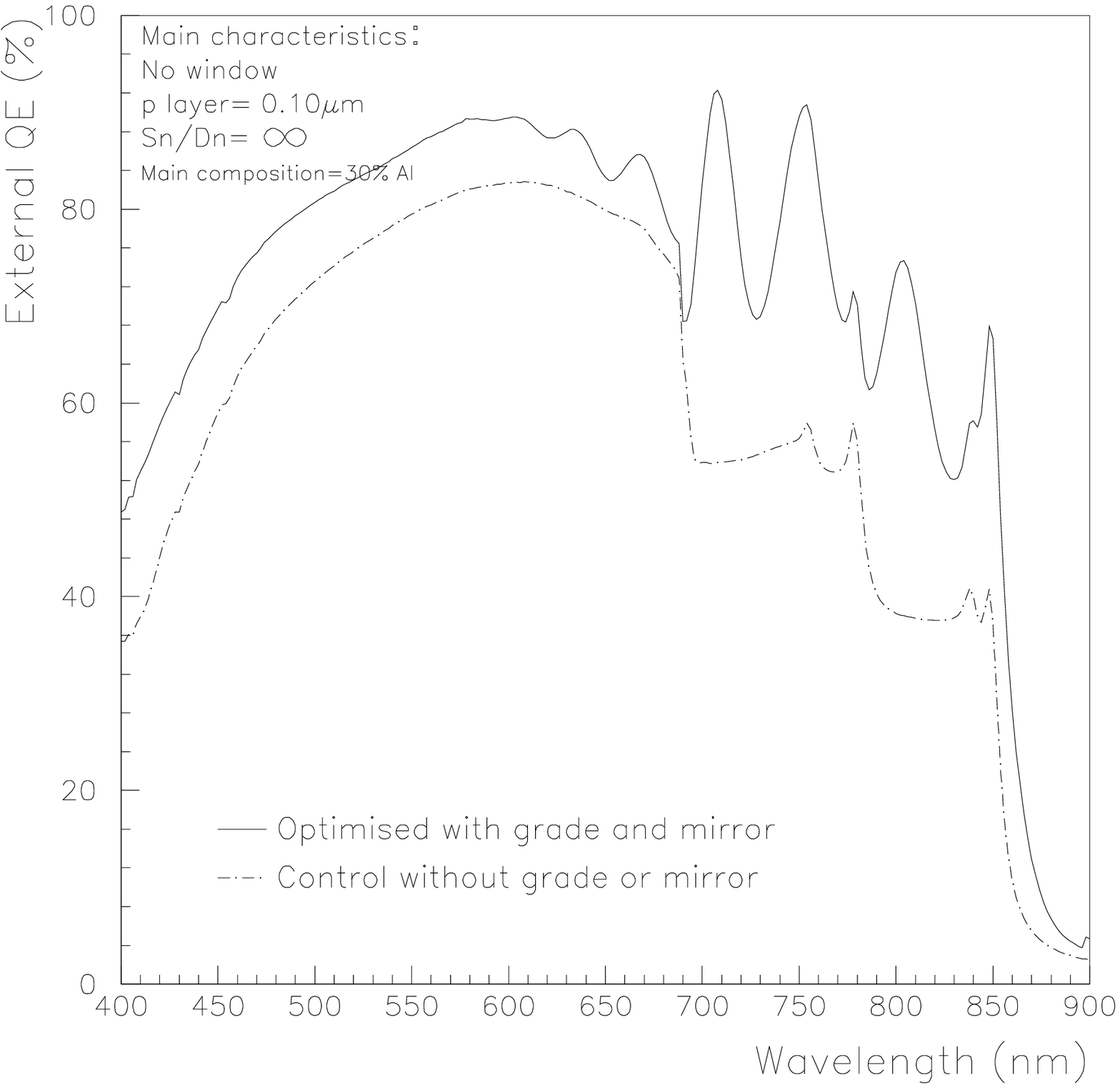}
{Theoretical prediction for optimised sample B combining the
Eindhoven reflectivity, a thin $0.10 \mu m$ grade with $X_{b1}=67\%$
and $X_{b2}=30\%$, no window, a negligible surface recombination and 
a back surface mirror. \label{figdangeroptimisation}}

\subsubsection{Material with More Efficient Transport}

Further efficiency enhancements are possible in material
with improved minority transport parameters. We saw in
section \ref{secqex} that a graded p layer may reduce the 
efficiency of samples with good material quality. 
Section \ref{secmodellinggrades} further showed that the $QE$ of graded
structures in particular is sensitive to the position dependence
of the diffusivity. Without a systematic
study of the compositional dependence of the minority carrier transport 
in such material, therefore, we cannot reliably make predictions
of the $QE$ and $J_{sc}$ enhancements expected in this system.

We therefore limit ourselves to short circuit current predictions of 
the best possible ungraded \algaas\ QWSC cell, using the highest reported 
literature values for the transport parameters. We will compare these
projected efficiencies with the results of the previous optimisation
section.

The paper by Ludowise \cite{ludowise84} mentioned in section
\ref{seclittreview} reports very high diffusion
lengths in a series of solar cell samples grown by MOVPE. These
values were derived from internal $QE$ spectra, using a method
similar to the one described in this thesis. 

An optimisation
based on the Ludowise diffusion length of $L_{n}\sim 1.4\mu m$
at $30\%$ aluminium composition yields a value of 
$J_{sc}=25.6 \rm mA cm^{-2}$. This corresponds to an efficiency
between $21\%$ and $24\%$ for low and high values of $F\!F$ and
$V_{oc}$ in AM1.5. These results apply to a
50 well sample with a back surface mirror, a $0.02\mu m$ window
of composition $90\%$ aluminium, no compositional grade and
an optimum p layer thickness of $0.15 \mu m$.

The Ludowise diffusion lengths, however, were estimated from
a set of n--p algaas samples. The structure of the samples is
not given in detail. The front n layer, however, is very thin.
This indicates that the diffusion lengths are estimated from
the $QE$ near the \algaas\ band-edge. The paper does not make
clear how the n and p contributions to the $QE$ in this regime
were separated, since the methods we have used in this work
do not apply to the Ludowise n--p structure. Loo \cite{loo88}
however has reported hole diffusion lengths of 2$\mu m$ at an
aluminium composition of 30\% and a minority carrier lifetime 
of 13ns, which is close to GaAs. The Loo paper combined with
data from Ludowise indicates that high minority carrier
transport parameters are achievable in \algaas.

More readily achievable minority carrier lifetimes are reviewed in
the review by Ahrenkiel \cite{ahrenkiel92}. These qualitatively
agree with the values we have derived for the Eindhoven samples
studied in section \ref{secgradediregion}, and
reproduce the functional form of the Hamaker parametrisation.
Using a value of $L_{n}\sim 0.4\mu m$ and a structure similar
to that described in the previous paragraph, we find an optimum
short circuit current of the order of 
$J_{sc}=24.2 \rm mA cm^{-2}$. This corresponds to lower and higher
limits to the estimated final efficiency of $20\%$ to $23\%$.

The second optimisation is more likely to be achievable, since it
is based on reproducible material quality. It should be noted that
the final efficiency estimates are subject to large errors because
of the lack of a reliable model for the $V_{oc}$ and the $F\!F$.

We note that the estimated efficiencies for optimised QWSC structures
without grades but with a back surface mirror are only about 5\% to 
10\% greater than those of the optimised graded structures presented
in the previous section for the \algaas\ material with significantly
poorer minority transport efficiency studied in this thesis.

\section{\label{secqeresultsconclusions}
Conclusions to chapter \ref{secqeresults}}

This chapter has demonstrated the accuracy of the theoretical
model for the $QE$ of mirrored and compositionally graded
QWSCs by comparison with experiment for a wide range of \algaas\
samples.

Results from a set of samples grown and processed in Eindhoven
University of Technology showed that the theoretical model is
capable of reproducing experimental data for QWSC and \pin\
structures with graded i layers. These samples further
demonstrated accurate fits in samples with good material.

MOVPE and MBE structures with and without quantum wells
and with ungraded neutral layers were characterised.
These are used to establish the dependence of the minority
carrier diffusion length $L_{n}$ on aluminium fraction. 
These samples have been successfully modelled, and values of
$L_{n}$ for both growth systems extracted. MOVPE values
are more variable than MBE ones but remain only fractionally
different to MBE ones. The fitting is most accurate for
intermediate aluminium fractions of $X_{b2}\simeq 30\%$.
The model has some difficulty reproducing experimental
$QE$ data near the band edge $E_{b2}$ for material with
low aluminium fractions. At higher aluminium fractions,
the well $QE$ could not be modelled in one MBE sample. This
is ascribed to a loss mechanism in the quantum wells.

Very high values of the surface recombination parameter 
$\cal S$ were required to fit some samples. The origin of
this problem has not been established. It is suggested
that it may be due to shortcomings in the way
the surface recombination is modelled. The presence of
a damaged layer near the surface of the cell may explain
this behaviour. Another candidate is dopant diffusion
in MBE \algaas, since this effect would extend the width of
the p layer, thereby reducing the $QE$ at short wavelengths.
More information is required to resolve this issue.

A wide range of QWSC and \pin\ designs with graded p layers
was then investigated. Samples grown by MOVPE have demonstrated
the $QE$ enhancement but have shown significant materials problems.
One optimised sample in particular cannot be modelled despite
extensive characterisation. This is ascribed to the poor material
quality indicated by high background doping levels and a very
low p layer $QE$.

Graded structures grown by MBE have shown a clearer picture. 
Good fits are obtained overall with values of the diffusivity
field $\cal D_{X}$ which are between 0 and the Hamaker value.
Samples requiring ${\cal D_{X}}\simeq 0$ are interpreted as
departing from the assumptions we have made regarding the variation
of $L_{n}$ with aluminium fraction, and, in some instances,
to the reproducibility of the SiN reflectivity.

An optimised sample which did not initially perform as
expected has been successfully modelled after supplementary
information was provided by SIMS and TEM measurements carried
out by collaborators. This information allowed us 
to correct the device specifications
to allow for dopant diffusion in MBE material. 
This has cast some doubt over the p layer thicknesses
used in modelling MBE samples from the same growth system.
It was concluded however that the influence on the modelled
transport parameters is less critical for other samples, because
they have thicker p layers. This however is reminiscent of
the conclusions we reached when discussing the surface recombination
parameter.

The optimised MBE sample was repeated on the MOVPE reactor. 
Experimental data from this sample cannot be modelled despite similar 
characterisation. The SIMS together with MIV
measurements, however, indicate poor material quality. This indicates
that the samples may be too highly contaminated by impurities to
operate successfully as solar cells.

Mirrored samples have been modelled accurately. One sample has
confirmed that the phase change upon reflection at the  back mirror
is a variable parameter. A significant and variable broadening
mechanism has been demonstrated by comparison of different devices
processed from the same wafer. In the case of a QT468a device,
this can be explained by an accidentally roughened back surface mirror
and consequent randomising of the relative phase of the incident
light.

The spacing of Fabry-Perot peaks on the wavelength axis has been
accurately reproduced. This indicates that $QE$ characterisation of
mirror backed samples may be of use in establishing the optical
thickness of samples. Finally, mirror backed samples are also useful
in determining the precise aluminium composition $X_{b2}$ of
QWSC samples, since the onset of Fabry-Perot oscillations is
a sensitive function of the former.

We have demonstrated that both grades and mirrors are effective
in increasing the $QE$ and hence the short circuit current of
QWSC structures in \algaas\ with poor minority transport
characteristics.

This information gives us sufficient confidence in the diagnostic
and predictive capabilities of the model to suggest further
optimisations. 

Optimisations using the model together with information from the
experimental data were presented. The projected efficiency for
the first of these optimisations is based on separate successful
design changes which were verified experimentally. This produces
a projected $J_{sc}$ which is significantly greater than published
good conventional \algaas\ solar cells with comparable aluminium 
fractions.

The second optimisation makes design changes which are an 
extrapolation of experimentally measured samples, but only
exceeds the conservatively optimised sample of the previous
paragraph slightly. A rough estimate of the efficiencies
of these projections suggests a value of the order of
19\% in AM1.5

Finally, these optimisations apply to available material. Further
optimisations of growth conditions are required for the QWSC
application in order to produce material with the minority
carrier transport efficiencies which have been observed in
Eindhoven material for example. We have mentioned the work
by Amano \cite{amano85} which started with fairly poor material 
but produced efficient minority carrier transport after a detailed 
growth optimisation study.

A set of calculations based on good material quality has predicted
efficiencies of between 20\% and 24\% in optimised QWSC structures 
without grades but with back surface mirrors. This study shows that
the material quality is the prime problem. A grade can compensate
for poor material quality to a certain extent, but is not expected
to improve good quality material.

We conclude however that the combination of a grade and a back surface
mirror can bring the $J_{sc}$ of QWSCs fabricated from poor material
quality to within a few percent of that achievable with the best material.

%% file: conclude.tex
\chapter{Conclusions}

This thesis has adressed the problem of increasing the
quantum efficiency ($QE$) of QWSC and \pin\ solar cells in 
the \gaasalgaas\ materials system. The approach to the problem 
consisted of joint theoretical and experimental investigations.

Chapter \ref{sectheory} reviewed previous models dealing with 
compositionally graded structures. These do not attempt to reproduce 
data. We have developed a theory based on previous work which
is capable of modelling a wide range of cells with
position dependent bandgaps in all regions of the solar
cell. In addition, the model is designed to model solar cells
with back surface mirrors. The internal reflections from quantum
well-barrier interfaces however was not treated.

We discussed properties of the analytical and numerical solutions
for the $QE$. We concluded that the wavelength dependence of the
$QE$ enables us to estimate the minority electron diffusion
length in certain samples. We further concluded that the recombination
velocity in the p layer is less significant in graded p layers.

Consequences of the modelling were discussed in chapter
\ref{secoptimisation}, together with an analysis of the
ideas underlying the optimisation program. The effect of
compositional grades in the p type emitter layer, and the
effect of adding a back mirror were investigated and illustrated
for a range of significant cell parameters.
Theoretical predictions of the $QE$ enhancement were 
presented separately for the compositional grading technique
and the mirror design.

After an introduction to the experimental techniques used
in the project, the preliminary characterisation of our
samples was presented. These results suggest a background
doping problem in multiple quantum well material. The problem
is more severe in MOVPE material. It was suggested that this
problem is due to a loss of compensation between carbon and
oxygen background impurities in the MOVPE QWSC structures.

The effect of the anti reflection coat on solar cell MIV
characteristics was considered. Parallels were drawn with
hydrogen passivated samples examined in the course of a
collaboration with Eindhoven University of Technology.

A set of samples with a range of i region compositional
profiles including quantum wells and grades
was characterised and modelled. The high minority carrier
efficiency predicted in these samples indicated good material
quality. These samples have confirmed the capacity of the model 
to reproduce the $QE$ of a range of complex graded i layers. They
also demonstrate successful modelling of the $QE$ of p and n layers
with efficient minority carrier transport.

We then examined a set of QWSC and \pin\ MBE samples with different 
aluminium  fractions. A tabulation of minority electron diffusion lengths 
$L_{n}$ in the p layer was estimated from the modelling. The
MOVPE samples grown for this project were also considered and
showed values of $L_{n}$ which are similar to the MBE material.
These values are significantly lower than those estimated
for Eindhoven material at a similar composition.
Furthermore, the values of $L_{n}$ are lower than the majority of
the published data for low aluminium fractions, and are weakly
dependent on composition overall. We concluded that the minority
carrier transport parameters can be improved in both materials
systems.

QWSC and \pin\ samples with compositionally graded p emitter layers 
were designed and characterised. Experimental enhancements were
observed in all these structures, except one MOVPE sample.
The QE measurements were modelled subject to the following
assumptions. We assumed that the variation of $L_{n}$ with 
aluminium composition in grades was similar to that estimated
from QE measurements of ungraded samples. We also assumed that
the diffusivity decreases exponentially with increasing aluminium
fraction. This allowed us to model the solar cells with a limited
number of parameters.

The experimental $QE$ data were successfully modelled for a range
of samples grown by MOVPE and MBE One device which could not
be modelled initially was characterised extensively by SIMS and
TEM. This additional information allowed accurate modelling of the
experimental $QE$ with corrected sample specifications.
Diffusion of the beryllium p type dopant into the i region was
identified as a problem. With this increased knowledge of the sample 
structure, a good fit was obtained. The companion MOVPE sample proved 
to suffer from poor material quality. Despite similar extensive
characterisation, this sample could not be modelled.

The study of ungraded samples has shown that the material used in
this work suffers from relatively poor minority carrier transport
efficiency.

The $QE$ data for graded structures concluded that significant $QE$ 
enhancements can be achieved by compositional grading of the p type 
emitter.

A series of mirror backed QWSC samples was examined. Significant
enhancements were seen in all devices. The front reflectivity
was used as a fitting parameter to adjust the amplitude of the
Fabry-Perot peaks observed in the data. One sample which shows no
Fabry-Perot peaks supports the suggestion that this decreased
amplitude is due to imperfectly smooth back surface mirrors. 

Two devices from a 50 quantum well sample showed evidence of
phase changes upon reflection from the back surface which differ
by approximately $\pi$. This variation supported the use of the
back surface phase change as a fitting parameter to model the
wavelength position of the Fabry-Perot peaks 

We then showed that the $QE$
model for mirrored samples is capable of accurately verifying both
the optical thickness of the sample and the aluminium composition
of the quantum well barriers. 
The origin of the overestimation of the Fabry-Perot peak amplitude
by the model should, however, be verified. A calculation of the effect
of light reflection at the quantum well-barrier interfaces would
contribute to better understanding of this point.
One sample matched the efficiency (about
14.2\% in AM1.5) of our previous best \algaas\ solar cell with benefitted 
from significantly better minority carrier transport.

The final part of this work presented two optimised designs incorporating
these two design features. The first is based on optimisations observed
separately in samples characterised during the course of the work. This
predicted a theoretical short circuit current of 22.3$\rm mA/cm^{2}$.
This represents an enhancement of 31\% over a control without grade or
back surface mirror. The projected lower limit to the efficiency 
in this case is 18.6\% for AM1.5.

The second optimised device extrapolated from the previous design by
thinning the p layer and removing the window layer. In this design,
the theoretical short circuit current prediction was 23.6$\rm mA/cm^{2}$.
This represents an enhancement of 20\% over a control, and a final
projected efficiency of about 19.7\%. These calculations demonstrate
enhancements possible relatively poor material. These efficiencies 
assume an unchanging $V_{oc}$ and $F\!F$, and therefore represent 
only a lower limit.

Preliminary calculations in section \ref{secqeresultsoptimisation}
compared the highest projected short circuit current obtainable
from graded QWSCs made from available material, and ungraded
QWSCs made from the best reported \algaas. Both designs were
coated with a back surface mirror. The QWSC structures
made from the best reported material have projected short
circuit current densities between 24.2$\rm mA/cm^{2}$ for
the most commonly reported higher limits to minority carrier transport,
and  25.6$\rm mA/cm^{2}$ for the highest reported values. The
projected efficiencies ranged from 21\% to 24\% for the most efficient
material, and from 20\% to 23\% for the standard, high quality material.
These calculations suggest that the compositional grading
and back surface mirror techniques can enhance 
the short circuit current density and efficiencies
of solar cells made from poor material to within
a few percent of cells made from the best material. All these
projections are significantly greater than the efficient, independently
verified 14.6\% achieved by Amano {\em et al.} \cite{amano85} and mentioned 
in chapter \ref{secqeresultsoptimisation}.

We conclude that both optimisation techniques have been successfully
demonstrated, and that significant enhancements in short circuit
current are possible. 

Future work will investigate the optimised designs presented in
chapter \ref{secqeresults}. The light intensity calculation should
be generalised to include internal reflections at interfaces other
than the front and back surfaces. The optimisation and modelling 
techniques can furthermore be generalised to other materials systems 
where material quality is an issue, and where compositional grades 
are attractive, such as CdHgTe or CuInGaSe.

%% file: app1.tex
\chapter{Analytical Solutions\label{secapp1}}

The analytical solutions for the quantum efficiency of
the neutral regions were described in section \ref{sectheory}.
This appendix discusses exact and approximate analytical
solutions to the differential equation for the minority carrier
concentration \ref{differentialeq} of chapter \ref{sectheory},
and uses the same definitions.

The solutions to equation \ref{differentialeq} are 
dependent on three materials and device dependent
transport parameters, which are the diffusion length $L_{n}$, the 
recombination velocity $S_{n}$ and the diffusivity $D_{n}$. 
The number of free parameters must be reduced to one before
experimental data can be modelled with confidence and optimised
structures designed.

This appendix examines these analytical solutions in
a number of limiting cases in order to determine conditions
in which the number of free parameters may be reduced
in order to allow single parameter fits.

The discussion quotes the results given in section \ref{sectheory},
and is similarly restricted to the case of a p type layer for
brevity. Where illustrative examples are given, the calculations apply
to a p type layer with an aluminium fraction of 30\%, which is the
design most commonly found in our structures.

The first section briefly restates the solutions obtained in
section \ref{sectheory} and states the exact general solution
subject to the usual boundary conditions.

The second section deals with an ungraded cell without
fields in the neutral regions. We examine the behaviour of the
analytical solution for the $QE$ in a number of limiting cases.

We then consider a cell with constant electric field 
in the neutral regions, but again with constant minority carrier 
transport parameters, and compare these results with the
results for the cell without fields.

\section{General Form of the Analytical Solution}

The general solution to equation \ref{differentialeq} can be written
\begin{equation}
\label{repeatgeneralsolution}
n(x)=a_{1}U_{1}+a_{2}U_{2}+U_{3}
\end{equation}
where functions $U_{1}$, $U_{2}$ and $U_{3}$ are described in the general
case in section \ref{secphotocurrent}.
We saw in section \ref{sectheorynonzeroe} that the effective field 
experienced by the minority electrons in the conduction band can be 
expressed in per metre as
\begin{equation}
\label{edefinition}
{\cal E}_{f}=\frac{q E}{2 k_{B}T}
\end{equation}
In order to further simplify the final expression for the
photocurrent, we introduce a surface boundary condition parameter
${\cal S}$ which includes both the recombination velocity and
electric field at the front of the layer
\begin{equation}
\label{sdefinition}
{\cal S}=2{\cal E}_{f}-\frac{S_{n}}{D_{n}}
\end{equation}
The surface boundary condition is then of the form
\begin{equation}
\label{bc1general}
{\cal S} n(0)+n'(0)=0
\end{equation}

The boundary condition at the interface between the neutral region
and the space charge region in the depletion approximation is that
the excess minority carrier concentration is zero
\begin{equation}
\label{bc2general}
n(z)=0
\end{equation}
where $z$ is the position of the depletion region, measured
from the front of the p layer.

Substituting for the expression for n and $n'$ into these two boundary
conditions allows us to express the general form of the constants
$a_{1}$ and $a_{2}$, which are
\begin{equation}
\label{generala1}
a_{1}=\frac
{-U_{3}(z)(U'_{2}(0)+{\cal S}U_{2}(0))
+U_{2}(z)(U'_{3}(0)+{\cal S}U_{3}(0))}
{U_{1}(z)(U'_{2}(0)+{\cal S}U_{2}(0))
-U_{2}(z)(U'_{1}(0)+{\cal S}U_{1}(0))}
\end{equation}

\begin{equation}
\label{generala2}
a_{2}=\frac
{U_{3}(z)(U'_{1}(0)+{\cal S}U_{1}(0))
-U_{1}(z)(U'_{3}(0)+{\cal S}U_{3}(0))}
{U_{1}(z)(U'_{2}(0)+{\cal S}U_{2}(0))
-U_{2}(z)(U'_{1}(0)+{\cal S}U_{1}(0))}
\end{equation}
These expressions define the general solution to equation
\ref{differentialeq}. It proves difficult however to find an
analytical expression for functions $U_{1}$ and $U_{2}$ in
most cases. We now consider the two simplest cases where this
is possible.

\section{Ungraded p Layers\label{secanalyhomosolutions}}

For an ungraded layer, the transport parameters
are constant and the pseudo-field field is zero.
We have seen in section \ref{sechomogeneoussolution} 
that the  functions $U_{1}$ and $U_{2}$ for an ungraded p 
layer are
\begin{equation}
\label{homodarksolutions}
\begin{array}{l}
U_{1}=e^{+kx} \\
U_{2}=e^{-kx} \\
\end{array}
\end{equation}
where $k$ is
\begin{equation}
\label{homok}
k=1/L_{n}
\end{equation}
The generation function $g$ defined in section \ref{seccurrentcontinuity}
takes the form
\begin{equation}
\label{homogenerationrate}
g(x,\lambda)=\frac{\alpha e^{-\alpha x}}{D_{n}}
\end{equation}
Function $U_{3}$ and its differential $U'_{3}$ are then found from
equation \ref{generalparticularsolution} and take the form
\begin{equation}
\label{homoparticular}
\begin{array}{l l}
U_{3}= & \frac{-\alpha e^{-\alpha x}}{D_{n}(\alpha^{2}-k^{2})} \\
U'_{3}= & \frac{\alpha^{2} e^{-\alpha x}}{D_{n}(\alpha^{2}-k^{2})} \\
\end{array}
\end{equation}
We note that both $U_{3}$ and $U'_{3}$ are inversely
proportional to the diffusivity $D_{n}$. 

Equations \ref{generala1} and \ref{generala2} for this case
yield the following forms for $a_{1}$ and $a_{2}$:
\begin{equation}
\label{homoa1}
a_{1}=\frac
{\alpha e^{-\alpha z}({\cal S}-k)+e^{-kz}(\alpha^{2}-\alpha{\cal S})}
{D_{n}(\alpha^{2}-k^{2})
\left[e^{+kz}({\cal S}-k)-e^{-kz}({\cal S}+k)\right]}
\end{equation}

\begin{equation}
\label{homoa2}
a_{2}=\frac
{-\alpha e^{-\alpha z}({\cal S}+k)-e^{+kz}(\alpha^{2}-\alpha{\cal S})}
{D_{n}(\alpha^{2}-k^{2})
\left[e^{+kz}({\cal S}-k)-e^{-kz}({\cal S}+k)\right]}
\end{equation}

Equations \ref{homodarksolutions} to \ref{homoa2} define the
excess minority carrier concentration n. We can then write down
the quantum efficiency using equations  \ref{qe} and 
\ref{pphotocurrent} from chapter \ref{sectheory}.
We choose to express the $QE$ for a unit incident light flux
in terms of the three following functions 
\begin{equation}
\label{homof1}
f_{1}=
\frac{\alpha ke^{+kz}}{k^{2}-\alpha^{2}}
\left[
\frac
{e^{-\alpha z}({\cal S}-k)+e^{-kz}(\alpha-{\cal S})}
{e^{+kz}({\cal S}-k)-e^{-kz}({\cal S}+k)}
\right]
\end{equation}

\begin{equation}
\label{homof2}
f_{2}=
\frac{\alpha ke^{-kz}}{k^{2}-\alpha^{2}}
\left[
\frac
{e^{-\alpha z}({\cal S}+k)+e^{+kz}(\alpha-{\cal S})}
{e^{+kz}({\cal S}-k)-e^{-kz}({\cal S}+k)}
\right]
\end{equation}

\begin{equation}
\label{homof3}
f_{3}=\frac{\alpha^{2} e^{-\alpha z}}{\alpha^{2}-k^{2}}
\end{equation}

such that the fractional quantum efficiency is 

\begin{equation}
\label{n'tof123}
QE(\lambda)=f_{1}+f_{2}+f_{3}
\end{equation}

We note that all three functions go through a singularity at
$k=\alpha$. It can be shown that $f_{1}$ is continuous.
Although functions $f_{2}$ and $f_{3}$ are both discontinuous,
a numerical study shows that the sum $f_{2}+f_{3}$ is small
compared to $f_{1}$ and continuous. This is illustrated in
figures \ref{figf1f2f3loln} and \ref{figf1f2f3hiln} for
high and low values of the diffusion length and high
recombination velocity.

The figures show that $f_{1}$ dominates the general form
the QE, whereas the sum $f_{2}+f_{3}$ constitutes a
correction which has a varying significance in different 
physical limits.

\smalleps{htbp}{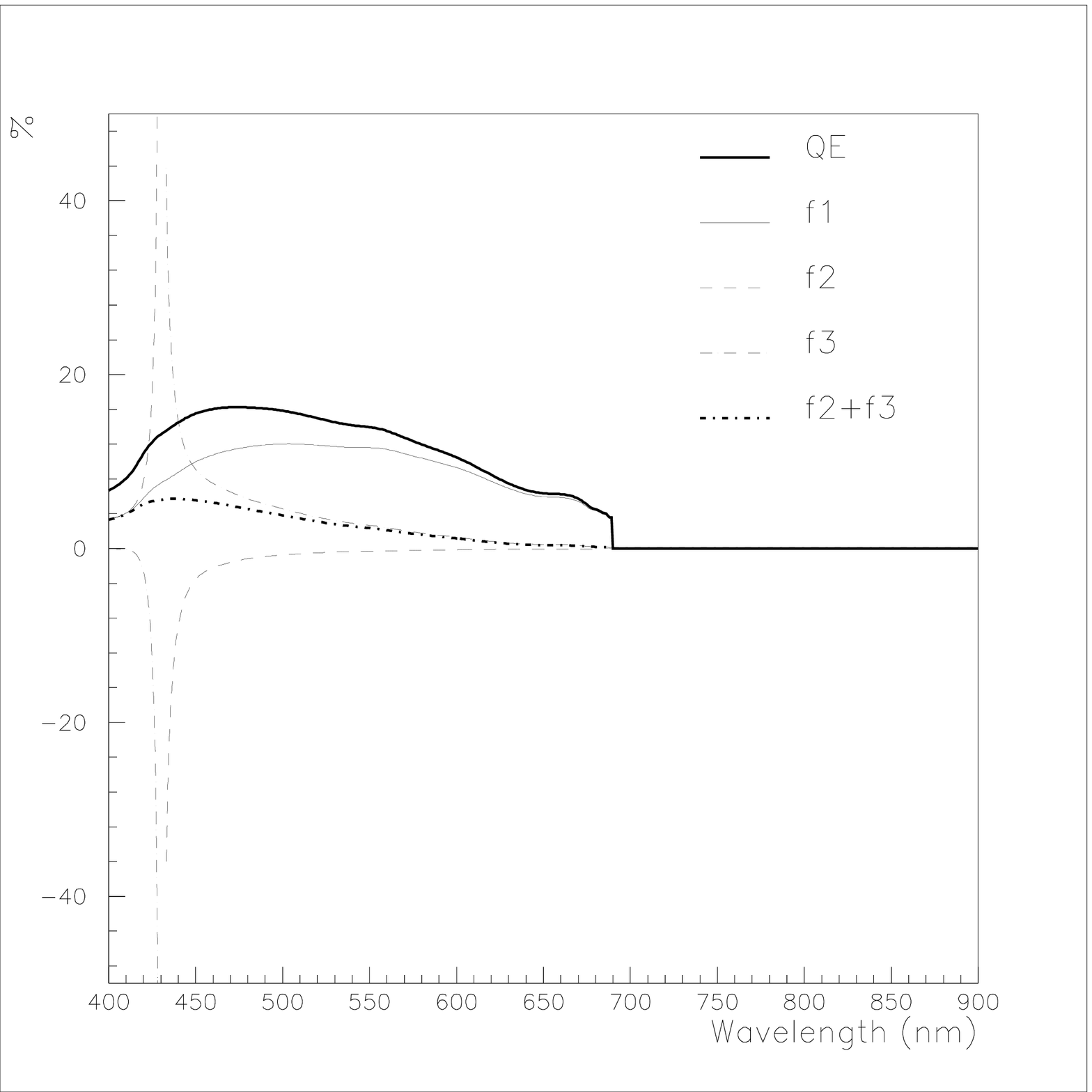}
{Components of the quantum efficiency for small diffusion
length\label{figf1f2f3loln}}

\smalleps{htbp}{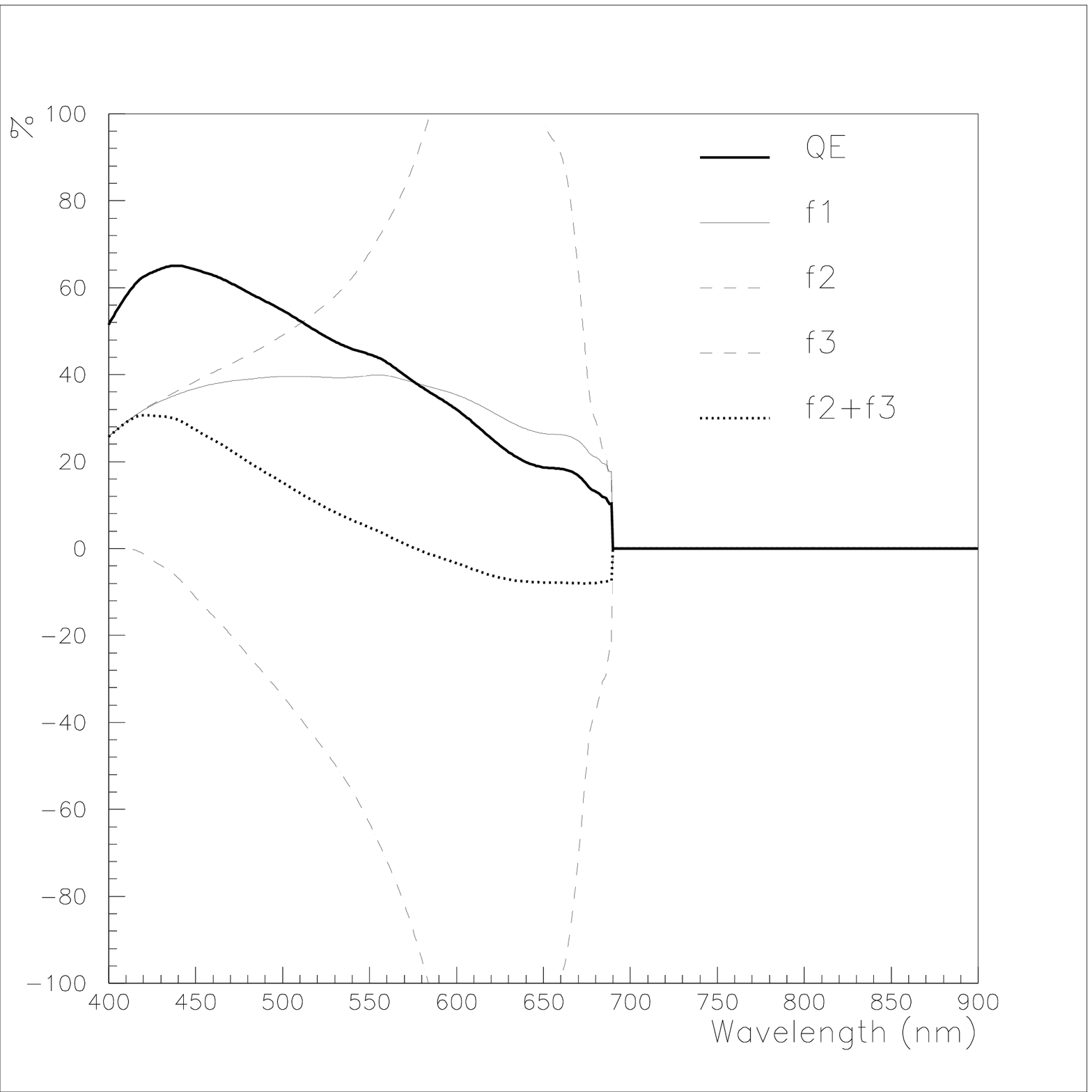}
{Components of the quantum efficiency for large diffusion
length\label{figf1f2f3hiln}}

\subsection{Case of Dominant Bulk Recombination\label{sechomobulkrecomb}}

An examination of equations \ref{homof1} to \ref{homof2} shows
that the surface boundary condition parameter $\cal S$
occurs in terms ${\cal S} \pm k$ and in the
term $\alpha^{2}-\alpha \cal S$.
For absorption coefficients greater than $\sim 10^{6}$, the latter
is dominated by $\alpha^{2}$. This is true in particular of the
absorption coefficient from $400nm$ to approximately $600nm$ in
the typical 30\% \algaas layer we are considering as an example.

We will see in the next section that for longer wavelengths and
lower absorption coefficients, the effect of the recombination velocity
is small if the diffusion length is small compared with $z$.

We find that the $QE$ will be
insensitive to both the recombination velocity and the
diffusivity if $\mid {\cal S}\pm k \mid \simeq k$. Expressed
in in terms of the materials parameters, this gives the
condition
\begin{equation}
\label{homobulkdominated}
\frac{1}{L_{n}}\gg \frac{S_{n}}{D_{n}}
\end{equation}
This is trivially true if the recombination velocity is
very low in which case the only free parameter is $L_{n}$.

More significantly, this condition will hold if the
diffusion length is very small. In this case, the number of minority
carriers lost at the surface is dwarfed by recombination in the bulk,
and $L_{n}$ is the main parameter determining the $QE$.
Further discussion of the balance between these parameters is given 
in chapter \ref{secqeresults}.


For $\cal S$ much greater than $k$ and $\alpha$,
functions $f_{1}$ and $f_{2}$ both tend to
limits which are independent of $\cal S$. This is equivalent
to a vanishingly small minority carrier concentration at the front of
the p layer. We conclude that the $QE$ is again independent of
the magnitude of $S_{n}$ and $D_{n}$ for $\cal S$ much larger than
$k$ and $\alpha$.

\subsection{QE Near the Band Edge \label{sechomolowabsqe}}

This section investigates simple limits for the behaviour
of the quantum efficiency near the band edge. In this regime,
the absorption coefficient is small compared to $z$ and
the generation rate is approximately constant. Generation rates
in this regime are illustrated in chapter \ref{secoptimisation}.

\subsubsection{Small Diffusion Length\label{seclowalphalowln}}

If the diffusion length is small compared to the p layer
thickness $z$, the product $kz$ becomes large. Function $U_{1}$ 
is much greater than $U_{2}$:

\begin{equation}
e^{-kz} \ll e^{+kz}
\end{equation}

This condition is usually true in the QWSC designs we consider,
function $U_{1}$ being typically one or two orders of
magnitude greater than $U_{2}$.

Of the three functions contributing to the $QE$, functions $f_{1}$
and $f_{2}$ have prefactors $e^{kz}$ and $e^{-kz}$ respectively.
Neglecting the contribution from function $f_{2}$ and terms in
$e^{-kz}$ in function $f_{1}$ gives the following expression
for the $QE$ near the band edge

\begin{equation}
QE \simeq 
\frac{ \alpha^{2} e^{-\alpha z}+k\alpha^{2} e^{-\alpha z}}
{k^{2}-\alpha^{2}}
\end{equation}

Cancelling by $\alpha+k$ and assuming $\alpha \ll k$ in
the resulting denominator gives the following result, expressed in
terms of the diffusion length:

\begin{equation}
\label{lowalphaandlnqe}
QE\simeq L_{n}\alpha e^{-\alpha z} 
\end{equation}

The $QE$ in this regime is independent of the
surface boundary condition and hence of the diffusivity. 
We are left with only one fitting parameter, which is the 
diffusion length.

An identical result is reached if we assume that all carriers
generated within a diffusion length of the depletion region
are collected. The quantum efficiency in this picture is
determined by the number of carriers generated in a region
of thickness $L_{n}$ adjacent to the depletion layer.
For low absorption and a constant generation
rate, this is simply the product of the generation rate and the
diffusion length, which is the result we obtain in equation
\ref{lowalphaandlnqe}.

\smalleps{htbp}
{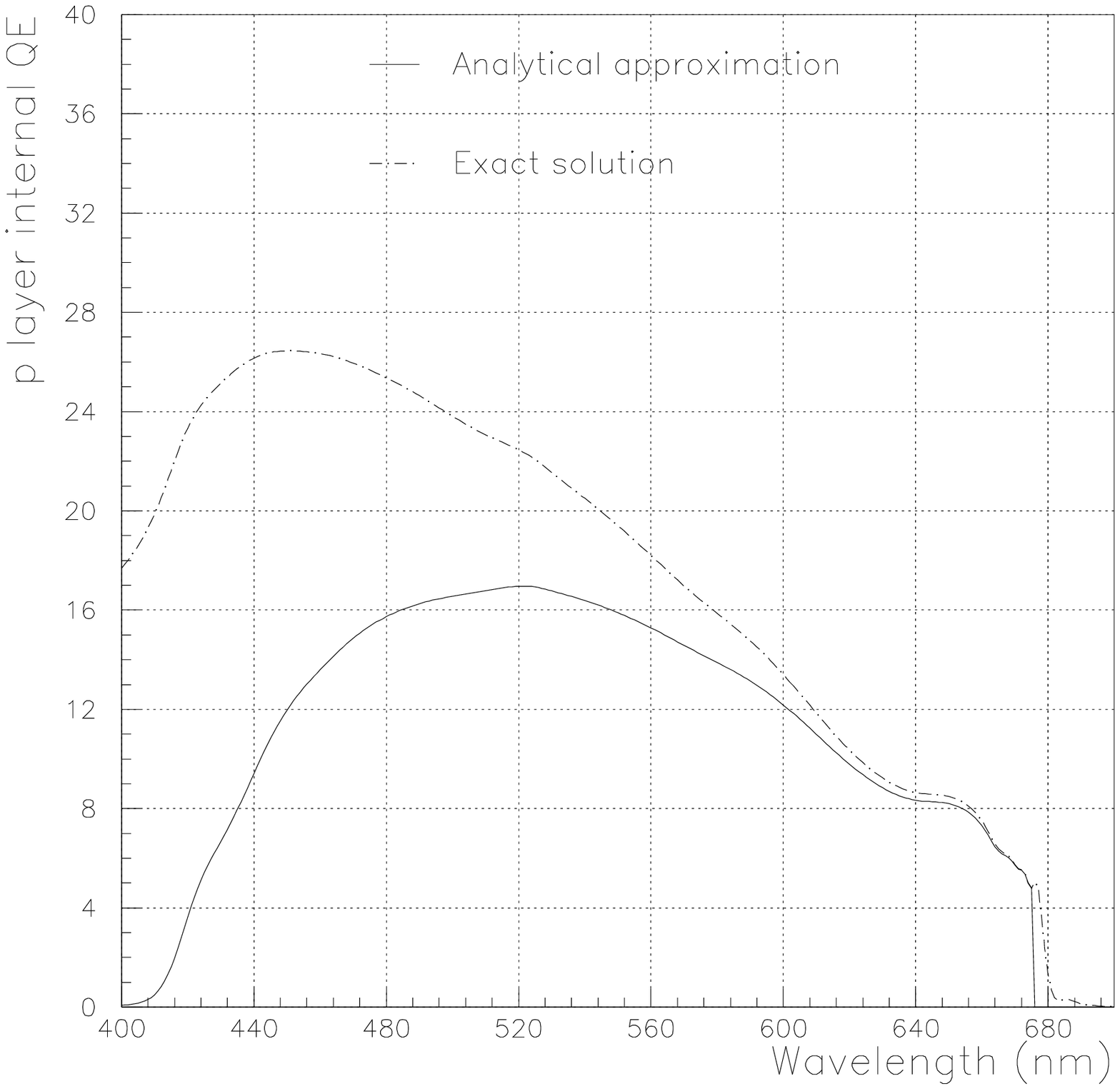}
{\label{figlowalphaandlnqe}Comparison of exact and 
approximate solutions to the QE for low absorption and low diffusion
length $L_{n}=0.07\mu m$ for $x_{wp}=0.148\mu m$}

Figure \ref{figlowalphaandlnqe} illustrates the agreement between
this approximation and the exact result for our standard 30\%
Al p layer of thickness $0.15\mu m$ for a standard diffusion
length of $0.08 \mu m$ and a high surface recombination velocity
of $500m s^{-1}$. The approximation is remarkably accurate
for wavelength within a few tens of nanometers of the band-edge.

For samples obeying this limit, modelling can indicate values
of the diffusion length from the magnitude of the $QE$ near
the band-edge, since the absorption coefficient is well known.
The $QE$ near the surface can then be modelled in terms of the
surface recombination velocity. This is a useful diagnostic
tool, which has already been proposed by Hovel \cite{hovel75}
and used by Ludowise \cite{ludowise84}.

At higher energies the generation rate becomes strongly position
dependent and the expression overestimates the minority carrier
concentration at the edge of the depletion region, leading to
overestimated values for the $QE$.

If the diffusion length is not small compared to $z$, minority
carriers can diffuse to the surface and recombine. The $QE$
across the whole wavelength range becomes sensitive to the
surface recombination.

\subsection{High Absorption}

Figures \ref{figf1f2f3loln} and \ref{figf1f2f3hiln} show that
function $f_{3}$ tends to zero at short wavelengths. This is
because the term $e^{-\alpha z}$ becomes small. To a good
approximation, the $QE$ in this wavelength range is approximately
given by
\begin{equation}
QE \simeq f_{1}+f_{2}
\end{equation}
Neglecting terms in $e^{-\alpha z}$ in functions $f_{1}$ and
$f_{2}$ then yields
\begin{equation}
\label{highalphaqe}
QE\simeq 
\frac
{\alpha k(\alpha-{\cal S})}
{(k^{2}-\alpha^{2})({\cal S}\sinh (kz)-k\cosh(kz))}
\end{equation}

\subsubsection{Low Diffusion Length}

For low diffusion length, $\sinh(kz)\sim\cosh(kz)\sim e^{kz}/2$.
Equation \ref{highalphaqe} becomes
\begin{equation}
QE\simeq \frac{2\alpha k(\alpha-{\cal S})e^{-kz}}
{(k^{2}-\alpha^{2})({\cal S}-k))}
\end{equation}
The short wavelength $QE$ is sensitive to both the surface recombination
velocity and the diffusion length

Since, however, section \ref{seclowalphalowln} has shown that a low
$L_{n}$ can be estimated by modelling experimental data in the long 
wavelength, low absorption regime, parameter $\cal S$ can be estimated
seperately by modelling the short wavelength response.

\subsubsection{High Diffusion Length}

If the diffusion length is large compared with $z$,
$\sinh(kz)\sim kz$ and $\cosh(kz)\sim 1$. Furthermore,
$k$ is small, and $\alpha \gg k$.
Equation \ref{highalphaqe} takes the simple form
\begin{equation}
\label{highalphahighlnqe}
QE\simeq 
\frac
{\alpha-{\cal S}}
{\alpha^{2}({\cal S}z-1)}
\end{equation}
This expression is accurate to within about 5\% of the $QE$ for
a range of about 10nm below 400nm, for typical structures.

The $QE$ in this case largely determined by the surface recombination
parameter. This means it is sensitive to a combination of recombination
velocity and minority carrier diffusivity. Recombination losses in the 
bulk become negligible compared to surface losses, and the $QE$ reaches a 
maximum which is limited by surface recombination.

\section{Homogeneous Cells with Electric Fields}

We now examine the hypothetical case of a cell with a non-zero
effective field in the doped regions but with constant
transport parameters. Models of this type are frequently used in
theoretical investigations of compositionally varying structures
(see section \ref{seclittreview}).

The parameters of equation for this type of structure are
are $b=2{\cal E}_{f}$ and $c=-1/L_{n}$, where definitions and
solutions are as given in section \ref{sectheorynonzeroe},
but are repeated here for convenience and take the form
\begin{equation}
\label{ak1withfield}
k_{1}=  +\sqrt{{\cal E}_{f}^{2}+L^{-2}}-{\cal E}_{f} 
\end{equation}
\begin{equation}
\label{ak2withfield}
k_{2}=  -\sqrt{{\cal E}_{f}^{2}+L^{-2}}-{\cal E}_{f}
\end{equation}
The two functions making up the restricted solution to the minority 
carrier equation were shown to be
\begin{equation}
U_{1}=e^{k_{1}x}
\end{equation}
\begin{equation}
U_{2}=e^{k_{2}x}
\end{equation}

Although the boundary condition at the edge of the depletion region
remains unchanged, a field term appears in the surface recombination
boundary condition, as explained in \ref{secpphotocurrent}. The
surface recombination parameter ${\cal S}$ of equation
\ref{generala1} is replaced with
\begin{equation}
\label{snparameterwithfield}
{\cal S}=2{\cal E}_{f}-S_{n}/D_{n}
\end{equation}
In a well designed cell, the field is negative,
causing carriers to drift towards the space charge region.
This study deals with fields $E \sim -10^{6}V m^{-1}$,
such that ${\cal E}_{f}\sim -10^{7} m^{-1}$ at room temperatures. 
This corresponds to a moderate grade from 40\% Al at the front 
surface of the layer to 30\% at the top interface of the 
depletion layer.

The presence of a field removes some of the symmetry we found in
the solutions for an ungraded layer. Function $U_{2}$ is
no longer the inverse of $U_{1}$. It becomes necessary to consider the
relative magnitudes of ${\cal E}_{f}$, $L_{n}$ and $S_{n}/D_{n}$.

The presence of a field term in the surface boundary condition
increases the magnitude of the surface boundary parameter
$\cal S$. This indicates a higher gradient in the 
minority carrier concentration $n'(0)$ (subject to the magnitude
of $n(0)$). This seemingly contradictory increase in the value
of $\cal S$ is due to the fact that minority carriers are removed
away from the surface by the effective field and swept towards
the depletion layer. The minority carrier profile as a whole is
shifted in a positive direction, increasing the gradient and hence
the $QE$ at the edge of the depletion layer. This is illustrated
in chapter \ref{secoptimisation}.

The solution for the $QE$ in this case is of a similar form to
that for an ungraded cell, but with the new expressions
\ref{k1withfield} to \ref{snparameterwithfield} substituted.
Accordingly, we rewrite equations \ref{homof1} to \ref{homof3}
in terms of three new functions:

\begin{equation}
\label{fielde1}
e_{1}=
\frac{\alpha}{(\alpha+k_{1})(\alpha+k_{2})}
\left[
\frac
{k_{1}e^{k_{1}z}(e^{-\alpha z}({\cal S}+k_{2})+e^{k_{2}z}(\alpha-{\cal S}))}
{e^{k_{1}z}({\cal S}+k_{2})-e^{k_{2}z}({\cal S}+k_{1})}
\right]
\end{equation}

\begin{equation}
\label{fielde2}
e_{2}=
\frac{\alpha}{(\alpha+k_{1})(\alpha+k_{2})}
\left[
\frac
{-k_{2}e^{k_{2}z}(e^{-\alpha z}({\cal S}+k_{1})+e^{k_{1}z}(\alpha-{\cal S}))}
{e^{k_{1}z}({\cal S}+k_{2})-e^{k_{2}z}({\cal S}+k_{1})}
\right]
\end{equation}

\begin{equation}
\label{fielde3}
e_{3}=\frac{\alpha^{2} e^{-\alpha z}}{(\alpha+k_{1})(\alpha+k_{2})}
\end{equation}

such that the fractional quantum efficiency is 

\begin{equation}
\label{n'toe123}
QE(\lambda)=e_{1}+e_{2}+e_{3}
\end{equation}

\subsection{Surface Recombination Effect with Field
\label{secfieldbulkrecomb}}

In this section we consider analogue to the balance
between bulk and surface recombination which was discussed
in section \ref{sechomobulkrecomb}.
The discussion of this case is more restricted
than for an ungraded cell because the spatial variation
of materials parameters forbids many of the simplifications
we were able to make in the previous section. We restrict the
discussion principally to the effect of the surface
recombination.

In section \ref{sechomobulkrecomb} we saw that the relative magnitudes
of $k$ and $\cal S$ determine the dominant recombination effect.
Equations \ref{fielde1} and \ref{fielde2} show that the $QE$ in this
case is insensitive to the surface recombination velocity if
$k_{1}+{\cal S}$ and $k_{2}+{\cal S}$ are both weakly dependent
on $S_{n}/D_{n}$

If the diffusion length is very small such that 
${\cal E}_{f} \ll L_{n}^{-2}$, we return to the situation
described in section \ref{sechomobulkrecomb}. A calculation shows
that this limit applies only for diffusion lengths
below $0.01 \mu m$ for the field strengths mentioned above.

More significantly, functions $e_{1}$and $e_{2}$ will be insensitive
to the surface recombination velocity if $\cal S$ is dominated by
$2{\cal E}_{f}$, that is, if
\begin{equation}
\label{fieldbulkdominated}
\mid {\cal E}_{f}\mid  \gg \frac{S_{n}}{2D_{n}}
\end{equation}

This condition is less stringent than equation \ref{homobulkdominated}
of section \ref{sechomobulkrecomb} and is fulfilled even the moderate
values of ${\cal E}_{f}$ of $\sim -10^{7} m^{-1}$. 

%
%
%


\subsection{QE Near the Band Edge for Small Diffusion Length with Field
\label{seclowalphalowlnwithe} \label{secfieldlowabsqe}}

Because ${\cal E}_{f}$ is negative, we see that the positive $k_{1}$
is greater than $k$ whereas the magnitude of $k_{2}$ is
smaller than $k$. The condition of section \ref{sechomobulkrecomb}
is more readily fulfilled. That is,
\begin{equation}
e^{k_{1}z} \gg e^{k_{2}z}
\end{equation}
This remains true even for very large diffusion lengths, for which
$k_{1}\simeq +\mid {\cal E}_{f} \mid $ and 
$k_{2}\simeq -\mid {\cal E}_{f} \mid $.  

We follow a similar argument to that given in section \ref{seclowalphalowln}.
Neglecting terms in $e^{k_{2}z}$ and assuming $\alpha \ll \mid k_{2} \mid$
gives
\begin{equation}
\label{fieldlowalphalowln}
QE \simeq 
\frac
{\alpha e^{-\alpha z}}
{\sqrt{{\cal E}_{f}^{2}+L_{n}^{-2}}+{\cal E}_{f}}
\end{equation}
This expression is in fact a restatement of equation \ref{lowalphaandlnqe}
in the presence of a field.

\section{Conclusions}

\subsection{Ungraded Solar Cells
\label{sechomocellconclusions}}

In section \ref{secanalyhomosolutions} we established that the
analytical expression for the $QE$ of a p layer can be reduced
to simple analytical expressions in a number of limiting cases.

For high absorption coefficient at short wavelengths, the $QE$ is
in general sensitive to both surface recombination and diffusion
length, and tends to a constant value.

For low absorption whithin a few tens of nm of the bandedge,
the $QE$ is independent of the surface recombination for the
structures we are studying, because the diffusion length is
smaller than the width of the p layer. The only free parameter
in this regime is the diffusion length.

Values of the diffusion length can therefore be estimated from
the $QE$ at long wavelengths. The short wavelength response is
then modelled independently in terms of the recombination velocity.

Cases where very large values of $\cal S$ are required indicate
a poor surface and the existance of a dead layer and a zero minority
carrier concentration at the front of the p layer. We have seen
however that the $QE$ in the low absorption regime remains principally
determined by the diffusion length.

\subsection{Solar Cells With Fields but Constant Materials Parameters
\label{secfieldcellconclusions}}

A first approximation to the $QE$ for a compositionally graded
cell is the example of an ungraded cell where an effective
field is present in the doped regions. Fewer conclusions can
drawn from this case than for an ungraded cell because the
spatial variation of materials parameters is not negligible.

The main conslusion we reach is that even moderate grades generate 
fields of sufficient magnitude to reduce the effect of the 
recombination velocity to negligible levels.

This model however overestimates the $QE$ for a graded structure
for two reasons. The constant absorption coefficient overestimates
the amount of light absorbed, particularly at long wavelengths.
Secondly, comparison between numerical and analytical calculations
shows that the gradient in the transport parameters is important.
The gradient in $D_{n}$, and hence the minority carrier mobility,
contributes a field-like term which is opposite to the band gap
gradient field.

The $QE$ in a real cell therefore remains insensitive to the
magnitude of the recombination velocity for realistic values
of the latter, but is significantly lower than the predictions
of the simplified analytical model.